\documentclass[12pt,a4paper]{article}

\usepackage[utf8x]{inputenc}

\usepackage{amsmath,amssymb}
\usepackage{amsfonts,latexsym}

\usepackage{titlesec}
\titleclass{\subsubsubsection}{straight}[\subsection]
\newcounter{subsubsubsection}[subsubsection]
\renewcommand\thesubsubsubsection{\thesubsubsection.\arabic{subsubsubsection}}
\titleformat{\subsubsubsection}
  {\normalfont\normalsize\bfseries}{\thesubsubsubsection}{1em}{}
\titlespacing*{\subsubsubsection}
{0pt}{3.25ex plus 1ex minus .2ex}{1.5ex plus .2ex}
\makeatletter
\def\toclevel@subsubsubsection{4}
\def\l@subsubsubsection{\@dottedtocline{4}{7em}{4em}}
\makeatother
\usepackage{graphicx}
\usepackage[bookmarks=true,colorlinks=true,citecolor=blue,linkcolor=blue,urlcolor=magenta]{hyperref}

\usepackage[left=1.5cm,right=1.5cm,top=2cm,bottom=2cm]{geometry}

\usepackage{har2nat} 
\usepackage[]{natbib}

\usepackage{xcolor}

\usepackage{soulutf8}
\definecolor{lightred}{rgb}{1,.80,.80}
\definecolor{lightgreen}{rgb}{0.9,1.0,0.9}
\definecolor{lightblue}{rgb}{.80,.90,1}
\definecolor{lightyellow}{rgb}{1.0,1.0,0.7}
\definecolor{lightorange}{rgb}{1.0,0.8,0.5}
\sethlcolor{lightyellow}

\title{\vspace{-0.5cm}\large{\textbf{Thermomagnetic Ettingshausen-Nernst effect in tachocline,\\
       magnetic reconnection phenomenon in lower layers,\\
       axion mechanism of solar luminosity variations,\\
       coronal heating problem solution and\\
       mechanism of asymmetric dark matter variations around black hole}}\vspace{0.5cm}}

\author{V.D.~Rusov$^1$\footnote{Corresponding author: Vitaliy D. Rusov, siiis@te.net.ua},
        M.V.~Eingorn$^2$,
        I.V.~Sharph$^1$,
        V.P.~Smolyar$^1$,
        M.E.~Beglaryan$^3$}

\date{}

\begin{document}

\maketitle

\begin{center}
$^1$\textit{Department of Theoretical and Experimental Nuclear Physics, \\Odessa National Polytechnic University, Odessa, Ukraine}

\vspace{0.5cm}

$^2$\textit{CREST and NASA Research Centers, North Carolina Central University,\\Durham, North Carolina, U.S.A.}

\vspace{0.5cm}

$^3$\textit{Department of Computer Technology and Applied Mathematics,\\Kuban State University, Krasnodar, Russia}
\end{center}

\begin{abstract}

It is shown that the holographic principle of quantum gravity (in the hologram
of the  Universe, and therefore in our Galaxy, and of course on the Sun!), in
which the conflict between the theory of gravitation and quantum mechanics
disappears, gives rise to the Babcock-Leighton holographic mechanism. Unlike the
solar dynamo models, it generates a strong toroidal magnetic field by means of
the thermomagnetic Ettingshausen-Nernst (EN) effect in the tachocline. Hence, it can
be shown that with the help of the thermomagnetic EN~effect,
a simple estimate of the magnetic pressure of an ideal gas in the tachocline of
e.g. the Sun can indirectly prove that by using the holographic principle of
quantum gravity, the repulsive toroidal magnetic field of the tachocline 
($B_{tacho}^{Sun} = 4.1 \cdot 10^7 ~G = - B_{core}^{Sun}$) precisely
``neutralizes'' the magnetic field in the Sun core, since the projections of
the magnetic fields in the tachocline and the core have equal values but
opposite directions.
The basic problem is a
generalized problem of the antidynamo model of magnetic flux tubes (MFTs), where
the nature of both holographic effects (the thermomagnetic EN~effect and Babcock-Leighton holographic mechanism), including magnetic cycles, manifests
itself in the modulation of asymmetric dark matter (ADM) and,
consequently, the solar axion in the Sun interior.

The general laws of the theory of virtually empty anchored flux tubes with
$B \sim 10^7 ~G$, which are born on the interface between the tachocline and
the overshoot layer, are developed. These flux tubes are pushed out by means of the effective increase in their magnetic buoyancy -- from the tachocline to the
surface of the Sun. The appearance of MFTs on the surface of the
Sun and the observed sunspots are identical. This means that the formation of
the magnetic cycles of flux tubes coincides with the observational data of the
Joy's Law, and both effects are the manifestation of dark matter (DM) -- solar axions
in the core of the Sun, the modulation of which is predetermined by the 
anticorrelation modulation of ADM.

It is shown that the hypothesis of the axion mechanism of solar luminosity
variations suggesting that the solar axion particles are born in the core of
the Sun and may be efficiently converted back into $\gamma$-quanta in the
magnetic field of the solar overshoot tachocline is physically relevant. As a
result, it is also shown that the intensity variations of the $\gamma$-quanta
of axion origin, induced by the magnetic field variations in the tachocline via
the thermomagnetic EN~effect, directly cause the Sun
luminosity variations and eventually characterize the active and quiet states
of the Sun. Within the framework of this mechanism estimations of the strength
of the axion coupling to a photon 
($g_{a \gamma} = 4.4 \cdot 10^{-11} ~GeV^{-1}$) and the hadronic axion
particle mass ($m_a \sim 3.2 \cdot 10^{-2} ~eV$) have been obtained. It is
also shown that the claimed axion parameters do not contradict any known
experimental and theoretical model-independent limitations.

The theoretical solution of the problem of coronal heating is suggested
involving the photons of axion origin, which are produced in the core of the
Sun. Our physical understanding of coronal heating and the coincidence of the theoretical and experimental photon spectra of the corona are connected with the appearance of a magnetic flux and simultaneously the flux of axion-origin photons in the outer layers of the Sun. On the other hand, it is connected with the basic mechanism of formation of sunspots and active regions, being an integral part of the solar cycle, which determines the corresponding variation in the release of energy in the corona and flares.

It is assumed that the modulation of the ADM density as a ``clock''
that anticorrelatively regulates the rate of the solar cycle, the photons of
axion origin, the solar abundance and sunspots, and the mechanism of ADM density
variations around the black hole (BH), is a consequence of the DM
modulation in the isolated group of galaxies (with the active galactic nuclei
(AGNs)), one of which is our Galaxy -- the Milky Way. The main result of our
model is the following one: the modulation of DM halo density at the
Galactic Center (GC), which is closely correlated with the modulation of baryon
matter density near supermassive black hole (SMBH), determines the appearance of the DM
modulation by S-stars found at the GC.
Based on empirical evidence, we make the following natural conclusion. If the
modulations of the DM halo at the GC give rise to the modulations of
the DM halo density at the surface of the Sun (through vertical
density waves from the disk to the solar neighborhood), then there is an ``experimental''
anticorrelational relation between the DM density modulation in
solar interior and the sunspot number. This is true for the relation between
the periods of S-stars and sunspot number cycles.

\end{abstract}

\section{Introduction}

A hypothetical pseudoscalar particle called axion is predicted by the theory
related to solving the CP-invariance violation problem in QCD. The most
important parameter determining the axion properties is the energy scale $f_a$
of the so-called U(1) Peccei-Quinn symmetry violation. It determines both the
axion mass and the strength of its coupling to fermions and gauge bosons
including photons. However, in spite of the numerous direct experiments, axions
have not been discovered so far. Meanwhile, these experiments together with the
astrophysical and cosmological limitations leave a rather narrow band for the
permissible parameters of invisible axion (e.g.
$10^{-6} eV \leqslant m_a \leqslant 10^{-2} eV$~\citep{ref01,ref02}), which is
also a well-motivated cold dark matter (CDM) candidate in this mass region
\citep{ref01,ref02}.

A whole family of axion-like particles (ALPs) may exist
along with axions, having the similar Lagrangian structure relative to the
Peccei-Quinn axion, as well as their own distinctive features. If they exist, the connection between their mass and their
constant of coupling to photons may be highly weakened, as opposed to the
axions. It should be also mentioned that the phenomenon of photon-ALP mixing in
the presence of the electromagnetic field not only leads to the classic
neutrino-like photon-ALP oscillations, but also causes the change in the
polarization state of the photons propagating in the strong enough magnetic fields (the $a \gamma \gamma$ coupling acts like a polarimeter \citep{ref03}). It
is generally assumed that there are light ALPs coupled only to photons,
although the realistic models of ALPs with couplings both to photons and to
matter are not excluded \citep{ref04}. Anyway, they may be considered a
well-motivated CDM candidate \citep{ref01,ref02} under certain
conditions, just like axions.

It is interesting to note that the photon-ALP mixing in magnetic fields of
different astrophysical objects including active galaxies, clusters of
galaxies, intergalactic space and the Milky Way, may be the cause of the
remarkable phenomena like dimming of stars luminosity (e.g. supernovae in the
extragalactic magnetic field \citep{ref06,Mirizzi2005}) and ``light  shining through
a wall'' (e.g. light from very distant objects, traveling through the Universe
\citep{ref03,ref05}). In the former case the luminosity of an astrophysical
object is dimmed because some part of photons transforms into axions in the
object's magnetic field. In the latter case photons produced by the object are
initially converted into axions in the object's magnetic field, and then after
passing some distance (the width of the ``wall'') are converted back into
photons in another magnetic field (e.g. in the Milky Way), thus emulating the
process of effective free path growth for the photons in astrophysical medium
\citep{ref08,ref09}.

For the sake of simplicity let us hereinafter refer to all such particles as
axions if not stated otherwise.

In the present paper we consider the possible existence of the axion mechanism
of Sun luminosity variations\footnote{Let us point out that the axion mechanism of Sun
luminosity used for estimating the axion mass was described for the first time
in 1978 by \cite{ref10}.} based on the ``light shining through a wall'' effect. To be
more precise, we attempt to explain the axion mechanism of Sun luminosity variations by the
``light shining through a wall'', when the photons born mainly in the solar
core are at first converted into axions via the Primakoff effect \citep{ref11}
in its magnetic field, and then are converted back into photons after passing
the solar radiative zone and getting into the magnetic field of the overshoot
tachocline (see e.g. Fig.~1 in \cite{Zioutas2009}).

It gives rise to the very intriguing questions: ``What is the nature of the
tachoclines on the Sun and other stars? If the existence of the tachocline is
predetermined by the holographic principle of quantum gravity, then why, in
contrast to the solar dynamo, the magnetic field of the tachocline is much
stronger than $10^5~G$? What is the strength of the magnetic field in the 
Sun core? What is it in the Sun core that controls the solar cycle? Can
axions and/or DM particles in the Sun core control the solar cycle
of the photons of axion origin, the abundance of Sun interior and sunspots?''

Based on the solution of such nontrivial physics, we estimate this magnetic
field within the framework of the thermomagnetic EN~effect
(as a consequence of the fundamental properties of the holographic principle of
quantum gravity). In addition, we obtain the consistent 
estimates for the axion mass ($m_a$) and the axion coupling constant to photons
($g_{a \gamma}$), based on this mechanism, and verify their values against the
axion model results and the known experiments including CAST, ADMX, RBF.

\section{Photon-axion conversion and the case of maximal mixing}

Let us give some implications and extractions from the photon-axion
oscillations theory which describes the process of the photon conversion into
an axion and back under the constant magnetic field $B$ of the length $L$. It
is easy to show \citep{ref05,Raffelt-Stodolsky1988,Mirizzi2005,Hochmuth2007} that in
the case of the negligible photon absorption coefficient
($\Gamma _{\gamma} \to 0$) and axions decay rate ($\Gamma _{a} \to 0$) the
conversion probability is
\begin{equation}
P_{a \rightarrow \gamma} = \left( \Delta_{a \gamma}L \right)^2 \sin ^2 \left( \frac{ \Delta_{osc}L}{2} \right) \Big/ \left( \frac{ \Delta_{osc}L}{2}
\right)^2 \label{eq01}\, ,
\end{equation}
where the oscillation wavenumber $\Delta_{osc}$ is given by
\begin{equation}
\Delta_{osc}^2 = \left( \Delta_{pl} + \Delta_{Q,\perp} - \Delta_{a} \right)^2 + 4 \Delta_{a \gamma} ^2
\label{eq02}
\end{equation}
while the mixing parameter $\Delta _{a \gamma}$, the axion-mass parameter
$\Delta_{a}$, the refraction parameter $\Delta_{pl}$ and the QED dispersion
parameter $\Delta_{Q,\perp}$ may be represented by the following expressions:
\begin{equation}
\Delta _{a \gamma} = \frac{g_{a \gamma} B}{2} = 540 \left( \frac{g_{a \gamma}}{10^{-10} GeV^{-1}} \right) \left( \frac{B}{1 G} \right) ~~ pc^{-1}\, ,
\label{eq03}
\end{equation}
\begin{equation}
\Delta _{a} = \frac{m_a^2}{2 E_a} = 7.8 \cdot 10^{-11} \left( \frac{m_a}{10^{-7} eV} \right)^2 \left( \frac{10^{19} eV}{E_a} \right) ~~ pc^{-1}\, ,
\label{eq04}
\end{equation}
\begin{equation}
\Delta _{pl} = \frac{\omega ^2 _{pl}}{2 E_a} = 1.1 \cdot 10^{-6} \left( \frac{n_e}{10^{11} cm^{-3}} \right) \left( \frac{10^{19} eV}{E_a} \right) ~~ pc^{-1},
\label{eq05}
\end{equation}
\begin{equation}
\Delta _{Q,\perp} = \frac{m_{\gamma, \perp}^2}{2 E_a} .
\label{eq06}
\end{equation}

Here $g_{a \gamma}$ is the constant of axion coupling to photons; $B$ is the
transverse magnetic field; $m_a$ and $E_a$ are the axion mass and energy;
$\omega ^2 _{pl} = 4 \pi \alpha n_e / m_e$ is an effective photon mass in terms
of the plasma frequency if the process does not take place in vacuum, $n_e$ is
the electron density, $\alpha$ is the fine-structure constant, $m_e$ is the
electron mass; $m_{\gamma, \perp}^2$ is the effective mass square of the
transverse photon which arises due to interaction with the external magnetic
field.

The conversion probability (\ref{eq01}) is energy-independent, when
$2 \Delta _{a \gamma} \approx \Delta_{osc}$, i.e.
\begin{equation}
P_{a \rightarrow \gamma} \cong \sin^2 \left( \Delta _{a \gamma} L \right)\, ,
\label{eq07}
\end{equation}
or whenever the oscillatory term in (\ref{eq01}) is small
($\Delta_{osc} L / 2 \to 0$), implying the limiting coherent behavior
\begin{equation}
P_{a \rightarrow \gamma} \cong \left( \frac{g_{a \gamma} B L}{2} \right)^2\, .
\label{eq08}
\end{equation}

It is worth noting that the oscillation length corresponding to (\ref{eq07})
reads
\begin{equation}
L_{osc} = \frac{\pi}{\Delta_{a \gamma}} = \frac{2 \pi}{g_{a \gamma} B} \cong 5.8 \cdot 10^{-3}
\left( \frac{10^{-10} GeV^{-1}}{g_{a \gamma}} \right)
\left( \frac{1G}{B} \right)  ~pc
\label{eq13}
\end{equation}
\noindent assuming a purely transverse field. In the case of the appropriate
size $L$ of the region a complete transition between photons and axions is
possible.

From now on we are going to be interested in the energy-independent case
(\ref{eq07}) or (\ref{eq08}) which plays the key role in determination of the parameters
for the axion mechanism of Sun luminosity variations hypothesis (the axion coupling
constant to photons $g_{a \gamma}$, the transverse magnetic field $B$ of length
$L$ and the axion mass $m_a$).

\section{Axion mechanism of Sun luminosity variations}

Our hypothesis is that the solar axions, which are born in the solar core
\citep{ref01,ref02} through the known Primakoff effect \citep{ref11}, may be
converted back into $\gamma$-quanta in the magnetic field of the solar
tachocline (the base of the solar convective zone). The magnetic field
variations in the tachocline cause the converted $\gamma$-quanta intensity
variations in this case, which in their turn cause the variations of the Sun
luminosity known as the active and quiet Sun states. Let us consider this
phenomenon in more detail.

As we noted above, the expression (\ref{eq01}) for the probability of the
axion-photon oscillations in the transversal magnetic field was obtained for
the media with quasi-zero refraction, i.e. for the media with a negligible
photon absorption coefficient ($\Gamma_{\gamma} \to 0$). It means that in order
for the axion-photon oscillations to take place without any significant losses,
a medium with a very low or quasi-zero density is required, which would
suppress the processes of photon absorption almost entirely.

Surprisingly enough, it turns out that such ``transparent'' media can take
place, and not only in plasmas in general, but straight in the convective zone
of the Sun. Here we generally mean the so-called MFTs, the
properties of which are examined below.

\subsection{Ideal photon channeling conditions inside the magnetic flux tubes}

\label{sec-channeling}

\begin{figure*}
\begin{center}
\includegraphics[width=12cm]{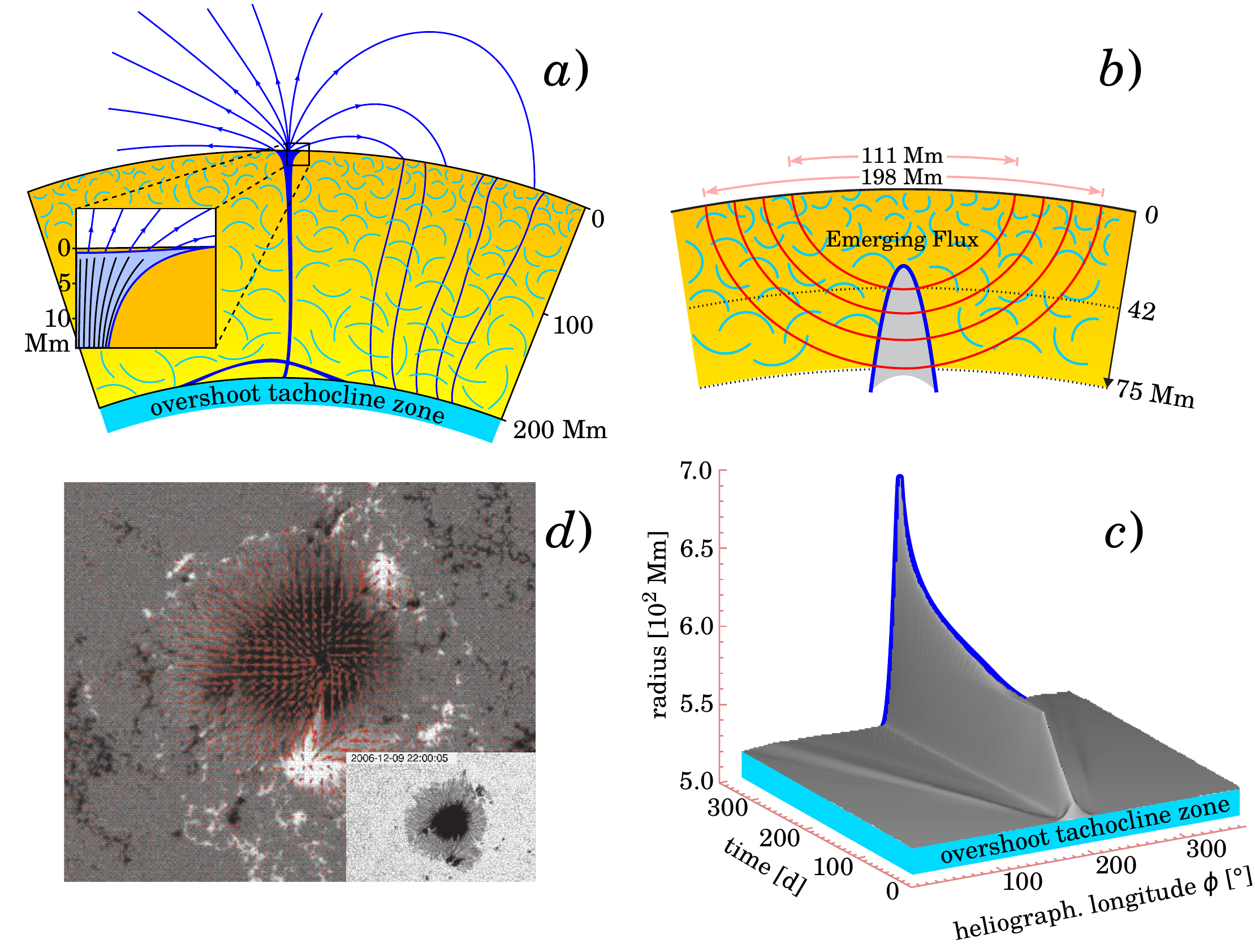}
\end{center}
\caption{(a) Vertical cut through an active region illustrating the connection
between a sunspot at the surface and its origins in the toroidal field layer at
the base of the convection zone. Horizontal fields stored at the base of the
convection zone (the overshoot tachocline zone) during the cycle. Active
regions form from sections brought up by buoyancy (one shown in the process of
rising). After eruption through the solar surface a nearly potential field is
set up in the atmosphere (broken lines), connecting to the base of the
convective zone via almost vertical flux tube. Hypothetical small scale
structure of a sunspot is shown in the inset (adopted from
\cite{ref18}
and
\cite{ref15}).
(b) Detection of emerging sunspot regions in the solar interior~\citep{ref18}.
Acoustic ray paths with lower turning points between 42 and 75 Mm
(1 Mm = 1000 km) crossing a region of emerging flux. For simplicity, only four
out of total of 31 ray paths used in this study (the time-distance
helioseismology experiment) are shown here. Adopted from~\cite{Ilonidis2011}.
(c) Emerging and anchoring of stable flux tubes in the overshoot tachocline
zone, and the time evolution in the convective zone. Adopted from \cite{ref20}.
(d) Vector magnetogram of the white light image of a sunspot (taken with the Solar Optical Telescope (SOT) on
a board of the Hinode satellite -- see inset) showing in red the direction of
the magnetic field and its strength (length of the bar). The movie shows the
evolution in the photospheric fields that has led to an X class flare in the
lower part of the active region. Adopted from~\cite{Benz2008}.}
\label{fig01}
\end{figure*}

The idea of the energy flow channeling along a fanning magnetic field has been
suggested for the first time by
\cite{ref12} as an explanation for
darkness of umbra of sunspots. It was incorporated in a simple sunspot model by
\cite{ref13}.
\cite{ref14} extended this suggestion to smaller
flux tubes to explain the dark pores and the bright faculae as well.
Summarizing the research of the convective zone magnetic fields in the form of isolated flux tubes,
\cite{ref15} suggested a simple mathematical model for the behavior of thin
MFTs, dealing with the nature of the solar cycle, sunspot
structure, the origin of spicules and the source of mechanical heating in the
solar atmosphere. In this model, the so-called thin tube approximation is used
(see \cite{ref15} and Refs. therein), i.e. the field is conceived to exist
in the form of slender bundles of field lines (flux tubes) embedded in a
field-free fluid (Fig.~\ref{fig01}). Mechanical equilibrium between the tube
and its surrounding is ensured by a reduction of the gas pressure inside the
tube, which compensates the force exerted by the magnetic field.  In our
opinion, this is exactly the kind of mechanism
\cite{Parker1955b} was thinking about when he wrote about the problem of flux
emergence: ``Once the field has been amplified by the dynamo, it needs to be
released into the convection zone by some mechanism, where it can be
transported to the surface by magnetic buoyancy''~\citep{ref17}.

In order to understand magnetic buoyancy, let us consider an isolated
horizontal flux tube in pressure equilibrium with its non-magnetic surroundings,
so that in cgs units

\begin{equation}
p_{ext} = p_{int} + \frac{\vert \vec{B} \vert^2}{8 \pi} ,
\label{eq21}
\end{equation}

\noindent where $p_{int}$ and $p_{ext}$ are the internal and external gas
pressures, respectively, while $B$ denotes the uniform field strength in the flux
tube. If the internal and external temperatures are equal, so that $T_{int} =
T_{ext}$ (thermal equilibrium), then since $p_{ext} > p_{int}$, the gas in the
tube is less dense than its surrounding ($\rho _{ext} > \rho _{int}$), implying
that the tube will rise under the influence of gravity.

In spite of the obvious, though turned out to be surmountable, difficulties of
the application to real problems, it was shown (see~\cite{ref15} and Refs.
therein) that strong buoyancy forces act in MFTs of the required
field strength ($10^4 - 10^5 ~G$~\citep{ref23}). Under their influence tubes
either float to the surface as a whole (e.g. Fig.~1 in \cite{ref24}) or they
form loops of which the tops break through the surface (e.g. Fig.~1
in~\cite{ref14}) and lower parts descend to the bottom of the convective zone,
i.e. to the overshoot tachocline zone. The convective zone, being unstable,
enhanced this process~\citep{Spruit1982,ref26}. Small tubes take longer to erupt
through the surface because they feel stronger drag forces. It is interesting
to note here that the phenomenon of the drag force, which raises the magnetic
flux tubes to the convective surface with the speeds about 0.3-0.6~km/s, was
discovered in direct experiments using the method of time-distance
helioseismology~\citep{Ilonidis2011}. Detailed calculations of the
process~\citep{MorenoInsertis1983} show that even a tube with the size of a very small spot,
if located within the convective zone, will erupt in less than two years. Yet,
according to~\cite{MorenoInsertis1983}, horizontal fields are needed in the overshoot tachocline
zone, which survive for about 11~yr, in order to produce an activity cycle.

A simplified scenario of MFTs birth and space-time
evolution (Fig.~\ref{fig01}a) can be presented as follows. A MFT is born in the
overshoot tachocline zone (Fig.~\ref{fig01}c) and rises up to the convective
zone surface (Fig.~\ref{fig01}b) without separation from the tachocline (the
anchoring effect), where it forms a sunspot (Fig.~\ref{fig01}d) or other
kinds of active solar regions when intersecting the photosphere.  There are
more fine details of MFT physics expounded in overviews by
\cite{ref17} and
\cite{ref24}, where certain fundamental questions, which need to be addressed
to understand the basic nature of magnetic activity, are discussed in detail:
How is the magnetic field generated, maintained and dispersed? What are its
properties such as structure, strength, geometry? What are the dynamical
processes associated with magnetic fields? \textbf{What role do magnetic fields
play in the energy transport?}

Dwelling on the last extremely important question associated with the energy
transport, let us note that it is known that thin MFTs can
support longitudinal (also called sausage), transverse (also called kink),
torsional (also called torsional Alfv\'{e}n), and fluting modes
(see e.g.~\cite{ref28,ref29,ref30,ref31,Stix2004}); for the tube modes supported by
wide MFTs, see
\cite{ref31}. Focusing on the longitudinal tube waves known to be an important
heating agent of solar magnetic regions, it is necessary to mention the recent
papers by
\cite{ref33}, which showed that the longitudinal flux tube waves are identified
as insufficient to heat the solar transition region and corona, in agreement
with the previous studies~\citep{ref34}.
In other words, \textbf{the problem of generation and transport of energy by
MFTs remains unsolved in spite of its key role in physics of
various types of solar active regions.}

It is clear that this unsolved problem of energy transport by magnetic flux
tubes at the same time represents another unsolved problem related to the
energy transport and sunspot darkness (see 2.2 in \cite{Rempel2011}). From a
number of known concepts playing a noticeable role in understanding of the
connection between the energy transport and sunspot darkness, let us consider
the most significant theory, according to our vision. It is based on the
Parker-Biermann cooling effect \citep{Parker1955a,Biermann1941,Parker1979b} and
originates from the works of~\cite{Biermann1941} and~\cite{Alfven1942}.

\begin{figure}[tb]
\begin{center}
\includegraphics[width=12cm]{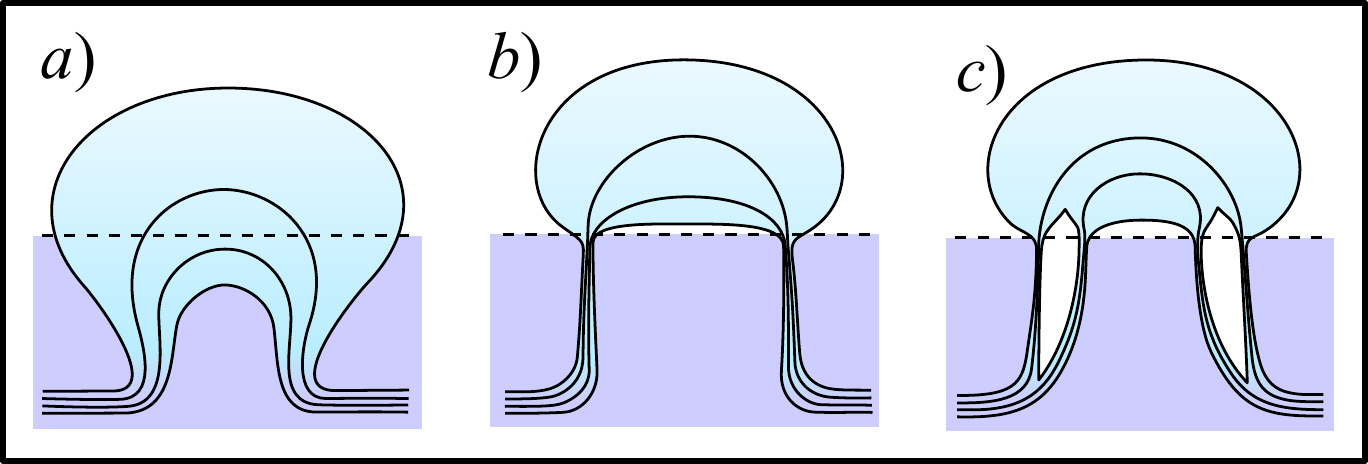}
\end{center}
\caption{The possible ways of the toroidal MFT development into a
sunspot.
(a) A rough representation of the form a tube can take after the rise to the
surface by magnetic buoyancy (adopted from Fig.~2a in \cite{Parker1955a});
(b) demonstrates the ``crowding'' below the photosphere surface because of
cooling (adopted from Fig.~2b in \cite{Parker1955a});
(c) demonstrates the tube splitting as a consequence of the inner region
cooling under the conditions when the tube is in thermal disequilibrium
with its surroundings and the convective heat transfer is suppressed
\citep{Biermann1941} above $\sim 0.71 R_{Sun}$. This effect as well as
the mechanism of the neutral atoms appearance inside the magnetic tubes
are discussed further in the text (see Fig.~\ref{fig-lampochka}a). 
Adopted from Fig.~2c in \cite{Parker1955a}.
} 
\label{fig04-3}
\end{figure}

The main point of the Parker-Biermann cooling effect is that the classical
mechanism of the magnetic tubes buoyancy (see e.g. Fig.~\ref{fig04-3}a,
\cite{Parker1955a}), emerging as a result of the shear flows instability
development in the tachocline, should be supplemented with the following
results of the~\cite{Biermann1941} postulate and the theory developed by
\cite{Parker1955a,Parker1979b}: the electric conductivity in the strongly ionized
plasma may be so high that the magnetic field becomes frozen into plasma and
causes the split magnetic tube (Fig.~\ref{fig04-3}b,c) to cool inside.

Biermann understood that the magnetic field within the sunspots might itself be
a reason of their darkness. Around the sunspots, the heat is transported up to
the surface of the Sun by means of convection (see 2.2.1 in~\cite{Rempel2011}),
while~\cite{Biermann1941} noted that such transport is strongly inhibited by
the nearly vertical magnetic field within the sunspot, thereby providing a
direct explanation for the reduced temperature at the visible surface. Thus,
the sunspot is dark because it is cooler than its surroundings, and it is
cooler because the convection is inhibited underneath.

Still, the missing cause of a very high conductivity in strongly ionized
plasma, which would produce a strong magnetic field ``frozen'' into this
plasma, has been the major flaw of the so-called~\cite{Biermann1941} postulate.

Let us show a solution to the known problem of the Parker-Biermann cooling
effect, which is defined by the nature of very large toroidal magnetic
fields in the tachocline (determined by the thermomagnetic EN~effect) and provides the physical basis for the photon channeling conditions 
inside the MFTs.

\subsubsection{Thermomagnetic Ettingshausen-Nernst effect and the nature of toroidal magnetic field in the tachocline}
\label{sec-nernst}

It is known~\citep{Schwarzschild1958} that the temperature dependence of the
thermonuclear reaction rate is proportional to T$^{4.5}$ in the 10$^7$ K
region. It means there is a sharp boundary between the hot region embracing the
majority of thermonuclear reactions and a cooler one with virtually no
thermonuclear reactions at all~\citep{Winterberg2015a}. This boundary between
the radiation zone and the convection zone is the tachocline. Because of the
great temperature gradient in the tachocline, the thermomagnetic
EN~effect~\citep{Ettingshausen1886,Sondheimer1948,Spitzer1956,Kim1969} produces
strong electric currents screening the intense magnetic fields
$\sim 5 \cdot 10^7$ G of the solar core~\citep{Fowler1955,Couvidat2003}.

Let us note that the thermomagnetic effect described by Eqs.~(5-49) and~(5-52)
in~\cite{Spitzer1956,Spitzer1962,Spitzer2006} is often called the Nernst effect.
This is not entirely a correct name, since the similar equations were
experimentally obtained earlier in 1886 by A.V. Ettingshausen, the teacher,
with his young student W.~Nernst helping him 
(see~\cite{Ettingshausen1886,Sondheimer1948}). Therefore let us hereinafter
refer to this effect as the thermomagnetic EN~effect
(see 3.1.1 in \cite{Rusov2015}).

The EN~effect theory and experiments are known to be the basis
for the quantum critical phenomena and kinetic fluctuations (see 
e.g.~\cite{Hertz1976,Coleman2005,Michaeli2009}) in the 2D materials under
magnetic fields. Let us recall e.g. the superconductors 
\citep{Xu2000,Wang2006}, the amorphous superconducting films
\citep{Michaeli2009}, semiconductor heterostructures and graphene
\citep{Bergman2010}, and what is not surprising taking into account the
``quantum conundrum'' \citep{Hertz1976,Zaanen2007} -- the dyonic BHs
\citep{Hartnoll2007}.

The bright idea of \cite{Winterberg2015a,Winterberg2016} to confine thermonuclear
plasma by means of the repulsive gravitational field in an ultracentrifuge
using the Einstein's general theory of relativity and the repulsive magnetic
field induced by the thermomagnetic effect prompts that it is
the repulsive quantum gravitation in 2D layer which gives rise to the repulsive
magnetic field induced by the thermomagnetic EN~effect. This
may happen e.g. in the tachocline of the Sun or white dwarfs 
(see \cite{Rusov2015}) or in dyonic BHs \citep{Hartnoll2007}).

A natural question arises here: can quantum gravity in 2D materials
(which is related to gravity in 3D anti-de Sitter (AdS) space (see 
\cite{Hartnoll2007})) remove the discrepancies between the general theory of
relativity and quantum mechanics (more exactly, the quantum field theory on the
boundary of the asymptotically AdS space) by means of the known
``holographic renormalization'' (see \cite{Skenderis2002,Papadimitriou2016}), and
give rise to the thermomagnetic EN~effect at the same time?
Let us roughly outline this below.

\subsubsubsection{Thermomagnetic Ettingshausen-Nernst effect of toroidal magnetic field
 in the tachocline}

The thermomagnetic current can be generated in the magnetized plasma under the
quasi-steady magnetic field in the weak collision approximation (the collision
frequency much less than the positive ion cyclotron
frequency)~\citep{Spitzer1962,Spitzer2006}. For the fully ionized plasma the
EN~effect yields the current density (see Eqs.~(5-49) and
(5-52) in~\cite{Spitzer1962,Spitzer2006}):

\begin{equation}
\vec{j} _{\perp} = \frac{3 k n_e c}{2 B^2} \vec{B} \times \nabla T
\label{eq06-01}
\end{equation}

\noindent where $n_e$ is the electron number density, $B$ is the magnetic
field, $T$ is the absolute temperature, $k$ and $c$ stand for the Boltzmann constant and the speed of 
light, respectively. When $n_e = \left[ Z / (Z+1)
\right] n$, where $n = n_e + n_i$, and $n_i = n_e / Z$ is the ion number
density for a $Z$-times ionized plasma,

\begin{equation}
\vec{j} _{\perp} = \frac{3 k n c}{2 B^2} \frac{Z}{Z+1} \vec{B} \times \nabla
T\, . \label{eq06-02}
\end{equation}

It exerts a force on plasma, with the force density $F$ given by

\begin{equation}
\vec{F} = \frac{1}{c} \vec{j} _{\perp} \times \vec{B} =
\frac{3 n k}{2 B^2} \frac{Z}{Z+1} \left( \vec{B} \times \nabla T \right)
\times \vec{B}\, ,
\label{eq06-03}
\end{equation}

or with $\nabla T$ perpendicular to $\vec{B}$

\begin{equation}
\vec{F} = \frac{3 n k}{2} \frac{Z}{Z+1} \nabla T\, ,
\label{eq06-04}
\end{equation}

which leads to the magnetic equilibrium (see Eq.~(4-1)
in~\cite{Spitzer1962}):

\begin{equation}
\vec{F} = \frac{1}{c} \vec{j} _{\perp} \times \vec{B} = \nabla p
\label{eq06-05}
\end{equation}

with $p = nkT$. By equating~(\ref{eq06-04})
and~(\ref{eq06-05}),

\begin{equation}
\frac{3 n k}{2} \frac{Z}{Z+1} \nabla T = nk \nabla T + kT \nabla n
\label{eq06-06}
\end{equation}

\noindent or

\begin{equation}
a \frac{\nabla T}{T} + \frac{\nabla n}{n} = 0,
~~~ where ~~ a = \frac{2 - Z}{2(Z+1)} ,
\label{eq06-06a}
\end{equation}

\noindent we obtain the condition

\begin{equation}
T ^a n = const .
\label{eq06-07}
\end{equation}

For the singly ionized plasma with $Z=1$,

\begin{equation}
T ^{1/4} n = const .
\label{eq06-08}
\end{equation}

For the doubly ionized plasma ($Z=2$) $n=const$. Finally, in the limit of large 
$Z$, $T^{-1/2}n = const$, and $n$ does not depend on $T$
strongly, as opposed to the case of the plasma at a constant pressure with
$Tn=const$. Thus, the thermomagnetic currents can change the pressure
distribution in the magnetized plasma considerably.

Choosing the Cartesian coordinate system with $z$ axis along $\nabla T$,
$x$ axis along the magnetic field and $y$ axis along the current,
and assuming the fully ionized hydrogen plasma with $Z=1$ in the tachocline,
we obtain

\begin{equation}
{j} _{\perp} = {j} _y = - \frac{3 n k c}{4 B} \frac{dT}{dz}. 
\label{eq06-09}
\end{equation}

From Maxwell's equation $4 \pi \vec{j}_{\perp}/ c = \nabla \times \vec{B}$, one has

\begin{equation}
{j} _y = \frac{c}{4 \pi} \frac{dB}{dz}, 
\label{eq06-10}
\end{equation}

\noindent
then by equating~(\ref{eq06-09}) and~(\ref{eq06-10}) we derive

\begin{equation}
2B \frac{dB}{dz} = -6 \pi k n \frac{dT}{dz}.
\label{eq06-11}
\end{equation}

From~(\ref{eq06-08}) one has

\begin{equation}
n = \frac{n _{tacho} T_{tacho}^{1/4}}{T^{1/4}},
\label{eq06-12}
\end{equation}

\noindent where the values $n = n_{tacho}$ and $T = T_{tacho}$ correspond to
the overshoot tachocline. Substituting~(\ref{eq06-12}) into~(\ref{eq06-11}), we
find

\begin{equation}
d\left(B^2\right) = -\frac{6 \pi k n _{tacho} T_{tacho}^{1/4}}{T^{1/4}} dT .
\label{eq06-13}
\end{equation}

\begin{figure}[tb]
\begin{center}
\includegraphics[width=10cm]{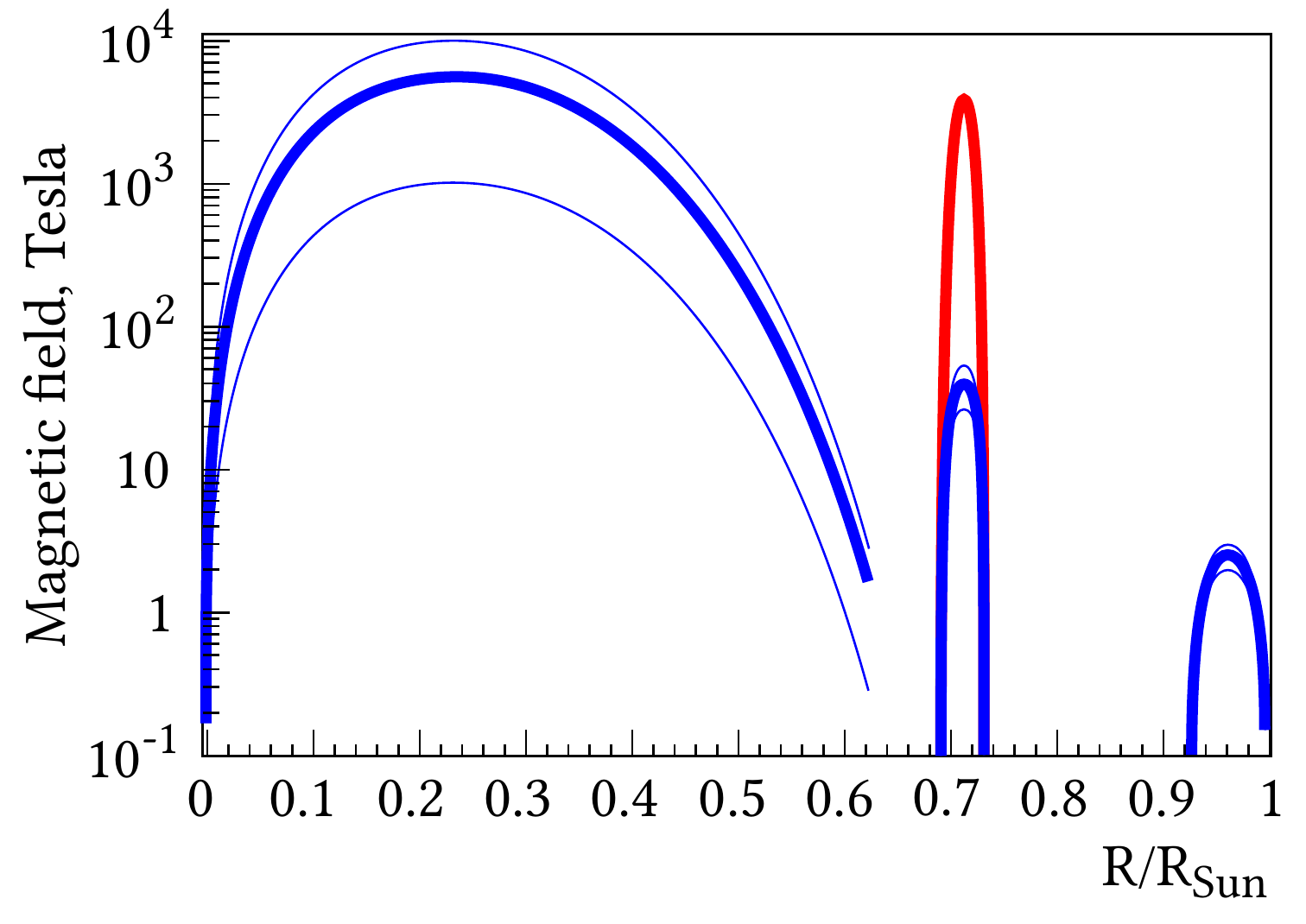}
\end{center}
\caption{The reconstructed solar magnetic field (in blue) simulation
from~\cite{Couvidat2003}: 10$^3$-10$^4$~Tesla (left), 30-50~Tesla (middle) and
2-3~Tesla (right), with temperature of $\sim$9~MK, $\sim$2~MK
and~$\sim$200~kK, respectively. The thin lines show the estimated range of
values for each magnetic field component. Internal rotation was not included in
the calculation. An additional axion production can modify both
intensity and shape of the solar axion spectrum (Courtesy Sylvaine 
Turck-Chi\`{e}ze (see Fig.~2 in~\cite{Zioutas2007})). The reconstructed solar 
magnetic field (in red) simulation from~(\ref{eq06-16}): $4 \cdot 10^3$~T in 
tachocline ($\sim0.7 R_{Sun}$).} 
\label{fig-R-MagField}
\end{figure}

\noindent As a result of integrating in the limits $[B_{tacho},0]$ on the
left and $[0,T_{tacho}]$ on the right,

\begin{equation}
\frac{B_{tacho}^2}{8 \pi} = n _{tacho} kT_{tacho}\, ,
\label{eq06-14}
\end{equation}

\noindent
expressing the fact that the magnetic field of the thermomagnetic current in
the overshoot tachocline ``neutralizes'' the magnetic field of the solar 
interior completely, so that the projections of the magnetic field in the
tachocline and in the core are equal but of opposite directions 
(see Fig.~\ref{fig-R-MagField}).

An intriguing question arises here of what forces are the cause of the
shielding of the strong magnetic fields of the solar core and the radiation
zone, which would be related to the enormous magnetic pressure in the overshoot
tachocline (see Fig.~\ref{fig-R-MagField}),

\begin{equation}
\frac{B_{tacho}^2}{8 \pi} = p_{ext} \approx 6.5 \cdot 10^{13} \frac{erg}{cm^3} ~~
at ~~ 0.7 R_{Sun}, 
\label{eq06-15}
\end{equation}

\noindent
where the gas pressure $p_{ext}$ in the solar tachocline ($\rho
\approx 0.2 ~g\cdot cm^{-3}$ and $T \approx 2.3 \cdot 10^6 K$
\citep{Bahcall1992} at~$0.7 R_{Sun}$) yields the toroidal magnetic field

\begin{equation}
B_{tacho} \simeq 4100 ~T = 4.1 \cdot 10^7 ~G\, .
\label{eq06-16}
\end{equation}

Propagating the question above even further, one may ask: ``What kind of
mysterious nature of the solar tachocline gives birth to the repulsive
magnetic field through the thermomagnetic EN~effect?''
And finally, ``If the thermomagnetic EN~effect exists in the
tachocline, what is its physical nature?''
The essence of this physics will be discussed below.

\subsubsubsection{Nature of toroidal magnetic field of the tachocline in the interior of the Sun}
\label{sec-toroidal-field}

The problem is devoted to the study of physics and the magnitude of the 
toroidal magnetic field in the tachocline. It is known that the radiation zone
of the Sun rotates approximately as a solid body, and the convection zone has
a differential rotation. This leads to the formation of a very strong shear 
layer between these two zones, which is called the tachocline. This gives rise
to a fairly simple and at the same time very complicated question: ``What is the
nature and the consequence of the existence of tachoclines on the Sun or other
stars?''

We propose a very unexpected answer that the existence of a two-dimensional
surface of the tachocline in solar interior is the manifestation of the 
holographic principle in the Universe and, therefore, on the Sun. It is important
to note that the holographic theory correlates the physical laws that act in 
some volume with the laws that act on the surface that limits this volume. The physics
at the boundary is represented by quantum particles that have ``colored'' 
charges and interact almost like quarks and gluons in standard particle
physics. The laws in the volume are a kind of string theory that includes the
force of gravity (see \cite{Maldacena2005}), which is difficult to describe in
terms of quantum mechanics. The main result of the holographic principle is the
fact that surface physics and physics in the volume are completely equivalent,
in spite of the completely different ways of describing them.

In most situations, the contradictory requirements of quantum mechanics and
general relativity are not a problem, because either quantum or gravitational
effects are so small that they can be neglected. However, with a strong
curvature of space-time, the quantum aspects of gravity become essential, and 
the conflict between the theory of gravity and quantum mechanics should disappear 
(see \cite{tHooft1993,Susskind1995,Maldacena1999,Hanada2014}). To create a
large curvature of space-time, a very large mass or density is
required. Some physicists believe that even the Sun is incapable of distorting
space-time so that the manifestations of quantum gravity effects become obvious.

Unlike other physicists, we believe that the fundamental holographic principle of quantum gravity predicts the experimental possibility of observational measurements of magnetic fields between the two-dimensional surface of the tachocline and the three-dimensional volume of the core on compact objects -- our Sun, magnetic white dwarfs, accreting neutron stars and BHs.

What does the equivalence of the two theories mean? First of all, for each 
object in one theory there must be an analog in the other (i.e. in quantum
mechanics and gravitation). Descriptions of objects can be completely
different: a certain particle within a volume can correspond to a whole set of
particles on its boundary, considered as a single entity. Second, the
predictions for the corresponding objects must be identical. For example, if
two particles in the volume collide with a probability of 40\%, then the
corresponding aggregate of particles on its boundary should also be confronted
with a probability of 40\% (see \cite{Maldacena2005}).

Hence, it is not difficult to show that with the help of the thermomagnetic 
EN~effect, a simple estimate of the magnetic pressure of an 
ideal gas in the tachocline of e.g. the Sun can indirectly prove that by using
the holographic principle of quantum gravity, the repelling toroidal magnetic
field of the tachocline 
($B_{tacho}^{Sun} = 4.1 \cdot 10^7 ~G = -B_{core}^{Sun}$; see 
Eq.~(\ref{eq06-16}) and \cite{Rusov2015}) exactly ``neutralizes'' the
magnetic field in the Sun's core,
where the projections of the magnetic fields of the
tachocline and the core have the equal value but the opposite directions.

\begin{figure}[tbp!]
\begin{center}
\includegraphics[width=15cm]{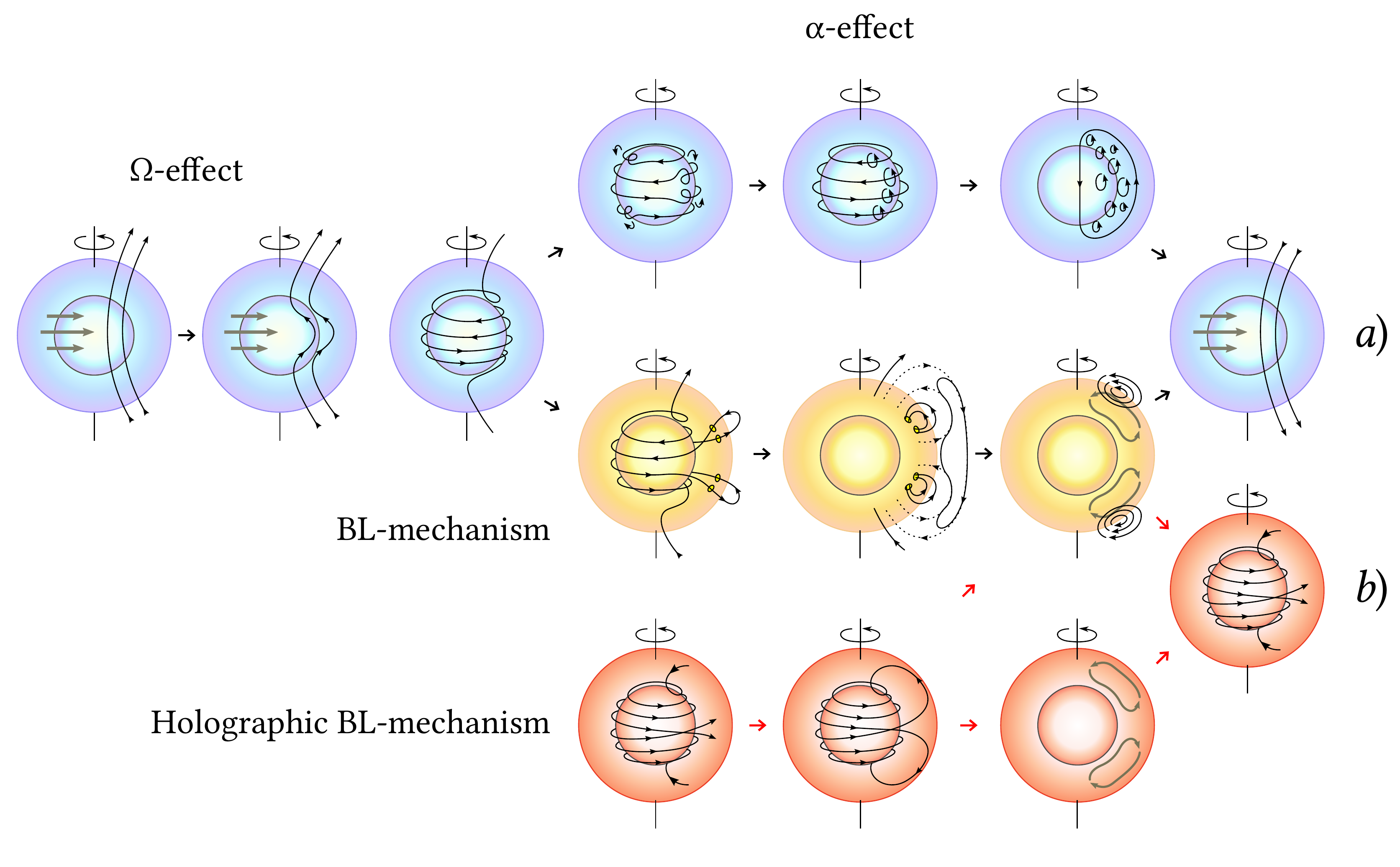}
\end{center}
\caption{An illustration of the main possible processes of a magnetically
active star of the Sun type.
(a) $\alpha$-effect, $\Omega$-effect and BL~mechanism as components of the
solar dynamo model. The $\Omega$-effect (blue) depicts the transformation of the
primary poloidal field into a toroidal field by differential rotation.
Regeneration of the poloidal field is then performed either by the
$\alpha$-effect (top) or by the BL~mechanism (yellow in the
middle). In case of $\alpha$-effect, the toroidal field at the base of the
convection zone is subject to cyclonic turbulence. In the BL~mechanism, the
main process of regeneration of the poloidal field (based on the 
$\Omega$-effect (blue)) is the formation of sunspots on the surface of the Sun
from the rise of floating toroidal flux tubes from the base of the convection
zone. The magnetic fields of these sunspots closest to the equator in each
hemisphere diffuse and join, and the field due to the spots closer to the
poles has a polarity opposite to the current that initiates rotation of the
polarity. The newly formed polar magnetic flux is transported by the meridional
flow to deeper layers of the convection zone, thereby creating a new
large-scale poloidal field. Adopted from \cite{Sanchez2014}.
(b) BL~mechanism and holographic BL~mechanism as components of our
solar antidynamo model. Unlike the component of the solar dynamo model (a), the
BL~mechanism, which is predetermined by the fundamental holographic principle
of quantum gravity, and consequently, the formation of the thermomagnetic
EN~effect (see \cite{Spitzer1962,Spitzer2006,Rusov2015}),
emphasizes that this process is associated with the continuous transformation
of toroidal magnetic energy into poloidal magnetic energy
($T \rightarrow P$ transformation), but not vice versa ($P \rightarrow T$). This
means that the holographic
BL~mechanism is the main process of regeneration of the primary toroidal field
in the tachocline, and thus, the formation of floating toroidal magnetic flux
tubes at the base of the convective zone, which then rise to the surface of the
Sun. The joint connection between the poloidal and toroidal magnetic fields is
the result of the formation of the so-called meridional magnetic field, which
goes to the pole in the near-surface layer, and to the equator at the base of
the convection zone. The connection between the theory of solar meridional
circulation and experimental data of the Joy's Law angle will be discussed in
Sect.~\ref{sec-tilt}.}
\label{fig-solar-dynamos}
\end{figure}

This means that on the one hand, the holographic Babcock-Leighton (BL) mechanism
(see Fig.~\ref{fig-solar-dynamos}), which we often refer to as the holographic
antidynamo mechanism, is caused by a remarkable example of the Cowling
antidynamo theorem. This theorem states that no axisymmetric magnetic field can
be maintained through the self-sustaining action of the dynamo by means of an
axially symmetric current~\citep{Cowling1933}. On the other hand, the BL~mechanism follows
our example of a thermomagnetic EN~effect, or the so-called
solar holographic antidynamo, in which the poloidal field originates directly
from the toroidal field, as shown in Fig.~\ref{fig-solar-dynamos}b, but not
vice versa (see Fig.~\ref{fig-solar-dynamos}a).

But from here we also understand that the remarkable properties of the
holographic BL~mechanism are the consequence of not only the fundamental
features of the holographic principle of quantum gravity, but also of DM. The existence and the true nature of DM are
predetermined, surprisingly enough, by the law of quantum-gravitational energy
conservation in the Universe, and therefore, in our Galaxy, and of course, in
the Sun, in which DM is actively captured by the solar gravitational field!

The main examples of the DM existence in the Sun and other stars in our Galaxy will be considered in our work.

\subsubsection{Parker-Biermann cooling effect, Rosseland mean opacity and\\ axion-photon oscillations in twisted magnetic tubes}

\label{parker-biermann}

Several local models are known to have been used with great success to 
investigate the formation of buoyant magnetic transport, which transforms 
twisted magnetic tubes generated through shear amplification near the base 
tachocline (see e.g.~\cite{Nelson2014}) and the structure and evolution of 
photospheric active regions (see e.g.~\cite{Rempel2011a}).

\begin{figure}[tb]
\begin{center}
\includegraphics[width=15cm]{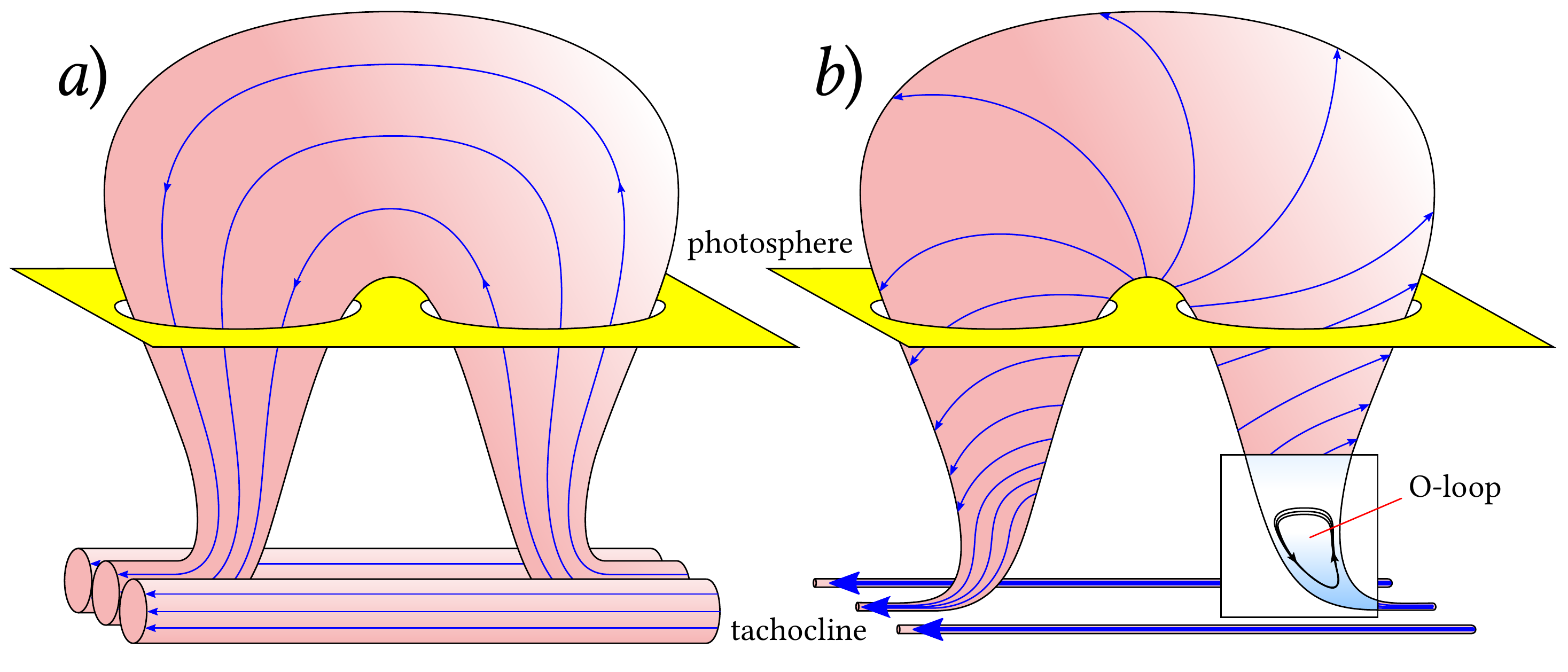}
\end{center}
\caption{An isolated and anchored in the tachocline (a) MFT (adopted
from~\cite{Parker1979a}) and (b) twisted MFT
(see e.g.~\cite{Stein2012}, Fig.~2 in~\cite{Gold1960}, Fig.~1 and Fig.~2
in~\cite{Sturrock2001}) bursting through the solar photosphere to form a
bipolar region. \textbf{Inset in (b)}: topological effect of the magnetic
reconnection in the magnetic tube (see~\cite{Priest2000}), where the
$\Omega$-loop reconnects across its base, pinching off the $\Omega$-loop to
form a free $O$-loop (see Fig.~4 in~\cite{Parker1994}). The buoyancy of the
$O$-loop is limited by the magnetic tube interior with Parker-Biermann
cooling.} 
\label{fig-twisted-tube}
\end{figure}

Because these models assume the anchored MFTs depending on the
poloidal field in the tachocline, it is not too hard to show that the magnetic
field $B_{tacho}$ reaching $\sim 4100$~T (see~(\ref{eq06-16})) may be at the same
time the reason for the Parker-Biermann cooling effect in the twisted magnetic
tubes (see~Fig.~\ref{fig-twisted-tube}b). The theoretical consequences of such
reasoning of the Parker-Biermann cooling effect are considered below.

First of all, we suggest that the classic mechanism of magnetic tubes buoyancy
(Fig.~\ref{fig-twisted-tube}a), appearing as a result of the shear instability
development in the tachocline, should be supplemented by the rise of the
twisted magnetic tubes in a stratified medium (Fig.~\ref{fig-twisted-tube}b;
see Fig.~1 and Fig.~2 in \cite{Sturrock2001}), where the magnetic field is
produced by antidynamo action throughout the convection zone, primarily by
stretching and twisting in the turbulent downflows (see~\cite{Stein2012}).

Second, the twisting of the magnetic tube can not only promote its splitting,
but also form a cool region under a certain condition 

\begin{equation}
p_{ext} = \frac{B^2}{8\pi}
\label{eq06v2-01}
\end{equation}

\noindent
when the tube (inset in~Fig.~\ref{fig-twisted-tube}b) is in thermal
disequilibrium with its surroundings and the convective heat transfer is
suppressed \citep{Biermann1941}.

\begin{figure*}[tbp]
  \begin{center}
    \includegraphics[width=14cm]{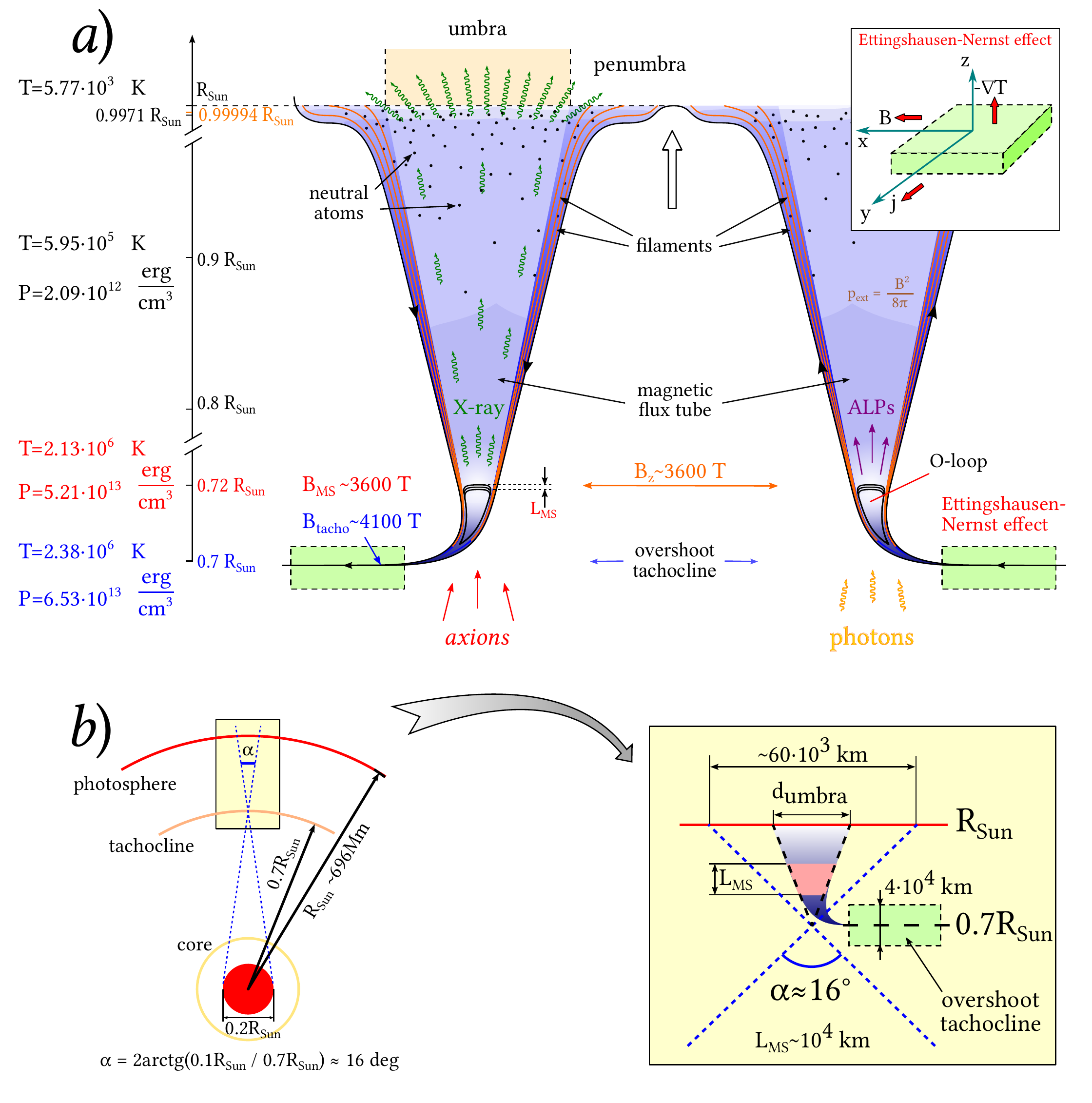}
  \end{center}
\caption{(a) Topological effects of magnetic reconnection inside the
magnetic tubes with the ``magnetic steps''. The left panel shows the
temperature and pressure change along the radius of the Sun from the tachocline
to the photosphere \citep{Bahcall1992}, $L_{MS}$ is the height of the magnetic
shear steps. At $R \sim 0.72~R_{Sun}$ the vertical magnetic field reaches $B_z
\sim 3600$~T, and the magnetic pressure $p_{ext} = B^2 / 8\pi 
\simeq 5.21 \cdot 10^{13}~erg/cm^3$ \citep{Bahcall1992}. The very cool regions 
along the entire convective zone caused by the Parker-Biermann cooling effect 
have the magnetic pressure (\ref{eq06v2-01}) in the twisted magnetic tubes.
\newline (b) All the axion flux, born via the Primakoff effect (i.e. the real
thermal photons interaction with the Coulomb field of the solar plasma), comes
from the region  $\leq 0.1 R_{Sun}$~\citep{Zioutas2009}. Using the angle
$\alpha = 2 \arctan \left( 0.1 R_{Sun} / 0.7 R_{Sun} \right)$ marking the
angular size of this region relative to tachocline, it is possible to estimate
the flux of axions distributed over the surface of the Sun. The flux of
X-rays (of axion origin) is defined by the angle
$\gamma = 2 \arctan \left( 0.5 d_{spot} / 0.3 R_{Sun} \right)$, where
$d_{spot}$ is the diameter of a sunspot on the surface of the Sun (e.g.
$d_{spot} \sim 11000~km$~\citep{Dikpati2008}).}
\label{fig-lampochka}
\end{figure*}

It is interesting to explore how the cool region stretching from the 
tachocline to the photosphere, where the magnetic tube is in thermal 
non-equilibrium~(\ref{eq06v2-01}) with its surroundings, deals with the 
appearance of neutral atoms (e.g. hydrogen) in the upper convection zone 
(see Fig.~\ref{fig-lampochka}a in contrast to Fig.~2c in \cite{Parker1955b}). In
other words, how does this very cool region prevent the neutral atoms to 
penetrate from the upper convection zone to the base of the convection zone, 
i.e. the tachocline?

It is essential to find the physical solution to the problem of solar convective zone which would fit the opacity experiments. The full calculation of solar opacities, which depend on the chemical composition, pressure and temperature of the gas, as well as the wavelength of the incident light, is a complex endeavor. The problem can be simplified by using the mean opacity averaged over all wavelengths, so that only the dependence on the gas physical properties remains (see e.g. \cite{Rogers1994,Ferguson2005,Bailey2009}). The most commonly used is the Rosseland mean opacity $k_R$, defined as:

\begin{equation}
\frac{1}{k_R} = \left. \int \limits_{0}^{\infty} d \nu \frac{1}{k_\nu} \frac{dB_\nu}{dT} \middle/ 
\int \limits_{0}^{\infty} d \nu \frac{dB_\nu}{dT} \right.\, ,
\label{eq06v2-02}
\end{equation}

\noindent
where $dB_\nu / dT$ is the derivative of the Planck function with respect to 
temperature, $k_{\nu}$ is the monochromatic opacity at frequency $\nu$ of the 
incident light or the total extinction coefficient,  including stimulated 
emission plus scattering. A large value of the opacity indicates strong 
absorption from beam of photons, whereas a small value indicates that the beam 
loses very little energy as it passes through the medium.

\begin{figure}[tbp!]
\begin{center}
\includegraphics[width=15cm]{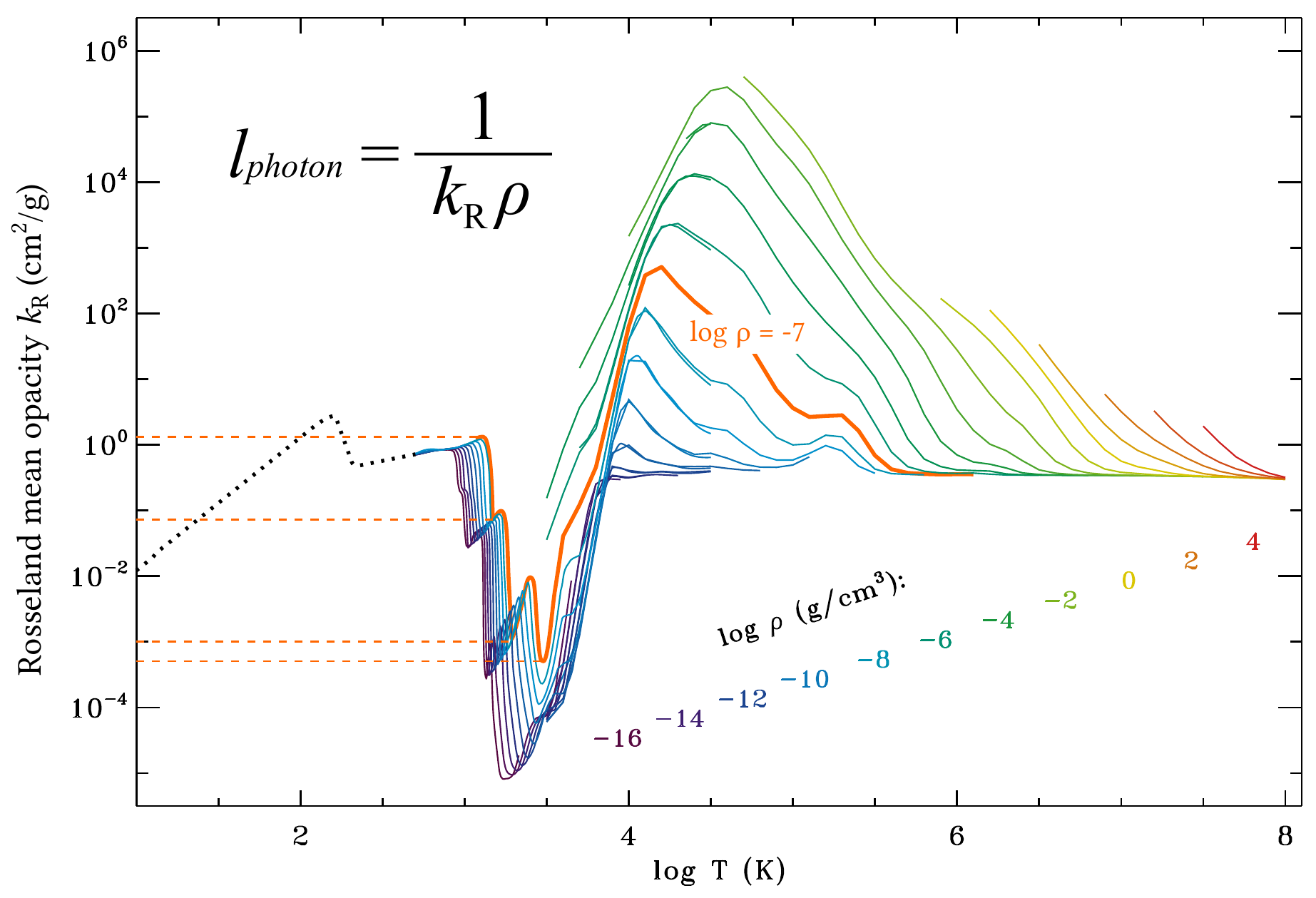}
\end{center}
\caption{Rosseland mean opacity $k_R$, in units of $cm^2 g^{-1}$, shown versus 
temperature (X-axis) and density (multi-color curves, plotted once per decade),
computed with the solar metallicity of hydrogen and helium mixture X=0.7 and 
Z=0.02. The panel shows curves of $k_R$ versus temperature for several 
``steady'' values of density, labeled by the value of $\log {\rho}$ (in 
$g/cm^3$). Curves that extend from $\log {T} = 3.5$ to 8 are from the Opacity 
Project (opacities.osc.edu). Overlapping curves from $\log {T} = 2.7$ to 4.5 
are from \cite{Ferguson2005}. The lowest-temperature region (black dotted 
curve) shows an estimate of ice-grain and metal-grain opacity from 
\cite{Stamatellos2007}. Adopted from \cite{Cranmer2015}.}
\label{fig-opacity}
\end{figure}

Note that the Rosseland opacity is the harmonic mean, in which the greatest 
contribution comes from the lowest values of opacity, weighted by a function 
that depends on the rate at which the blackbody spectrum varies with 
temperature (see Eq.~(\ref{eq06v2-02}) and Fig.~\ref{fig-opacity}), and the
photons are most efficiently transported through the ``windows'' where $k_\nu$ 
is the lowest (see Fig.~2 in \cite{Bailey2009}).

Taking the Rosseland mean opacities shown in Fig.~\ref{fig-opacity}, one may 
calculate, for example, four consecutive cool ranges within the convective 
zone (Fig.~\ref{fig-lampochka}a), where the internal gas pressure $p_{int}$ is 
defined by the following values:

\begin{equation}
p_{int} = n k T, ~where~ 
\begin{cases}
T \simeq 10^{3.48} ~K, \\
T \simeq 10^{3.29} ~K, \\
T \simeq 10^{3.20} ~K, \\
T \simeq 10^{3.11} ~K, \\
\end{cases}
\rho = 10^{-7} ~g/cm^3\, .
\label{eq06v2-03}
\end{equation}

Since the inner gas pressure~(\ref{eq06v2-03}) grows towards the tachocline so 
that

\begin{align}
p_{int} &(T = 10^{3.48} ~K) \vert _{\leqslant 0.85 R_{Sun}}  > 
p_{int} (T = 10^{3.29} ~K) \vert _{\leqslant 0.9971 R_{Sun}} > \nonumber \\
& > p_{int} (T = 10^{3.20} ~K) \vert _{\leqslant 0.99994 R_{Sun}} > 
p_{int} (T = 10^{3.11} ~K) \vert _{\leqslant R_{Sun}} ,
\label{eq06v2-04}
\end{align}

\noindent
it becomes evident that the neutral atoms appearing in the upper convection 
zone ($\geqslant 0.85 R_{Sun}$) cannot descend deep to the base of the 
convection zone, i.e. the tachocline (see Fig.~\ref{fig-lampochka}a).

Therefore, it is very important to examine the connection between the Rosseland 
mean opacity and axion-photon oscillations in a twisted magnetic tube.


Let us consider the qualitative nature of the $\Omega$-loop formation and
growth process, based on the semiphenomenological model of the magnetic
$\Omega$-loops in the convective zone.

\vspace{0.3cm}

\noindent $\bullet$ A high concentration azimuthal magnetic flux 
($B_{tacho} \sim 4100$~T, see Fig.~\ref{fig-lampochka}) in the overshoot 
tachocline develops through the shear flows instability.

An interpretation of such a link is related to the fact that helioseismology
places the principal rotation $\partial \omega / \partial r$ of the Sun in the
overshoot layer immediately below the bottom of the convective zone
\citep{Parker1994}. It is also generally believed that the azimuthal magnetic
field of the Sun is produced by the shearing $r \partial \omega / \partial r$
of the toroidal field $B_{tacho}$, from which it is generally concluded that the
principal azimuthal magnetic flux resides in the shear layer
\citep{Parker1955b,Parker1993}.

\vspace{0.3cm}

\begin{figure}[tb]
  \begin{center}
    \includegraphics[width=15cm]{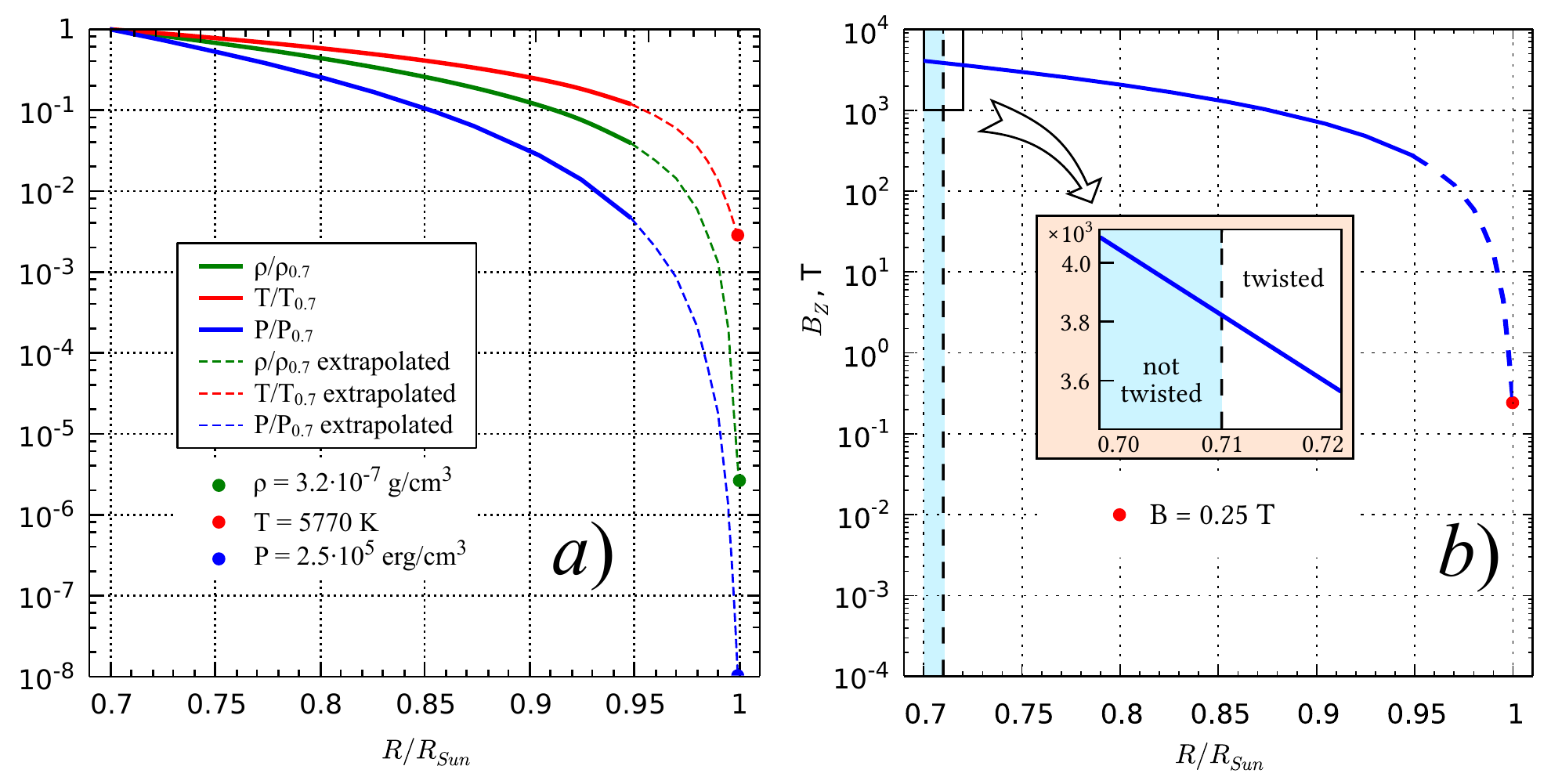}
  \end{center}
\caption{
(a) Normalized external temperature, density and gas pressure as functions of 
the solar depth $R/R_{Sun}$. The standard solar model with $He$ diffusion 
\citep{Bahcall1992} was used  for $R < 0.95 R_{Sun}$ (solid lines). The dotted 
lines mark extrapolated values.
(b) Variation of the magnetic field strength $B_z$  along the emerging
$\Omega$-loop as a function of the solar depth $R / R_{Sun}$ throughout the
convection zone. The solid blue line marks the permitted values for
the same standard solar model with $He$ diffusion \citep{Bahcall1992} starting at
the theoretical estimate of the magnetic field 
$B_{tacho} \approx 4100$~T. The dashed line is the continuation, 
according to the existence of the very cool regions inside the magnetic tube. 
The red point marks the up-to-date observations showing the mean magnetic field 
strength at the level $\sim 0.25~T = 2500 ~G$ \citep{Pevtsov2011,Pevtsov2014}.}
\label{fig-Bz}
\end{figure}

\noindent
$\bullet$ If some ``external'' factor of the local shear perturbation appears
against the background of the azimuthal magnetic flux concentration, such
an additional local density of magnetic flux can lead to the magnetic field
strength as high as e.g. $B_z \sim 3600$~T (see Fig.~\ref{fig-lampochka}a
and Fig.~\ref{fig-Bz}b). Of course, this brings up a question about the 
physics behind such an ``external'' factor and the local shear perturbation.

In this regard let us consider the superintense magnetic $\Omega$-loop
formation in the overshoot tachocline through the local shear caused by the
high local concentration of azimuthal magnetic flux. The buoyant force
acting on the $\Omega$-loop decreases slowly with concentration, so the vertical
magnetic field of the $\Omega$-loop reaches $B_z \sim 3600$~T at about 
$R / R_{Sun} \sim 0.72$ (see Fig.~\ref{fig-lampochka}a and Fig.~\ref{fig-Bz}b).
Because of the magnetic pressure (see analogous Eq.~(\ref{eq06-15}) and 
Fig.~\ref{fig-lampochka}a) 
$p_{ext} = B_{0.72 R_{Sun}}^2 / 8\pi = 5.21\cdot 10^{13}~erg/cm^3$
\citep{Bahcall1992}, this leads to significant cooling of the $\Omega$-loop
tube (see Fig.~\ref{fig-lampochka}a).

In other words, we assume the effect of the $\Omega$-loop cooling to be the
basic effect responsible for the magnetic flux concentration. It arises from
the well known suppression of convective heat transport by a strong magnetic
field~\citep{Biermann1941}. It means that although the principal azimuthal
magnetic flux resides in the shear layer, it predetermines the additional local
shear giving rise to significant cooling inside the $\Omega$-loop.

Thus, the ultralow pressure is set inside the magnetic tube as a result of the
sharp limitation of the magnetic steps buoyancy inside the cool magnetic tube
(Fig.~\ref{fig-lampochka}a). This happens because the buoyancy of the magnetic
flows requires finite \textbf{superadiabaticity} of the convection zone
\citep{Fan1996,Fan2009}, otherwise, expanding according to the magnetic
\textbf{adiabatic} law (with the convection being suppressed by the magnetic
field), the magnetic clusters may become cooler than their surroundings, which
compensates the effect of the magnetic buoyancy of superintense magnetic O-loop.

Eventually we suppose that the axion mechanism based on the X-ray
channeling along the ``cool'' region of the split magnetic tube
(Fig.~\ref{fig-lampochka}a) effectively supplies the necessary energy flux
``channeling'' in the magnetic tube to the photosphere while the convective heat transfer is heavily
suppressed.

In this context it is necessary to have a clear view of the energy transport by X-rays of axion origin, which are a primary transfer mechanism. The recent improvements in the calculation of the radiative properties of solar matter have helped to resolve several long-standing discrepancies between the observations and the predictions of theoretical models (see e.g. \cite{Rogers1994,Ferguson2005,Bailey2009}), and now it is possible to calculate the photon mean free path (Rosseland length) for Fig.~\ref{fig-opacity}:

\begin{equation}
l_{photon} = \frac{1}{k_R \rho} \sim 
\begin{cases}
2 \cdot 10^{10} ~cm  ~~ & for ~~ k_R \simeq 5 \cdot 10^{-4} ~cm^2/g, \\
10^{10} ~cm          ~~ & for ~~ k_R \simeq 10^{-3} ~cm^2/g, \\
1.5 \cdot 10^{8} ~cm ~~ & for ~~ k_R \simeq 6.7 \cdot 10^{-2} ~cm^2/g, \\
10^{7} ~cm           ~~ & for ~~ k_R \simeq 1 ~cm^2/g,
\end{cases}
~~ \rho = 10^{-7} ~g/cm^3\, ,
\label{eq06v2-05}
\end{equation}

\noindent
where the Rosseland mean opacity values $k_R$ and density $\rho$ are chosen so 
that the very low internal gas pressure $p_{int}$ (see Eq.~(\ref{eq06v2-04})) 
along the entire magnetic tube almost does not affect the external gas pressure
$p_{ext}$ (see (\ref{eq06v2-05}) and Fig.~\ref{fig-opacity}).

Let us now examine the appearance of the X-rays of axion origin, induced by the magnetic field variations near the tachocline (Fig.~\ref{fig-lampochka}a), and their impact on the Rosseland length (see~(\ref{eq06v2-05})) inside the cool region of the magnetic tubes.

Let us remind that the magnetic field strength $B_{tacho}\sim 4100$~T in the overshoot 
tachocline (see Fig.~\ref{fig-lampochka}a) and the 
Parker-Biermann cooling effect (see~(\ref{eq06v2-01})) lead to the corresponding 
value of the magnetic field strength $B(0.72 R_{Sun}) \sim 3600$~T
(see Fig.~\ref{fig-lampochka}a), which in its turn implies the virtually zero 
internal gas pressure of the magnetic tube.

As it is shown above (see~\cite{Priest2000}), the topological effect of the
magnetic reconnection inside the $\Omega$-loop results in the formation of the
so-called O-loops (Fig.~\ref{fig-twisted-tube} and Fig.~\ref{fig-lampochka}a)
with their buoyancy limited from above by the strong cooling inside the
$\Omega$-loop (Fig.~\ref{fig-lampochka}a). It is possible to derive the value
of the horizontal magnetic field of the magnetic steps at the top of the O-loop:
$B_{MS} \approx B(0.72 R_{Sun}) \sim 3600$~T.

So in the case of the large enough Rosseland length (see Eq.~(\ref{eq06v2-05})), 
X-rays of axion origin, induced by the horizontal magnetic field in O-loops, reach
the photosphere freely, while in the photosphere itself, according to the 
Rosseland length

\begin{equation}
l_{photon} \approx 100 ~km < l \approx 300 \div 400 ~km,
\label{eq06v2-06}
\end{equation}

\noindent
these photons undergo multiple Compton scattering
producing a typical directional pattern
(Fig.~\ref{fig-lampochka}a).

Aside from the X-rays of axion origin with mean energy of $4.2~keV$
\citep{Andriamonje2007,Zioutas2009}, there are
also $h \nu \sim 0.95 ~keV$ X-rays (originating from the interface between the
radiation zone and overshoot tachocline, according to a theoretical estimate by
\cite{Bailey2009}). Such
X-rays would produce the Compton-scattered photons with mean energy of
$\leqslant 0.95~keV$. These photons ``disappear'' by inverse X-ray transformation
into axions (see Fig.~\ref{fig-lampochka}, Fig.~\ref{fig-axion-compton}a,b and
Fig.~\ref{fig-lower-heating}a). This way the $\sim 0.95~keV$ X-rays do not
contradict the known measurements of the photons with mean energy of $\sim 4~keV$ 
(see Fig.~1 in \cite{Andriamonje2007} and \cite{Rieutord2014}) by involving the X-rays
of axion origin in O-loops (see Figs.~\ref{fig-lampochka} and 
\ref{fig-axion-compton}a,b).

And finally, let us emphasize that we have just shown a theoretical possibility
of the time variation  of the  sunspot activity to correlate with the flux 
of the X-rays of axion origin; the latter being controlled by the magnetic 
field variations near the overshoot tachocline. As a result, it may be 
concluded that the axion mechanism for solar luminosity variations 
based on the lossless X-ray ``channeling'' along the
magnetic tubes allows to explain the effect of the almost complete suppression of the
convective heat transfer, and thus to understand the known puzzling darkness of
the sunspots \citep{Rempel2011}.

\subsubsection{Effect of virtually empty magnetic flux tubes, magnetic reconnection phenomenon and emergence of reconnecting flux-rope dynamo}
\label{sec-empty-tubes}

The appearance of sunspots on the solar surface is one of the major
manifestations of solar activity demonstrating the cyclic behavior with the
period of about 11 years (see e.g. \cite{Hathaway2015}). A very high density
of the magnetic field is observed within the sunspots, which suppresses the
convective heat flow from the solar interior to the surface
\citep{Biermann1941,Cowling1953,Parker1955a}. That is why the sunspots are cooler
and darker against the background of the solar disk. More than a hundred years
ago~\cite{Hale1908} discovered the vertical ``vortices'' of the magnetic field in
the sunspots. About a year later British astronomer~\cite{Evershed1909} was
conducting the observations in Kodaikanal (Tamil Nadu, India) and found that in
spite of the Hale's vertical ``vortices'', the magnetic field in the penumbra is
radially divergent from the center of a sunspot. The mechanism of the
sunspot formation including the umbra and penumbra as well as the Evershed
effect are still the subject of active discussions and studies today, and lots
of fundamental questions remain unanswered 
\citep{Solanki2003,Borrero2011,Tiwari2015,Pozuelo2015}. We consider some
possible solutions to these problems below.

We are mostly interested in effects for which the current theories, assuming
that the sunspots are produced by the dynamo action at the bottom of the
convection zone, fail to provide the convincing proofs and explain the dynamo
action in the convection zone. This problem becomes even stronger with the
recent findings. The numerical simulations of the solar dynamo have not
revealed thin MFTs of the comparable strength so far 
(see e.g. \cite{Guerrero2011,Nelson2013,Kapyla2013}). The helioseismology
does not give any evidence of the upward MFTs existence either
\citep{Birch2013,Birch2016}. This is especially important because of the
recent helioseismological investigations which set the strict limitations on
the velocities of large-scale convection inside the Sun and demonstrated
the inconsistency with the existing global magnetoconvection
\citep[e.g.][]{ref34-3} modeling
\citep{Hanasoge2010,Hanasoge2012,Hanasoge2015,Gizon2012,Birch2013,Birch2016}.

Thus, our major question is: ``How are the sunspots generated by the strong
magnetic field at the base of the convection zone without any dynamo action?'',
or otherwise ``Which fundamental processes connect the sunspot cycle with the
large-scale magnetic field of the Sun?''

A possible answer to this question is that the properties of the
tachocline, bringing the most important thermomagnetic EN~effect, play the key role in the process of the strong magnetic field
variation (without any large-scale dynamo!), and this variation is at least
partially responsible for the solar activity cycles
\citep{Spiegel1980,Rosner1980,Glatzmaier1985,Weiss1994,Schou1998,
Zaqarashvili2010,Zaqarashvili2015,Guerrero2016}.

\begin{figure}[tbp!]
\includegraphics[width=16cm]{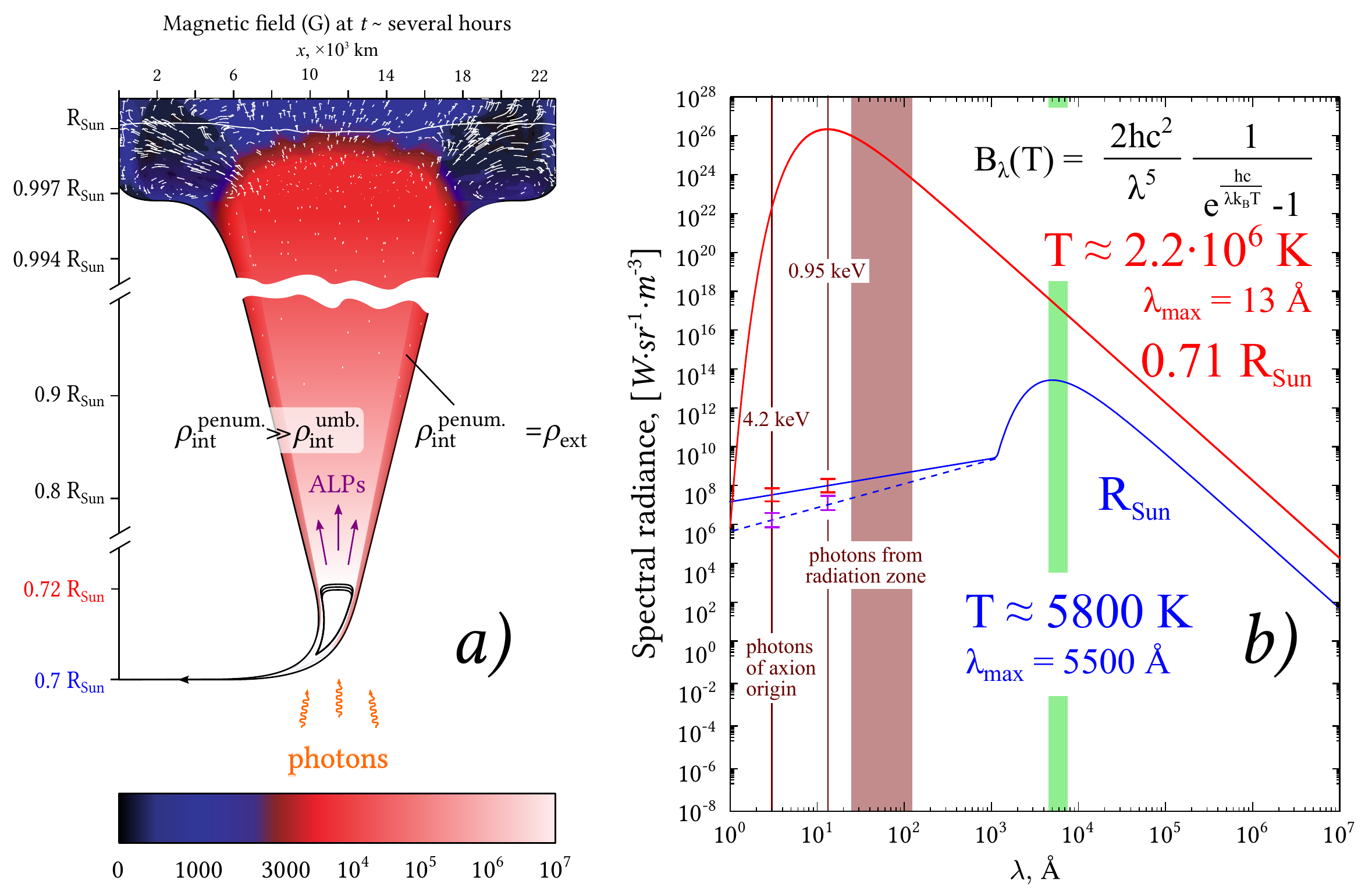}
\caption{\textbf{(a)} A virtually empty magnetic tube is born anchored to the
tachocline and lifted to the surface of the Sun by the neutral buoyancy
($\rho_{int}^{penum} = \rho_{ext}$). The significant convection suppression by
the magnetic field provokes the rapid decrease in temperature and density
($\rho_{int}^{penum} \gg \rho_{int}^{umbra}$), which leads to the significant
decrease of the gas pressure in the umbra. The upper part (photosphere, blue
color) shows the magnetic field strength and magnetic flux (light arrows).
Adopted from \cite{Heinemann2007} and \cite{Schussler2006}. At the boundary
layer between the overshoot tachocline and the underlying radiation zone, photons coming from the radiation zone and passing through the horizontal
fields of the O-loop (see Fig.~\ref{fig-axion-compton}a and 
Fig.~\ref{fig-lampochka}a), are converted into axions, thus almost
completely suppressing the radiative heating in the almost empty magnetic tube.
However, a small fraction of photons passes through the ``ring'' of the
magnetic tube (left and right of the O-loop in the figure (a)) and reaches the
penumbra.
\textbf{(b)} Spectrum of the $2.22 \cdot 10^6~K$ black body at
$0.71 R_{Sun}$ (see Fig.~\ref{fig-Bz} and \cite{Bahcall1992}). The X-ray
luminosity is determined by the thermal photons of average energy
$\sim 0.95~keV$ (in the tachocline \citep{Bailey2009}). These photons are
converted into axions in the ``magnetic steps'' at $\sim 0.72 R_{Sun}$, and
therefore they do not constitute the upper layers spectrum. A noticeable
amount of photons passes between the O-loop and the magnetic tube walls though
(see Fig.~\ref{fig-lampochka}a and Fig.~\ref{fig-axion-compton}a). They reach
the penumbra where there are a lot of visible photons along with the keV
ones (see the green line in Fig.~\ref{fig-axion-compton}b and 
Appendix~\ref{appendix-luminosity}). The blue line shows the spectrum of the
black body with the temperature of the solar surface.}
\label{fig-axion-compton}
\end{figure}

Of all the known concepts playing a noticeable role in understanding of the
link between the energy transfer and the darkness of sunspots, let us
consider the most significant one, in our opinion. It is based on the
Parker-Biermann cooling effect \citep{Parker1955a,Biermann1941,Parker1979b}
which explains the high density of the magnetic field suppressing the
convective heat transfer from inside the Sun to its surface. It also provides a
direct explanation of the lower temperature of the corresponding area on the
visible surface of the Sun (see e.g. Fig.~\ref{fig-lampochka}a and
Eq.~(\ref{eq06v2-04})). According to Fig.~\ref{fig-lampochka}a and
Eq.~(\ref{eq06v2-01}), the cool regions along the entire convection zone,
which are formed under the Parker-Biermann cooling effect, have strong
magnetic pressure in twisted magnetic tubes. This pressure can be estimated
from the condition for the cool umbra (Fig.~\ref{fig-axion-compton}a),

\begin{equation}
p_{ext} = \left[ p_{int} + \frac{B^2}{8 \pi} \right] _{penumbra} = 
\left[ \frac{B^2}{8 \pi} \right]_{umbra}\, .
\label{eq07-01}
\end{equation}

\noindent
The anchored magnetic tubes form a pair of sunspots with umbra and
penumbra on the solar surface. The neutral buoyancy of the MFT
means that when the external density is equal to the internal density
($\rho_{ext} = \rho_{int}$), the temperature inside the tube is lower than
that of the ambient medium (see Fig.~\ref{fig-lampochka}a and 
Fig.~\ref{fig-axion-compton}a). This leads to the magnetic tube lifting towards the
surface of the Sun. So the neutral buoyancy parameters, such as the densities

\begin{equation}
\rho_{ext} = (\rho_{int})_{penumbra} \gg (\rho_{int})_{umbra},
\label{eq07-02}
\end{equation}

\noindent the external and internal gas pressures

\begin{equation}
p_{ext} \gg (p_{int})_{penumbra} \geqslant (p_{int})_{umbra},
\label{eq07-03}
\end{equation}

\noindent plasma $\beta$-parameter (external to internal gas pressure ratio)

\begin{equation}
\left( \frac{B^2}{8 \pi p_{ext}} \right) \equiv \frac{1}{\beta} \sim 1 \gg
 \left( \frac{B_{eq}^2}{8 \pi p_{ext}} \right) \sim 10^{-7},
\label{eq07-04}
\end{equation}

\noindent and the toroidal magnetic field in the tachocline

\begin{equation}
\left( B \right) _{umbra} \gg B_{eq} \sim 10^4 ~G\, ,
\label{eq07-05}
\end{equation}

\noindent
characterize the virtually empty magnetic tube (see Fig.~\ref{fig-axion-compton}a).
$B_{eq}$ is the magnetic field strength in the tachocline at
which the magnetic energy density equals to the kinetic energy density
of the convective downflows.

Our results confirm (see Eq.~(\ref{eq07-01}) for the umbra) that the convective
energy transfer is hindered all along the convection zone from the tachocline
to the umbra inclusively (see Fig.~\ref{fig-axion-compton}) when the plasma 
$\beta$-parameter ($\beta = 8 \pi p_{ext}/B^2$) tends to 1, i.e. when the
magnetic pressure dominates the internal gas pressure. It is also noteworthy
that the convection suppression is more pronounced for the horizontal magnetic
field, since the Lorentz field counteracts the vertical magnetic flux in this
case. In other words, exceeding certain strength of the magnetic field (see
e.g. Eq.~(\ref{eq07-01}) and parameters~(\ref{eq07-05})) apparently leads to
global convection suppression 
\citep{Biermann1941,Chandrasekhar1952,Parker1955a,Valyavin2014,Tremblay2015}.

\begin{figure}[tb!]
\begin{center}
\includegraphics[width=16cm]{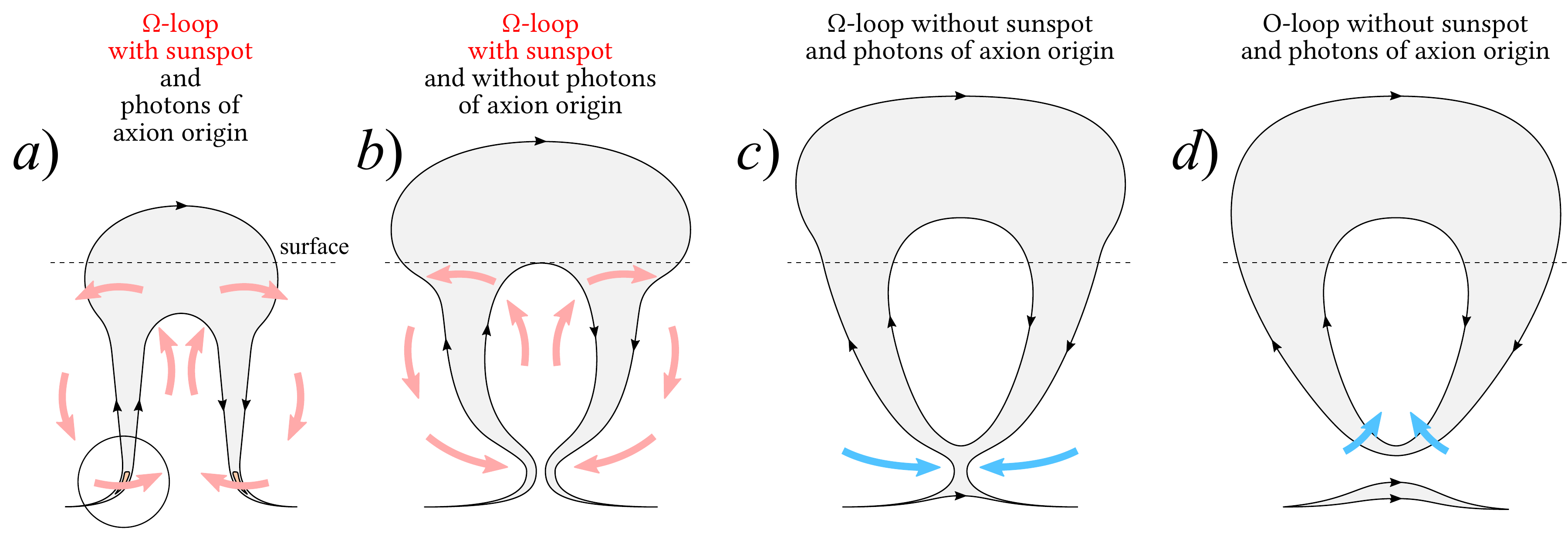}
\end{center}
\caption{A sketch of the magnetic reconnection near the tachocline.
\textbf{(a)} $\Omega$-loop forms the sunspot umbra (with photons of axion
origin) via the indirect thermomagnetic EN~effect
(Fig.~\ref{fig-lampochka}a);
\textbf{(b)} $\Omega$-loop with a sunspot (without photons of axion origin);
pink arrows show the upward convective flow between the ``legs'' of the
$\Omega$-loop during its rise from the tachocline to the visible surface;
\textbf{(c)} $\Omega$-loop with reconnection and without a sunspot;
\textbf{(d)} O-loop without a sunspot.
Going through the stages (a), (b), (c), (d) (left to right), the convection
around the rising $\Omega$-loop ``closes'' it at its base, then a free O-loop is
formed via reconnection, and the initial configuration of the azimuthal field
at the bottom of this region is restored. Blue arrows show the substance motion
leading to the loop ``legs'' connection.}
\label{fig-lower-reconnection}
\end{figure}

It is known (see e.g. \cite{Miesch2012,Miesch2015}) that the Lorentz force
feedback leads to significant suppression of the turbulent $\alpha$-effect
for a large Reynolds number $R_m$, which may be related to the near
(e.g. photosphere) magnetic field spirality. The generation of the spiral flows
gives birth to the small-scale spirality in the opposite ``direction'', which in
its turn can suppress the large-scale dynamo, provided it is not dissipated or
taken away by the magnetic spirality flux \citep{Miesch2012}. So, on the one
hand, the suppression of the convection from the tachocline to the photosphere
is interpreted as the large-scale dynamo suppression (see 
Fig.~\ref{fig-lampochka}a and Fig.~\ref{fig-lower-reconnection}a). On the other hand, with the
magnetoconvection in the penumbra (see Eq.~(\ref{eq07-01}) for the penumbra), it
connects (via the small-scale dynamo action) the sunspots and solar cycle to
the large-scale magnetic field of the Sun.

This solution explicitly depends on the lifetime of the magnetic tubes rising
from the tachocline to the solar surface. Therefore, because of the magnetic
reconnection in the lower layers (see Fig.~4 in \cite{Parker1994}), it is not
the final stage of the simulation. The $\Omega$-loop, forming the sunspot umbra
(Fig.~\ref{fig-lower-reconnection}a) through the convection suppression from the tachocline to
the photosphere, also gives rise to the convective upflow around the
$\Omega$-loop forming the sunspot penumbra. Because of the pre-reconnection
(Fig.~\ref{fig-lower-reconnection}a,b) when the ``legs'' of the $\Omega$-loop collide
(Fig.~\ref{fig-lower-reconnection}b), the convective flow is generated at the base of the
convection zone.

As described by \cite{Spruit1987,Wilson1990,Parker1994,Parker2009}, the upward
convection flow around the rising $\Omega$-loop brings its ``legs'' together in
such a way that the magnetic field reconnection occurs across this loop. This
cuts off the loop from the azimuthal magnetic field, turning it into an O-loop
(see Fig.~3 in \cite{Spruit1987} and Fig.~4 in \cite{Parker1994}). After that the
azimuthal magnetic field restores its initial configuration and becomes ready
for another process with $\Omega$-loop.

Let us now make some important remarks on the turbulent reconnection, the
$\Omega$-loop transformation into the O-loop by rapid ``legs'' closure, the
restoration of the initial azimuthal field and the preconditions for another
$\Omega$-loop formation in the same place. It is also necessary to explain the
physical interpretation of the overshoot process near the tachocline and
estimate the velocity $v_{rise}$ and time $\tau _{rise}$ of the magnetic tube
rise from the overshoot boundary layer -- starting with the azimuthal
magnetic flux strength of $B_{tacho} \sim 4 \cdot 10^7 ~G$
(see Eq.~(\ref{eq06-16})).

One ultimate goal of this section is to determine the general regularities in
the theory of MFTs, which are generated by the magnetic buoyancy
of virtually empty tubes rising from the tachocline to the surface of the Sun
(Fig.~\ref{fig-lampochka}a and Fig.~\ref{fig-axion-compton}a). Another one is
the physical interpretation of the process of MFTs reconnection
in the lower layers of the convection zone (Fig.~\ref{fig-lower-reconnection}).
Not only this is related to the magnetic cycles of flux tubes coinciding with
the observed Joy's law for the tilt angle, but both effects (surprisingly
enough) are induced by the existence of DM -- the solar axions
generated in the core of the Sun.

\subsubsubsection{Convective heating and the buoyant rise of magnetic flux tubes in the solar interior: the model of antidynamo flux tubes and the phenomenon of dark matter solar axions}
\label{sec-radiative-heating}

The first problem is devoted to the study of the effect of virtually empty
magnetic tubes and the phenomenon of DM of solar axions.

The assumption that the virtually empty magnetic tubes
(Fig.~\ref{fig-lampochka}a) are neutrally buoyant ($\rho_{int} = \rho_{ext}$
\citep{Spruit1982}) implies that the temperature inside these tubes is lower than
that of the ambient medium (Fig.~\ref{fig-lampochka} and Fig.~\ref{fig-lower-reconnection}a).
This leads to the heat inflow, and consequently, the flux tube rises up
(see \cite{Parker1975} or Sect.~8.8 in \cite{Parker1979a}). For
a horizontal tube with a cross-section of radius $a$ the rise velocity follows
from the Parker's analysis (see \cite{Parker1975}, Eq.~(60) in 
\cite{Ballegooijen1982}):

\begin{equation}
v_{rise} = 2 \frac{H_p}{\tau _d} \frac{B^2}{8 \pi p_{ext}}
\left( -\delta + 0.12 \frac{B^2}{8 \pi p_{ext}} \right)^{-1},
\label{eq07-39}
\end{equation}

\noindent
where $H_p = \Re T_{ext}/g = p_{ext}/g \rho _{ext} = 0.08 R_{Sun}$ 
\citep{BohmVitense1958,Spruit1977,Brun2011} is the pressure scale height at the
tachocline, $T_{ext}$ and $p_{ext}$ are the external gas temperature and
pressure, $\delta \equiv Y = \nabla _e - \nabla _{ad} = -c_p^{-1} dS / d \xi 
= -c_p^{-1} H_p dS / dz$ is the dimensionless entropy gradient 
(see \cite{Ballegooijen1982,Smolec2008,Smolec2010}),
$\nabla _e \equiv d \ln T_{ext} / d \ln p_{ext}$ and
$\nabla _{ad} \equiv (\partial \ln T / \partial \ln p)_s$ are the local and
adiabatic temperature gradients in external and internal plasma
\citep{Spruit1974,Ballegooijen1982,Christensen1995}, $s$ is the specific
entropy, $c_p$ is the heat capacity at constant pressure, and $\tau _d$ is the
radiation and/or convection diffusion time of the flux tube:

\begin{equation}
\tau_d = \frac{c_p \rho a^2}{k_e} \simeq 
c_p \rho a^2
\left[ \frac{c_p  F_{tot} }{g} 
\left( 1 + \frac{2 \ell _{ov}}{5 H_p} \right)^{\nu} \right]^{-1}.
\label{eq07-40}
\end{equation}

\noindent
where for the fully ionized gas $c_p = 2.5 \Re$ ($\Re$ is the gas constant
in the ideal gas law $p = \rho \Re T$), $T(z)$ and $\rho(z)$ are the mean
temperature and density; $k_e$ is the radiative heat conductivity
(see Eq.~(36) in \cite{Ballegooijen1982});
$\ell _{ov} \approx 0.37 H_p$ \citep{Ballegooijen1982,Christensen2011} is the
thickness of the overshoot layer; the total radiative energy flux $F_{tot} = L/(4 \pi r^2)$ depends on the Sun luminosity $L$; $g$ is the
gravitational acceleration.

Next we apply the condition of hydrostatic equilibrium, $dp / dz = \rho g$,
when the adiabatic temperature gradient $(dT/dz)_{ad} = g/c_p$ may be used,
and the neutral buoyancy of the flux tube in the overshoot zone 
$(\vert \delta T \vert /T_{ext})^{-1} \sim \beta \equiv 8\pi p_{ext}/B^2$. This way we are able to
estimate the time of the radiative and/or convective diffusion $\tau_d$
(see Eq.~(\ref{eq07-40})) of the flux tube:

\begin{equation}
\tau_d = \frac{c_p \rho a^2}{k_e} \approx
\vert \delta T \vert c_p \rho 
\frac{a^2}{(1.148)^{\nu} \delta z F_{tot} }\, ,
\label{eq07-41}
\end{equation}

\noindent where

\begin{equation}
\delta z \sim (1.148)^{-\nu} \left( \frac{a}{H_p} \right)^2
H_p \frac{\nabla _e}{\nabla _{rad}}, ~~ where ~~\nu \geqslant 3.5\, ,
\label{eq07-42}
\end{equation}

\begin{equation}
F_{tot} = \frac{L}{4 \pi R_{tacho}^2} = 
H_p \frac{\nabla _{rad}}{\nabla _e} \left( \frac{dQ}{dt} \right)_1 .
\label{eq07-43}
\end{equation}

\noindent
Here $\nabla _{rad} = (\partial \ln T_{ext} / \partial \ln p_{ext})_{rad}$ is the
radiative equilibrium temperature gradient; $(dQ/dt)_1$ is the rate of radiative heating, which only depends on the
thermodynamic parameters $k_e$ and $T_{ext}$ of the ambient plasma, depending only
on the radial distance from the Sun center \citep{Spruit1974,Ballegooijen1982}.

As a result, it is not difficult to show that the van Ballegooijen model
combining equations (\ref{eq07-39})-(\ref{eq07-43}) gives the final
expression for the rise time by radiation and/or convective diffusion from the
boundary layer of the overshoot to the solar surface,

\begin{equation}
\tau_d \approx \frac{2}{\beta} T_{ext} \left[ \frac{1}{c_p \rho_{ext}} 
\left( \frac{dQ}{dt} \right)_1 \right]^{-1},
\label{eq07-44}
\end{equation}

\noindent
and the lifting speed of the MFT from the overshoot boundary
layer to the surface of the Sun,

\begin{equation}
v_{rise} = H_p \nabla _{ad} \frac{1}{p_{ext}} \left( \frac{dQ}{dt} \right)_1
\left( -\delta + 0.12 \frac{B^2}{8 \pi p_{ext}} \right)^{-1}, ~~~ 
\nabla _{ad} = \nabla _e \simeq 0.4,
\label{eq07-45}
\end{equation}

\noindent
which are almost identical to the equations (29) and (30) of \cite{Fan1996}.

Hence, we understand that the van Ballegooijen model is a special case for
magnetic fields of $B_{tacho} \leqslant 10^5~G$, under which a magnetic dynamo
can exist. On the other hand, we know that based on the holographic BL~mechanism, generating (in contrast to dynamo!) the toroidal magnetic
field in the tachocline, the universal model of flux tubes predetermines the
existence of not only the fields of $B_{tacho} \leqslant 10^5~G$, but also
the strong magnetic fields of the order $B_{tacho} \sim 10^7 ~G$.

Unlike the van Ballegooijen model, we adopt the universal model of MFTs with

\begin{equation}
v_{rise} = 2 \frac{H_p}{\tau _d} \frac{B^2}{8 \pi p_{ext}}
\left( -\delta + 0.12 \frac{B^2}{8 \pi p_{ext}} \right)^{-1}, \nonumber
\end{equation}

\noindent where

\begin{equation}
\tau_d = \frac{c_p \rho a^2}{k_e} 
\left[ 1 + \left( \frac{dQ}{dt} \right)_2 
\middle/ \left( \frac{dQ}{dt} \right)_1 \right]^{-1}\, .
\label{eq07-46}
\end{equation}

\noindent
Here $(dQ/dt)_2$ represents the radiative diffusion across the
flux tube due to the temperature difference ($\delta T \equiv T - T_{ext}$)
between the tube and the external plasma (see \cite{Fan1996}).

Using simple calculations of equations (\ref{eq07-39}) and (\ref{eq07-46}) for
MFTs, it is easy to show that with the help of the total
expression

\begin{equation}
\frac{dQ}{dt} = \left( \frac{ dQ}{dt} \right)_1 + \left( \frac{ dQ}{dt} \right)_2
\label{eq07-47}
\end{equation}

\noindent
of the universal model

\begin{equation}
\tau_d \approx \frac{2}{\beta} T_{ext} 
\left[ \frac{1}{c_p \rho_{ext}} \frac{dQ}{dt} \right]^{-1}
\label{eq07-48}
\end{equation}

\noindent and

\begin{equation}
v_{rise} = H_p \nabla _{ad} \frac{1}{p_{ext}} \frac{dQ}{dt}
\left( -\delta + 0.12 \frac{B^2}{8 \pi p_{ext}} \right)^{-1}, ~~~ 
\nabla _{ad} = \nabla _e \simeq 0.4,
\label{eq07-49}
\end{equation}

\noindent
which is the general case of the so-called universal model of van Ballegooijen-Fan-Fisher (vanBFF model).

On the other hand, let us remind that on the basis of the BL
holographic mechanism, generating the toroidal magnetic field in the
tachocline, the universal model of flux tubes is predetermined by the
existence of strong magnetic fields of the order of $B_{tacho} \sim 10^7~G$.
Since the physics of the holographic BL~mechanism does not
involve a magnetic dynamo, we often refer to it as the universal antidynamo
vanBFF model. It is determined by the following total
energy rate per unit volume:

\begin{equation}
\frac{dQ}{dt} = 
\left( \frac{ dQ}{dt} \right)_1 + \left( \frac{ dQ}{dt} \right)_2 = 
\left( \frac{ dQ}{dt} \right)_1 \left[ 1 + \frac{\alpha _1 ^2}{\nabla _e} 
\left( \frac{H_p}{a}  \right)^2 \frac{1}{\beta} \right],
\label{eq07-50}
\end{equation}

\begin{equation}
\left( \frac{dQ}{dt} \right)_1 = -\nabla \vec{F}_{rad} = F_{tot}
\frac{\nabla _e}{\nabla _{rad}} \frac{1}{H_p} = k_e \nabla _e
\frac{T_{ext}}{H_p^2} ,
\label{eq07-51}
\end{equation}

\begin{equation}
\left( \frac{dQ}{dt} \right) _2 = -k_e \frac{\alpha _1^2}{a^2} (T - T_{ext}) ,
\label{eq07-52}
\end{equation}

\noindent
where 
we used the relation $\vert \delta T \vert /T_{ext} \sim 1/\beta$; the parameter $\alpha _1 ^2 \approx 5.76$
\citep{Fan1996,Weber2015}; $\vec{F}_{rad}$ is the radiative energy flux 
\citep{Spruit1974}; $\nabla _e \sim 1.287 \nabla _{rad}$
(see Table 2 in \cite{Spruit1974}); $H_p/a$ is a factor for
the lower convection zone (see \cite{Ballegooijen1982}).

As a result, we understand that the heating rate of MFTs
consists of the rate of radiative and convective heating (see 
Eqs.~(\ref{eq07-47})-(\ref{eq07-52}); Eq.~(10) in \cite{Fan1996} and
Eq.~(7) in \cite{Weber2015}):

\begin{equation}
\frac{dQ}{dt} = \rho T \frac{dS}{dt} \approx 
-div \vec{F}_{rad} - k_e \frac{\alpha _1^2}{a^2} (T - T_{ext}) ,
\label{eq07-53}
\end{equation}

\noindent
where $S$ is the entropy per unit mass. The first term $(dQ/dt)_1$ determines the mean temperature
gradient between the lower convection zone and the overshoot (see
Eq.~(\ref{eq07-51})). It deviates from the radiative equilibrium substantially,
implying the existence of the nonzero divergence of the heating radiation
flux. The second term $(dQ/dt)_2$ (see Eq.~(\ref{eq07-52})) represents the 
radiation diffusion through the flux tube because of the temperature difference
between the tube and the external plasma. Its effect is to reduce the
temperature difference.

Hence, it is clear that with strong toroidal magnetic fields in the
tachocline (of the order $> 10^5 ~G$) the second term $(dQ/dt)_2$
(in contrast to \cite{Ballegooijen1982,Fan1996,Weber2015,Weber2016})
is the dominant source of convective heating. 
For the toroidal magnetic field 
$B_{tacho} \sim 4 \cdot 10^7 ~G$ in the tachocline, our estimate of
the second term

\begin{equation}
\left( \frac{dQ}{dt} \right)_2 = \frac{\alpha _1^2}{\nabla _e} 
\frac{1}{\beta} \left( \frac{H_p}{a} \right)^2 \left( \frac{dQ}{dt} \right)_1
\geqslant 5 \cdot 10^{4} ~erg \cdot cm^{-3} s ^{-1} ~~ at ~~ \beta \simeq 1
\label{eq07-54}
\end{equation}

\noindent
is related to the first term 
$(dQ/dt)_1 \approx 29.7~erg \cdot cm^{-3} \cdot s^{-1}$
(see Eq.~(\ref{eq07-43}); see also Eq.~(18) in \cite{Fan1996}) and the following parameters:
$\alpha _1^2 \approx 5.76$ \citep{Fan1996,Weber2015}, 
$\nabla _e \approx 0.4$ (see Table~2 in \cite{Spruit1974,Fan1996}),
$1 / \beta = B_{tacho}^2 / 8 \pi p_{ext} = 1$,
$\nabla _e / \nabla _{rad} \approx 1.287$, 
$a = (\Phi / \pi B_{tacho})^{1/2} \leqslant 0.1 H_p$ (see \cite{Fan1993})
with an average value of magnetic flux $\Phi \sim 10^{21}~Mx$ (see e.g. 
\cite{Zwaan1987}).

Taking into account the consequences of the non-local theory of mixing
length \citep{Spruit1974,Spruit1982}, it can be shown that thin,
neutrally buoyant flux tubes are stable in the stably stratified medium,
provided that its field strength $B$ is smaller than a critical value $B_c$
\citep{Ballegooijen1982}, which is approximately given by

\begin{equation}
\frac{B _c^2}{8\pi p_{ext}} = - \gamma \delta = - \frac{5}{3} \delta .
\label{eq07-55}
\end{equation}

So, for the maximum value of the toroidal magnetic field
($B_{tacho} \sim 4 \cdot 10^7 ~G$) the estimate of the rise time of the
radiation and/or convection diffusion of the flux tube (see 
Eq.~(\ref{eq07-48}))

\begin{equation}
\tau _d \approx \frac{2}{\beta} T_{ext} 
\left[ \frac{1}{c_p \rho _{ext}} \left( \frac{dQ}{dt} \right)_2\,\right] ^{-1}
\geqslant 10^4 ~year
\label{eq07-56}
\end{equation}

\noindent
and the lifting speed of an almost empty magnetic tube (see Eqs.~(\ref{eq07-54}), (\ref{eq07-55}))

\begin{equation}
v_{rise} \simeq \frac{H_p \nabla _e}{p_{ext}}  \left( \frac{dQ}{dt} \right)_2
\frac{1}{0.72} \sim 2.6 \cdot 10^{-8} ~km/s
\label{eq07-57}
\end{equation}

\noindent
show that the existence of MFTs on the surface of the Sun is
meaningless.

There is one more beautiful problem which is associated with our problem of
almost total suppression of radiative heating in virtually empty magnetic
tubes (see Fig.~\ref{fig-axion-compton}). Let us remind that photons going from the
radiation zone through the horizontal field of the O-loop near the tachocline
(see Fig.~\ref{fig-axion-compton}a and Fig.~\ref{fig-lampochka}a) are turned into
axions, thus almost completely eliminating the radiative heating in the
virtually empty magnetic tube. Some small photon flux can still pass
through the ``ring'' between the O-loop and the tube walls (see
Fig.~\ref{fig-lampochka}a and Fig.~\ref{fig-axion-compton}a) and reach the penumbra.
Let us denote the area of the magnetic tube ``ring'' by $\pi a _{axion}^2$.

The axion-photon interactions may be described by the incoherent
\cite{ref11} process, when the axions are born in e.g. the core of the
Sun, or the coherent process in the presence of the external magnetic
field, when the axions can be turned into photons and back. This is similar
to neutrino oscillations, and the external magnetic field is necessary to
compensate the spin violation in the case of photon-axion oscillations
\citep{Raffelt-Stodolsky1988,Mirizzi2005,Hochmuth2007,ref05,Marsh2016}.

\begin{figure}[tpb!]
\centering
\includegraphics[width=8cm]{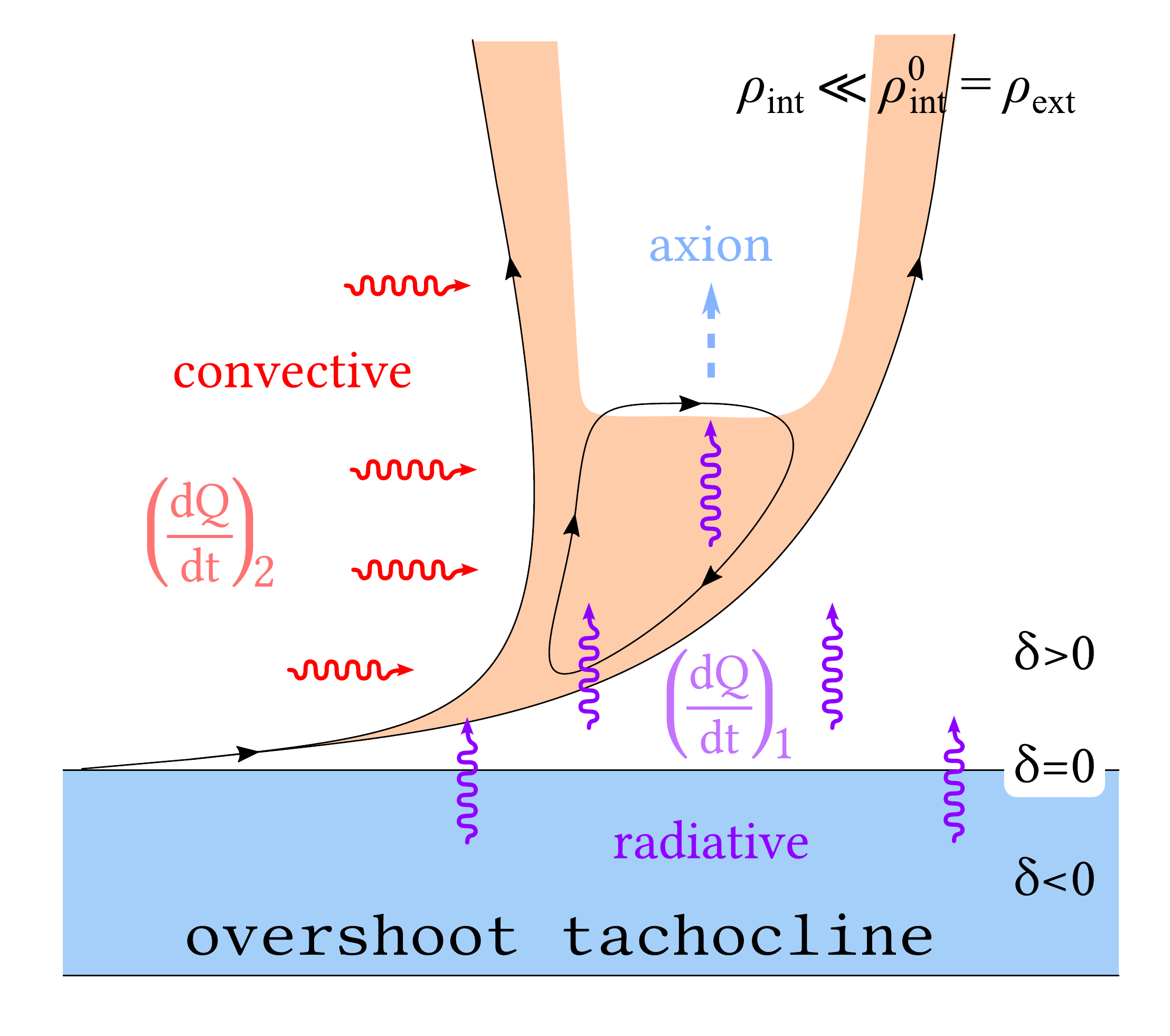}
\caption{\textbf{(a)} A sketch of the magnetic tube born anchored to the
tachocline and risen to the solar surface by the neutral buoyancy
($\rho _{ext} = \rho _{int}^0$). The strong convection suppression inside the
tube leads to the abrupt decrease of temperature and density
($\rho_{int}^0 \gg \rho_{int}$), which in its turn leads to the significant
decrease in gas pressure in the umbra. At the top of the overshoot tachocline
the first term $(dQ/dt)_1$ characterizes the radiative heating which depends on
the thermodynamic quantities $k_e$ and $T_{ext}$ of the external plasma, changing
with the distance from the center of the Sun only (see Eq.~(\ref{eq07-51})). The
second term $(dQ/dt)_2$ represents the diffuse radiation through
the flux tube because of the temperature difference between the tube and the
ambient plasma (see Eq.~(\ref{eq07-52})). The keV photons 
(see Fig.~2 in \cite{Bailey2009}) coming from the radiation zone are turned into
axions in the horizontal magnetic field of the O-loop (see Fig.~\ref{fig-axion-compton}a
and Fig.~\ref{fig-lampochka}a). Therefore, the radiative heating almost vanishes
in the virtually empty magnetic tube. The base of the convection zone is defined
as a radius at which the stratification switches from almost adiabatic
($\delta = \nabla _e - \nabla _{ad} = 0$) to sub-adiabatic
($\delta = \nabla _e - \nabla_{ad} < 0$). Meanwhile, the external plasma turns
from sub-adiabatic to super-adiabatic ($\delta = \nabla_e - \nabla _{ad} > 0$).}
\label{fig-lower-heating}
\end{figure}

We are interested in the axion-photon oscillations for the photons, coming 
from the radiation zone near the tachocline and passing through the horizontal
magnetic field of the O-loop (see Fig.~\ref{fig-lower-heating}a, Fig.~\ref{fig-axion-compton}a and
Fig.~\ref{fig-lampochka}a). They are intensively converted into axions,
thus almost exhausting the radiative heating in the virtually empty
magnetic tube.

So, it is clear that the radiative heating coming from the overshoot boundary
layer can only pass through the cross-section of the magnetic tube ``ring''.
As a result, assuming the mean cross-sectional radius of the magnetic tube

\begin{equation}
a_{axion} \sim 3.7 \cdot 10^{-4} ~H_p,
\label{eq07-58}
\end{equation}

\noindent
we apply a new analysis of the universal vanBFF model
(see Eq.~(\ref{eq07-49}) and (\ref{eq07-47}) (analogous to 
\cite{Ballegooijen1982}) or (\ref{eq07-49}) and (\ref{eq07-55}) (analogous to
Eq.~(29) in \cite{Fan1996})), where the calculated values such as the magnetic
flux $\Phi$ and the rise speed $(v_{rise})_{axion}$ of the MFT
to the surface of the Sun, do not contradict the known observational data:

\begin{itemize}

\item the value of the magnetic flux of the tube

\begin{equation}
\Phi = \pi a_{axion} ^2 B_{tacho} \approx 5.7 \cdot 10^{12} T \cdot m^2 \equiv 
5.7 \cdot 10^{20} Mx ,
\label{eq07-59}
\end{equation}

\noindent
which is in good agreement with the observational data of \cite{Zwaan1987};

\item the time of radiation diffusion of the flux tube (\ref{eq07-48})-(\ref{eq07-50})

\begin{equation}
(\tau_d)_{axion} = 2T_{ext} \left \lbrace \frac{1}{c_p \rho_{ext}} 
\left( \frac{dQ}{dt} \right)_1
\left[ 1 + \frac{\alpha _1 ^2}{\nabla_e} 
\left( \frac{H_p}{a_{axion}} \right)^2 
\right] \right \rbrace ^{-1} \approx 1.1 \cdot 10^5 ~s ~ \sim 1.3 ~ day
\label{eq07-60}
\end{equation}

\noindent
and, as a consequence, the magnitude of the lifting speed of the MFT (see Eq.~(\ref{eq07-39}))

\begin{equation}
(v_{rise})_{axion} = \frac{2.77 H_p}{(\tau_d)_{axion}} \sim 1.4 ~km/s,
\label{eq07-61}
\end{equation}

\noindent
which are almost identical to the observational data of the known works by
\cite{Ilonidis2011,Ilonidis2012,Ilonidis2013} and \cite{Kosovichev2016}.
\end{itemize}

In this context, it is known that \cite{Ilonidis2011,Ilonidis2012,Ilonidis2013}
and \cite{Kosovichev2016} recently detected significant magnetic perturbations
at a depth of about 42−75~Mm (i.e. $\sim 0.9 R_{Sun}$) and showed that these
perturbations were associated with magnetic structures that emerged with an
average speed of 0.3−1.4~km/s and appeared at the surface several hours -- 
2~days after the detection of the perturbations. Interestingly, the results of
several attempts to detect emerging magnetic flux prior to its appearance in the
photosphere can be compared to our theoretical estimates
(\ref{eq07-60})-(\ref{eq07-61}) and demonstrate surprising agreement as
regards the velocity and time of the magnetic tube rise.

Hence, a very important question arises as to how the values of the rise time
and speed of the MFTs, which are predetermined by the radius of
the magnetic tube (see Eq.~(\ref{eq07-58})), do not contradict the
observational data of the Joy's law.
If this is the case, under strong magnetic fields in the
tachocline, which are generated by the holographic principle of quantum gravity
(see Fig.~\ref{fig-solar-dynamos}b), DM of solar axions causes
the appearance of a ``ring'' around a MFT.
Photons of the near-surface radiation of the overshoot regions passing
through the horizontal magnetic field of the O-loop (see 
Figs.~\ref{fig-lampochka}, \ref{fig-axion-compton} and \ref{fig-lower-heating})
are converted into axions, thus completely suppressing the radiation
heating $(dQ/dt)_1$ in some cross-section of the magnetic tube (see~(\ref{eq07-58})).

On the other hand, the most interesting point is that if the axions, which are
born in the Sun core, are directly converted into X-rays near the tachocline,
then the axion-photon oscillations predetermine the appearance of magnetic
sunspot cycles (see Sect.~\ref{sec-osc-parameters} and \cite{Rusov2015}).

It means that if the rising speed of the buoyant magnetic tube, which is
determined by the values of the magnetic field $\sim 4.1 \cdot 10^7 ~G$ and the
cross-section radius $\sim 3.7 \cdot 10^{-4} ~H_p$, causes the appearance
of MFTs in the form of sunspots, then the parameters
of the universal vanBFF model, such as the magnetic
cycles, are almost identical to the observational data of the tilt angle of Joy's law.

\subsubsubsection{Magnetic reconnection of magnetic tubes in lower layers and the observed features of the tilt angle of Joy's Law}
\label{sec-tilt}

The problem is devoted to physics of the magnetic reconnection of a
magnetic tube in the lower layers, which is associated with the so-called
reconnecting dynamo and the observed features of the tilt angle of Joy's law.

Let us consider the physics of turbulent reconnection in the lower
layers of a magnetic tube (see Fig.~\ref{fig-lower-reconnection}), which, as we
discussed above, determines the connection (Fig.~\ref{fig-lower-reconnection}b)
between the global dynamo and the sunspot cycle.
According to the prevalent model, the solar dynamo forms thin MFTs with a strength of $\sim 10^5 ~G$ near the tachocline \citep{DSilva1993},
and some of these tubes can become magnetically buoyant and rise to the
surface, forming sunspots and active regions
\citep[][and Refs. therein]{Parker1955b,Parker1979a,Choudhuri1995,Fan1996,Fan2009}.
However, until now it has been believed 
\citep[][and Refs. therein]{Jabbari2016,JabbariEtAl2016} that numerical and
observational studies do not support this scenario. This is due to the fact
that the theoretical model of the magnetic field vector tilt in sunspots and
bipolar regions \citep{DSilva1993} cannot describe the known observed
latitudinal dependence of the magnetic loops tilt (often called Joy's law \citep{Hale1919}), where the latitudinal distribution of the mean
magnetic field on the surface (and sunspots) as a function of time has the form
of a butterfly diagram (see e.g. Fig.~4 in \cite{Ossendrijver2003} and Fig.~17
in \cite{Hathaway2015}).

It is consequently understandable why the physics of virtually empty magnetic
tubes (see Fig.~\ref{fig-lampochka}a, Fig.~\ref{fig-axion-compton}a and
Fig.~\ref{fig-lower-reconnection}a), which depend on a very strong toroidal
magnetic field on the basis of the EN~effect, does not
contradict the possibility of the existence of neutrally
buoyant magnetic tubes, emerging from the boundary layer of the overshoot
tachocline to the solar surface, as well as the formation of the
small-scale dynamo, which is connected to the sunspot cycle
(Fig.~\ref{fig-axion-compton}a and Fig.~\ref{fig-lower-heating}a).

Hence, our previous question can be clarified as follows: what is the mechanism of the magnetic reconnection in the lower layers of a neutrally floating tube, which at the beginning (see Fig.~\ref{fig-lower-reconnection}a) has a practically empty $\Omega$-loop without convection (see Fig.~\ref{fig-lower-heating}a)? In other words, how does the scenario of the model described by well-known physicists (see \cite{Spruit1987,Wilson1987,Parker1994,Parker2009}) transform the turbulent $\Omega$-pumping with the so-called turbulent reconnection \citep{Loureiro2009,Huang2010,Beresnyak2017}, in which turbulent plasma pumping from the azimuthal field is repeated in the $\Omega$-loop again and again (see Fig.~\ref{fig-lower-reconnection} and Fig.~\ref{fig-lower-reconnection2})?

\textbf{\textit{Below is the brief analysis and summary of the answer.}}
As known, magnetic reconnection is the most striking and amazing
observation of the Sun, which causes solar flares. Observations estimate the
reconnection speed as small, though not too small, fraction of the
Alfv\'{e}n speed of the so-called rapid reconnection (see e.g. 
\cite{Beresnyak2017} and Refs. therein). Given the remarkable properties of
fast turbulent reconnection, we understand why the mechanism of sunspot
formation and corona heating are very strongly associated with the
phenomenon of magnetic reconnection (see \cite{Loureiro2009,Huang2010,
Beresnyak2013,Beresnyak2017,JabbariEtAl2016}), which changes the topology of the
magnetic field and leads to the conversion of magnetic energy into thermal
energy, kinetic energy, and even particle acceleration. Let us discuss the
known ``unanswered'' basic questions related to the nature of dynamo and
the origin of sunspots and active regions, together with our respective
answers:

\begin{figure}[tbp]
\begin{center}
\includegraphics[width=16cm]{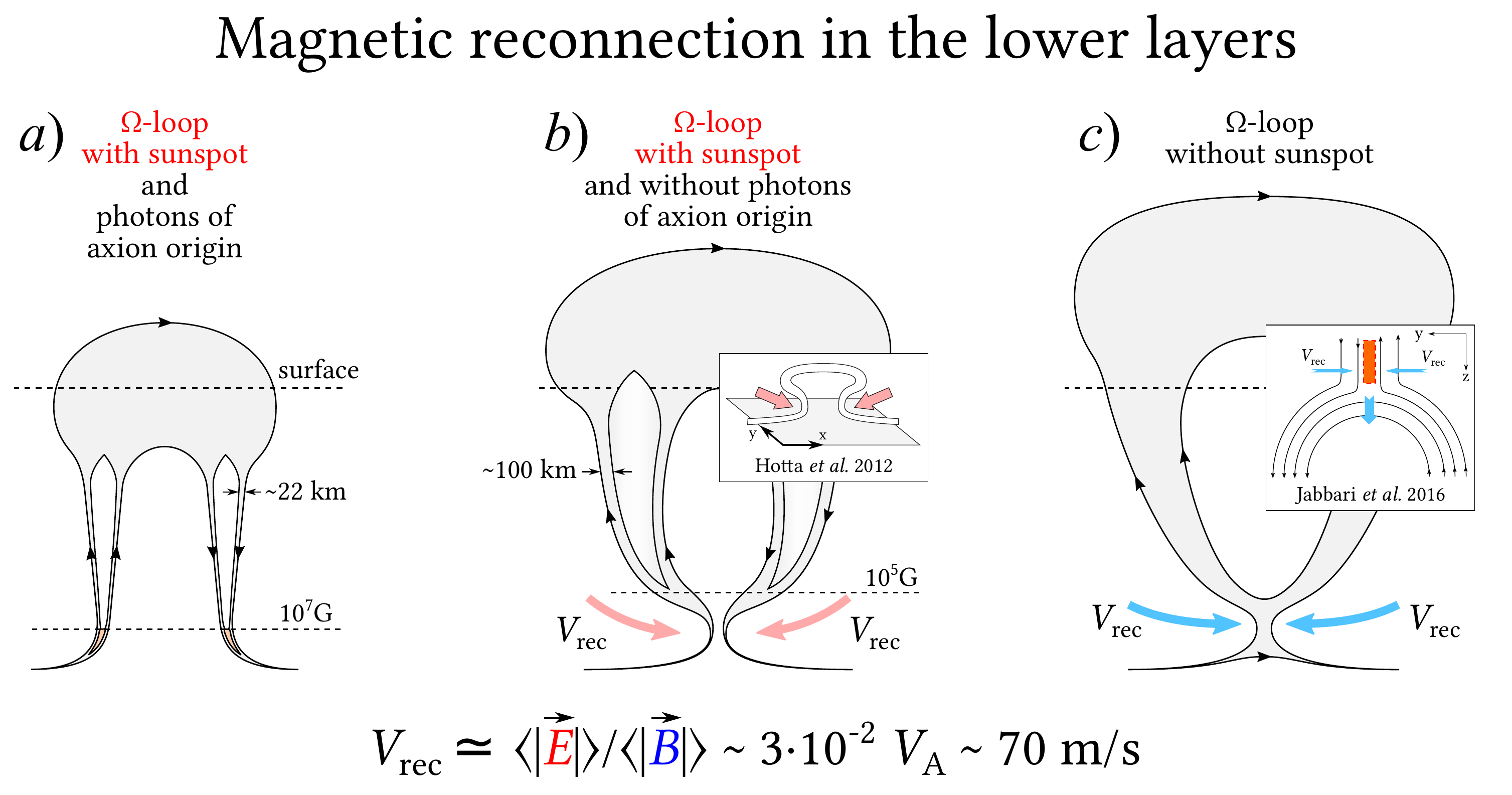}
\end{center}
\caption{A sketch of magnetic reconnection near the tachocline (see analogous Fig.~\ref{fig-lower-reconnection}). Understanding the essence of a virtually empty tube, which (a) is first born without any dynamo, is related to the physics (b)-(c) of the turbulent reconnection of magnetic bipolar structures (see Eq.~(8) in \cite{JabbariEtAl2016}), and, as a consequence, a very rare model of fluctuation dynamo caused by a multiscale turbulence model (see \cite{Baggaley2009}), but necessarily with the help of the so-called turbulent reconnection \citep{Loureiro2009,Huang2010,Beresnyak2017}, at which the turbulent pumping of the plasma from the azimuthal field is repeated in the $\Omega$-loop again and again (see Fig.~\ref{fig-lower-reconnection}).}
\label{fig-lower-reconnection2}
\end{figure}

$\bullet$ \textbf{Is there a phenomenon of magnetic reconnection in the lower layers of magnetic tubes? If they appear near the tachocline, then why isn't there a large-scale dynamo in the virtually empty magnetic tubes?}

\begin{figure}[tbp]
\begin{center}
\includegraphics[width=16cm]{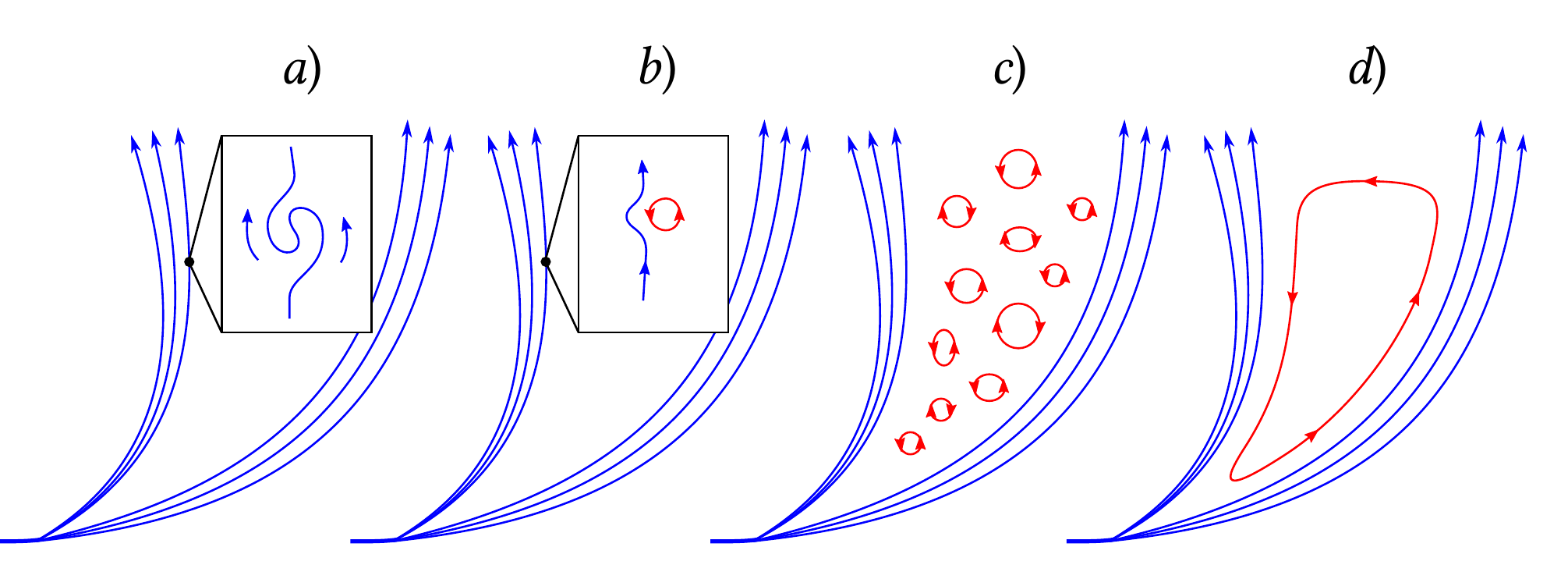}
\end{center}
\caption{Turbulent cascade \citep{Kolmogorov1941,Kolmogorov1968,Kolmogorov1991} and magnetic reconnection in the lower layers inside the magnetic tube. Common to these various turbulent systems is the presence of the inertial range of Kolmogorov, through which the energy is cascaded from large to small scales, where dissipative mechanisms (as a consequence of magnetic reconnection) overcome the turbulent energy in plasma heating. \textbf{(a)} When the magnetic field of the unipolar part of the tube is strong (see the inset), it can be moved as a passive scalar, and its spectrum will simulate the Kolmogorov turbulence. On the one hand, the inset (a) contains the ``description'' of the turbulent-vortex cascade on a very small scale, and on the other hand, the inset \textbf{(b)} represents a description of a small magnetic island of fluctuations that is born by means of reconnection. \textbf{(c)} As the magnetic field increases, the Alfv\'{e}n velocity $V_A = B/(4 \pi \rho )^{1/2}$ accelerates the inverse cascade, when a group of magnetic islands with different dimensions of the loop merge to form larger ones. \textbf{(d)} The extremely large O-loop will always be smaller than the initial magnetic field of the current tube, since the emerging O-loop predetermines the generation of losses by turbulent magnetic energy, which by reconnection is converted into plasma heat.}
\label{fig-Kolmogorov-cascade}
\end{figure}

We have already shown (see Figs.~\ref{fig-lampochka}a and~\ref{fig-axion-compton}b) that the topological effects of magnetic reconnection inside magnetic tubes near the tachocline form the O-loop (Fig.~\ref{fig-Kolmogorov-cascade}), the ``magnetic steps'' of which (see $L_{MS}$ in Fig.~\ref{fig-lampochka}a) participate in the formation of photons of axion origin (see Fig.~\ref{fig-lampochka}a).

\begin{figure}[tbp]
\begin{center}
\includegraphics[width=16cm]{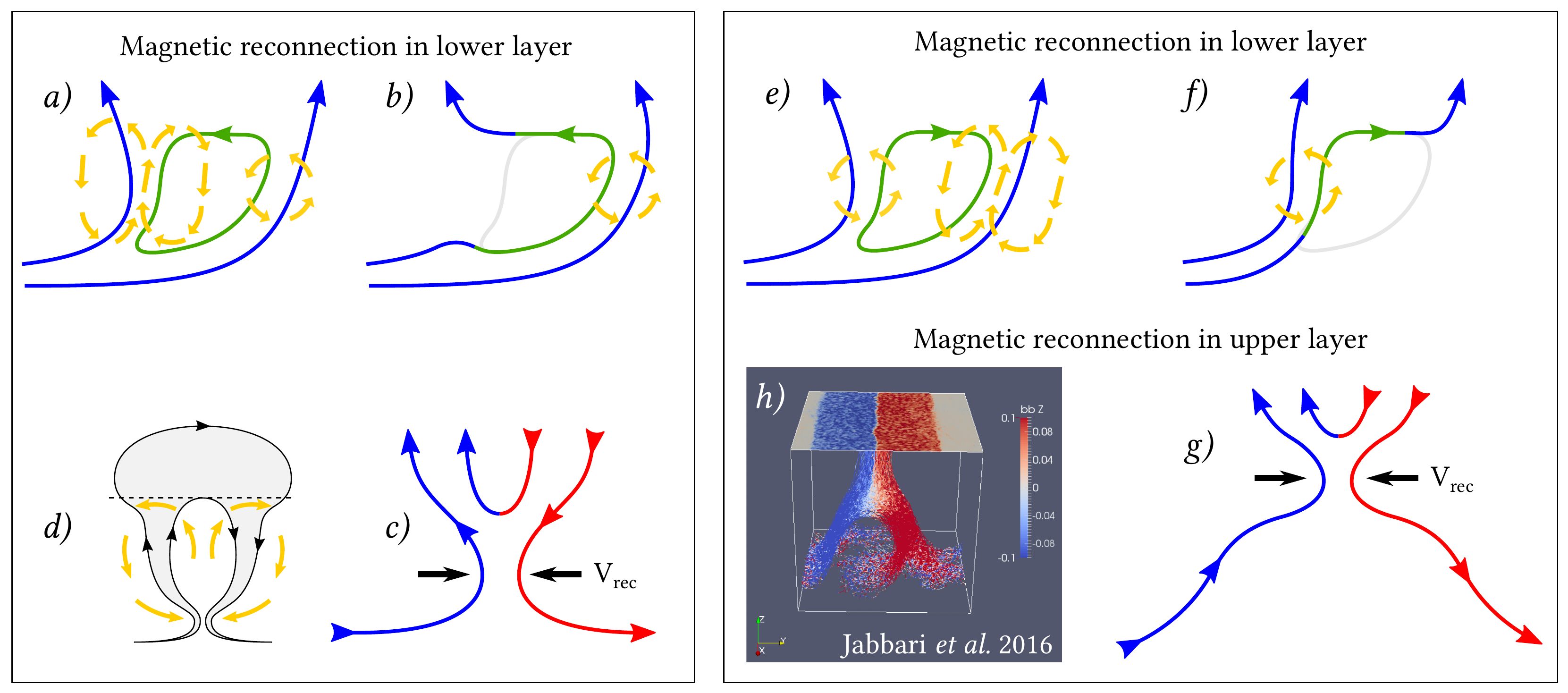}
\end{center}
\caption{Topological effects of magnetic reconnection in the lower (left) or upper (right) layers of the magnetic tube. Here, the unipolar part of the $\Omega$-loop is rebuilt on its base, compressing the $\Omega$-loop (blue lines) to form a free O-loop (green lines) (Fig.~\ref{fig-lampochka}a). The yellow lines show the movement of the substance leading to the connection of the ``leg'' loop (see analogous Fig.~4 in \cite{Parker1994}). If the O-loop (green lines) can randomly have different directions of magnetic fields, then the magnetic reconnection can generate loop ``legs'' in different layers, for example, in lower layers (Fig.~\ref{fig-upper-reconnection}c and also Fig.~\ref{fig-upper-reconnection}d as an analog of Fig.~\ref{fig-lower-reconnection}b) and upper layers (Fig.~\ref{fig-upper-reconnection}g and also Fig.~\ref{fig-upper-reconnection}h as an analog of Fig.~4 in \cite{JabbariEtAl2016}).}
\label{fig-upper-reconnection}
\end{figure}

In Fig.~\ref{fig-upper-reconnection} one can see the conditions for the magnetic reconnection between the O-loop (green lines) and the unipolar part of the $\Omega$-loop (blue lines) that can organize them in the lower layer (Fig.~\ref{fig-upper-reconnection}, left) or in the upper layer (Fig.~\ref{fig-upper-reconnection}, right), thereby showing the appearance of bipolar magnetic tubes in various versions (Fig.~\ref{fig-upper-reconnection}c,d and g,h).

The O-loop mentioned above may have different directions of magnetic fields, or may not exist within the $\Omega$-loop at all. When there is no O-loop near the tachocline in the $\Omega$-loop, the absence of the horizontal magnetic field of the O-loop (see Figs.~\ref{fig-axion-compton}a and ~\ref{fig-lampochka}a) makes it impossible to convert axions (with average energy $\sim 4.2 ~keV$) into photons (see Fig.~\ref{fig-lampochka}a). At the same time it passes thermal photons with average energy $\sim 0.95 ~keV$ (in the tachocline \citep{Bailey2009}), which are strongly scattered near the tachocline (see Fig.~\ref{fig-axion-compton}a). This means that due to the strong scattering of $0.95~keV$ photons near the tachocline, already visible photons reaching the photosphere and above generate only ``bright'' sunspots, i.e. the visible photons form ``invisible'' sunspots (see Fig.~\ref{fig-lower-reconnection2}), which exist inside the convective zone, but are not observed optically!

\begin{figure}[tbp]
\begin{center}
\includegraphics[width=16cm]{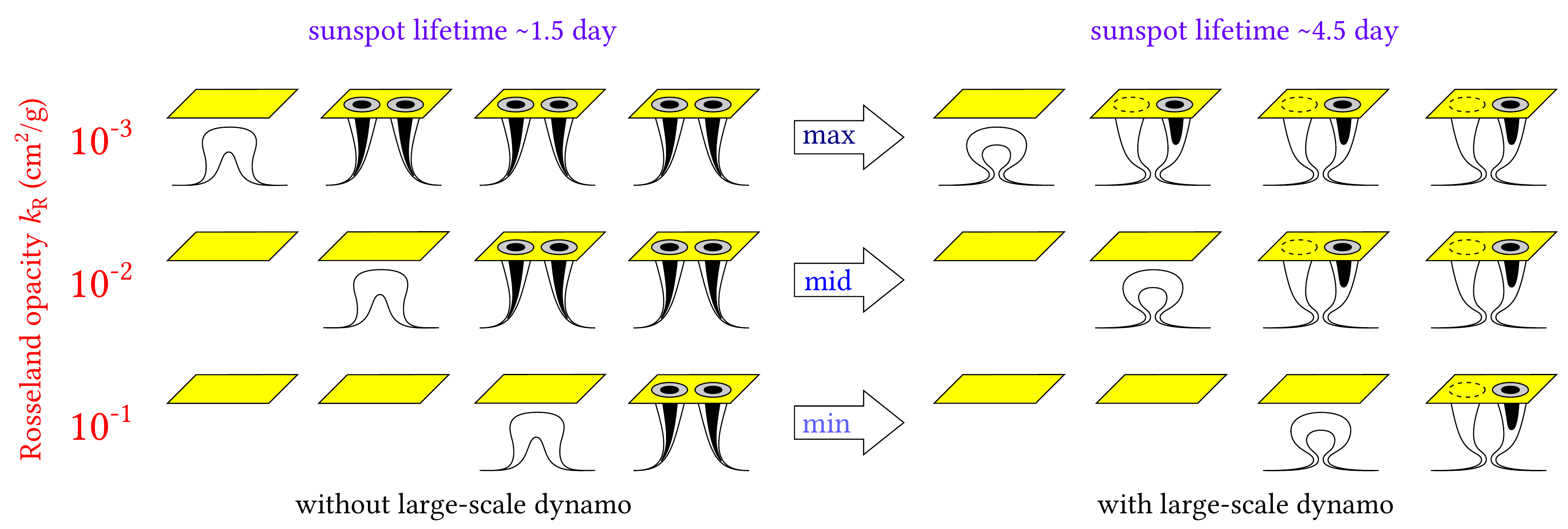}
\end{center}
\caption{The physical nature of the cycle of sunspots as a consequence of the modulation of MFTs, rising from the tachocline to the surface of the Sun. Left: ``one'' of the initial bipolar magnetic tubes is ``invisible'' due to the absence of a free O-loop and, thus, the absence of photons of axion origin in the $\Omega$-loop, which, as a consequence, rises super-slowly to the solar surface: here the rise time of the magnetic tube from the boundary layer of the overshoot to the surface of the Sun is many orders of magnitude greater than the lifetime of the sunspot (see \cite{Petrovay1997,Solanki2003} and vanBFF model (Sect.~\ref{sec-radiative-heating})). Right: simplifying the drawing, we drew different unipolar magnetic tubes that actually correspond to visible or ``invisible'' bipolar magnetic tubes, cyclically (from the maximum to the minimum of activity) appearing on the solar surface. The rough values of the Rosseland opacity $k_R$ can be estimated using Eq.~(31) and Fig.~6 in~\cite{Rusov2015}.}
\label{fig-sunspots-vanish}
\end{figure}

Finally, let us note that we have two types of magnetic reconnection inside the $\Omega$-loop near the tachocline, one of which is manifested by the existence of a free O-loop (red lines in Fig.~\ref{fig-Kolmogorov-cascade}, green lines in Figs.~\ref{fig-upper-reconnection}a and ~\ref{fig-upper-reconnection}e), and the other one practically does not exist. In other words, magnetic reconnection inside the $\Omega$-loop may or may not accidentally give birth to a free O-loop near the tachocline (see Fig.~\ref{fig-Kolmogorov-cascade}). This means that if a free O-loop is created in the $\Omega$-loop near the tachocline, classical sunspots appear on the surface (see black tubes, Fig.~\ref{fig-sunspots-vanish}), and the ``transparent'' bipolar tubes on the surface appear as ``optically invisible'' spots otherwise (see white bipolar tubes, Fig.~\ref{fig-sunspots-vanish}).

In this case we are interested in the relationship between cycles of sunspots and a large-scale dynamo. On the one hand, we know that one of the main difficulties with the dynamo of the convective zone is that magnetic buoyancy rather displaces any magnetic flux from the convective zone, without allowing sufficient time to amplify the dynamo (see \cite{Parker1975,MorenoInsertis1983,DSilvaChoudhuri1993}). On the other hand, we know that according to \cite{Choudhuri1987}, the magnitude of the effect of magnetic buoyancy is determined by the strength of the magnetic field. This dictates the reduction in density inside the tube compared to its surroundings at a given level. This difference in density $\rho$ can be estimated by considering a magnetic tube in the thermal state, i.e. in complete equilibrium of pressure and temperature with its surroundings (see \cite{Choudhuri1987,Choudhuri1989,DSilvaChoudhuri1993}). It is known that if the magnetic field at these levels is supported by a dynamo and the magnetic fields are diffuse (see, for example, Fig.~\ref{fig-lower-heating}a), then the field strength is unlikely to be greater than $\sim 10^4 ~G$, since large fields will greatly suppress the dynamo action \citep{Deluca1986}. Since we know that the azimuthal magnetic fields of the tube represent very strong fields near the tachocline ($\sim 10^7 ~G$), the strong decrease in internal pressure of the gas, which is predetermined by the thermomagnetic EN~effect (see Eq.~(25) in \cite{Rusov2015}), yields

\begin{equation}
\Delta \rho / \rho_{ext} = B^2 / 8 \pi p_{ext} = 1 / \beta \cong 1 .
\label{eq07-62}
\end{equation}

Let us remind that photons coming from the radiation zone through the horizontal magnetic field of the O-loop from the $\Omega$-loop near the tachocline (see Figs.~\ref{fig-axion-compton}a and~\ref{fig-lampochka}a) are converted into axions, virtually eliminating the radiation heating in the empty MFT (see Fig.~\ref{fig-lower-heating}a). Hence, the most intuitive result of this solution is the fact that the magnetic cycles of sunspots (see Fig.~\ref{fig-sunspots-vanish}) are a consequence of the practically empty flux tubes (see Figs.~\ref{fig-axion-compton}a, ~\ref{fig-lampochka}a and ~\ref{fig-lower-heating}a), which, reaching the photosphere, initially do not have the effect of secondary reconnection in the lower or upper layer of the magnetic tube (see Fig.~\ref{fig-sunspots-vanish}). If a large-scale dynamo appears with the spots present, this means that the emerging dynamo with the help of the secondary reconnection near the tachocline (see Figs.~\ref{fig-lower-reconnection}b,c, ~\ref{fig-lower-reconnection2} and ~\ref{fig-sunspots-vanish}) leads to the MFT cut-off (see Fig.~\ref{fig-lower-reconnection}c) and, as a consequence, the sunspots leaving the surface of the Sun towards the corona.

So, the final connection between the cycles of sunspots and the large-scale dynamo is the fact that the formation of sunspots and their cycles does not depend on the large-scale dynamo, which, strangely enough, is predetermined only by the process of the sunspots liftoff from the surface of the Sun!

$\bullet\bullet$\textbf{ How do the primary and secondary magnetic reconnection in the lower layers of flux tubes (with magnetic field strength $\sim 10^7 ~G$ (see Fig.~\ref{fig-lower-reconnection2}a) and $\sim 10^5 ~G$ (see Fig.~\ref{fig-lower-reconnection2}b), respectively) explain the physics and theoretical estimates of the buoyant tubes -- the rise time and the speed of rising 
to the surface of the Sun, and the tendency of the tilt angle of Joy's law, which does not contradict the known experimental data?}

\begin{figure}[tbp]
\begin{center}
\includegraphics[width=14cm]{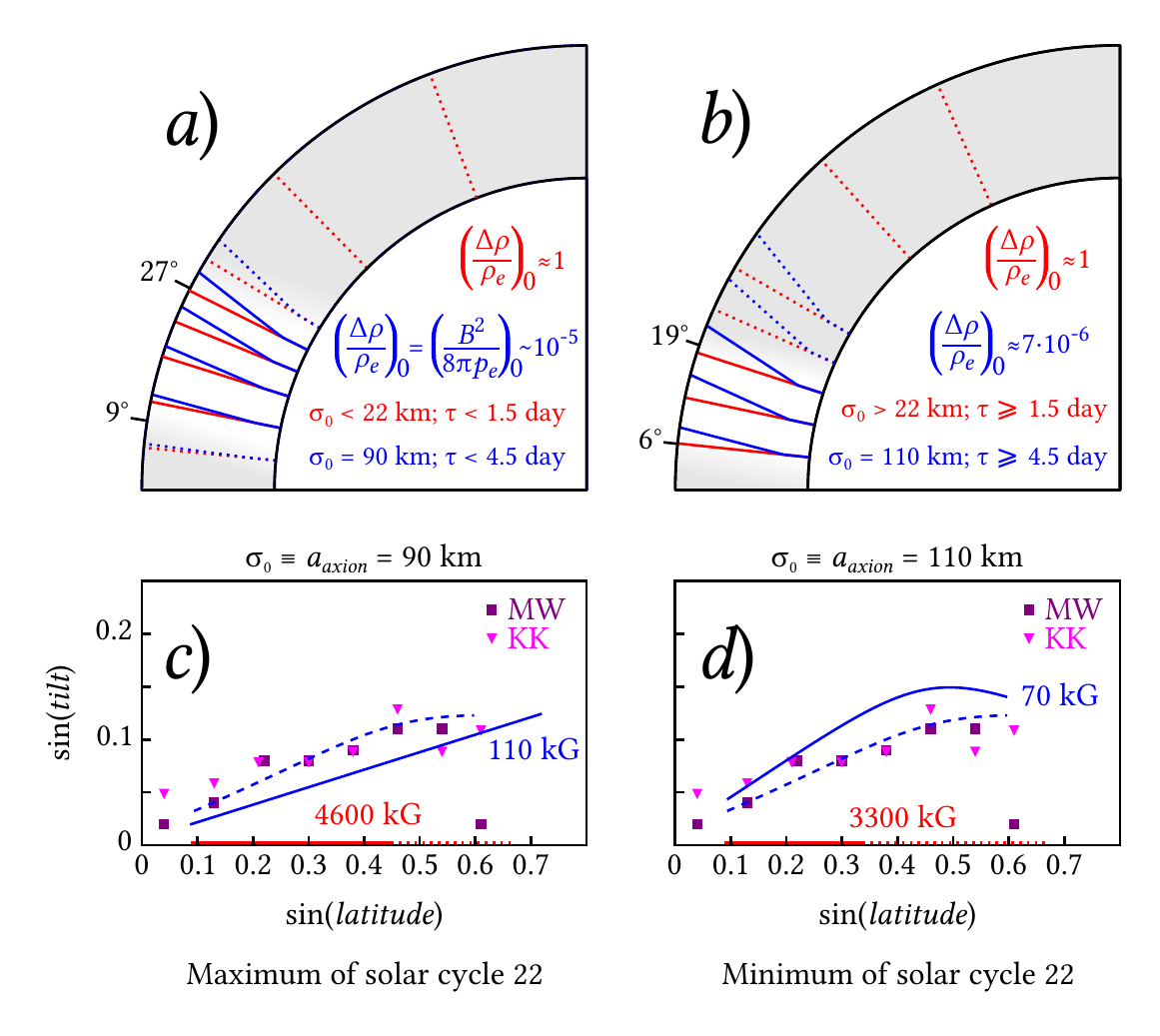}
\end{center}
\caption{\textbf{(a)}-\textbf{(b)} Flux tubes without drag of adiabatic flux ring in the superadiabatic convection zone (red lines: based on the \cite{Choudhuri1987} and \cite{Choudhuri1989} MEQ model (see also (10) in \cite{DSilvaChoudhuri1993}), as well as on the vanBFF MEQ model (see Sect.~\ref{sec-tilt})), and the trajectories of flux rings in thermal equilibrium (TEQ) incorporating drag (blue lines: based on the TEQ model of \cite{Choudhuri1987} and \cite{Choudhuri1989}; see also (5)-(10) in \cite{DSilvaChoudhuri1993}). In (a) we use the values for the maximum of the magnetic cycle (red lines) of $(\Delta \rho / \rho_{ext})_0 = B_0^2 / 8 \pi p_{ext,0} \approx 1$ and $B_0 \approx 4.6 \cdot 10^7 ~G$ for the $\Omega$-loop with a sunspot and photons of axion origin (see Figs.~\ref{fig-lampochka}a and~\ref{fig-lower-reconnection}a) and (blue lines) of $(\Delta \rho / \rho_{ext})_0 \approx 10^{-5}$ and $B_0 \approx 1.1 \cdot 10^5 ~G$ for the $\Omega$-loop with a sunspot and without photons of axion origin (see Fig.~\ref{fig-lower-reconnection}b). In (b) the values correspond to the minimum of the magnetic cycle: (red lines)  $(\Delta \rho / \rho_{ext})_0 \approx 1$ and $B_0 \approx 3.3 \cdot 10^7 ~G$ for the $\Omega$-loop with a sunspot and photons of axion origin (see Fig.~\ref{fig-lower-reconnection}a) and (blue lines) $(\Delta \rho / \rho_{ext})_0 \approx 2 \cdot 10^{-6}$ and $B_0 \approx 10^7 ~G$ for the $\Omega$-loop with a sunspot and without photons of axion origin (see Fig.~\ref{fig-lower-reconnection}b). In (a) and (b) we use red and blue dotted lines of trajectories that depend on large (see Eqs.~(17)-(19) in \cite{Choudhuri1987}, where the diffusion coefficient depends on the density and $k_R$) and small values of the Rosseland mean opacity $k_R$ (see (\ref{eq06v2-02}) and Fig.~\ref{fig-opacity}). \textbf{(c)}-\textbf{(d)} Dependence of $\sin (tilt)$ on $\sin (latitude)$ in different theoretical and experimental data series. At an average value of $4.1 \cdot 10^7 ~G$ in the tachocline, the maximum ($\sim 4.6 \cdot 10^7 ~G$) and minimum ($\sim 3.6 \cdot 10^7 ~G$) of the magnetic field of flux tubes are predetermined by two bound estimates, for example, the observational data on the variations of the magnetic field of tubes on the solar surface 
(see \cite{Pevtsov2011,Pevtsov2014}) and the theoretical estimates of magnetic variations in the tachocline with the EN~effect (see Eqs.~(\ref{eq06-08}) and~(\ref{eq06-14})). At an average value of $\sim 10^5 ~G$ near the tachocline, the maximum ($\sim 1.1 \cdot 10^5 ~G$) and minimum ($\sim 7.0 \cdot 10^4 ~G$) of the magnetic field of flux tubes are predetermined by two analogous bound estimates. The dashed lines show the linear regressions of the average slopes of sunspot groups in the latitude range of five degrees for Mount Wilson (MW) and Kodaikanal (KK) observatories. The solid blue lines show theoretical calculations of Joy's law (see \cite{Ivanov2012} and more explanation in the text).}
\label{fig-magtube-tilt}
\end{figure}

Based on the MEQ model of \cite{Choudhuri1987} and vanBFF MEQ model (see Sect.~\ref{sec-radiative-heating}), we showed that the results of the simulation of flux tube trajectories without adiabatic ring flux drag in the super-adiabatic zone (see Fig.~\ref{fig-lower-heating}a) are well represented in Fig.~\ref{fig-magtube-tilt}a,b (red lines). The most intriguing results of the simulation of flux tube trajectories are a very strong basis not only for understanding the complex physics, but also for understanding the theoretical estimates of buoyant MFTs, like the rise time and speed of rising to the Sun surface (see (\ref{eq07-60})-(\ref{eq07-61}), and also analogous Fig.~1 and Fig.~5 in \cite{Browning2016}), and the explanation of the tendency of the tilt angle of Joy's law (see Fig.~\ref{fig-magtube-tilt}c,d), which do not contradict the known experimental data, for example, the well-known works of
\cite{DasiEspuig2010,DasiEspuig2013,Ivanov2012,McClintock2013,Pevtsov2014,
Tlatova2015,Pavai2015,Baranyi2015,Wang2015,Wang2017}.

A brief explanation of the complex physics is as follows. It is known that the mechanical equilibrium (MEQ) of the flux tube is characterized by neutral buoyancy and a balance of curvature and Coriolis forces due to azimuthal flux along the tube. Indeed, an $\Omega$-loop with magnetic field of the order of $10^7 ~G$ is an ascending development of the azimuthal stream trapped under the bottom of the convective zone. This is because the sharp reduction of pressure in the lower layers of the considered $\Omega$-loop produces a gas that not only enhances the concentration of the azimuthal flux of magnetic bundles, but also promotes the buoyancy of the magnetic tube. This means that the magnetic buoyancy of the tube is not only associated with azimuthal flux bundles that are in magnetostatic equilibrium along its entire length, but also is a very important consequence of the thermomagnetic EN~effect creating strong toroidal flux tubes on the overshoot tachocline. It is not surprising, but this also means that the solar magnetic fields are generated not by the dynamo action in the lower part of the convective zone, but with the help of the thermomagnetic EN~effect in the tachocline (see Fig.~\ref{fig-solar-dynamos}b). 

Hence, the question arises as to how and when the large-scale dynamos are born in the flux tubes. It is known that turbulence is the defining feature of magnetized plasma in the cosmic, astrophysical and, of course, solar environments, which are almost always characterized by very large magnetic Reynolds numbers (see e.g. \cite{Loureiro2017}). In this case the fast dynamo problem is related to the generation of magnetic fields in the limit of the infinite magnetic Reynolds number $Re_{m} \rightarrow \infty$, and has been extensively studied, for example, by \cite{Childress1995}. Among the very well known models of turbulent dynamo, we single out a very rare but remarkable idea of the dynamo fluctuation model 
(see \cite{Blackman1996,Archontis2003b,Archontis2003a,Baggaley2009,Baggaley2010}) in which the magnetic field is condensed by thin ropes caused by the multiscale model of turbulence (see \cite{Baggaley2009}). This model may be considered as an implementation of the asymptotic limit $Re_{m} \rightarrow \infty$ for the continuous magnetic field, where the magnetic dissipation, which is predetermined by a very sparse, hot plasma and high Alfv\'{e}n velocity, is strongly localized in small regions of strong field gradients, for example, in the lower layers of MFTs. This means that in the initially homogeneous plasma $\beta_{average} \equiv (p_{int} / p_{mag}) \ll 1$, the nonlinear Alfv\'{e}n compression waves split the material on the energy scale containing the turbulent vortices, with density increase of the order of $1 / \beta$, where $p_{int}$ and $p_{mag}$ are the average internal gas and magnetic pressures. It is noteworthy that the balance between the Lorentz work (e.g. in the virtually empty tube) and the Joule dissipation occurs mainly from small regions where strong magnetic flux structures are concentrated (see \cite{Archontis2003b}). Hence, it becomes clear that since the Alfv\'{e}n velocity associated with the rarefied regions is large, as a consequence, the magnetic dissipation occurs only by reconnection in the lower and/or upper layers of the flux tubes (see Fig.~\ref{fig-upper-reconnection}), namely with the help of the large-scale dynamo or the so-called reconnection dynamo (reconnecting flux-rope dynamo \citep{Baggaley2009}).

Here the question arises as to what is the purpose of the so-called reconnecting dynamo. In this context, we briefly consider the injection of the magnetic and turbulent energy at some large scales inside a magnetic tube. As a consequence, the energy density on a large scale is cascaded down to smaller scales through the Kolmogorov inertial range (see \cite{Kolmogorov1941} and e.g. \cite{Goldreich1995}) and, therefore, at different scales leads to turbulent vortices that mostly move in the direction, corresponding to the local magnetic field, than in the perpendicular direction \citep{Goldreich1995}. It is very important that in the lower part of the cascade there is a dissipation range, where the gradients in the flow are large enough for effective dissipation, in which the only mechanism of magnetic dissipation is the reconnection of magnetic lines carried out in the straightforward manner. Thus, the reconnection more effectively transforms the magnetic energy of the plasma flow into the thermal one, in our case by the dynamo action, which is predetermined by the so-called reconnecting dynamo (see \cite{Baggaley2009}).

This means that, surprisingly, the reconnecting dynamo is never a source of magnetic field generation, except for the magnetic reconnection source, which is a consequence of the mechanism of magnetic dissipation. The intriguing moment of the reconnecting dynamo is the fact that the dynamo never generates magnetic fields and, unlike magnetic diffusion (see \cite{Priest2000}), is sensitive only to the nature of magnetic dissipation, the reconnection of which in the final stage releases the heat energy from the flux tube to e.g. the solar corona. This is the goal of forming a large-scale dynamo, which by means of reconnection organizes the volatilization of a free O-loop with the energy of heating (and without sunspots, see Figs.~\ref{fig-lower-reconnection}d and~\ref{fig-lower-reconnection2}) from the surface of the Sun to the corona. The known experiments suggest that, according to \cite{Baggaley2009,Baggaley2010}, the theoretical estimates of the probability distribution function of the energy density released in the reconnecting dynamo have a power-law form (see Fig.~27 in \cite{Baggaley2010}), which does not contradict the observed data, for example, heating the solar corona with nanoflares (see e.g. \cite{Charbonneau2001,Benz2008,Testa2014}).

Further, for evaluating other experimental data, for example, the Joy's law slope data 
(see e.g. \cite{DasiEspuig2010,Ivanov2012,McClintock2013,Pevtsov2014,
Tlatova2015,Pavai2015,Isik2015,Karak2017}), we are interested in the second stage of modeling the trajectory of flux tubes (see Fig.~\ref{fig-magtube-tilt}), when a magnetic tube that reaches the solar surface above the photosphere changes its shape and structure with the topological effect of magnetic reconnection in the lower layers of the magnetic tube, see Figs.~\ref{fig-lower-reconnection}b and~\ref{fig-lower-reconnection2}. Moreover, we believe that the secondary magnetic reconnection of the flux tubes leads to the real decrease in the magnetic field to $B \sim 10^5 ~G$ at $\sim 0.8~R_{Sun}$ (see Fig.~\ref{fig-Bz-diffusivity}), at which the evolution of the tubes (in thermal equilibrium (TEQ) with the surroundings it is often referred to as magnetic buoyancy \citep{Parker1975}) is controlled by the latitudinal pressure gradient in magnetic layers on the overshoot tachocline that allows a balance between nonzero buoyancy force, curvature force and pressure force in the absence of azimuthal flow (see bottom panel in Fig.~2 in \cite{Schussler2002}; see also review and Fig.~5b in \cite{Fan2009}), which generally allows a balance of TEQ between four main forces: nonzero buoyancy, magnetic tension, aerodynamic drag, and Coriolis force.

\begin{figure}[tbp]
\begin{center}
\includegraphics[width=16cm]{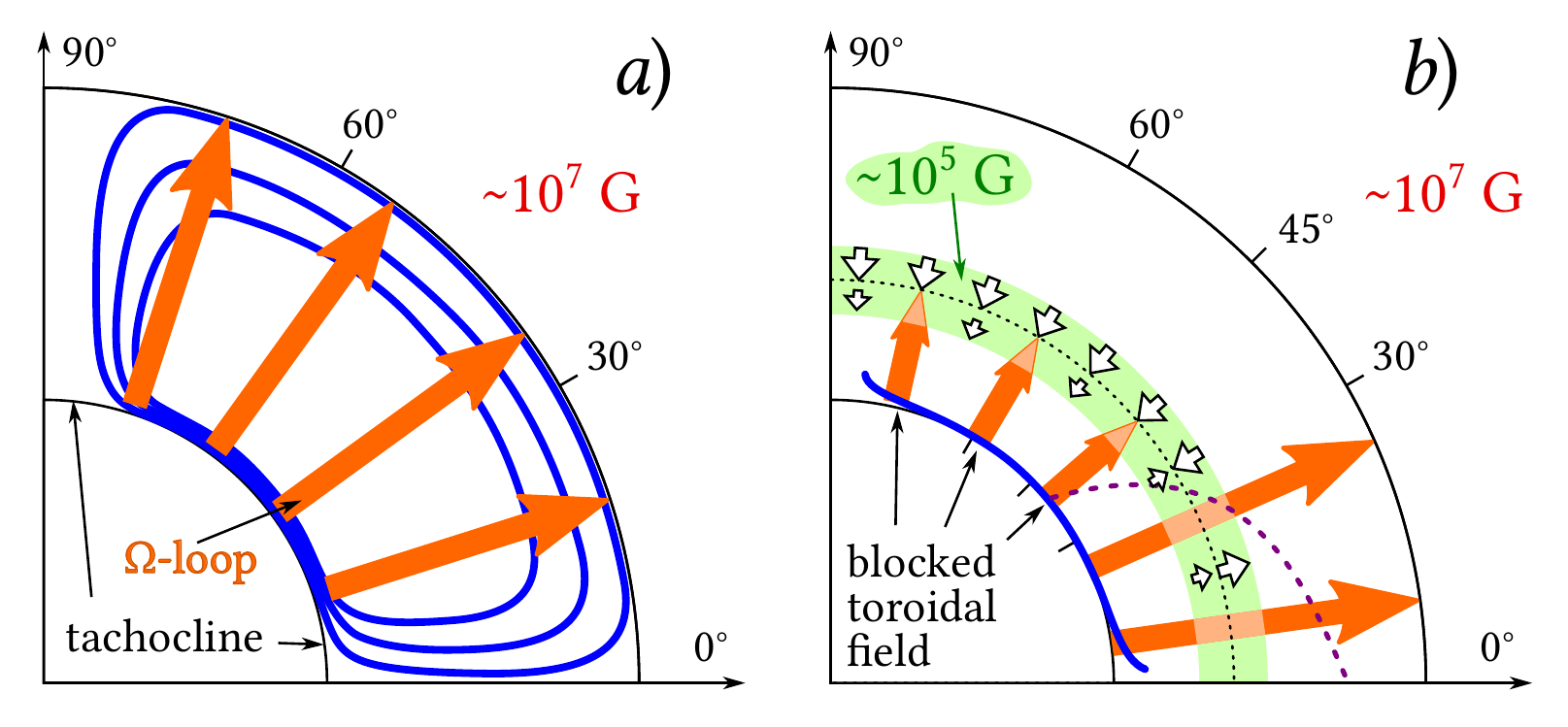}
\end{center}
\caption{Scheme of turbulent reconstruction of the toroidal magnetic field in the convective zone: \textbf{(a)} meridional circulation (closed blue poloidal field lines), which is generated by the toroidal field in the tachocline (thin black lines), and the magnetic buoyancy of the flux tubes (red arrows) with magnetic field $\sim 10^7 ~G$;
\textbf{(b)} the joint interactions of magnetic buoyancy (red arrows) and the rotating magnetic $\nabla \rho$-pumping (short white arrows) generate the total buoyancy of magnetic tubes in which the toroidal magnetic field $\sim 10^5 ~G$ predetermines the appearance of magnetic buoyancy on the surface of the Sun only at low, and to a lesser extent at middle latitudes (see analogous Fig.~5 in \cite{Krivodubskij2005}).}
\label{fig-meridional-cut}
\end{figure}

\begin{figure}[tbp]
\begin{center}
\includegraphics[width=16cm]{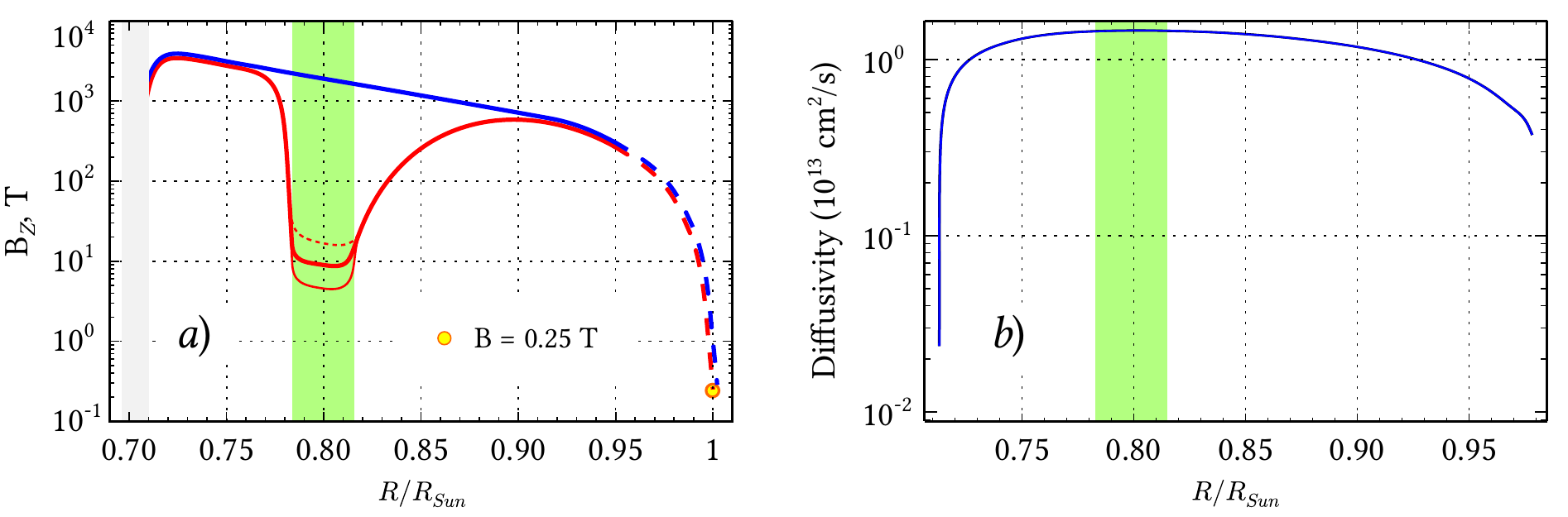}
\end{center}
\caption{\textbf{(a)} Change in the magnetic field strength $B_z$ along the rising $\Omega$-loop as a function of the Sun depth $R/R_{Sun}$ in the convective zone. The blue line (see also Fig.~\ref{fig-Bz}) denotes admissible values ​​for the standard solar model with diffusion of helium \citep{Bahcall1992}, with the initial value of the theoretical estimate of the magnetic field $B_z \approx  4 \cdot 10^7 ~G$ (see Eq.~(\ref{eq06-16})). The red line
corresponds to the cool areas inside the magnetic tube, which by means of the reconnection of the $\Omega$-loop keeps the emerging magnetic diffusivity (green band). \textbf{(b)} Radial profile of turbulent magnetic diffusivity in the solar convection zone based on the \cite{Stix1990} model (see also ~\cite{Stix2004}). Diamagnetic pumping must be very strong near the base of the convective zone, where diffusivity almost jumps by orders of magnitude. Its gradient is the rate of descending diamagnetic pumping (see the beginning of the pumping (green band)).}
\label{fig-Bz-diffusivity}
\end{figure}

Despite our complicated theoretical calculations of the trajectories (Fig.~\ref{fig-magtube-tilt}a,b) of the flux tubes without drag of adiabatic flux ring in the superadiabatic convection zone (red lines: based on the MEQ model by \cite{Choudhuri1987} and the vanBFF MEQ model (see Sect.~\ref{sec-radiative-heating})) and involving the flux ring resistance in thermal equilibrium (blue lines: based on the TEQ model by \cite{Choudhuri1987}), below we will describe the scenario, which explains the simple physics of why the magnetic tubes emerging on the solar surface, can only be at low, and to a lesser extent at middle latitudes (see Figs.~\ref{fig-magtube-tilt}a,b and~\ref{fig-meridional-cut}b).

The first part of the scenario consists in discussing the physics of magnetic flux buoyancy and, as a consequence, estimating the rise speed of an almost empty magnetic tube, $v_{rise}$, based on the vanBFF model (see Sect.~\ref{sec-radiative-heating}), which simultaneously coincides with the known expressions for the buoyancy speed, $v_{B}$ (see e.g. \cite{Parker1975}; see also Eq.~(17) and Fig.~2 in \cite{Ballegooijen1988}; Eq.~(34) in \cite{Fan1993}; Eq.~(30) in \cite{Khaibrakhmanov2017}):

\begin{equation}
(v_{rise})_{axion} \equiv (v_{B})_{axion} \approx v_A 
\left( \frac{\pi}{C_D} \frac{a_{axion}}{H_p} \frac{z}{H_p} \right) ^{1/2} ,
\label{eq07-63}
\end{equation}

\noindent where $v_A \equiv B / (4 \pi \rho_{int})^{1/2}$ is the Alfv\'{e}n speed of the magnetic field (see e.g. \cite{Roberts2001}), $C_D \approx 1$ is the drag coefficient, $\rho_{int}$ is the internal density of the gas in the MFT. It is assumed that the magnetic tube is formed inside the disk at an altitude $z / H_p \approx 1$.

For the neutral buoyancy condition ($\rho_{int} = \rho_{ext} \approx 0.2 ~g/cm^3$; see Figs.~\ref{fig-Bz}a and~\ref{fig-lower-heating}a) and the strong toroidal field of the magnetic tube, $B_{tacho}^{Sun} = 4.1 \cdot 10^7 ~G$, as well as the transverse radius of the ``thin'' ring $a_{axion} \sim 3.7 \cdot 10^{-4} ~H_p$ (see Eq.~(\ref{eq07-58})) between the O-loop and the walls of the magnetic tube (see Fig.~\ref{fig-axion-compton}a), it is not difficult to show that the estimate of the Alfv\'{e}n speed (see Eq.~(\ref{eq07-63}))

\begin{equation}
v_A \cong 4.1 \cdot 10^6 ~cm/s
\label{eq07-64}
\end{equation}

\noindent allows us to estimate the analytical coincidence of the rise speed $v_{rise}$ (see Eq.~(\ref{eq07-61})) and the magnetic buoyancy speed $v_{B}$:

\begin{equation}
(v_{rise})_{axion} \equiv (v_{B})_{axion} \approx 1.4 \cdot 10^5 ~cm/s ,
\label{eq07-65}
\end{equation}

\noindent
at which for such large magnetic fields there is a significant number of rising tubes at all latitudes (see also the red lines in Figs.~\ref{fig-meridional-cut}a and~\ref{fig-magtube-tilt}a,b).

So, for the considered case, the secondary magnetic reconnection of the flux tubes leads to the real decrease in the magnetic field to $B \sim 10^5 ~G$ at $\sim 0.8 R_{Sun}$ (see Figs.~\ref{fig-magtube-tilt}a,b and~\ref{fig-Bz-diffusivity}a), and thereby reveals the nonzero magnetic buoyancy (see the blue lines in Fig.~\ref{fig-magtube-tilt}a,b). This means that for the condition of nonzero buoyancy ($\rho _{ext} \approx 0.09 ~g/cm^3$) and the toroidal magnetic field of the flux tube, as well as the transverse radius of the ``thin'' ring $\sigma _0 \equiv a_{axion} \sim 100 ~km \approx 4.5 \times 3.7 \cdot 10^{-4} H_p $ (see Fig.~\ref{fig-magtube-tilt}c,d), the estimate

\begin{equation}
v_A \cong 1.1 \cdot 10^4 ~cm/s
\label{eq07-66}
\end{equation}

\noindent
allows us to estimate the analytical coincidence of the rise speed $v_{rise}$ (see Eq.~(\ref{eq07-61})) and the magnetic buoyancy speed $v_{B}$:

\begin{equation}
(v_{rise})_{axion} \equiv (v_{B})_{axion} \approx 7.2 \cdot 10^2 ~cm/s ,
\label{eq07-67}
\end{equation}

\noindent
where we assume that MFTs are formed inside the disk at a height $z \sim H_p$ (see e.g. \cite{Khaibrakhmanov2017}). Below we consider the reason for the rising magnetic buoyancy
only at low and middle latitudes.

The second part of the scenario, which is based on the remarkable idea of \cite{Kichatinov1991} (see also Fig.~5 in \cite{Krivodubskij2005}), includes the generation of the magnetic field near the bottom of the convective zone (see Fig.~\ref{fig-solar-dynamos}b for the holographic BL~mechanism) and transfer of the toroidal field from the deep layers to the solar surface, where the efficiency of the magnetic buoyancy transfer is predetermined by the participation of two processes: macroscopic turbulent diamagnetism 
(see \cite{Zeldovich1957,Radler1968a,Radler1968b,Vainshtein1980,
VainshteinKichatinov1983,Stix1989,Kichatinov1992,Kitchatinov2016}) and rotational $\nabla \rho$-pumping 
(see \cite{Drobyshevski1974,Vainshtein1983,VainshteinKichatinov1983,
Stix1989,Kichatinov1991,Ossendrijver2002,Kitchatinov2012}), which are also associated with the process of meridional circulation 
(see e.g. \cite{Ballegooijen1982,Spruit1982,Dudorov1985,Ballegooijen1988,
Wang1991,Choudhuri1995,Caligari1995,Nandy2002} and also \cite{Khaibrakhmanov2017}).

Let us note some important properties of macroscopic turbulent diamagnetism. It is known that \cite{Zeldovich1956,Zeldovich1957} and \cite{Spitzer1956} discovered the diamagnetism of inhomogeneously turbulent conducting liquids, in which the inhomogeneous magnetic field moves as a single whole. In this case the turbulent fluid, for example, with nonuniform effective diffusivity $\eta _T \approx (1/3) v l$ (see Fig.~1 in ~\cite{Kitchatinov2008}; see also Fig.~\ref{fig-Bz-diffusivity}b) behaves like a diamagnetic one and carries the magnetic field with the effective velocity

\begin{equation}
\vec{v}_{dia} = - \frac{1}{2} \nabla \eta_T\, ,
\label{eq07-68}
\end{equation}

\noindent
where $l$ is the mixing length of turbulent pulsations, and $v = \sqrt{\langle v^2 \rangle}$ is the root-mean-square velocity of turbulent motion. The minus sign on the right in Eq.~(\ref{eq07-68}) shows the meaning of turbulent magnetism: it is not paramagnetic magnetism, so magnetic fields repel from regions with relatively high turbulent intensity. In other words, macroscopic turbulent plasma diamagnetism and, as a consequence, the so-called macroscopic diamagnetic effect (see \cite{Radler1968a,Radler1968b}) in the physical sense is the displacement of the averaged magnetic field $B$ from regions with increased intensity of turbulent pulsations to regions with less developed turbulence \citep{Vainshtein1980,Krause1980}.

However, there is an interesting problem of the diamagnetic process caused by inhomogeneous turbulent intensity with allowance for the total nonlinearities in the magnetic field. This is due to the fact that up to the present time analytical estimates have been obtained only for the limiting cases of weak and strong magnetic fields. For example, at strong magnetic fields of flux tubes, at $B \gg B_{eq}$, the diamagnetic effect becomes almost negligible, in particular, strong magnetic damping of diamagnetism $\sim B^{-3}$ is obtained for super-equipartitions of fields \citep{Kichatinov1992}, when turbulence is close to two-dimensional \citep{Zeldovich1957}. On the other hand, for very weak fields the diamagnetic pumping, which is predetermined by the intensity of turbulence at $B \ll B_{eq}$, is a very effective process \citep{Kichatinov1992}.

Between the known limiting cases of weak and strong magnetic fields we are interested in the average toroidal magnetic field of a flux tube, that is $B _{tacho}^{Sun} \sim 10^5 ~G$, when $B \geqslant B_{eq}$. This is due to the fact that the secondary magnetic reconnection of the flux tubes (see Figs.~\ref{fig-upper-reconnection}b,d,f,g) leads to the real decrease of the magnetic field to $B \sim 10^5 ~G$ at $\sim 0.8~R_{Sun}$ (see the blue lines in Fig.~\ref{fig-magtube-tilt}a,b). This means that the real decrease in the toroidal magnetic field of the flux tube is a consequence of the formation of the reconnecting dynamo (see \cite{Baggaley2009}), as well as the important appearance of two ``anti-buoyancy'' effects: the downwardly directed turbulent diamagnetic transfer and the rotational effect of the magnetic $\nabla \rho$-pumping 
\citep{VainshteinKichatinov1983,Krivodubskij2005}.

In this regard, we consider turbulence with quasi-isotropic spectral tensor \citep{Kichatinov1987}, which is certainly the simplest representation for inhomogeneous turbulence (see Eq.~(2.12) in \cite{Kichatinov1992}. As a consequence, the information on the spectral properties of turbulence (given by Eqs.~(2.12)-(2.16) from \cite{Kichatinov1992}) is sufficient to reduce the expression for the average electromotive force $\varepsilon$ (see Eq.~(2.1) in \cite{Kichatinov1992}) to its traditional form, where only integrations over the wave number $k$ and frequency $\omega$ remain. After such shortening it is possible to get (see \cite{Kichatinov1992})

\begin{equation}
\vec{F} = (\vec{v}_{dia} + \vec{v}_{dens}) \times \vec{B}
\label{eq07-69}
\end{equation}

\noindent
with the speed of turbulent diamagnetic transfer

\begin{equation}
\vec{v}_{dia} = -\nabla \int \limits _{0} ^{\infty} \Re _{dia} (k, \omega, B)
\frac{\eta k^2 q (k, \omega, x)}{\omega ^2 + \eta ^2 k^4} dk d\omega\, ,
\label{eq07-70}
\end{equation}

\noindent
where $q$ stands for the local velocity spectrum, and the rate of rotational magnetic advection caused by the vertical heterogeneity of the fluid density in the convective zone, i.e. the magnetic $\nabla \rho$-pumping effect,

\begin{equation}
\vec{v}_{dens} = \frac{\nabla \rho}{\rho} \int \limits _{0} ^{\infty} \Re _{dens} (k, \omega, B)
\frac{\eta k^2 q (k, \omega, x)}{\omega ^2 + \eta ^2 k^4} dk d\omega\, .
\label{eq07-71}
\end{equation}

\noindent
The effective speeds $v_{dens}$ and $v_{dia}$ are consequences of the non-uniformity of density and of turbulence intensity, respectively, where the latter is attributed to the known diamagnetic pumping.

We are interested in the problem of reconstructing a strong toroidal field $\sim 10^7 ~G$ of flux tubes (see Figs.~\ref{fig-meridional-cut}a and~\ref{fig-Bz-diffusivity}a), which by the reconnecting dynamo \citep{Baggaley2009} transform regions of the mean magnetic field $\sim 10^5 ~G$ in the convective zone (see Figs.~\ref{fig-meridional-cut}b and ~\ref{fig-Bz-diffusivity}a) and, thereby, allow the organization of the amazing balance between the magnetic buoyancy, turbulent diamagnetism, and the rotationally modified $\nabla \rho$-effect. It can be shown that for the parameters $\beta = B/B_{eq} \sim 1$ and $\cos \varphi$ (see Eq.~(3.5) in \cite{Kichatinov1992}), the speeds~(\ref{eq07-70}) and~(\ref{eq07-71}), which depend on the magnetic field $B$ through the kernels $\Re _{dens}(\beta, \varphi)$ and $\Re _{dia}(\beta, \varphi)$ (see Eqs.~(3.6) and~(3.7) in \cite{Kichatinov1992}), have the following estimates:

\begin{equation}
v_{dia} ^{black} \sim 1.3 \cdot 10 ~cm/s
\label{eq07-72}
\end{equation}

\noindent and

\begin{equation}
v_{dens} ^{light} \sim 7.2 \cdot 10^2 ~cm/s .
\label{eq07-73}
\end{equation}

This raises the question of how a certain balance appears between the speeds of magnetic buoyancy (see Eq.~(\ref{eq07-67}) and Fig.~\ref{fig-meridional-cut}), diamagnetic pumping (see Eq.~(\ref{eq07-72})), and rotating densely stratified $\nabla \rho$-pumping (see Eq.~(\ref{eq07-73}) and Fig.~\ref{fig-meridional-cut}b),

\begin{equation}
(v_B ^{red})_{axion} + v_{dia}^{black} + v_{dens}^{light} \cong 
(v_B ^{red})_{axion} + v_{dens}^{light} = ?
\label{eq07-74}
\end{equation}

In order to consider the balance Eq.~(\ref{eq07-74}), it is necessary to apply the widely used approximation of the mixing length (see e.g. \cite{BohmVitense1958,Bradshaw1974,Gough1977a,Gough1977b,Barker2014,Brandenburg2016}), which, according to \cite{Kichatinov1991}, fully satisfies this goal. This approximation will be understood as the replacement of nonlinear terms along with time derivatives in the equations for fluctuating fields by means of $\tau$-relaxation terms, i.e. instead of equations (3.1) and (3.9) from \cite{Kichatinov1991}, we now have the equation of the radial speed of the toroidal field in the convection zone

\begin{equation}
v_{dens}^{light} = \tau \langle u^2 \rangle ^{\circ}
\frac{\nabla \rho}{\rho} \left[ \phi_2 (\hat{\Omega}) - \cos ^2 \theta \cdot 
\phi_1 (\hat{\Omega}) \right] \approx
\label{eq07-75}
\end{equation}

\begin{equation}
\approx 6 v_p \left[ \phi_2 (\hat{\Omega}) - \cos ^2 \theta \cdot 
\phi_1 (\hat{\Omega}) \right] =
\label{eq07-76}
\end{equation}

\begin{equation}
= - \frac{3 \kappa g}{(\gamma - 1)c_p T}
\left[ \phi_2 (\hat{\Omega}) - \cos ^2 \theta \cdot \phi_1 (\hat{\Omega}) \right]\, ,
\label{eq07-77}
\end{equation}

\noindent
where $\tau \approx l / (\langle u^2 \rangle ^{\circ})^{1/2}$ is a typical lifetime of a convective eddy; $l$ is the mixing length; $\langle u^2 \rangle ^{\circ}$ is the mean intensity of fluctuating velocities for original turbulence; $\theta$ is the latitude; $e_r \nabla \rho / \rho = -e_r g / [(\gamma - 1) c_p T]$ , where $e_r$ is the radial unit vector; $T \cong 1.35 \cdot 10^6 ~K$ is the temperature at $0.8 R_{Sun}$; $g = g_0 (R_{Sun}/r)$ is the gravitational acceleration, where $g_0 = 2.74 \cdot 10^4 ~cm/s^2$ is the surface value; $c_p = 3.4 \cdot 10^8 ~cm^2 s^{-2} K^{-1}$ (fully ionized hydrogen) is the specific heat at constant pressure; $\gamma = 5/3$ is the ratio of specific heats $c_p / c_V$; $3 \kappa = \tau \langle u^2 \rangle ^{\circ}$, where $\kappa \equiv \eta _T \approx 5 \cdot 10^{13} ~cm^2 \cdot s^{-1}$ is turbulent diffusivity supplied by the model of non-rotating convection zone 
\cite[][see also Fig.~1 in \cite{Kitchatinov2008}]{Spruit1974,Gough1976,
Stix1990,Parker2009,Karak2014} and the mixing length relation $\langle u^2 \rangle ^{\circ} = - \nabla \Delta T l^2 g / (4 T)$, where the $\nabla \Delta T$ is superadiabatic temperature gradient; $v_p = (1/6) \tau \langle u^2 \rangle ^{\circ} (\nabla \rho / \rho)$ is the velocity of the magnetic field transfer caused by the density gradient (see Eq.~(36) in \cite{VainshteinKichatinov1983}); $\hat{\Omega} = Co = 2 \tau \Omega$ is the Coriolis number (reciprocal of the Rossby number), where $\Omega$ is the rotation speed, $\tau$ is the turnover time, and the functions

\begin{equation}
\phi _n (\hat{\Omega}) = (1/8) I_n (\Omega, k, \omega)
\label{eq07-78}
\end{equation}

\noindent
(see $I_1$ and $I_2$ in Eqs.~(3.12) and~(3.21) in \cite{Kichatinov1991}) are

\begin{equation}
\phi _1 (\hat{\Omega}) = \frac{1}{4 \hat{\Omega}^2} \left[
-3 + \frac{\hat{\Omega}^2 + 1}{\hat{\Omega}} \arctan \hat{\Omega} \right]\, ,
\label{eq07-79}
\end{equation}

\begin{equation}
\phi _2 (\hat{\Omega}) = \frac{1}{8 \hat{\Omega}^2} \left[
1 + \frac{\hat{\Omega}^2 - 1}{\hat{\Omega}} \arctan \hat{\Omega} \right]
\label{eq07-80}
\end{equation}

\noindent
(see also analogous Eq.~(19) and Fig.~2 in \cite{Kitchatinov2016}), which describe the rotational effect on turbulent convection.

According to \cite{Kapyla2014}, the rotational effect on the flow can be measured with the local Coriolis number $\hat{\Omega} = Co = 2 \tau \Omega$, especially if $\tau$ is estimated on the basis of the theory of mixing length, which predicts values of $\hat{\Omega}$ reaching more than 10 in the deep layers (see e.g. \cite{Ossendrijver2003,Brandenburg2005,Kapyla2011,KapylaEtAl2011}). However, on the other hand, according to \cite{Kapyla2014}, the question of whether there are such deep layers of solar and stellar convection zones or not is still open.

Below we show that the estimate of the Coriolis number for solar convection in the deep layer should be $\hat{\Omega} \approx 20$. As a consequence of~(\ref{eq07-78})-(\ref{eq07-80}), the following values are assumed for this quantity:

\begin{equation}
\phi_1 \cong 0.0171 , ~~~ \phi_2 \cong 0.0098 ,
\label{eq07-81}
\end{equation}

\noindent
at which the radial velocity~(\ref{eq07-75}) of toroidal field transport changes the sign at the latitude $\theta ^{*} = \arccos \sqrt{\varphi_2 / \varphi_1} \cong 41^{\circ}$, being negative (downward) for $\theta > \theta ^{*}$ and positive (upward) for $\theta < \theta ^{*}$. Using~(\ref{eq07-77}), we find that the value of the radial velocity in the convective zone $v_{dens}^{light}$ (see Eqs.~(\ref{eq07-75})-(\ref{eq07-77})) near low latitudes (e.g. $\theta ^{*} = \arccos (0.985) \cong 10^{\circ}$;  see also Fig.~\ref{fig-meridional-cut}b) almost completely coincides with the value of the speed (\ref{eq07-73}), which was previously calculated with the help of the magnetic field $B$ in the kernel $\Re_{dens} (\beta, \varphi)$.

This means that the balance of the magnetic buoyancy (see Eq.~(\ref{eq07-67}) and Fig.~\ref{fig-meridional-cut}) and rotating density-stratified $\nabla \rho$-pumping (see Eq.~(\ref{eq07-73}) and Fig.~\ref{fig-meridional-cut}b) provides both the process of blocking the magnetic buoyancy at high latitudes,

\begin{equation}
\uparrow (v_B ^{red})_{axion} + \downarrow v_{dens}^{light} \leqslant 0
~~~ \text{at high latitudes} ,
\label{eq07-82}
\end{equation}

\noindent
and the process of lifting MFTs from the base of the convective zone to the solar surface at low latitudes,

\begin{equation}
\uparrow (v_B ^{red})_{axion} + \uparrow v_{dens}^{light} > 0
~~~ \text{at lower latitudes} ,
\label{eq07-83}
\end{equation}

\noindent
which are simultaneously predetermined by the following rise time value:

\begin{equation}
(\tau_B ^{red})_{axion} \sim \frac{z_0}{(v_B ^{red})_{axion} + v_{dens}^{light}}
\leqslant 4.5 ~~day,
\label{eq07-84}
\end{equation}

\noindent
where $z_0 \approx 0.1 ~H_p$ is the length of the magnetic line at $0.8 R_{Sun}$ (see the green band in Fig.~\ref{fig-Bz-diffusivity}). Here we must remember that magnetic $\nabla \rho$-pumping of plasma from the azimuthal field through the formation of the $\Omega$-loop is repeated again and again (see Figs.~\ref{fig-lower-reconnection} and \ref{fig-lower-reconnection2}; see also Fig.~4 in \cite{Parker2009}). Moreover, our theoretical estimates of the rise time of the $\Omega$-loop are in good agreement with the experimental observations by \cite{Gaizauskas1983}, who mention the repeated appearance of new $\Omega$-loops with an interval of 5-8 days, which simultaneously indicates the continuing convective pumping of plasma.

We have shown above that the sunspots -- the dark magnetic regions occurring at low latitudes on the surface of the Sun (see Fig.~\ref{fig-meridional-cut}) -- are indicators of the magnetic field generated not with the help of the dynamo mechanism (this is very important!), which does not exist here, but with the help of the holographic BL~mechanism as components of the model of solar antidynamo (see Fig.~\ref{fig-opacity}b). Hence, the question arises as to how the sunspots originating from magnetic buoyancy on the surface only at low and, to a lesser extent, middle latitudes do not contradict the known observational data of the tilt angle of Joy's law (see Fig.~\ref{fig-magtube-tilt}c,d).

The common notion about solar magnetic fields is that they are created by the dynamo action. Although the solar dynamo theory currently faces many serious difficulties (see e.g. \cite{Karak2014}), there has been no satisfactory alternative theory so far \citep{Cowling1981,Gilman1986,Choudhuri1989}. Avoiding an alternative theory, most physicists believe that it is necessary to study more deeply the known physical problems of various aspects of the theory of solar dynamo (see \cite{Karak2014}) before passing a verdict on it.

Conversely, if the question arises whether there is a satisfactory alternative theory against the action of the dynamo, then our short answer is ``yes''. We understand that a strong toroidal magnetic field and a tachocline are generated not by the dynamo action, but by the holographic BL~mechanism (see Fig.~\ref{fig-solar-dynamos}b), which is a consequence of the holographic principle of quantum gravity in the Universe and, therefore, in the Sun. This means that the holographic principle predetermines the formation of the fundamental properties of a nearly two-dimensional surface of the tachocline based on the three-dimensional volume of the Sun below the tachocline, and thus the formation of the thermomagnetic EN~effect 
\citep[see][and also Fig.~\ref{fig-lampochka}a]{Spitzer1962,Spitzer2006,Rusov2015}. Hence, it is easy to show that with the help of the thermomagnetic EN~effect a simple estimate of the magnitude of magnetic pressure of an ideal gas in the tachocline can indirectly prove that the repulsive toroidal magnetic field of the tachocline precisely ``neutralizes'' the magnetic field in the Sun core (see also Eq.~(\ref{eq06-16}) and \cite{Rusov2015}). This is the main and at the same time fundamental result of the existence of the holographic BL~mechanism, which is a consequence of the remarkable properties of the holographic principle of quantum gravity on the Sun.

\begin{figure}[tbp]
\begin{center}
\includegraphics[width=14cm]{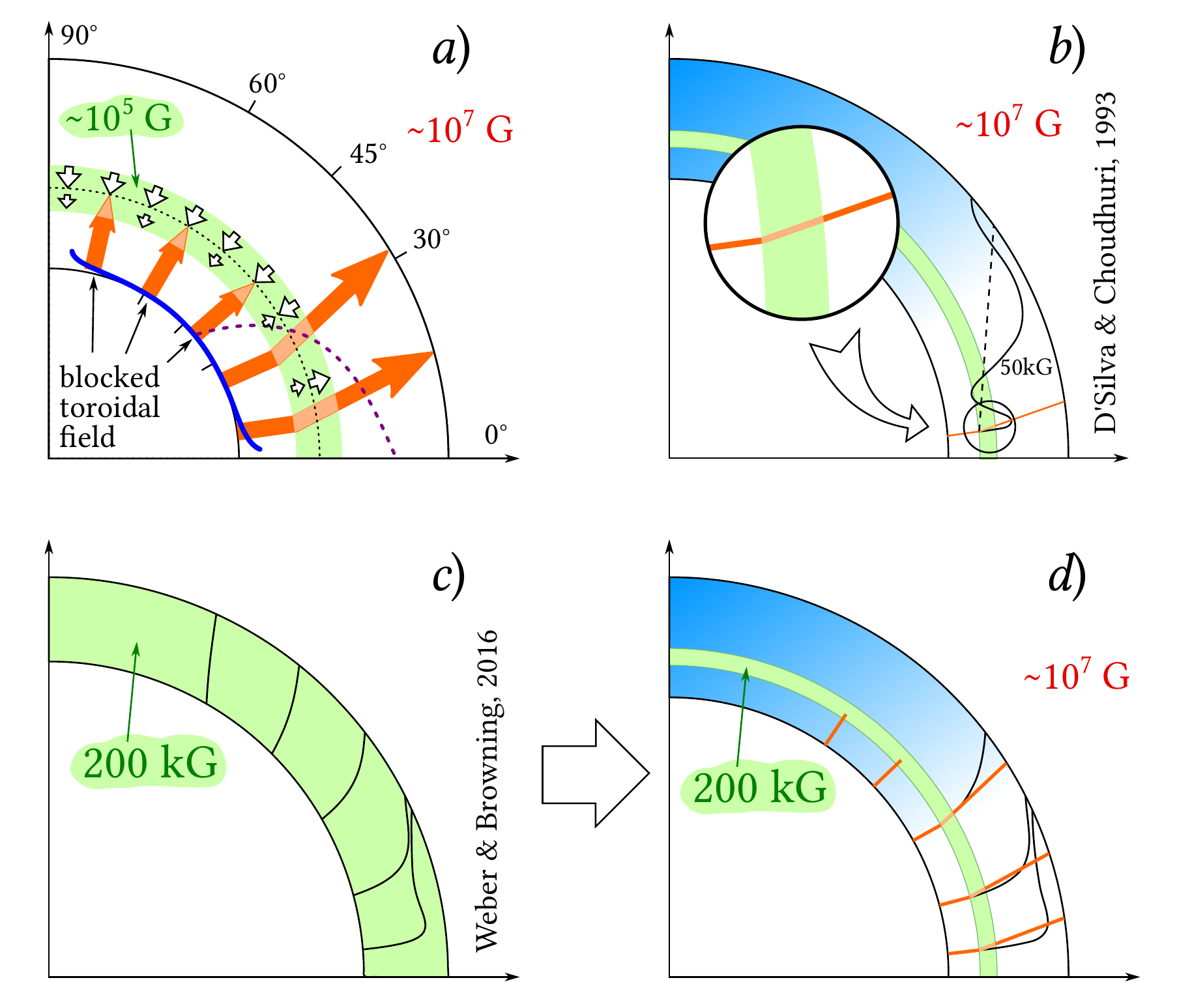}
\end{center}
\caption{\textbf{(a)} Interactions of magnetic buoyancy (red arrows) and rotating magnetic $\nabla \rho$-pumping (short white arrows) generate the total buoyancy of magnetic tubes in which the toroidal magnetic field $\sim 10^7 ~G$ predetermines the appearance of magnetic buoyancy by using the dominant Coriolis force in $\nabla \rho$-pumping with $\sim 10^5 ~G$ (see Fig.~\ref{fig-meridional-cut}b), which ultimately generates a curved upward loop on the surface of the Sun with a slope in the lower and, to a lesser extent, middle latitudes (see Fig.~\ref{fig-magtube-tilt}a,b). 
\textbf{(b)} According to \cite{DSilvaChoudhuri1993}, the trajectory of stream loops with 50~kG in the $\xi - \theta$ plane realized at latitude $5^{\circ}$ in the lower part of the convective zone. The dashed line shows the contour of constant angular momentum. A streaming ring of 50~kG, realized at $5^{\circ}$, also oscillates around this contour and comes out to a high latitude. The streaming ring ``hugs'' the contour of constant angular momentum, which is almost parallel to the axis of rotation. When we use Fig.~\ref{fig-meridional-cut-tilt}b, which is the modified Fig.~14 from \cite{DSilvaChoudhuri1993}, the inset shows the trajectory of the tubes with $\sim 10^7 ~G$, which, ultimately, by $\nabla \rho$-pumping with $\sim 5 \cdot 10^4 ~G$ (green line) generates a curved ascending loop on the solar surface with a tilt in the lower latitudes. At the same time, the flow ring ``hugs'' the contour of constant angular momentum, which is not parallel to the axis of rotation, which, of course, will be in good agreement with the observations of the slope angle of Joy's law (see Fig.~\ref{fig-magtube-tilt}c,d).
\textbf{(c)} According to \cite{Weber2016}, the trajectories of tubes initiated at $0.75 R_{Sun}$ (black lines) can cause (by means of induced Coriolis forces) tubes to move horizontally outwards to smaller layers at lower latitudes, and deeper layers at higher latitudes, caused by the direction of the flow backward.
\textbf{(d)} When we use Fig.~\ref{fig-meridional-cut-tilt}c, which is the modified Fig.~6 from \cite{Weber2016}, the trajectory of flux tubes with $\sim 10^7 ~G$ ultimately, by $\nabla \rho$-pumping with $\sim 5 \cdot 10^4 ~G$ (green line, see also Fig.~\ref{fig-meridional-cut-tilt}d) generates a curved ascending loop on the solar surface with a tilt in the lower latitudes.}
\label{fig-meridional-cut-tilt}
\end{figure}

Hence, it is very interesting that the holographic BL~mechanism, which is a satisfactory alternative theory against the action of the dynamo, predetermines not only the generation of sunspots themselves on the surface at low and middle latitudes (see Fig.~\ref{fig-meridional-cut}), but their coincidence with the observed slope angle of Joy's law, where, as a consequence, the average angle of inclination of bipolar sunspots increases with latitude (see Fig.~\ref{fig-magtube-tilt}c,d). This is due to the fact that strong toroidal magnetic fields in the overshoot tachocline are generated by the holographic BL~mechanism (see Fig.~\ref{fig-solar-dynamos}b), and bipolar magnetic tubes are created by lifting $\Omega$-loops caused by magnetic buoyancy. As a result, these ascending $\Omega$-loops, according to \cite{Choudhuri1989}, will be bent (towards the pole; see also Fig.~\ref{fig-meridional-cut-tilt}) by the Coriolis force, so that they eventually appear on the surface of the Sun with the tilt angle (see Fig.~\ref{fig-magtube-tilt}).

Using the magnetic field strength near the bottom of the convective zone of the order of $10^5 ~G$ at $0.8 R_{Sun}$ (see Figs.~\ref{fig-meridional-cut}b and \ref{fig-Bz-diffusivity}a), we find that the Coriolis force plays a dominant role, and the MFTs, starting from the bottom at low latitudes, deviate by $0.8 R_{Sun}$ (see Figs.~\ref{fig-magtube-tilt}a,b and \ref{fig-meridional-cut-tilt}) and appear on the surface of the Sun at low and middle latitudes, located in the direction of the poles as sunspots. It is obvious that since the time of radiation diffusion of the flux tube (see Fig.~\ref{fig-magtube-tilt}a) is related to Eq.~(\ref{eq07-84}),

\begin{equation}
(\tau_B ^{red})_{axion} \equiv (\tau_d)_{axion} \leqslant 4.5 ~~day,
\label{eq07-85}
\end{equation}

\noindent
using the equation of the universal model (\ref{eq07-48}),

\begin{equation}
(\tau_d)_{axion} \approx \tau_d \cdot a_{axion}^2 / a^2\, ,
\label{eq07-86}
\end{equation}

\noindent
it is possible to estimate the radius of the cross-section of the ring of the MFT:

\begin{equation}
a_{axion} \sim 100 ~km .
\label{eq07-87}
\end{equation}

Here we are interested in the relationship between the speed of magnetic buoyancy, $(v_B^{red})_{axion}$, and the radius of the cross-section of the ring, $\sigma_0$, which is identical to $a_{axion}$ based on the vanBFF model (see Sect.~\ref{sec-radiative-heating}). Using Eq.~(23) from \cite{Choudhuri1987}

\begin{equation}
\frac{(v_B ^{red})_{axion}}{4.3 \cdot 10^4 ~cm/s} \equiv
u_t ' = \left( \frac{4 \pi}{C_D} \cdot \frac{\sigma_0}{R_{Sun}} \cdot 
\frac{(\Delta \rho / \rho_{ext})_0}{2 \cdot 10^{-6}} \right)^{1/2} ,
\label{eq07-88}
\end{equation}

\noindent
which connects the radius of the cross-section $\sigma_0 \equiv a_{axion}$ (see Eq.~(\ref{eq07-51})) with the thermal velocity in dimensionless coordinates (see analogous Table~1 in \cite{Choudhuri1987}), it is not difficult to show that for $(\Delta \rho / \rho_{ext})_0 \sim 10^{-5}$ (or equivalently at $B_0 = v_A(4 \pi \rho)^{1/2} \sim 10^5 ~G$) the magnetic buoyancy values $(v_B^{red})_{axion} / 4.3 \cdot 10^4 ~cm/s \equiv u_t ' \approx 1.7 \cdot 10^{-2}$ and $\sigma _0 \equiv a_{axion} \sim 100 ~km$ remarkably coincide with the corresponding values of Eqs.~(\ref{eq07-67}) and~(\ref{eq07-87}), respectively.

As a consequence of~(\ref{eq07-87})-(\ref{eq07-88}), from the understanding of physics of the flux tubes trajectory (i.e. the radius of the cross-section $\sigma_0 \equiv a_{axion} \sim 100 ~km$ and strong magnetic fields $B \sim 10^5 ~G$), the question arises: in what way the theoretical calculations of the trajectory of flux tubes with the participation of the drag force (see Fig.~\ref{fig-magtube-tilt}a,b) obey Joy's law, when the values of the tilt of flux loops are very close to the observed values ​​(see Fig.~\ref{fig-magtube-tilt}c,d), and thereby do not contradict the known experimental data?

In the beginning, according to \cite{DSilvaChoudhuri1993}, we provide our calculations of the flux tube trajectories using the dynamic equation of a thin flux tube \citep{Spruit1981,Roberts1979,FerrizMas1989}. The basic equation for a non-axisymmetric ring using spherical coordinates was discussed by \cite{Choudhuri1989}. Here we use the equation by \cite{Choudhuri1989} (see Eq.~(1); also see Eqs.~(5)-(9) in \cite{DSilvaChoudhuri1993}) with an additional term of drag $D_n$ (per unit length):

\begin{equation}
\frac{D_n}{\pi \sigma_n ^2 \rho_e (\xi _n)} = - \frac{C_D}{2 \pi \sigma _n}
\left[ v_n ^2 - \left(\vec{v}_n \cdot \hat{\vec{l}}\,\right)^2 \right]^{1/2}
\left[ \vec{v}_n - \left(\vec{v}_n \cdot \hat{\vec{l}}\,\right)\hat{\vec{l}}\,\right] ,
\label{eq07-89}
\end{equation}

\noindent
where $\xi _n = r_n / R_{Sun}$, $r_n$ is the position of the $n^{th}$ Lagrange marker, $\rho_e (\xi_n)$ is the density, and $\sigma _n$ is the radius of the cross-section. For the velocity $\vec{v}_{k,n}$ perpendicular to the local tangent at the $n^{th}$ point we have $\vec{v}_{k,n} = \hat{\vec{l}} \times (\vec{v}_n \times \hat{\vec{l}})$, where $\vec{v}_n$ is the velocity and $\hat{\vec{l}}$ represents the unit vector along the tangent at the $n^{th}$ point. The expression for $\hat{l}$ was given by \cite{Choudhuri1989}. Hence, as a result, the term of drag, as shown in Eqs.~(5)-(9) in \cite{DSilvaChoudhuri1993}, will increase with decreasing the size of the flux tube and vice versa.

It is very important to note here that the flux tube trajectories always appear in the overshoot tachocline with the help of the strong toroidal magnetic field $\sim 10^7 ~G$ (see Figs.~\ref{fig-lower-reconnection}a,b and ~\ref{fig-lower-reconnection2}), but if by means of magnetic reconnection the initial trajectory (see Fig.~\ref{fig-lower-reconnection}a) switches over to the following trajectory (see Fig.~\ref{fig-lower-reconnection}b), then it depends on the appearance of the azimuthal magnetic field $\sim 10^5 ~G$, whose radial length is equal to the radial length ($z_0 \approx 0.1 ~H_p$ at $0.8 R_{Sun}$) of the meridional field in the lower layer near the tachocline (see green bands in Figs.~\ref{fig-meridional-cut}b and \ref{fig-Bz-diffusivity}a).

The main result of the theoretical estimation of the flux tube trajectory parameters with $B_0 \sim 10^7 ~G$ is predetermined by the fact that the parameters of the initial trajectory (see red lines in Fig.~\ref{fig-magtube-tilt}a,b) and the secondary trajectory by magnetic reconnection with $B_0 \sim 10^5 ~G$  (see blue lines in Fig.~\ref{fig-magtube-tilt}a,b) are parts of one common trajectory of the flux tubes, in which the motion of the flux ring is mainly due to four basic properties:

\begin{enumerate}
\item
The existence of anchored flux tubes with $10^7 ~G$ in the overshoot tachocline is a consequence of the fundamental properties of the holographic principle of quantum gravity, one of which (unlike the dynamo action!) generates a strong toroidal field in the tachocline with the help of the holographic BL~mechanism (see Fig.~\ref{fig-solar-dynamos}b).

\item
The existence of the magnetic buoyancy of flux tubes on the surface of the Sun is a consequence of the formation of a secondary trajectory by magnetic reconnection with $B_0 \sim 10^5 ~G$ (see blue lines in Fig.~\ref{fig-magtube-tilt}a,b), at which, on the one hand, the formation of the rotating density-stratified $\nabla \rho$-pumping with $\sim 10^5~G$ near the tachocline (see green bands in Figs.~\ref{fig-meridional-cut}b and ~\ref{fig-Bz-diffusivity}a) provides the process of blocking magnetic buoyancy at high latitudes, and on the other hand, it predetermines the existence of the dominant Coriolis force in $\nabla \rho$-pumping with $\sim 10^5 ~G$, which ultimately generates an arched ascending loop on the surface of the Sun with a tilt in the lower and, to a lesser extent, middle latitudes (see Figs.~\ref{fig-magtube-tilt} and ~\ref{fig-meridional-cut-tilt}a).

\item
The basic properties of an almost empty magnetic tube (see Figs.~\ref{fig-lampochka}a, ~\ref{fig-axion-compton} and ~\ref{fig-lower-heating}), characterizing thin radii of the cross-section ($\sigma_0 \equiv a_{axion} \sim 100~km$) and strong magnetic fields $\sim 10^7 ~G$, do not contradict the existence of a thin flux tube, since a virtually empty magnetic tube, unlike thin flux tubes, has not only strong magnetic fields, but also a thin radius of the ring between the surface of the thick O-loop and the wall of the tube (see Figs.~\ref{fig-axion-compton} and~\ref{fig-lower-heating}). It means that it is not the thin flux tubes with $\sim 10^5 ~G$ near the tachocline (see green bands in Figs.~\ref{fig-meridional-cut}b and~\ref{fig-Bz-diffusivity}a) that serve as a basis for the role of the Coriolis force at a certain latitude of the rising tilt in the direction of the active region. Therefore, some cross-section of the tube (see Fig.~\ref{fig-magtube-tilt}c,d) and the arched ascending loop of the magnetic field lines (see Fig.~\ref{fig-meridional-cut-tilt}) are allowed.

\item
The averaged theoretical estimates of the magnetic cycle of flux tubes are practically identical to the observational (averaged) data of the tilt angle of Joy's law (see Fig.~\ref{fig-magtube-tilt}c,d).
\end{enumerate}

Since the study of the tendencies of the tilt angle of Joy's law is very important for understanding the evolution of the solar magnetic field, then unlike the heavy calculations of theoretical estimates of the averaged angle tilt of sunspots with increasing latitude, we can summarize our results (see Eqs.~(5)-(9) in \cite{DSilvaChoudhuri1993} and also Eq.~(\ref{eq07-58}) for Fig.~\ref{fig-magtube-tilt}c,d), which can simultaneously be expressed in the form of simple and understandable physics of Joy's law. It is known that, according to \cite{DSilvaChoudhuri1993} and \cite{Fan1993}, the Coriolis force is proportional to the buoyancy speed $(v_B)_{axion}$, which can be estimated taking into account the balance between the buoyancy force (left term) and the drag force (right term):

\begin{equation}
\frac{B^2}{8 \pi H_p} = C_D \frac{\rho_{ext} (v_B^2)_{axion}}{\pi a_{axion}}\, ,
\label{eq07-90}
\end{equation}

\noindent or

\begin{equation}
(v_B)_{axion} \propto B^{3/4} (\pi a_{axion}^2 B)^{1/4} , 
\label{eq07-91}
\end{equation}

\noindent
where in our case the value of the buoyancy speed $(v_B)_{axion}$ is identical to the known expressions for the rising speed $v_r$ of the loop (see \cite{DSilvaChoudhuri1993,Fan1993}). Taking into account Eqs.~(18) and~(19) by \cite{Fan1994},

\begin{equation}
a \propto B^{-1/2} (\pi a_{axion}^2 B)^{1/2} v_{B}^{-1} \sin \theta\, ,
\label{eq07-92}
\end{equation}

\noindent
we can obtain by means of~(\ref{eq07-91}) the following equation:

\begin{equation}
a \propto B^{-5/4} (\pi a_{axion}^2 B)^{1/4} \sin \theta ,
\label{eq07-93}
\end{equation}

\noindent
where, using the theoretical values of the slope angle $\alpha$, the latitude angle $\theta$, and the radial parameter $\xi = r / R_{Sun} \sim 0.8 R_{Sun}$, we can write the simplified equation (see analogous Eq.~(\ref{eq07-93})) in the following form:

\begin{equation}
\sin (tilt) \propto \frac{\sqrt{a_{axion} (\xi , \theta)}}{B} \sin (latitude) .
\label{eq07-94}
\end{equation}

It is very important to understand that, based on the properties of the trajectories of anchored flux tubes with $10^7 ~G$, which are a consequence of the holographic BL~mechanism (see Fig.~\ref{fig-solar-dynamos}b), our studies show that the strength of the toroidal magnetic field is $\sim 10^5 ~G$ (near the tachocline see the ``green'' term of the trajectory in Fig.~\ref{fig-meridional-cut}b and ~\ref{fig-Bz-diffusivity}a) and the corresponding radius of the cross-section ($\sigma _0 \equiv a_{axion} \sim 100~km$) appears by means of reconnection of the flux tubes (see Figs.~\ref{fig-lower-reconnection}b and~\ref{fig-lower-reconnection2}b). This means that the anchored flux tubes with $10^7 ~G$ after magnetic reconnection become the flux tubes with $10^5 ~G$. So it is not difficult to understand why the Coriolis force (see Eq.~(\ref{eq07-90})) can explain the tilt angle described by Joy's law (see Fig.~\ref{fig-magtube-tilt}c,d) and Eq.~(\ref{eq07-94}).

Hence, it becomes clear that if all properties of the trajectories of anchored flux tubes with $10^7 ~G$ are considered to be a consequence of the fundamental holographic BL~mechanism (see Fig.~\ref{fig-solar-dynamos}b), then this mechanism generates oscillations of the magnetic field in the Sun core 
(see \cite{Gough1988,Dicke1988}) (again, not the dynamo action!), which not only are the result of the formation of magnetic cycles, but both effects, not surprisingly, are caused by the existence of DM -- solar axions in the core of the Sun. (see Sect.~\ref{sec-radiative-heating}).

Let us remind that there is practically no satisfactory alternative theory against the action of the dynamo  
(see e.g. \cite{Cowling1981,Gilman1986,Dicke1978,Dicke1982,Dicke1988,
Gough1980,Gough1981,Gough1988,Choudhuri1989,Karak2014}). At the same time, the problem of the dynamo action is not yet solved to the end! Although it was believed that 
``...To sum up, the dynamo theory of the Sun’s magnetic field is subject to a number of unresolved objections, but alternative theories advanced so far are to much greater objections'' \citep{Cowling1981}. In this sense, we still remember the words of \cite{Choudhuri1989}: ``...It is perhaps fair to say that a major uncertainty remains at the present time regarding the question of where exactly the dynamo process takes place. If we assume the dynamo to operate within the main body of the convection zone, then the magnetic buoyancy poses a serious problem.  On the other hand, until we understand how the magnetic flux can get out of the clutches of the Coriolis force, \textbf{the hypothesis that the dynamo operates at the bottom of the convection zone remains at best a far-fetched speculation}.  One may even take an extreme point of view and raise the question \textbf{whether the dynamo theory itself is the correct theory for the generation of solar magnetic fields. There have been some recent claims that the solar cycle may involve oscillations  penetrating to the core of the Sun} \citep{Gough1988}. If these claims turn out to be true, then it will be necessary to understand their implications for the dynamo theory. More theoretical and observational work will certainly be needed before the fate of the  dynamo theory can be decided''.

From here we understand that the remarkable properties of the holographic BL~mechanism are a consequence of not only the fundamental properties of the holographic principle of quantum gravity, but also DM, the existence of the true nature of which is predetermined, not surprisingly, by the conservation law of quantum-gravitational energy in the Universe, and therefore in our Galaxy, and of course in the Sun!

\sethlcolor{lightyellow}

\subsection{Coronal heating problem solution by means of photons of axion origin}
\label{sec-osc-parameters}

To estimate the hadron axion-photon coupling constant, we focus on the
conversion probability

\begin{equation}
P_{a \rightarrow \gamma} = \frac{1}{4} \left( g_{a \gamma} B_{MS} L_{MS} \right)^2 \sim 1\, ,
\label{eq3.31}
\end{equation}

\noindent
where the complete conversion between photons and axions is possible by means
of estimating the axion coupling constant to photons.

\begin{figure}[tbp!]
  \begin{center}
    \begin{minipage}[h]{0.44\linewidth}
      \includegraphics[width=7.6cm]{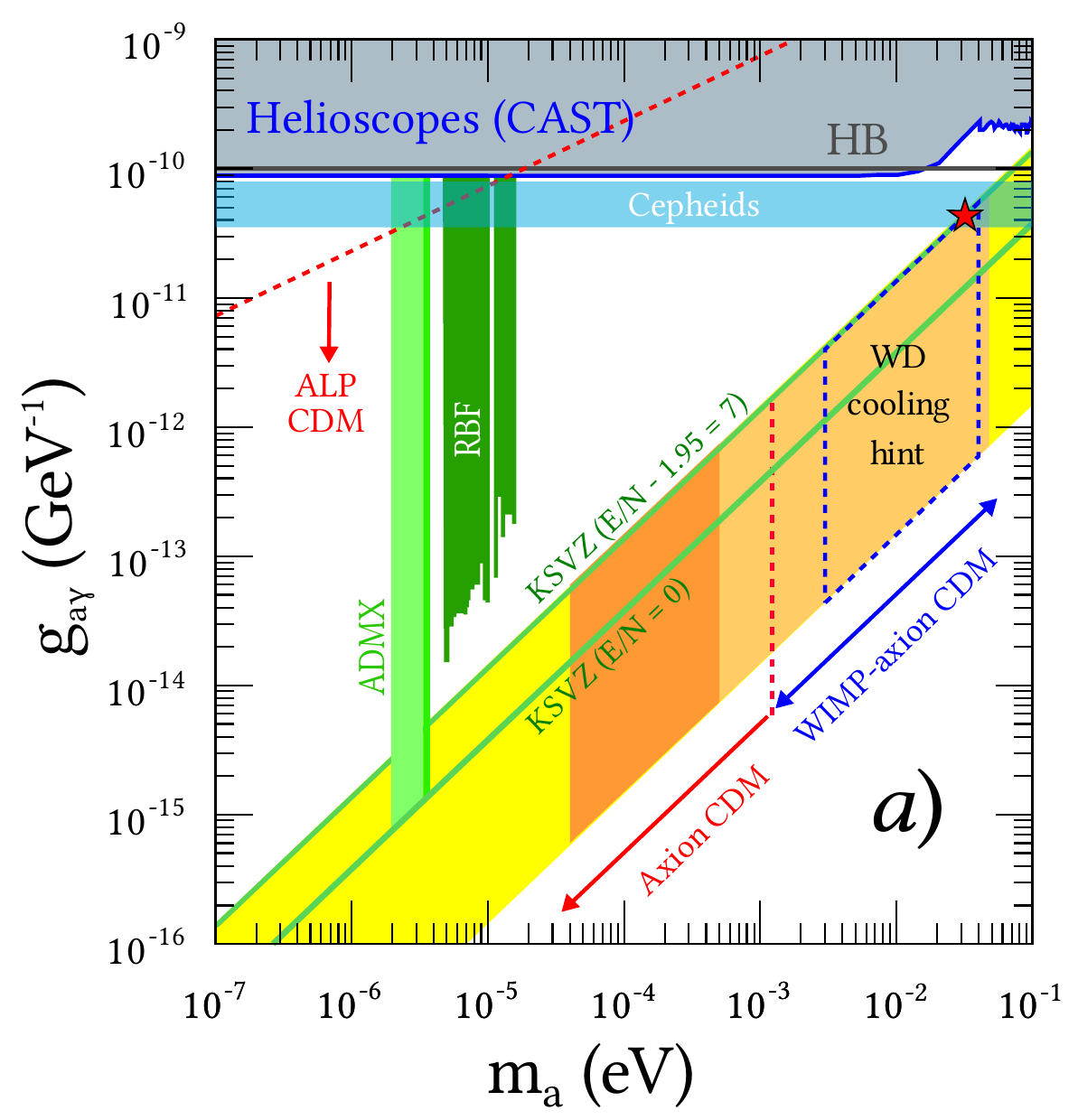}
    \end{minipage}
    \hfill
    \begin{minipage}[h]{0.53\linewidth}
      \includegraphics[width=8.8cm]{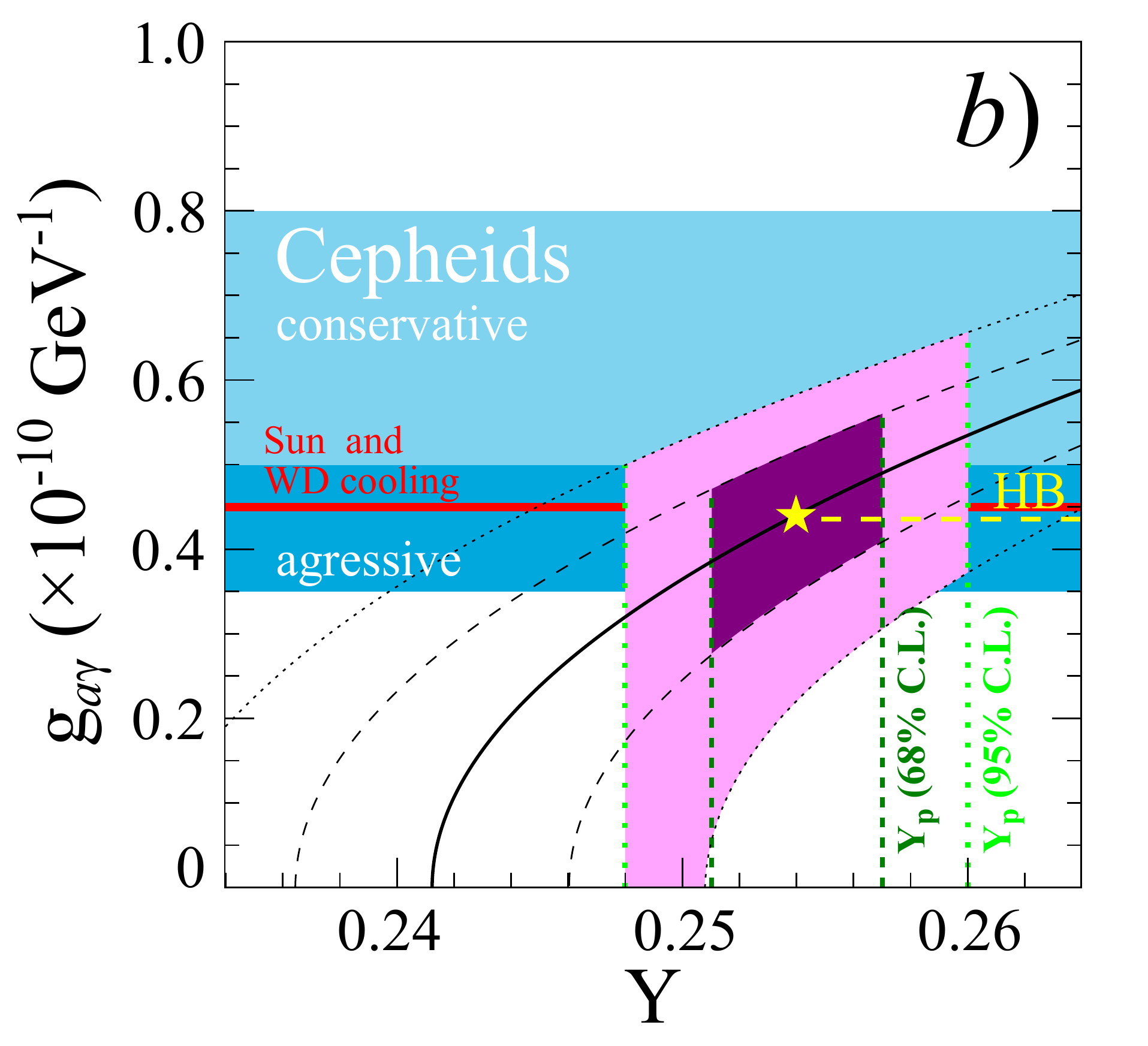}
    \end{minipage}
  \end{center}
\caption{\textbf{(a)} Summary of astrophysical, cosmological and laboratory
constraints on axions and ALPs. Comprehensive axion/ALP
parameter space, highlighting two main front lines of direct detection
experiments: helioscopes (CAST~\citep{Andriamonje2007,ref72,CAST2011,Arik2013}) and
haloscopes (ADMX~\citep{ref50} and RBF~\citep{ref51}). The astrophysical bounds
from horizontal branch and massive stars are labeled ``HB''~\citep{ref02} and
``Cepheids''~\citep{Carosi2013}, respectively. The QCD motivated models
(KSVZ~\citep{ref46,ref46a} and DFSZ~\citep{ref47,Dine1981}) for axions lay in
the yellow diagonal band. The orange parts of the band correspond to
cosmologically interesting axion models: models in the ``classical axion
window'' possibly composing the totality of DM (labeled ``Axion CDM'') or a
fraction of it (``WIMP-axion CDM''~\citep{Baer2011}). For more generic ALPs,
practically all allowed space up to the red dashed line may contain valid ALP
CDM models~\citep{Arias2012}. The region of axion masses invoked in the white dwarf
cooling anomaly is shown by the blue dashed line~\citep{Irastorza2013}. The red
star marks the values of the axion mass $m_a \sim 3.2 \cdot 10^{-2} eV$ and the
axion-photon coupling constant $g_{a\gamma} \sim 4.4 \cdot 10^{-11} GeV^{-1}$
chosen in the present paper on the basis of the suggested relation between the
axion mechanisms of the Sun and the white dwarf luminosity variations.
\newline
\textbf{(b)} $R$ parameter constraints on $Y$ and $g_{a \gamma}$ (adopted from
\cite{Ayala2014}). The dark purple area delimits the 68\%~C.L. for $Y$ and
$R_{th}$ (see Eq.~(1) in \cite{Ayala2014}). The resulting bound on the axion
($g_{10} = g_{a \gamma \gamma}/(10^{-10} ~GeV^{-1})$) is somewhere between
rather conservative $0.5 < g_{10} \leqslant 0.8$ and most aggressive $0.35 <
g_{10} \leqslant 0.5$ \citep{Friedland2013}. The red line marks the value of
the axion-photon coupling constant $g_{a \gamma} \sim 4.4 \cdot 10^{-11}
~GeV^{-1}$ chosen in the present paper.
The blue shaded area represents the bounds from Cepheids
observation. The yellow star corresponds to $Y$=0.254 and the bounds from HB
lifetime (yellow dashed line).}
\label{fig05}
\end{figure}

The $\sim 4100$~T magnetic field in the overshoot
tachocline and the Parker-Biermann cooling effect can 
produce the O-loops with the horizontal magnetic field
$B_{MS} \approx B(0.72 R_{Sun}) \sim 3600$~T
stretching for about $L_{MS} \sim 1.28 \cdot 10^4 ~km$ (see Eq.~(\ref{eq13})),
 and surrounded by virtually
zero internal gas pressure of the magnetic tube (see Fig.~\ref{fig-lampochka}a,b).
Since $P_{a \rightarrow \gamma} \sim 1$, we obtain the following parameters of
the hadron axion (see Fig.~\ref{fig05}).

\begin{equation}
g_{a \gamma} \sim 4.4 \cdot 10^{-11} ~ GeV^{-1}, ~~~ m_a \sim 3.2 \cdot 10^{-2} ~eV.
\label{eq3.30}
\end{equation}

The choice of these values is also related to the observed solar luminosity
variations in the X-ray band (see (\ref{eq3.34})). The theoretical estimate and
the consequences of such choice are considered below.

Thus, it is shown that the hypothesis about the possibility for the solar
axions born in the core of the Sun to be efficiently converted back into
$\gamma$-quanta in the magnetic field of the magnetic steps of the O-loop
(above the solar overshoot tachocline) is relevant. Here the variations of the
magnetic field in the solar tachocline are the direct cause of the converted
$\gamma$-quanta intensity variations. The latter in their turn may be the cause
of the overall solar luminosity variations known as the active and quiet Sun phases.

It is easy to show that the theoretical estimate for the part of the axion
luminosity $L_a$ in the total luminosity of the Sun $L_{Sun}$ with respect to
(\ref{eq3.30}) is~\citep{Andriamonje2007}

\begin{equation}
\frac{L_a}{L_{Sun}} = 1.85 \cdot 10 ^{-3} \left( 
\frac{g_{a \gamma}}{10^{-10} GeV^{-1}} \right)^2 \sim 3.6 \cdot 10^{-4} .
\label{eq3.32}
\end{equation}

\begin{figure*}
  \begin{center}
    \includegraphics[width=14cm]{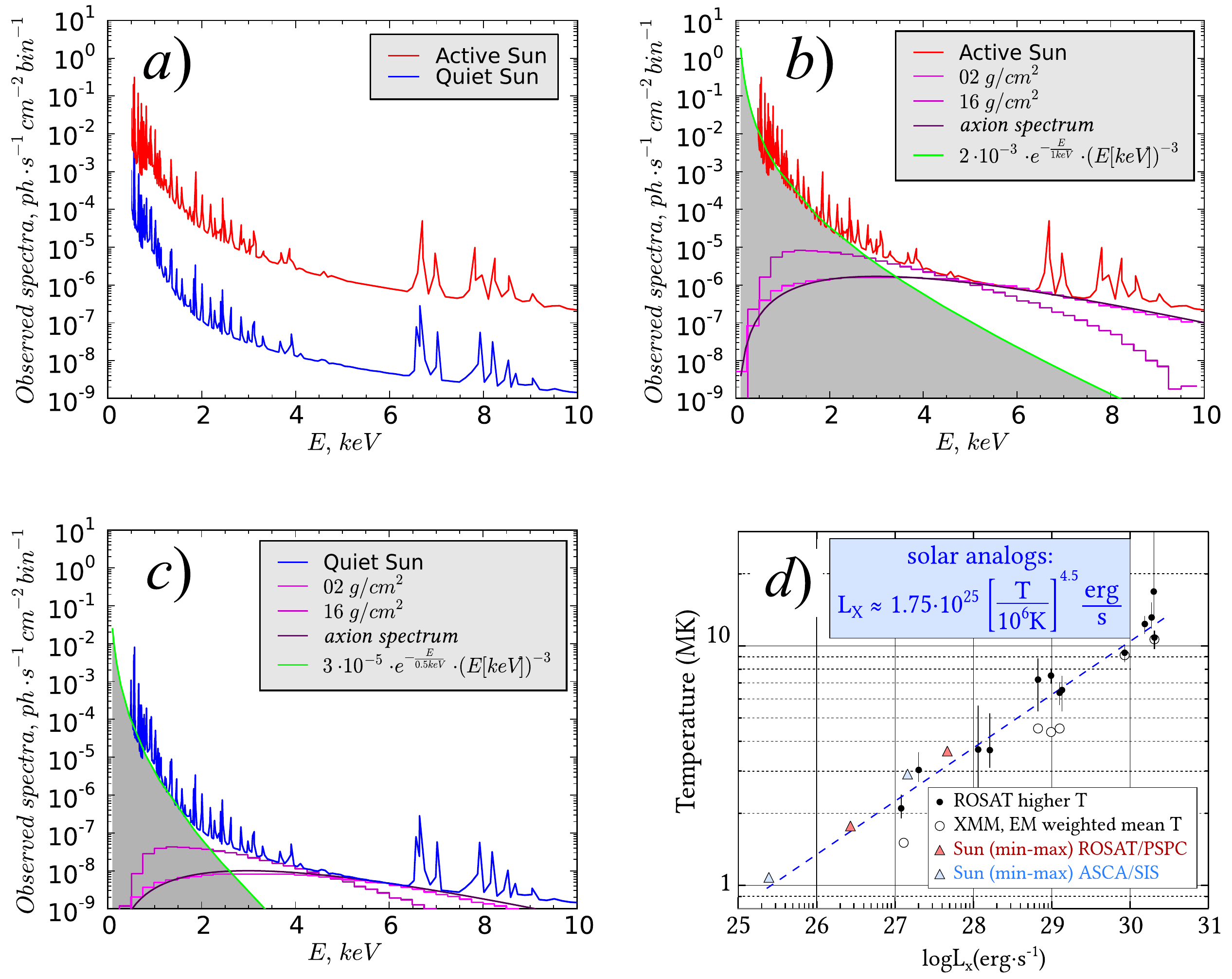}
  \end{center}

\caption{\textbf{(a)} Reconstructed solar photon spectrum in the 0.5\,-\,10~keV band from 
the active
Sun (red line) and quiet Sun (blue line) from accumulated observations
(spectral bin is 6.1~eV wide). Adopted from~\cite{Peres2000}.
\newline
\textbf{(b)} Reconstructed solar photon spectrum fit in the active phase of the Sun by
the quasi-invariant soft part of the solar photon spectrum (grey shaded area;
see \mbox{Eq.~(\ref{eq06-34})}) and three spectra (\ref{eq3.33}) degraded to
the Compton scattering for column densities above the initial conversion place
of 16~$g / cm^2$ \citep{Zioutas2009} and 2~$g / cm^2$ (present paper).
\newline
\textbf{(c)} Similar curves for the quiet phase of the Sun (grey shaded area 
corresponds to \mbox{Eq.~(\ref{eq06-35})}).
\newline
\textbf{(d)} 
Corona temperature against the X-ray luminosity of the solar
analogs (see ROSAT, XMM-Newton, adopted from \cite{Gudel2004}). The data from
\textit{ROSAT/PSPC} and \textit{ASCA/SIS} are shown as
compared to the maximum and minimum of the solar cycle (see Table 2 in
\citet{Peres2000}).}
\label{fig06}
\end{figure*}

As opposed to the classic mechanism of the Sun modulation, the
axion mechanism is determined by the magnetic tubes rising to the photosphere,
and not by the over-photosphere magnetic fields. In this case the solar
luminosity modulation is determined by the axion-photon oscillations in the
magnetic steps of the O-loop causing the formation and channeling of $\gamma$-quanta inside the almost empty magnetic $\Omega$-tubes (see
Fig.~\ref{fig-twisted-tube} and Fig.~\ref{fig-lampochka}a). When the magnetic
tubes cross the photosphere, they ``open'' (Fig.~\ref{fig-lampochka}a), and the
$\gamma$-quanta are ejected to the photosphere, where their comfortable journey
along the magnetic tubes (without absorption and scattering) ends. As the
calculations by \cite{Zioutas2009} show, the further destiny of the $\gamma$-quanta
in the photosphere may be described by the Compton scattering, which actually
agrees with the observed solar spectral shape (Fig.~\ref{fig06}b,c).

Considering the above remarks, we believe that it is necessary to calculate the
effect of axion-originated photons in active and quiet phases of the Sun on the
solar corona. To do this, in Sec.~\ref{sec-sunspot-cycles} we first make the
preliminary assessment of the magnetic cycles of sunspots using magnetic flux
tubes and photons of axion origin on the corona. 
In Sec.~\ref{sec-coronal-heating} the theoretical photon spectra of the corona
are calculated using inverse Compton scattering (ICS) of isotropic photons on
isotropic (moderately relativistic) electrons.
In Sec.~\ref{sec-sun-chronometer} we discuss whether there is a dark matter
chronometer hidden deep in the Sun core.

\subsubsection{Preliminary estimation of magnetic cycles of sunspots with the help of photons of axion origin}
\label{sec-sunspot-cycles}

From the axion mechanism point of view the solar spectra during
the active and quiet phases (i.e. during the maximum and minimum solar
activity) differ from each other by the smaller or larger part of the Compton
spectrum, the latter being produced by the $\gamma$-quanta of axion origin
ejected from the magnetic tubes into the photosphere (see Fig.~4 in
\cite{ChenF2015}).

A natural question arises at this point: ``What are the real parts of the
Compton spectrum of axion origin in the active and quiet phases of the Sun,
and do they agree with the experiment?'' Let us perform the
mentioned estimations being based on the known experimental results by ROSAT/PSPC,
where the Sun coronal X-ray spectra and the total luminosity during the
minimum and maximum of the solar coronal activity were obtained~\citep{Peres2000}.

Apparently, the solar photon spectrum below 10~keV of the active and quiet Sun
(Fig.~\ref{fig06}a) reconstructed from the accumulated ROSAT/PSPC and ASCA/SIS
observations can be described by two Compton spectra for different column
densities rather well (Fig.~\ref{fig06}b,c). This gives grounds for the
assumption that the hard part of the solar spectrum is mainly determined by the
axion-photon conversion efficiency:

\begin{align}
\left( \frac{d \Phi}{dE} \right)^{(*)} \approx
\left( \frac{d \Phi}{dE} \right)^{(*)}_{CS} +
\left( \frac{d \Phi _{\gamma}}{dE} \right)^{(*)}_{axions} ,
\label{eq06-33}
\end{align}

\noindent where ${d \Phi}/{dE}$ is the observed solar spectra during the
active (red line in Fig.~\ref{fig06}a,b) and quiet (blue line in
Fig.~\ref{fig06}a,c) phases, $\left({d \Phi}/{dE} \right)_{CS}$
represents the power-like theoretical spectra
of the Compton-scattered photons of axion origin at 0.5\,-\,2.0 keV,

\begin{equation}
\left( \frac{d \Phi}{dE} \right)_{CS} \sim E^{-(1+\alpha)} e^{-E/E_0} ,
\label{eq06-33a}
\end{equation}

\noindent
where the power-law decay with the ``semi-heavy tail'' takes place in practice 
\citep{Lu1993} instead of the so-called power laws with heavy tails 
\citep{Lu1991,Lu1993} (see e.g. Figs.~3 and~6 in \cite{Uchaikin2013}). 
Consequently, the observed corona spectra ($0.5 ~keV < E < 2.0 ~keV $)
(shaded area in Fig.~\ref{fig06}b)

\begin{align}
\left( \frac{d \Phi}{dE} \right)^{(active)}_{CS} \sim
5 \cdot 10^{-3} \cdot (E~[keV])^{-3} \cdot \exp{\left(-\frac{E}{1 keV} \right)} 
~~for~the~active~Sun
\label{eq06-34}
\end{align}

\noindent and (shaded area in Fig.~\ref{fig06}c)

\begin{align}
\left( \frac{d \Phi}{dE} \right)^{(quiet)}_{CS} \sim
1 \cdot 10^{-4} \cdot (E~[keV])^{-3} \cdot \exp{\left(-\frac{E}{0.5 keV} \right)} 
~~for~the~quiet~Sun ;
\label{eq06-35}
\end{align}

\noindent $\left( {d \Phi _{\gamma}}/{dE} \right)_{axions}$ is the
reconstructed solar photon spectrum fit ($2 ~keV < E < 10 ~keV$) built up from
three spectra (\ref{eq3.33}) degraded to the Compton scattering for different
column densities (see Fig.~\ref{fig06}b,c for the active and quiet phases of
the Sun, respectively).

As is known, this class of flare models (Eqs.~(\ref{eq06-34})
and~(\ref{eq06-35})) is based on the recent paradigm in statistical physics
known as self-organized criticality
\citep{Bak1987,Bak1988,Bak1989,Bak1996,Aschwanden2011}. The basic idea is that
the flares are a result of an ``avalanche'' of small-scale magnetic
reconnection events cascading \citep{Lu1993,Charbonneau2001,Aschwanden2014} 
through the highly intense coronal magnetic structure \citep{Shibata2011} driven
at the critical state by the accidental photospheric movements of its magnetic
footprints. Such models thus provide the natural and computationally convenient
basis for the study of Parker's hypothesis of the coronal heating by nanoflares
\citep{Parker1988}.

Another significant fact discriminating the theory from practice, or rather 
giving a true understanding of the measurements against some theory, should be 
recalled here (see e.g. Eq.~(\ref{eq06-33a}); also see Eq.~(5) in \cite{Lu1993}). The 
nature of power laws is related to the strong connection between the consequent
events (this applies also to the ``catastrophes'', which  in turn give rise to
a spatial nonlocality related to the appropriate structure of the medium (see 
page 45 in \cite{Uchaikin2013})). As a result, the ``chain reaction'', i.e. the
avalanche-like growth of perturbation with more and more resource involved,
leads to the heavy-tailed distributions. On the other hand, obviously, none of 
the natural events may be characterized by the infinite values of the mean and 
variance. Therefore, the power laws like (\ref{eq06-33a}) are approximate and
must not hold for the very large arguments. It means that the power-law decay
of the probability density rather corresponds to the average asymptotics, and
the ``semi-heavy tails'' must be observed in practice instead.

In this regard we suppose that the application of the power-law distributions 
with semi-heavy tails leads to a soft attenuation of the observed corona 
spectra (which are not visible above $E > 2 ~keV$), and thus to a close
coincidence between the observed solar spectra and $\gamma$-spectra of axion 
origin (Fig.~\ref{fig06}), i.e.

\begin{equation}
\left( \frac{d \Phi}{dE} \right)^{(*)} \approx
\left( \frac{d \Phi _{\gamma}}{dE} \right)^{(*)}_{axions} 
~~~ \text{for energies} ~~ E > 2 ~keV.
\label{eq06-35a}
\end{equation}

It means that the physics of the formation and ejection of the $\gamma$-quanta 
above $2 ~keV$ through the sunspots into corona is not related to the 
magnetic reconnection theory by e.g. \cite{Shibata2011} (Fig.~\ref{fig06}d),
and may be of axion origin.

With this in mind, let us suppose that the part of the differential solar axion
flux at the Earth~\citep{Andriamonje2007}
\begin{align}
\frac{d \Phi _a}{dE} = 6.02 \cdot 10^{10} \left( \frac{g_{a\gamma}}{10^{10} GeV^{-1}} \right)^2 E^{2.481} \exp \left( - \frac{E}{1.205} \right) ~~cm^{-2}
s^{-1} keV^{-1} ,
\label{eq3.33}
\end{align}
\noindent characterizing the differential $\gamma$-spectrum of axion
origin $d \Phi _{\gamma} / dE$
(see $[ d \Phi _{\gamma} / dE ]_{axions}$ in (\ref{eq06-33}) and 
(\ref{eq06-35a}))

\begin{align}
\frac{d \Phi _{\gamma}}{dE}  = P_{\gamma} \frac{d \Phi _{a}}{dE}
~~ cm^{-2} s^{-1} keV^{-1} \approx
6.1 \cdot 10^{-3} P_{\gamma} \frac{d \Phi _{a}}{dE}
~ ph\cdot cm^{-2} s^{-1} bin^{-1}\, ,
\label{eq3.34}
\end{align}

\noindent
where the spectral bin width is 6.1~eV (see Fig.~\ref{fig06}a);
the probability $P_{\gamma}$ describing the relative portion of $\gamma$-quanta
(of axion origin) channeling along the magnetic tubes can be defined, according
to~\cite{Peres2000}, from the observed solar luminosity variations in the X-ray
band, recorded in ROSAT/PSPC experiments (see Fig.~6 and Table~2 in \cite{Peres2000}):
$\left(L_{corona}^X \right) _{min} \approx 2.7 \cdot 10^{26} ~erg/s$ at minimum
and $\left( L_{corona}^X \right) _{max} \approx 4.7 \cdot 10^{27} ~erg/s$ at
maximum (see Fig.~\ref{fig06}d), and in ASCA/SIS experiments (see Fig.~9 and Table~2 in \cite{Peres2000} and also
Figs.~\ref{fig06} and~\ref{fig-lower-heating}b): 
$\left(L_{corona}^X \right) _{min} \approx 2.3 \cdot 10^{25} ~erg/s$ at minimum
and $\left( L_{corona}^X \right) _{max} \approx 1.5 \cdot 10^{27} ~erg/s$ at
maximum (see Fig.~\ref{fig06}d).

Consequently, the probability $P_\gamma$ describing the relative fraction of
the sunspot area on the surface over the magnetic flux tubes during the
maximum of Sun luminosity can be determined as:

\begin{equation}
(P_{\gamma})_{max} = \left( P_{a \rightarrow \gamma} \right)_{max} \cdot \dfrac{2 \left \langle sunspot ~area \right \rangle _{max}}
{(4/3) \pi R_{Sun})^2} \sim 0.74 \cdot 10^{-2},
\label{eq7.5-sep-11}
\end{equation}

\noindent and during the minimum of Sun luminosity as:

\begin{equation}
(P_{\gamma})_{min} = \left( P_{a \rightarrow \gamma} \right)_{min} \cdot \dfrac{2 \left \langle sunspot ~area \right \rangle _{min}}
{(4/3) \pi R_{Sun})^2} \sim 5.2 \cdot 10^{-4},
\label{eq7.5-sep-12}
\end{equation}

\noindent where $\langle sunspot ~area \rangle _{max} \approx 7.5 \cdot 10^9 ~km^2 \approx 2470 ~ppm$, $\langle sunspot ~area \rangle _{min} \approx 7.5 \cdot 10^8 ~km^2 \approx 247 ~ppm$ of visible hemisphere is the sunspot area (over the visible hemisphere~\citep{Dikpati2008,Gough2010}) for the cycle 22 experimentally observed by the Japanese X-ray telescope Yohkoh (1991) (see \cite{Zioutas2009}); $R_{Sun} = 6.96 \times 10^5 ~km$; $P_{a \rightarrow \gamma} \approx 1$ at maximum of solar luminosity, and $P_{a \rightarrow \gamma} \approx 0.7$ at minimum (see Eqs.~(\ref{eq07-117a}) and~(\ref{eq07-117b})).

The product of the total fraction of axions originating from the Sun core and the fraction of the sunspot area (see \citet{Dikpati2008,Gough2010}) yields the total fraction of the corona luminosity, or more precisely, the fraction of photon luminosity of axion origin in the corona.

\begin{equation}
\frac{L_{corona}^X}{L_{Sun}} = \frac{L_a}{L_{Sun}} \cdot P_{\gamma} ,
\label{eq7.5-sep-13}
\end{equation}

\noindent where $L_{corona}^X / L_{Sun}$ is the fraction of the corona luminosity in the total Sun luminosity,
$L_a / L_{Sun}$ is the fraction of the axion ``luminosity'', $L_{Sun} = 3.8418 \cdot 10^{33} ~erg/s$ is the solar luminosity~\citep{Bahcall2004}.

The axions luminosity fraction $L_a$ at the maximum and minimum of the Sun luminosity

\begin{equation}
\frac{(L_a)_{max}}{L_{Sun}} \sim 3.6 \cdot 10^{-4},
\label{eq7.5-sep-14}
\end{equation}

\begin{equation}
\frac{(L_a)_{min}}{L_{Sun}} \sim 3.6 \cdot 10^{-5},
\label{eq7.5-sep-15}
\end{equation}

\noindent
are determined by the 11-year ADM variations, which are gravitationally captured in the solar interior (see Sect.~\ref{sec-sun-chronometer}). An important note is that this can lead to the absence of self-annihilation today, which allows large amounts of ADM to accumulate in stars like the Sun~\citep{Vincent2015a}. The ADM particles absorb energy in the hottest, central part of the core, they then travel to a cooler, more peripheral, area before the scattering again and redistribute their energy \citep{GouldRaffelt1990a}.
This reduces the contrast of temperature across the core region and reduces the central temperature. A colder core produces less neutrinos from the temperature-sensitive fusion reactions, and also less axions from the temperature-sensitive reactions of electron-nucleus collisions (see e.g.~\citet{Raffelt1986}). So the fluxes of $^7 Be$ and $^8 B$ neutrino, and axions (see Eqs.~(\ref{eq7.5-sep-14}) and (\ref{eq7.5-sep-15})) can be significantly reduced. This is accompanied by a slight change in the $pp$ and $pep$ fluxes, as required by the constancy of the solar luminosity.

Taking into account the known observed variability of the corona luminosity in the X-ray range recorded in the ROSAT/SPC experiments,

\begin{equation}
\frac{(L_{corona}^X)_{max}}{L_{Sun}} \sim 1.22 \cdot 10^{-6}, ~~~
\frac{(L_{corona}^X)_{min}}{L_{Sun}} \sim 7.0 \cdot 10^{-8},
\label{eq7.5-sep-17}
\end{equation}

\noindent
we can compare the ``experimental'' (on the left; also (\ref{eq7.5-sep-17})) and theoretical equations (on the right; also Eqs.~(\ref{eq7.5-sep-11}) - (\ref{eq7.5-sep-12}) and (\ref{eq7.5-sep-14})-(\ref{eq7.5-sep-15}))

\begin{equation}
\frac{(L_{corona}^X)_{max}}{L_{Sun}} = \frac{(L_a)_{max}}{L_{Sun}} \cdot (P_{\gamma})_{max} \sim 2.7 \cdot 10^{-6},
\label{eq7.5-sep-18}
\end{equation}

\begin{equation}
\frac{(L_{corona}^X)_{min}}{L_{Sun}} = \frac{(L_a)_{min}}{L_{Sun}} \cdot (P_{\gamma})_{min} \sim 2.0 \cdot 10^{-8},
\label{eq7.5-sep-19}
\end{equation}

\noindent which fit well enough!

At the same time, we must keep in mind that the ``experimental'' data (on the left: Eq.~(\ref{eq7.5-sep-17})) come from the spectra of the corona, synthesized with the MEKAL spectral code directly from the $EM (T)$ distribution (see Sec.~\ref{sec-sunspot-cycles}) derived with spectral fitting of ROSAT/PSPC spectra, as reported by \citet{Peres2000} (see Table~2 and thick solid lines in Fig.~6a,b). Therefore, the values of the ``experimental'' data (on the left: Eq.~(\ref{eq7.5-sep-17})) are rather rough, while the theoretical data are more accurate.

As a result, in contrast to the variability of corona luminosity (on the left Eqs.~(\ref{eq7.5-sep-20a}) and (\ref{eq7.5-sep-20b})), according to the ``experimental'' ASCA/SIS data (see Fig.~\ref{fig06}d)),

\begin{equation}
\frac{(L_{corona}^X)_{max}}{L_{Sun}} \sim 0.39 \cdot 10^{-6} \neq \frac{(L_a)_{max}}{L_{Sun}} \cdot (P_{\gamma})_{max} \sim 2.7 \cdot 10^{-6} ,
\label{eq7.5-sep-20a}
\end{equation}

\begin{equation}
\frac{(L_{corona}^X)_{min}}{L_{Sun}} \sim 0.6 \cdot 10^{-8} \neq \frac{(L_a)_{min}}{L_{Sun}} \cdot (P_{\gamma})_{min} \sim 2.0 \cdot 10^{-8},
\label{eq7.5-sep-20b}
\end{equation}

\noindent
the strongest results is the theoretical variability of corona luminosity (on the right: Eqs.~(\ref{eq7.5-sep-18})-(\ref{eq7.5-sep-19})), which practically coincide with the variations of the corona luminosity according to ROSAT/PSPC (on the left: Eqs.~(\ref{eq7.5-sep-18})-(\ref{eq7.5-sep-19})).

The preliminary result is as follows. If the hadronic axions found in the Sun are the same particles that are associated with the known axion coupling to photons (see (\ref{eq3.30}) and Fig.~\ref{fig06}a,b,c), it is quite natural that the independent observations of the hard part of the solar photon spectrum are mainly determined by the axion-photon conversion efficiency, and the theoretical estimate for the luminosity fraction from the axions produced in the solar core, the fraction of sunspot area (see \citet{Dikpati2008,Gough2010}), i.e. from the magnetic cycles (see Eqs.~(\ref{eq7.5-sep-18})-(\ref{eq7.5-sep-19})) of the photons of axion origin.

\subsubsection{Asymmetric dark matter and axion from the solar data global fit and coronal heating problem}
\label{sec-ADM-SolData}

An interesting problem arises here related to the heating of the practically empty magnetic tube (see Figs.~\ref{fig-lampochka}a, \ref{fig-lower-heating} and \ref{fig-lower-reconnection2})
 and heating of the solar corona, which has long been unresolved (see \cite{DeMortel2015,DeMortel2016}). There are many assumptions about the unusually high temperature in the corona (see e.g. $T_{average} \approx (1.5 - 3) \cdot 10^6 ~K$ in \cite{Gudel2009} and \cite{Kariyappa2011,Cirtain2013,Peter2014,Reale2014,Laming2015,Peter2015,Morgan2017}) compared to the chromosphere and photosphere. It is known that energy comes from the underlying layers, including, in particular, the photosphere and chromosphere. Here are just some of the elements, possibly involved in the heating of the corona: 
magneto-acoustic and Alfv\'{e}n waves 
(see \cite{Alfven1947,Schatzman1949,Schatzman1962,Parker1964,Callebaut1994,
Jess2009,Jess2016,Ballegooijen2011,Ballegooijen2014,Arregui2015,Morton2016,
Laming2017,Vigeesh2017}), 
magnetic reconnections (see e.g. \cite{Shibata1999,Watanabe2011,Archontis2014,Xue2016,Sun2015,Huang2018}), 
nano-flares (see e.g. \cite{Aulanier2013,Testa2014,Brosius2014,Cargill2015,Klimchuk2015}), 
Ellerman bombs (see e.g. \cite{Watanabe2011,Vissers2013,Vissers2015,Libbrecht2017}). 
It is considered (see e.g. \cite{Jess2009}) that the possible mechanism of corona heating is the same as for the chromosphere: convective cells in the form of granulation rising from the depth of the Sun and appearing in the photosphere (see e.g. \cite{Ballegooijen2011,Ballegooijen2014,Dudik2014,Cranmer2015}) lead to a local imbalance in the gas, which leads to the propagation of magneto-acoustic and Alfv\'{e}n waves (see e.g. Fig.~13 in \cite{Laming2015}) moving in different directions. In this case, a chaotic change in the density, temperature and velocity of the substance in which these waves propagate leads to a change in the speed, frequency and amplitude of the magneto-acoustic waves and can be so large that the gas becomes supersonic. Shock waves appear (see \cite{Solanki2003,Grib2014,Santamaria2016}), which lead to the heating of the gas and, as a consequence, to the heating of the corona.

\begin{figure}[tbp]
\begin{center}
\includegraphics[width=16cm]{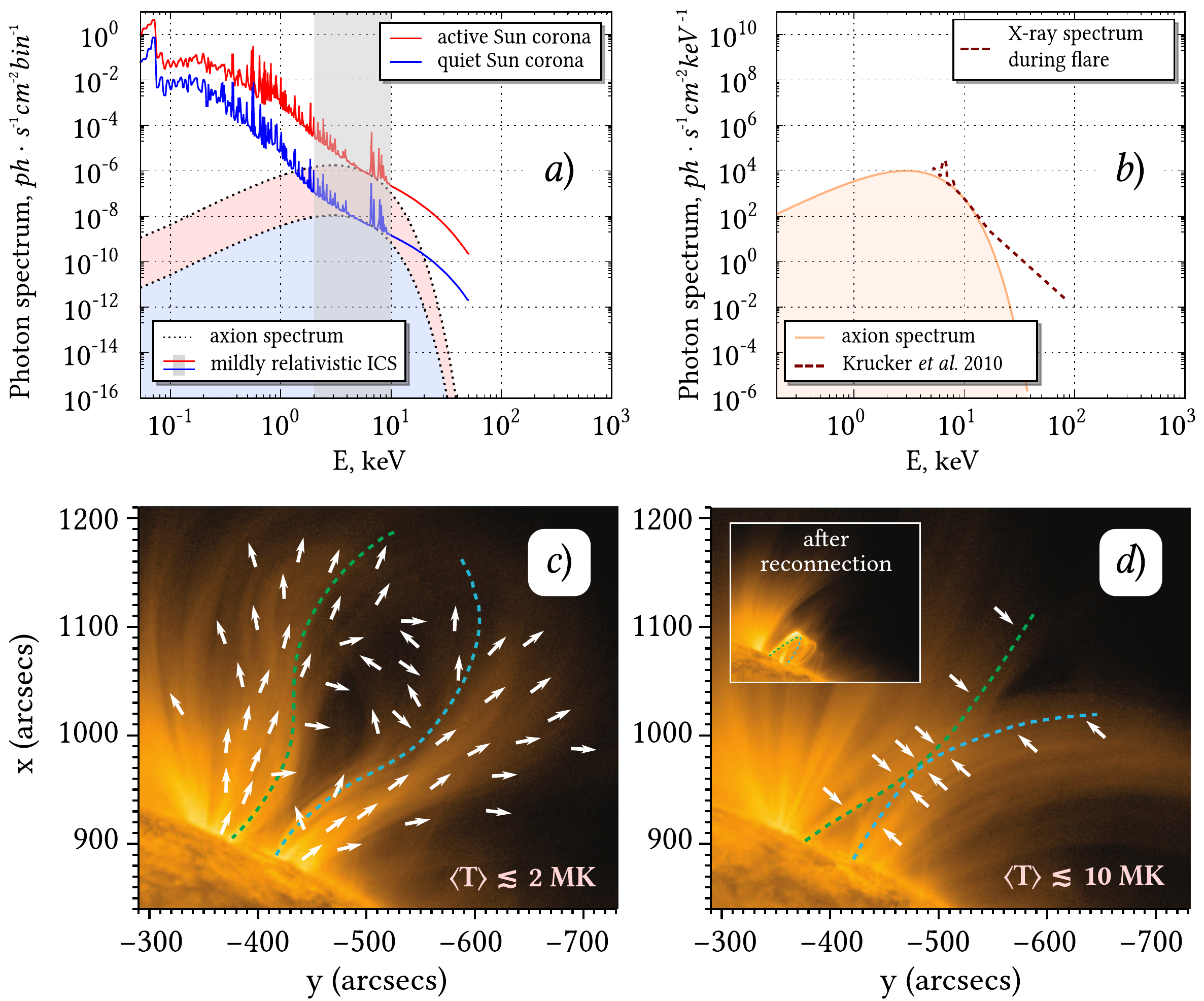}
\end{center}
\caption{\textbf{Top:} Synthesized photon spectra of the corona and flares.
\textbf{(a)} The coronal total luminosity of the Sun (see Table~2 in
\cite{Peres2000}) in the ROSAT/PSPC band ranges from 
$\approx 2.7 \cdot 10^{26} ~erg/s$ at minimum to 
$\approx 4.7 \cdot 10^{27} ~erg/s$ at maximum. Red line (ICS in the shaded
band) marks the solar maximum. Blue line (ICS in the shaded band) marks the
solar minimum. The theoretical photon spectra of the corona (see shaded band)
were calculated using inverse Compton scattering (ICS) of isotropic photons on
isotropic electrons (see analogous Fig.5~b in \citet{Chen2012}).
\textbf{(b)} Typical X-ray spectrum of the ``full sun'' (spatially integrated)
during flares. Spectral fitting during a hard X-ray peak (see the partially
disk-occulted solar flare of 2007 December 31 in \citet{Krucker2010}), 
originating from the electrons scattered on the photons of axion origin.
Although the experimental data on 11-year variations of the over-coronal flares
luminosity have not yet been obtained, we can describe the nature of e.g.
the minimum and maximum luminosity of the over-coronal flares. It may be calculated
via the product of the fraction of axions, originating from the solar core, and the
fraction of the flare-spots area (see analog \citet{Song2016}), and
thus obtain the luminosity fraction of photons of axion origin over the corona
(see e.g. analogous Eq.~(\ref{eq7.5-sep-13})).
\textbf{Bottom:} The origin of magnetic flux and the associated dynamic coronal phenomenon on the Sun \citep{Sun2015} due to the appearance of a flux of photons of axion origin. The blue and green dashed curves show the selected coronal loops representing the two lines of magnetic fields involved in the process (c) and (d). The sequence of extreme ultraviolet images clearly shows that two groups of oppositely directed and non-coplanar magnetic loops (c) gradually approach each other, sometimes (see \cite{Archontis2012,Sun2015}) forming a separator \citep{Parnell2010a,Parnell2010b} or a quasi-separator \citep{Aulanier2005} (d), causing a magnetic reconnection (inset). As a consequence, the free energy stored by photons of axion origin (white arrows in (c) and (d)) in the magnetic field (before, during and after reconnection) is quickly released and converted to the heating and volumetric plasma motions and acceleration of nonthermal particles.}
\label{fig-corona-compton}
\end{figure}

On the other hand, we believe that the main effective mechanism for heating the solar corona is  the emission of axions from the solar core with an energy spectrum with the maximum of about 3~keV and the average of 4.2~keV. These axions are supposed to convert into soft X-rays in very strong transverse magnetic field of an almost empty tube at the base of the convective zone (see Figs.~\ref{fig-lampochka}a and \ref{fig-lower-heating}). As a result, X-rays, passing through the photosphere at high speed and scattering in the Compton process, reach the transition region between the chromosphere and corona. The bulk of soft X-rays dissipates in the corona (see Fig.~\ref{fig-corona-compton}c,d).

The total power radiated by X-rays of axion origin is only about one millionth of the total solar luminosity,
so there is sufficient energy on the Sun to heat the corona.

It should be recalled here that the energy distribution of the emitted axions is far from being a blackbody spectrum because the spectrum of the incident photons is modulated by the frequency dependence of the cross-section. For the typical solar spectrum, the maximum of the differential flux of axions occurs at $E_a / T \approx 3.5$, whereas the average axion energy is $\langle E_a / T \rangle \approx 4.4$ \citep{Raffelt1986}. This means that the average energy of the photon of axion origin can generate a temperature of the order of $T_a \sim 1.11 \cdot 10^7 ~K$ under certain conditions of coronal substances \citep{Priest2000}, which is close to the temperature $T_{core} \sim 1.55 \cdot 10^7 ~K$ of the Sun core (see e.g. \cite{Fiorentini2001,Fiorentini2002,Bahcall1992,Bahcall1995}).

This raises an intriguing question about the paradoxical heating of the corona 
(see e.g. \cite{Edlen1943,Alfven1947,Parker1958,Gibson1973,Withbroe1977,
Parker1988,Klimchuk2006,Tomczyk2007,Aschwanden2007,Erdelyi2007,Golub2009,
Pontieu2011,Parnell2012,Reale2014,AschwandenEtAl2014,AschwandenEtAl2015,
AschwandenEtAl2016,AschwandenEtAl2017,Tan2014,DeMortel2015,Klimchuk2015,
Barnes2016,Morton2016}): how do the soft X-rays of axion origin heat the solar corona and flares  to a temperature of more than two and, correspondingly, three orders of magnitude higher than in the photosphere?

Below we show why the solar corona is so hot with the help of photons of axion origin.

\subsubsubsection{Solar axion and coronal heating problem solution}
\label{sec-coronal-heating}

It is known \citep{Priest2000} that traditionally in the atmosphere of the Sun there are three types of eruptions, such as coronal mass ejections, prominence eruptions and eruptive flares, and they are considered bound and are the result of the same physical process. Coronal mass ejections (CMEs) are large-scale mass ejections and magnetic flux from the lower corona to the interplanetary space. It is believed that they should be created by the loss of equilibrium in the structures of the coronal magnetic plasma, which causes sharp changes in the magnetic topology. A typical CME carries approximately $10^{23} ~Mx$ of flux and $10^{13} ~kg$ of plasma into space \citep{Priest2000}. During the active phase of the solar cycle, CME can occur more often than once a day. The intermittent appearance of a new magnetic flux from the convective zone (which originates from twist in flux tubes 
(see \cite{Archontis2012,Schmieder2014,Pontieu2014}) in the corona is the most important process for the dynamic evolution of the coronal magnetic field 
\citep{Galsgaard2007,Fang2010,ArchontisHood2012,ArchontisHood2013}, in which the rearrangement of the intersection of closed coronal lines of force causes the accumulation of coronal field stress. When the stress exceeds a certain threshold, the stability of the magnetic field configuration is broken and erupts (see e.g. Fig.~\ref{fig-corona-compton}c; Fig.~2 in \cite{Sun2015}). This model is called a storage model \citep{Yamada2010}, although it is unfortunately known that the question of how magnetic fields rise from the tachocline to the convective zone of the Sun and exit through the photosphere and chromosphere into the corona has not yet been resolved (see e.g. \cite{Archontis2008,Bushby2012,Schmieder2015}).

Nevertheless, this plausible explanation is associated not only with the appearance of a magnetic flux, but also necessarily with the appearance of photons of axion origin from an empty magnetic tube (see Figs.~\ref{fig-lampochka}a, \ref{fig-lower-reconnection}a, \ref{fig-lower-heating}a), in which, according to our theoretical and experimental observations (see Fig.~\ref{fig-corona-compton}a), the simultaneous occurrence of a magnetic flux and the flux of photons of axion origin in the outer layers of the Sun is the main mechanism of the formation of sunspots and active regions, being the integral part of the solar cycle, and also of the high energy release in the corona and flares. The physical solution of this problem will be shown below.

First, let us consider the work and method of \cite{Peres2000}, who use the X-ray image of the Sun, collected by Yohkoh/SXT, to obtain the corresponding full disk coronal distribution of the emission as a function of temperature -- $EM(T)$. From $EM(T)$ the spectrum for wide X-ray spectral range with the corresponding resolution is calculated. In this case, using the summation of the obtained photon spectra of the corona with the available spectral sensitivity of non-solar X-ray observatories, such as ROSAT and ASCA, a synthesized spectra is obtained from the histogram that corresponds to the focal plane of the photon spectrum of the solar corona.

Brief stages of the method of \cite{Peres2000} are as follows. From the Yohkoh/SXT images made using two different filters, the authors get the temperature $(T)$ and the emission measure $(EM)$ in pixels of the images, which then, by sorting out the emission measure values, allows the construction of a histogram of $EM$ vs. $T$.

Let us note that the broadband filters are used which provide reliable thermal diagnostics, since they depend weakly on details of atomic physical models, for example, on the presence of unknown or not very well known spectral lines for the choice of the prevalence of elements (see \cite{Reale2014}). The filter ratio map (see e.g. Fig.~6 in \cite{Reale2014}) provides information on the spatial distribution of temperature and the emission measure (see e.g. \cite{Vaiana1973}). The emission of the optically thin isothermal plasma when measured in the j-th filter passband is

\begin{equation}
I_j = EM \times G_j (T) ,
\label{eq07-114}
\end{equation}

\noindent
where $T$ is the temperature and $EM$ is the emission measure defined as

\begin{equation}
EM = \int \limits _{V} n_e ^2 dV = \int \limits _{\Delta T} DEM (T) dT ;
\label{eq07-115}
\end{equation}

\noindent
the distribution of the differential emission measure (see \cite{Rosner1978})

\begin{equation}
\frac{dEM (T)}{dT} \equiv DEM (T) = n_e ^2 \frac{dV}{dT} ~cm^{-3} K^{-1} ,
\label{eq07-116}
\end{equation}

\noindent
where $n_e$ is the electron density, and $V$ is the volume of the plasma. The ratio $R_{ij}$ of emission in two different filters $i$, $j$ is independent of the density, and depends only on the temperature function:

\begin{equation}
R_{ij} = \frac{I_i}{I_j} = \frac{G_i (T)}{G_j (T)} .
\label{eq07-117}
\end{equation}

The inversion of this relationship gives the temperature value, based on the isothermal assumption.

It should be noted that the calculation of $EM(T)$ differs from $DEM (T)$ mentioned in \cite{Peres2000}, because it is the line-of-sight-averaged emission measure distributed throughout the solar corona (but outside the peaks of flares! (see Sect.~5. Summary in \cite{Peres2000}) and differs also from the temperature distribution of the emission measure within one pixel's line of sight.

Following the steps in the method of \cite{Peres2000}, on the basis of the histogram of $EM$ vs. $T$, the emitted spectrum is synthesized using one of the available spectral codes, for example, the MEKAL spectral code (see \cite{Mewe1985,Mewe1986}). Converting the spectrum with device response, and also using the observational consequences (photon noise, distance from the source, exposure time, etc.), the authors get a realistic simulation of the solar corona, observed as if it were a nearby star. Hence, for these spectra, the standard analysis of data taking into account the response of the instrument can be used.

So the ``experimental'' photon spectra of the corona, synthesized with the MEKAL spectral code directly from the $EM (T)$ distribution (see Fig.~\ref{fig-corona-compton}a), agree with the solar cycle, since, on the one hand, the total coronal luminosity during the solar cycle 22 in the ROSAT/PSPC band ranges from $\sim 2.7 \cdot 10^{26} ~erg \cdot s^{-1}$ at minimum to $\sim 4.7 \cdot 10^{27} ~erg \cdot s^{-1}$ at maximum (see Table 2 in \cite{Peres2000}), and on the other, from the maximum to the minimum of solar activity, the temperature of the maximum emissivity measure (see $EM$ vs. $T$ from Eq.~(\ref{eq07-114})-(\ref{eq07-115})) shifts from $\sim 2 ~MK$ to $\sim 1 ~MK$ (see Fig.~3 in \cite{Peres2000}; Fig.~\ref{fig-corona-compton}c).

To compare the ``experimental'' photon spectra of the corona \citep{Peres2000} with the theoretical photon spectra of the corona, we consider the behavior of soft X-ray emissions (SXR) of axion origin (see Fig.~\ref{fig-corona-compton}a) generated inside the magnetic tube (see Figs.~\ref{fig-lampochka}a, \ref{fig-axion-compton}a, \ref{fig-lower-heating}a), which propagate through the convective zone and the photosphere to the surface of the transition region, between the chromosphere and the corona, where the bulk of the X-rays reaches the corona (see Fig.~\ref{fig-corona-compton}c).

The theoretical spectrum of the scattered photons in the corona ($photons$ per
$keV$ per $second$ per $electron$) are calculated using inverse Compton
scattering (ICS) of isotropic photons on isotropic (mildly relativistic)
electrons (see Fig.~3a in \citet{Chen2012};
\citet{Blumenthal1970,Tucker1975,Brunetti2000,Aharonian1981,Krucker2008}),
where the energy distributions of electrons with different spectral indices are
$\delta = 1$ (based on the kinetic energy of the electron, which has a
power-law form $\sim (\gamma - 1)^{-\delta}$ \citep{Chen2012}.
The energy of the incident EUV/SXR photon of axion origin is 4.2~keV
(Fig.~\ref{fig-corona-compton}a; see analogous Fig.~5 in \citet{Chen2012}) with
a photon density of $10^7 ~cm^{-3}$, respectively. The limits of the energy
spectrum of electrons are $\gamma = 10$ ($\approx 4.6 ~MeV$). The results are
normalized to one electron above $\sim 0.5~MeV$ as was done in Sec.~2.1 of
\citep{Chen2012}.

This means that the fraction of the photon spectra of the corona (see 
Fig.~\ref{fig-corona-compton}a), which are calculated using the inverse Compton
scattering (ICS) by means of isotropic photons on isotropic (moderately
relativistic) electrons, almost completely coincides with the product of the
total fraction of axions ($L_a / L_{Sun}$) originating from the core of the
Sun, and the fraction of the sunspot area ($P_{\gamma}$).

\begin{equation}
\frac{(L_{corona}^X)_{max}}{L_{Sun}} \sim 2.7 \cdot 10^{-6} = \frac{(L_a)_{max}}{L_{Sun}} \cdot (P_{\gamma})_{max} \sim 2.7 \cdot 10^{-6} ,
\label{eq7.5-sep-24}
\end{equation}

\begin{equation}
\frac{(L_{corona}^X)_{min}}{L_{Sun}} \sim 1.2 \cdot 10^{-8} \approx \frac{(L_a)_{min}}{L_{Sun}} \cdot (P_{\gamma})_{min} \sim 2.0 \cdot 10^{-8} ,
\label{eq7.5-sep-25}
\end{equation}

\noindent where $(L_{corona}^X)_{max} \sim 1.04 \cdot 10^{28} ~erg/s$
and $(L_{corona}^X)_{min} \sim 4.61 \cdot 10^{25} ~erg/s$.

A more accurate explanation is as follows. The Sun as an X-ray star describes
solar luminosity at minimum and maximum (see ROSAT/PSPC data in 
Fig.~\ref{fig06}d). Therefore, the fraction $(L_{corona}^X)_{max}/L_{Sun}$
describing the spectra of photons of axion origin, may be equal to the maximum
and minimum of corona luminosity (on the left: 
Eq.~(\ref{eq7.5-sep-24})-(\ref{eq7.5-sep-25})). On the other hand, the spectra
of the photons of axion origin may be calculated from the product of the axion
fraction and the sunspot area at the top of magnetic tubes. It is not
surprising that the left-hand side of 
Eqs.~(\ref{eq7.5-sep-24})-(\ref{eq7.5-sep-25})
depend only on moderately relativistic electrons, which interact with the
photons of axion origin near 4.2~keV, and produce a wide range of photons from the 
ultraviolet/soft X-rays to $\gamma$-energies, and this is in the corona of the
solar disk!

So on the basis of photons of axion origin, we obtain two significant results.
First, the solar luminosity of the corona is absolutely identical to the
luminosity of the photosphere (which is confirmed by 
Eqs.~(\ref{eq7.5-sep-18})-(\ref{eq7.5-sep-19}) and
(\ref{eq7.5-sep-24})-(\ref{eq7.5-sep-25})), since the number of photons of
axion origin, which exit through the magnetic flux tubes of the photosphere,
despite the change in their density, is almost equal to the number of photons
in the low-density ambient plasma of the corona (see e.g. Fig.~\ref{fig06}a).
As a result, the experimental ratio $L_{corona}^X / L_{Sun}$ (see 
Fig.~\ref{fig-corona-compton}a) is the same as the theoretical ratio 
$L_a / L_{Sun} \times P_{\gamma}$ (see e.g. 
Eqs.~(\ref{eq7.5-sep-18})-(\ref{eq7.5-sep-19}) and
(\ref{eq7.5-sep-24})-(\ref{eq7.5-sep-25})) in the minimum and maximum solar
luminosity.

Second, the energy distribution of the emitted axions is not like a blackbody
spectrum, since the spectrum of the incident photons of axion origin is
modulated by the frequency dependence of the cross section. As we have already
shown, for a typical solar spectrum, the maximum of the axion differential flux
occurs at $E_a / T \approx 3.5$, while the average axion energy 
$\langle E_a / T \rangle \approx 4.4$ \mbox{\citep{Raffelt1986}}. This means
that the energy of the average photon of axion origin can generate a
temperature of the order of $T_a \sim 1.11 \cdot 10^7 ~K$ (see e.g. 
Fig.~\ref{fig-corona-compton}a,c) under certain conditions of coronal
substances, which is close to the temperature $T_{core} \sim 1.55 \cdot 10^7~K$
of the solar core! Under some conditions, the photons of axion origin with very
powerful flares in the corona can partially generate energy,
which corresponds to a temperature of the order of 
$T_a \leqslant 10^8~K = 10~MK$ (see e.g. Fig.~\ref{fig-corona-compton}b,d).

\begin{figure}[tb!]
\begin{center}
\includegraphics[width=18cm]{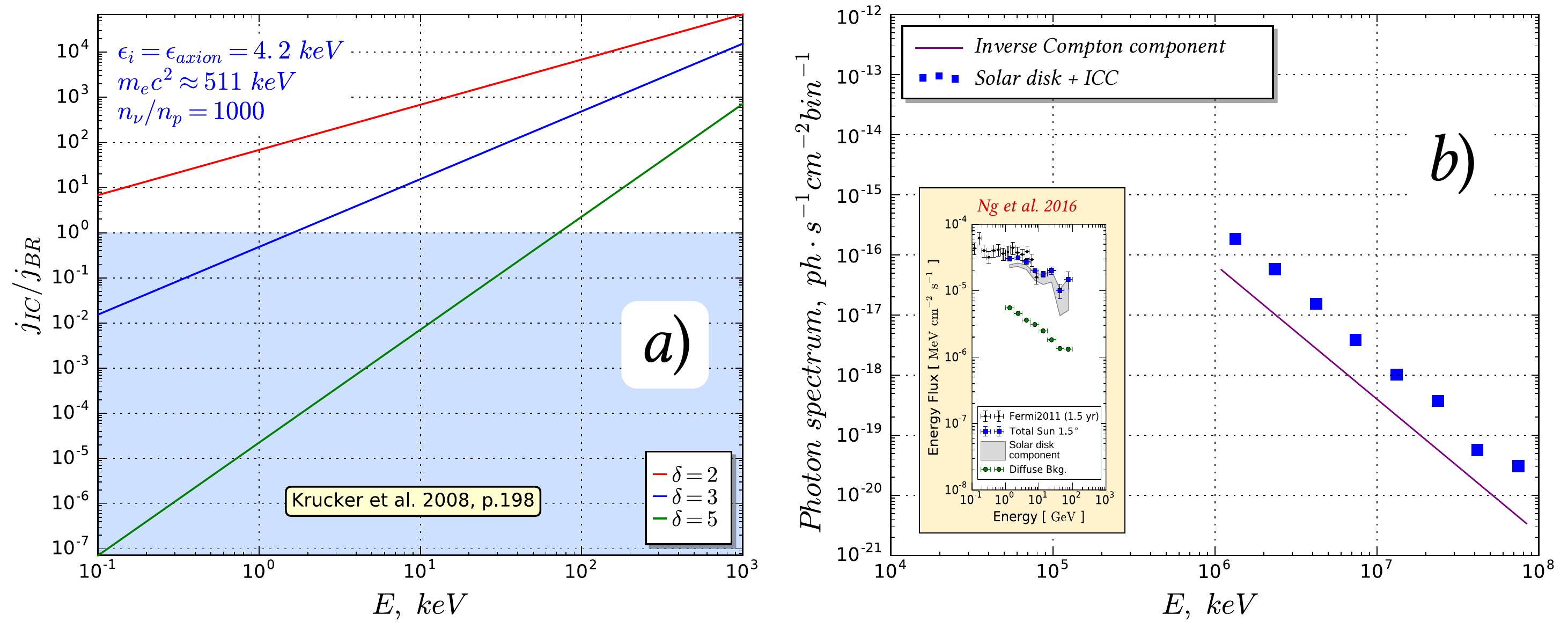}
\end{center}
\caption{Emission mechanism of ICS revisited. 
\textbf{(a)} Observing the photon energy $\varepsilon \ll m_e c^2$, we consider the region in which the electrons have the energy distribution $N( E ) \approx E^{-\delta}$ extending to arbitrarily high energies. Vertical shows the relative values of inverse Compton $(j_{IC})$ and bremsstrahlung $(j_{BR})$ contributions according to Eq.~(5) of Sect.~5.2 and Eq.~(7) of Sect.~5.3 from \cite{Krucker2008}. Inverse Compton radiation, with its harder photon spectrum, eventually dominates over bremsstrahlung. With the ambient coronal density of $10^9 ~cm^{-3}$, $n_\nu / n_p = 1000$ and this ratio may approach or exceed unity, in the 10-100~keV photon energy range, for the most likely hard energy distributions (e.g. $\delta = 2,3,5$).
\textbf{(b)} Energy spectrum of gamma rays from the Sun. Blue squares are the total solar gamma-ray flux (solar disk + IC) within (see inset) 1.5$^{\circ}$ of the Sun with only statistical error bars. Inset: Black dots are the solar-disk-only component from \cite{FermiLAT2011}; the gray band shows the solar-disk-only component found by \cite{Ng2016}. Green circles are the estimated diffuse background within 1.5$^{\circ}$ of the Sun.}
\label{fig-inverse-compton}
\end{figure}

As a result, according to our physical understanding, the coincidence of the theoretical and experimental photon spectra of the corona 
(see e.g. Eqs.~(\ref{eq7.5-sep-18})-(\ref{eq7.5-sep-19}) and
(\ref{eq7.5-sep-24})-(\ref{eq7.5-sep-25}))
is connected, on the one hand, with the appearance of the magnetic flux and simultaneously the flux of photons of axion origin in the outer layers of the Sun, and on the other hand, with the manifestation of the basic mechanism of formation of sunspots and active regions correlated with the solar cycle, which determines formation of the corresponding variation in the energy release in the corona and flares.

Thus, the next question is: ``What controls the solar cycle? Can axions and/or DM particles control the solar cycle?'' If they exist, how are the magnetic cycles of stars around a BH in the center of the galaxy controlled, for example, by modulating the density of ADM particles?

The first two questions seem to us understandable. Below is the idea of using axions and ADM particles to explain the problem of solar cycles.

\subsubsubsection{Is there a chronometer of dark matter hidden deep in the Sun core?}
\label{sec-sun-chronometer}

Here we consider the simplified physics of the chronometric model of DM, in which the modulation of the DM density within the solar interior affects the so-called solar modulation of axions. Let us show how the chronometric model of DM predetermines the relationship between helioseismological observables and predictions of solar models, which at the same time correlates with the modulation of magnetic sunspot cycles (see Fig.~\ref{fig06}d and \ref{fig-corona-compton}a) and anticorrelates with the cosmic ray (CR) intensity modulation here in the Solar System and, particularly, in the solar-disk component (see e.g. Figs.~6 and~7 in \cite{Ng2016}; see also Fig.~\ref{fig-inverse-compton}b).

We already know that the thermomagnetic EN~effect (see Sect.~\ref{sec-nernst}) is the basis of the modulation mechanism for the toroidal magnetic field in the tachocline (see Eq.~(\ref{eq06-14})),

\begin{equation}
\frac{B_{tacho}^2}{8 \pi} = n_{tacho} k T_{tacho}\, ,
\nonumber
\end{equation}

\noindent
by means of the condition (see Eq.~(\ref{eq06-08})) for the fully ionized plasma at $Z = 1$

\begin{equation}
T_{tacho}^{1/4} n_{tacho} = const .
\nonumber
\end{equation}

If we assume that the magnetic field in the tachocline changes in proportion to the magnetic field on the surface of the solar photosphere, for example, for the maximum and minimum of the 22$^{nd}$ solar cycle (see e.g. Fig.~1 in \cite{Dikpati2008}), then

\begin{equation}
(B_{tacho})_{max} = \langle B_{tacho} \rangle 
\frac{(B_{photosphere})_{max}}{B_{photosphere}} \approx 4592 ~T,
\label{eq07-117a}
\end{equation}

\begin{equation}
(B_{tacho})_{min} = \langle B_{tacho} \rangle 
\frac{(B_{photosphere})_{min}}{B_{photosphere}} \approx 3280 ~T,
\label{eq07-117b}
\end{equation}

\noindent
where the average magnetic field is $\langle B_{tacho} \rangle \approx 4.1 \cdot 10^7 ~G = 4100 ~T$ (see Eq.~(\ref{eq06-16})), the magnetic field of the photosphere ($B_{photospere}$) at the maximum is $2800 ~G = 0.28 ~T$, and at the minimum it is $2000 ~G = 0.2 ~T$ for the 22$^{nd}$ solar cycle (see \cite{Pevtsov2014}). Checking the same formulas for the input data of \cite{Bahcall1992}, which we now consider as averaged data, gave the density

\begin{equation}
(n_{tacho})_{max} \approx 1.902 \cdot 10^{23} cm^{-3} ,
\label{eq07-118a}
\end{equation}

\begin{equation}
(n_{tacho})_{min} \approx 2.381 \cdot 10^{23} cm^{-3} 
\label{eq07-118b}
\end{equation}

\noindent and temperature

\begin{equation}
(T_{tacho})_{max} \approx 3.2 \cdot 10^6 K ,
\label{eq07-119a}
\end{equation}

\begin{equation}
(T_{tacho})_{min} \approx 1.3 \cdot 10^6 K .
\label{eq07-119b}
\end{equation}

On the basis of the above-mentioned variations of the magnetic field (see Eqs.~(\ref{eq07-117a}, \ref{eq07-117b})), we showed earlier (Sect.~\ref{sec-ADM-SolData} and Fig.~\ref{fig-magtube-tilt})  the theoretical probability of the temporal variations of solar activity, that correlate with the flux of X-rays of axion origin (see Fig.~\ref{fig-lower-reconnection}); the latter being controlled by variations of the magnetic field near the overshoot tachocline.

Consequently, let us  recall that the axion mechanism of solar luminosity variations, which determines the effect of the ``channeling'' of X-rays of axion origin along magnetic tubes (see Fig.~\ref{fig-lampochka}a),
allows to explain the effect of practically complete suppression of convective thermal heating and, thus, understand the famous puzzling darkness of sunspots (see e.g. \cite{Rempel2011}), which simultaneously predetermine the physics of variations in solar cycles (see also Joy's law in Fig.~\ref{fig-magtube-tilt}c,d).

Hence, it is clear that the source of the cyclic variations of the initial magnetic field and, for example, the axions which are born by electron-nucleus collisions, $\gamma + (e^{-}, Ze) \rightarrow (e^{-}, Ze) + a$ (see e.g. \cite{Raffelt1986}), should be within the radius of the Sun core. On the other hand, it is known that helioseismology, the diagnostics of which depends on the temperature $(T)$, average molecular mass $(\bar{\mu})$ and their gradients, and consequently, the speed of solar sound as $\delta c_s / c_s \sim (1/2)[\delta T / T - \delta \bar{\mu} / \bar{\mu}]$ (see \cite{Christensen2002,Vincent2015a}), is unable to measure the temperature or density in this deepest central region below $0.2 ~R_{Sun}$ (see e.g. \cite{Garcia2007,Serenelli2016}), but can only indirectly estimate these parameters using known solar neutrino experiments. Since neutrinos are formed in several neutrino nuclear reactions of the proton-proton chain and carbon-nitrogen-oxygen cycle 
\citep{Bethe1939,Blanch1941,Bahcall1990,Gruzinov1998,Haxton2013,BOREXINO2014,
Haxton2014}, which take place at different radii of the Sun core (see e.g. \cite{Bahcall2006,LopesSilk2012}), in this case neutrinos, in particular, $^8 B$ neutrino fluxes (see \cite{Gando2012}) can provide an almost precise determination not only of locally averaged temperatures and densities, but also of their possible variations in the core, and as a consequence, of solar cyclic variations in the convective envelope (see e.g. $g$- and $p$-modes in \cite{Wolff2009}).

So after we understood the origin of the first-order flux of solar neutrinos, which was proclaimed a major breakthrough more than a decade ago (see e.g. \cite{Akhmedov2009,Fogli2015,Smirnov2016}), we can focus on one of the second-order terms, for example, the variation of solar neutrino fluxes due to the temperature and density variations in the core. Except for several works by 
\cite{Grandpierre1990,Grandpierre1996a,Grandpierre1996b,Grandpierre2000,
Grandpierre2010,Grandpierre2015,Grandpierre2005}, such physics is little known \citep{Wolff2010,Grandpierre2015}.

In order to find the physical connection between the solar activity and the nuclear processes of the solar core, Attila Grandpierre studied the possible relationship between the local thermonuclear instability and the physical conditions associated with energy and neutrinos produced in the solar core 
(see e.g. \cite{Gough1996,Morel1996,Bludman1996,Bludman1999,Wolff2009,
Wolff2010}). At the same time, the presence of thermonuclear micro-instabilities generated by magnetic instabilities (see \cite{Spruit2002,Braithwaite2017,Brun2017}) and, as a result, the formation of hot bubbles in the solar core \citep{Grandpierre1990,Grandpierre1996a,Grandpierre1996b} causes a significant deviation from the thermal equilibrium and changes the Maxwell-Boltzmann distribution of plasma particles, since the dynamic system is already far from thermodynamic equilibrium \citep{Grandpierre2000}. Such a complex dynamic Sun ceases to be a closed system, since the production of its energy is partially regulated by tiny external influences, such as planetary tides (see e.g. \cite{Javaraiah1995} and Refs. therein; 
\cite{Jager2005,Wilson2008,Scafetta2012a,Scafetta2012b,Cionco2012,
Charbonneau2013,Cionco2015,Cionco2017,Liu2016,Cionco2018}). Using a non-Maxwellian distribution, for example, the Tsallis distribution, which was applied by means of the Fokker-Planck equation kinetic approach (see \cite{Kaniadakis1996,Kaniadakis2001,Ribeiro2017}), such a modification of the system leads to the increase in the temperature of the solar core, which at a given solar luminosity can compensate for non-standard cooling, and thereby, through the generalized Tsallis statistics, maintain the temporal behavior of the neutrino flux (see \cite{Kaniadakis1996,Grandpierre2004,Grandpierre2005}) associated with a known estimate of the modulation of the solar neutrino flux in the deep interiors of the Sun (see e.g. \cite{Wolff2009,Khondekar2012,Sturrock2016}).

Omitting some of the interesting challenges of \cite{Grandpierre2015} -- What is the reason for the mixed core or, in other words, how much excess of temperature is needed for the bubbles to reach the solar shell from the core? What is the strength of the magnetic field in the Sun core? Is there a mass flow in the kernel? -- we will consider the fundamental problem of the existence of hot bubbles, not only within the solar core, but in general in the solar-stellar interiors
(see e.g. \cite{Grandpierre2005,Grandpierre2010,Grandpierre2015,Wolff2007,
Wolff2009,Wolff2010}).

First, it is not difficult to show that the gravitational energy of the Solar System, supplied by the planetary tides near the center of the Sun, is very small, and less than $10^{22} ~erg/s$ (see Table~1 in \cite{Grandpierre2004}; p.~18 in \cite{Grandpierre2015}). But the most important, though surprisingly simple, is the fact that the Sun orbital motion is a state of free fall; therefore, aside from very small tidal motions, the associated particle velocities do not vary as a function of position on or within the body of the Sun (see e.g. Fig.~1 in \cite{Shirley2006}). It is necessary to prevent the inappropriate use of rotational motion equations for particle modeling due to orbital rotation (an example can be found in Sect.~2 in \cite{Jager2005}) and the repetition of future errors of this type (see \cite{Shirley2006}). In this case, the solution of the fundamental physics of the interconnection between the orbital and rotational angular momenta of the Sun (see \cite{Shirley2006,Shirley2017a,Shirley2017b}), i.e. the solution of the principal stumbling block for dynamical spin-orbit coupling hypotheses is considered below.

Second, because the gravitational energy of the Solar System, supplied by the planetary tides near the center of the Sun, is very small, such small heating in the core cannot initiate the mass flow at all, but is dissipated in the form of thermal and Alfv\'{e}n waves (see \cite{Grandpierre2005}). Hence, we can understand why the famous physicists \cite{Wolff2010} incorrectly argue that planetary gravitational forcing can cause a mass flow inside the Sun that could carry fresh hydrogen fuel at deeper levels, including the solar core and, as a consequence, increase the rate of solar nuclear fusion.

\begin{figure}[tbp]
\begin{center}
\includegraphics[width=16cm]{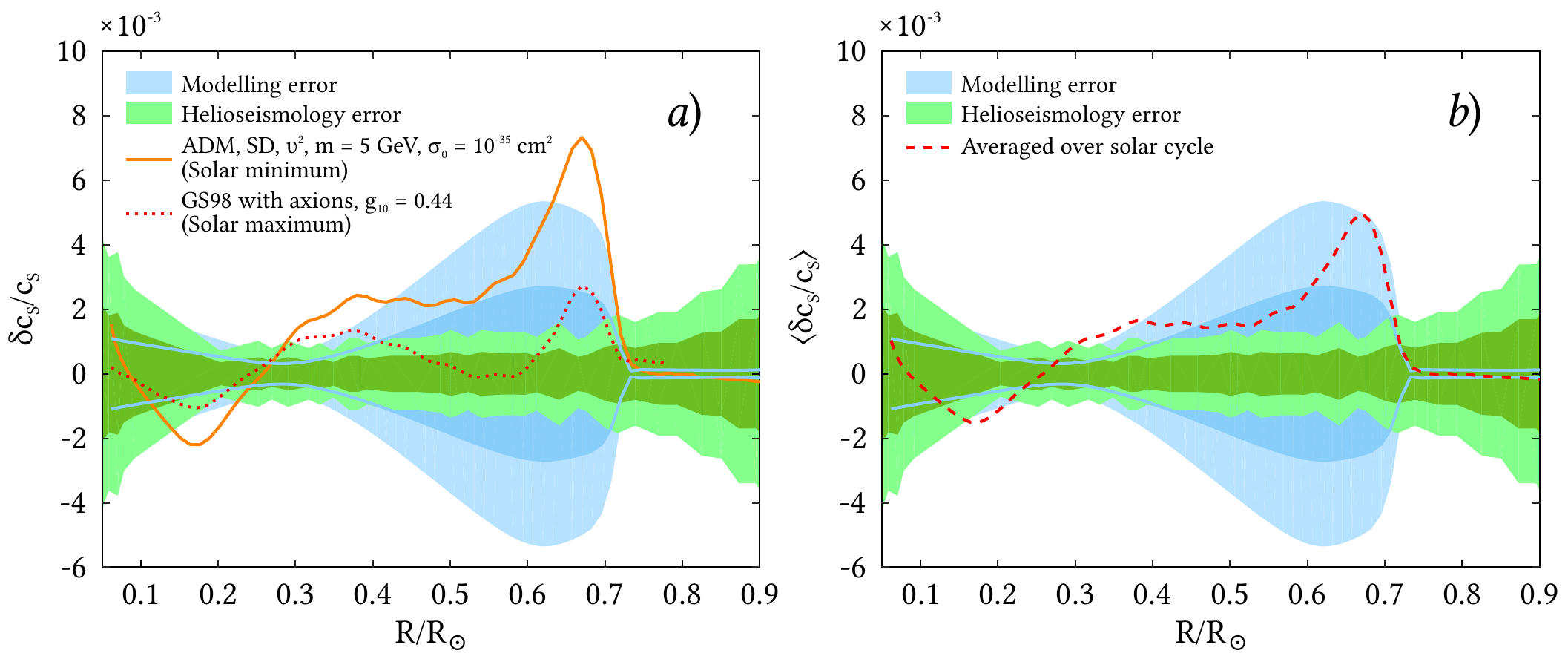}
\end{center}
\caption{\textbf{(a)} Deviation of radial sound speed profile (Sun − model)/Sun in the solar interior from values of \cite{Vincent2016} (solid line)  and \cite{Grevesse1998}
(GS98) with axions (dotted line) using two solar models. \cite{Vincent2016} show the best fit for a constant cross-section in black, and two best fits that are also allowed by CDMSlite data \citep{Agnese2016}, the spin-dependent (SD) $v^2$ and $q^2$ models (indicated by daggers in Table~1 in \cite{Vincent2016}. Red dots from values of GS98 with axions almost coincide with values of \cite{Vinyoles2015}. Colored regions indicate 1$\sigma$ and 2$\sigma$ errors in modeling \citep{Vincent2015a,Vincent2015b} (thick blue band) and on helioseismological inversions  \citep{DeglInnoccenti1997,Fiorentini2001} (thinner green band). 
\textbf{(b)} Deviation of radial sound speed profile from values averaged over the solar cycle, which is easily predetermined by (a) with the solar minimum and solar maximum, is in good agreement with the nonstandard solar models.}
\label{fig-helioseismology}
\end{figure}

The results of \cite{Wolff2010} may be sharply changed in the physically powerful and beautiful way if there is a mass flow, for example, in the solar interior, but an invisible mass flow of DM like ADM, that correlates with baryonic matter. They have remarkable properties (see Figs.~\ref{fig-inverse-compton} and~\ref{fig-helioseismology}a) for understanding the evolution of the Sun within the halo of DM of the Milky Way (see e.g. \cite{Spergel1985,Gould1987,Gould1990,GouldRaffelt1990a,GouldRaffelt1990b,LopesSilk2002,
LopesSilk2010,Frandsen2010,Lopes2014,Vincent2015a,Vincent2015b,Vincent2016}).

In the current cosmological scenario, we are interested in understanding the evolution of the Sun within the halo of DM. The DM problem is an empirical proof of the reality of physics outside the Standard Model \citep{Bertone2004}. Here we demonstrate that the existence of the weakly interacting ADM in our problem (see the fraction of WIMP-axion CDM in Fig.~\ref{fig05}a) is a motivation for experimental observations in which the density of DM is about 5 times greater than the density of visible matter (VM):

\begin{equation}
\Omega _{DM} = \Omega _{ADM} + \Omega _{axion} \approx 5 \Omega _{VM} ,
\label{eq07-120}
\end{equation}

\noindent
where $\Omega$ as usual denotes the density of the given component with respect to the critical density (see \cite{Petraki2013,Zurek2014}). At the same time, multiple-GeV ADM can explain most anomalies if (and only if -- see, for example, \cite{Vincent2015a,Vincent2015b}) the interaction force between DM and nucleons depends on the momentum exchanged between them.

Unlike weakly interacting massive particles (WIMPs), the motivation for
asymmetric dark matter (ADM, see \citet{Petraki2013,Zurek2014}) comes from the
baryonic sector of the Standard Model and is based on the initial asymmetry between DM and anti-DM to produce the correct relic abundance. As a consequence, these ADM conditions can lead to the absence of self-annihilation today (see Fig.~\ref{fig-helioseismology}a), which allows large amounts of ADM to accumulate in the center of the Sun due to their capture by its gravitational field \citep{Gould1987}. In this case, ADM captured during the evolution of the Sun is in the central part of the nuclear-reacting core of the Sun, where the radius of the ADM core is inversely proportional to the square root of the ADM mass \citep{Spergel1985}. When the weakly interacting DM particles absorb energy in the hottest, central part of the nucleus, they then travel to a cooler, more peripheral area, where before the scattering again and again additionally accumulate energy \citep{GouldRaffelt1990a}. This reduces the contrast of temperature across the core region and reduces the central temperature by a few percent (see e.g. \cite{LopesSilk2002,LopesSilk2010,LopesSilk2012,Lopes2014}).

It is known that the strong dependence of the rate of nuclear reactions on temperature allows the use of neutrino nuclear reactions, such as $pp$, $pep$, $^7 Be$ and $^8 B$, which occur at various locations in the nuclear region, including the radial temperature distribution in the nucleus. When the weakly interacting DM absorbs energy, it cools the solar core and, of course, lowers the neutrino flux, which is a very sensitive probe of central temperature (see e.g. \cite{LopesSilk2012}). On the other hand, helioseismic observations cannot directly determine the temperature inside the Sun core, since no one can determine the gas temperature from knowledge of the sound velocity if the chemical composition of the core is not known (see \cite{Garcia2007}). In this case, neutrino and helioseismic information thus complement each other, especially when the presence of particles such as ADM in the Sun, affecting the heat transfer in the solar interior and, consequently, neutrino fluxes, can simultaneously solve a very difficult ``solar composition problem'' or otherwise called ``solar abundance problem'' (see e.g. 
\cite{VillanteRicci2010,Villante2010,Frandsen2010,Taoso2010,Cumberbatch2010,
LopesSilk2012,Villante2014,Vincent2015a,Vincent2016,Vincent2015b,Serenelli2016}), as well as the surprising problem of the variation of luminosity 
(see e.g. \cite{Sheldon1969,Spergel1985,Faulkner1985,Gilliland1986,
Wilson1987,GouldRaffelt1990a,GouldRaffelt1990b,Wolff2007,Wolff2009,Wolff2010,
Khondekar2012}).

Despite the tremendous success in predicting neutrino fluxes and describing the bulk structure of the Sun, the Standard Solar Model (SSM) still cannot reproduce the key observables related to helioseismology. In order to bring the solar simulations into line with these precise observables (see \cite{Vinyoles2017,Serenelli2016} and Refs. therein), we try the solutions beyond the SSM with the help of e.g. ADM (see e.g. \cite{Vincent2016}).

At first, it became clear that the accurate solar models, which include the capture and transfer of heat from standard spin-dependent or spin-independent ADM, can be constructed to satisfy, for example, the solar radius, age and luminosity \citep{Taoso2010,Cumberbatch2010,LopesSilk2012}. Further, the improvement in the quality of the solar model was advanced by \cite{Vincent2015a,Vincent2015b} who found that a light asymmetric particle with momentum and velocity-dependent interactions could be captured in a sufficiently large amount on the Sun and conduct heat in such a way that helioseismic observables could be brought in full agreement with the data. Although this led to the decrease in the predicted neutrino fluxes, a significant improvement in sound velocity and convective zone depth was sufficient to produce a noticeable improvement over the SSM of 6 standard deviations and a potential solution to the solar composition problem. However, only after the limits of direct detection of recoil spectra for DM particles have been well defined (see the experiments of CRESSTII in \cite{Angloher2016} and CDMSlite in \cite{Agnese2016}), the best solution is spin-dependent $v^2$ scattering with a reference cross-section of $10^{-35} cm^2$ (at a reference velocity $v_0 = 220 ~km \cdot s^{-1}$), and a DM mass of about 5~GeV (\cite{Vincent2016}).

Here arises a question of how the deviation of the radial sound speed profile
$\delta c_s / c_s$ in the solar interior depends on the values obtained from
helioseismological data for the SSM and the ADM model.

It can be shown that the deviation of the radial sound speed profile
$\langle \delta c_s / c_s \rangle$ from values averaged over the solar cycle
(Fig.~\ref{fig-helioseismology}b), determined by the solar minimum (model of
ADM with axions in Fig.~\ref{fig-helioseismology} (see analogous Fig.~6 in 
\cite{Vincent2016})) and solar maximum (GS98 with axions in 
Fig.~\ref{fig-helioseismology}a (see analogous Fig.~5 in \cite{Vinyoles2015}
and Fig.~3 in \cite{Schlattl1999})), demonstrates a good agreement with the
non-standard solar model only.

It is clear from the fits (see e.g. the orange line in Fig.~\ref{fig-helioseismology}a; see also \cite{Vincent2016}) that the obtained nonstandard solar model is good, but not the best, and it may never become the best if the solar abundance problem is not really solved as a consequence of the fundamental physics of the Sun. In this sense, as very deeply and interestingly expressed by \cite{Vincent2016}, “It is clear from the fits we present here that the addition of heat transport by DM in the Sun can improve the SSM significantly. The improvement is well beyond what one would naively expect from adding just two degrees of freedom. Even if DM is not the solution, the fact that the true solution can be accurately mimicked by DM would seem to indicate something fundamental about the physical processes behind the solution...”

Hence, it becomes clear that instead of the averaged deviation of the radial
sound speed profile $\langle \delta c_s / c_s \rangle$ over the solar cycle, 
which is consistent with the simplified nonstandard solar model 
(Fig.~\ref{fig-helioseismology}b), it is necessary to model the exact
variations of the nonstandard solar model, which, including heat capture and
transport from standard spin-dependent ADM, can be constructed not just to
satisfy the solar radius, age and luminosity (see e.g. 
\cite{Taoso2010,Cumberbatch2010,LopesSilk2012}), but by varying the ADM density
corresponding to the variations of the solar radius (see 
\cite{Meftah2015,Meftah2016,Meftah2017,Rozelot2015,Rozelot2016}) and luminosity
of the Sun (see \cite{Willson1988,Foukal2006,Ulrich2010,Chapman2013}) and, as a
consequence, modulations of solar abundances that have low metallicity at the
Sun maximum (see e.g. \cite{Asplund2009}) and high metallicity at the Sun
minimum (see e.g. \cite{Grevesse1998}), against different observables including
solar neutrinos, surface helium, the depth of the convective shell, and the
sound speed profile (see \cite{Vinyoles2017}).

Surprisingly, the existence of ADM can be a remarkable result of the fact that
\citet{Ng2016} (Fig.~\ref{fig-inverse-compton}b), who for the first time
recently discovered a significant time variation in the gamma radiation flux of
the solar disk, believes that anticorrelation with solar activity may be the
result of the interaction of ADMs arising from the solar core with cosmic rays
in the solar atmosphere.

Curiously enough, we have understood that the fundamental physics of the Sun, as if recalling a forgotten, but physically profound question, raised long ago by R.H.~Dicke, ``Is there a chronometer hidden deep in the Sun?'' \citep{Dicke1978,Dicke1979,Dicke1988}, suggests the possibility of the existence of ADM (as a result of the energy conservation law with the Galactic frame velocity, density and dispersion) in the Sun core, with which, for example, variations in the ADM density and, as a consequence, variations in the neutrino flux, the solar cycle, solar luminosity, sunspots and other solar activities, seem to be paced by an accurate ``clock'' inside the Sun.

The question is very simple: If the ADM density variation controls the ``clock''
inside the Sun, then what is the physics behind this process?

\sethlcolor{lightred}
\section{How the ADM density variation around the black hole controls the ADM density variation inside the Sun}
\label{sec-dark-matter}

Let us try to find an answer to the important, yet unconventional question: is there an observable connection between the 11-year variations of ADM density in the solar interior and the periods of S-stars revolution around the SMBH at the center of the Milky Way (see e.g. \cite{Genzel2010,Dokuchaev2015,Mapelli2016})?

In order to answer this question, let us first consider all unexpected and intriguing implications of the 11-year modulations of ADM density in the solar interior and ADM around the BH. We will start with a short overview of our current understanding of the Sun and its mysterious dynamo action, which should be supplemented with the more consistent comparison with other stars. Hopefully this will lead us to the future understanding of the synergy between studies of the stars and the Sun \citep{Thompson2014}.

Let us recall that the major puzzle about the dynamo is the fact that “for the understanding of the solar dynamo, the Sun is not enough” \citep{Judge2014}. The attempts to understand this behavior often refer to the combination of the differential rotation, convection and meridional convection flow as the cause of the global field modulation (solar activity cycle) through the magnetic dynamo (see e.g. \cite{Rempel2006}). It is known however that the existing models of solar dynamo suffer from a number of disadvantages. Some of them are highly idealized mathematical or computational models which may explain certain principles but do not correspond to the real behavior of the Sun \citep[see][]{Thompson2014}. Others may contain some special parameters which correspond to the observed large-scale behavior (like the sunspot number of butterfly diagram), but fail to provide any predictions. As a result, none of them describes the features of the Sun and Sun-like stars correctly, leaving a lot of room for the further grounds and explanations (see e.g. \cite{Judge2014,Thompson2014,Metcalfe2017,Wargelin2017}).

It is also known (see e.g. \cite{Vinyoles2015}) that the Standard Model of particles is not a complete theory, and unfortunately it does not determine the existence of the solar cycle. Indeed, some extensions are necessary to address the open questions of fundamental physics like: the nature of DM; matter-antimatter asymmetry in the Universe; origin of neutrino masses; strong CP-violation problem. As noted by \citet{Vinyoles2015}, 
``In order to solve these problems, physics beyond the Standard Model is needed and the existence of new particles or non-standard properties of known particles are generally invoked'', e.g. ADM and axion.

We have already examined the fundamental solutions of the problems related to the existence of DM -- ADM, which comes from the baryon sector, and axion -- in the solar interior:

\textbf{1. The problem of dynamo, sunspots and coronal heating.} 
In the beginning, we note the BL~mechanism and the holographic BL~mechanism as components of our model of solar antidynamo. In contrast to the component of the solar dynamo model (see Fig.~\ref{fig-solar-dynamos}b), the BL~mechanism, which is predetermined by the fundamental holographic principle of quantum gravity and, as a consequence, the formation of the thermomagnetic EN~effect in the tachocline  (see \cite{Spitzer1962,Spitzer2006,Rusov2015}; see also Sect.~\ref{sec-nernst}), emphasizes that this process is associated with the continuous transformation of toroidal magnetic energy into poloidal magnetic energy, but not vice versa. On this basis, the general theory of a buoyant magnetic tube is constructed. The tube is born on the boundary between the tachocline and the overshoot layer. This means that the existence of an anchored flux tube with $10^7 ~G$ in the overshoot tachocline is a consequence of the fundamental properties of the holographic principle of quantum gravity, one of which (unlike the known dynamo action!) generates a strong toroidal field in the tachocline by the BL holographic mechanism. Hence, the final problem is a generalized problem of the antidynamo model of the MFTs, where the nature of both effects (the EN~effect and holographic BL~mechanism), including magnetic cycles, manifests itself precisely in the existence of modulation of ADM and, as a consequence, the solar axion in the Sun interior.

In Sect.~\ref{sec-ADM-SolData} we showed that the supposition of the periodic ADM variations in the solar interior, and consequently the axion flux, became possible because of the theoretical correlation of the time variation of the sunspot activity with the flux of the X-rays of axion origin in magnetic tubes (Fig.~\ref{fig-lower-heating}a) up to the photosphere (see Sect.~\ref{sec-channeling}) and in the corona (Fig.~\ref{fig-corona-compton}). Though the most remarkable result of the ADM variations in the solar interior is the fact that they can serve as a fundamental solution (see Sect.~\ref{sec-osc-parameters}) of the coronal heating problem (see Fig.~\ref{fig-corona-compton}) and an intriguing assessment (see the hint of Fig.~\ref{fig-helioseismology}b) of the deviation of radial sound speed profile from the values averaged over the solar cycle, which is easily explained by the solar minimum and maximum, providing a good agreement with non-standard solar models.

The key point here is that the magnetic cycles of sunspots (see Figs.~\ref{fig-axion-compton}, \ref{fig-lower-heating}a and Fig.~\ref{fig-lampochka}a) are the consequence of the effect of virtually empty MFTs, reaching the photosphere and having no influence of the lower layer reconnection near the tachocline at the first stage (see Figs.~\ref{fig-lower-reconnection}a, ~\ref{fig-lower-heating}a). Surprisingly, this means that the magnetic sunspot cycle does not depend on any magnetic dynamo! When dynamo does appear (the so-called reconnecting dynamo (see \cite{Baggaley2009} and e.g. Fig.~\ref{fig-upper-reconnection})), it cuts off the MFTs at the tachocline via the reconnection, thus leading to the sunspot volatilization from the solar surface (see Fig.~\ref{fig-lower-reconnection}b,c).

\textbf{2. A complete theory of the Sun with ADM and axion.} Earlier in Sect.~\ref{sec-ADM-SolData} we showed that the deviations of radial sound speed profile (Fig.~\ref{fig-helioseismology}a) from values of \cite{Vincent2016} with ADM (solid line) and \cite{Grevesse1998} (GS98) with axions (dotted line) using two solar models correspond to the solar minimum and solar maximum, and the average of them demonstrates the total correspondence of the non-standard solar model with the helioseismic data (see Fig.~\ref{fig-helioseismology}b).

Put simply, when the ADM density at the solar minimum causes the temperature and pressure decrease in the core, the matter density increases, which leads to the changes in several key nuclear reactions rate (see e.g. Sect.~2.1 in \cite{Vinyoles2015}). These changes directly impact the predictable solar neutrino fluxes (see detailed results for all neutrino fluxes in Table~6 in \cite{Vinyoles2015}). The most important fact here is that $^7 Be$- and $^8 B$-neutrino fluxes, which are very sensitive probes of the impacts of DM on the core of the Sun, weaken significantly. Meanwhile, these phenomena are simultaneously accompanied by a slight change in the $pp$ and $pep$ fluxes, in accordance with the requirement of reducing the solar luminosity by $\sim 10^{-3} ~L_{Sun}$ \citep{Willson1988,Sofia2004,Foukal2006,Ulrich2010,Chapman2013}, which is the main result of the solar cycle modulation data on the total solar irradiance \citep{Willson2014,Dewitte2016,Kopp2016,Coddington2016,deWit2017,Lean2018,deWit2018}.

And vice versa, when the lower ADM density during the solar maximum rises the temperature and pressure in the core, the decrease in matter density is accompanied by the higher fluxes of all sorts of solar neutrinos. The $^7 Be$- and $^8 B$-neutrino fluxes increase significantly together with the corresponding changes in $pp$ and $pep$ fluxes, associated with the solar irradiance growth to $L_{Sun}$  (see also \cite{Foukal2006,Frohlich2013,Pelt2017}).

One of the major indirect evidences of the 11-year variations in ADM density in the solar interior is the fact of almost complete coincidence of the helioseismological observations with the predictions of the new non-standard solar model (see Fig.~\ref{fig-helioseismology}b), which in addition to the  spectral determination of solar abundances takes into account the impact on the temperature, density and chemical composition of the solar core.

Let us remind that when the ADM is accumulated near the center of the Sun, it may provide an additional mechanism for the energy transfer from the solar core. The presence of such particles leads to a change in local solar luminosity of about 0.1\%, which is accessible for the solar seismic experiments. We use the typical ADM with spin-dependent $\upsilon ^2$ scattering with a reference cross-section $\sigma _0 \sim 10^{-35} \div 10^{-37}~cm^2$ and the DM mass of about $m_{\chi} \sim 10 ~GeV$ (see e.g. Fig.~6 in \cite{Vincent2016}, Fig.~4 in \cite{Rott2013} or our Fig.~\ref{fig-helioseismology}a).

As a result, it can be shown that the ADM density variations in the solar interior, which provide feedback to the solar cycle, would also be anticorrelated to the variations of $^7 Be$- and $^8 B$-neutrino fluxes (see analogous values in Table~6 in \cite{Vinyoles2017}), axions and photons of axion origin (see Sect.~\ref{sec-empty-tubes}) and magnetic fields (see thermomagnetic EN~effect, Sect.~\ref{sec-nernst}) near the solar core, tachocline or photosphere of the Sun. Despite the small population of DM (less than $\sim$1 DM particle per $10^{10}$ baryons, see e.g. \cite{Vincent2016}), these little corrections to the temperature gradient may have a serious impact on the Sun. They affect the solar structure itself -- including the speed of sound $c_s (r)$ and the convection zone radius $r_{CZ}$ -- and the neutrino fluxes from thermonuclear processes because of their strong dependence on the temperature in the core.

\begin{figure*}[tbp]
  \begin{center}
    \includegraphics[width=17cm]{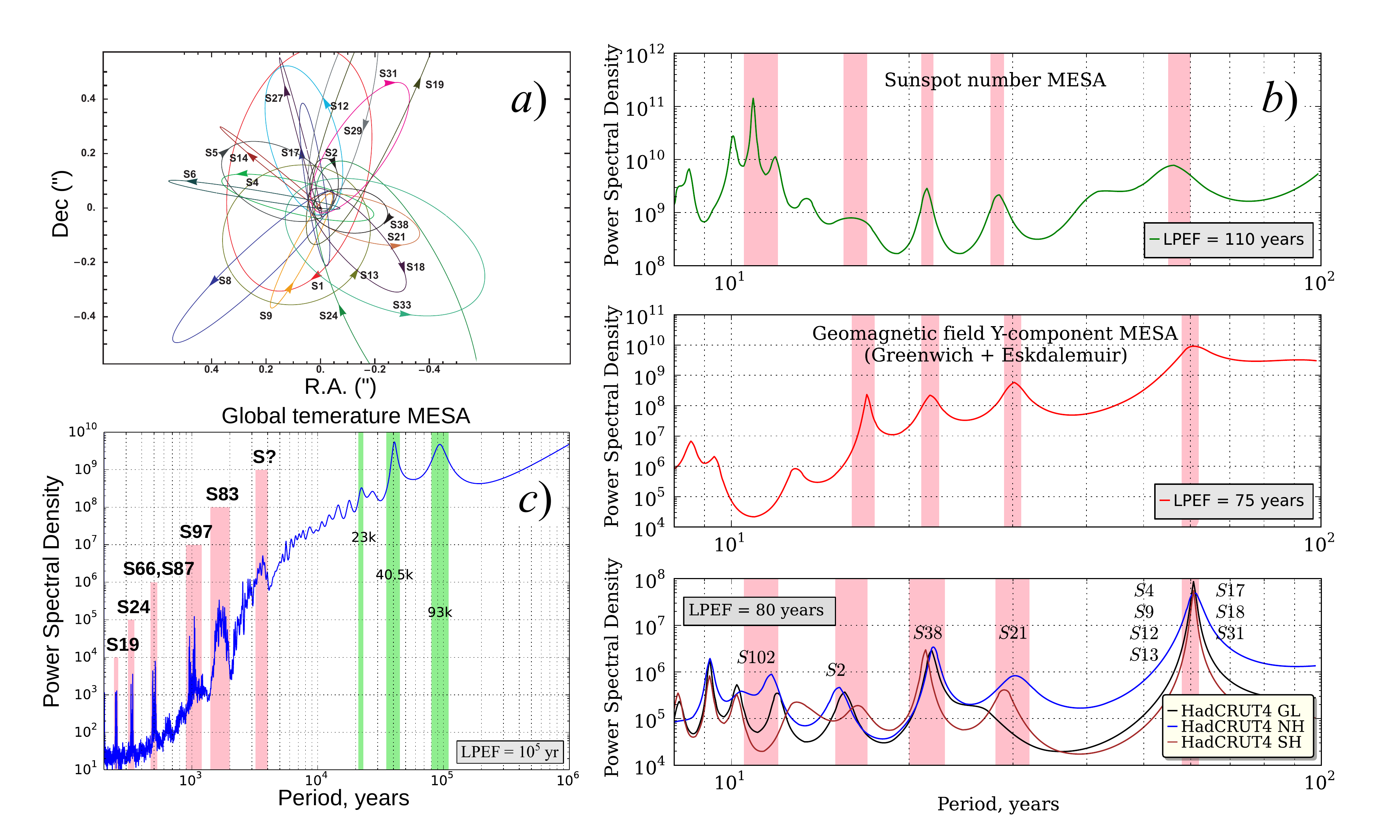}
  \end{center}
\caption{\textbf{(a)} Stellar orbits at the GC in the central arcsecond (declination Dec(”) as a function of time for the stars and red ascension R.A.(”)) \citep{Gillessen2009,Genzel2010}. The coordinate system is chosen so that Sgr A* (the SMBH with the mass $\sim 4.3 \cdot 10^6 ~M_{Sun}$ ) is at rest. 
\textbf{(b)} Power spectra of the sunspots (1874-2010 from Royal Greenwich Observatory), geomagnetic field Y-component (from Greenwich and Eskdalemuir observatories, \cite{WDC2007}) and HadCRUT4 GST (1850-2012) (black), Northern Hemisphere (NH) and Southern Hemisphere (SN) using the maximum entropy spectral analysis (MESA); red boxes represent major astronomical oscillations associated to the major heliospheric harmonics associated to the orbits of the best known short-period S-stars (S0-102, S2, S38, S21, S4-S9-S12-S13-S17-S18-S31) at the GC \citep{Gillessen2009,Genzel2010,Gillessen2013,Gillessen2017} and to the solar cycles (about 11-12, 15-16, 20-22, 29-30, 60-61 years).
\textbf{(c)} Power spectra of the global temperature as reconstructed by \cite{Bintanja2008}; red boxes represent the major astronomical oscillations associated to the major heliospheric harmonics associated to the orbits of the best known non-short-period S-stars (S19, S24, S66-S87, S97, S83, S?) at the GC \citep{Gillessen2009,Genzel2010} and to the solar cycles (about 250, 330, 500, 1050, 1700, 3600 years); green boxes represent the major temperature oscillations presumably associated with the variations of the Earth orbital parameters: eccentricity ($\sim$93~kyr), obliquity ($\sim$41~kyr) and axis precession ($\sim$23~kyr).}
\label{fig-stellar-orbits}
\end{figure*}

\textbf{3. ADM modulation and periodic cycles of S-stars at the GC.} On basis of the experimental data (see Fig.~\ref{fig-stellar-orbits}) on the oscillations of the sunspot number, the geomagnetic field Y-component and the global temperature we explicitly demonstrate that their periods coincide with revolution periods of S-stars orbiting a SMBH at the GC of the Milky Way. It is absolutely obvious that such a fine coincidence cannot be random. Then the next quite natural question arises: how do the solar and terrestrial observables ``know'' about motion of S-stars? And a hypothesis inevitably comes to mind: the ``carrier'' is none other than DM. More specifically, S-stars can modulate DM flows in our galaxy and, consequently, cause variations of DM space and velocity distributions, in particular, at the Sun and Earth positions. Further, these variations may cause the corresponding variations of the Solar System observables by means of some mechanism, e.g. the interaction of DM particles, which correlate with baryonic matter, with the cores of the Sun and the Earth. Such a probable mechanism is a subject of our Sect.~\ref{sec-who-generates}. Here our aim is to stress that the available experimental data indicate the frequency transfer from the center of our galaxy to the Solar System. This fact can serve as an indirect evidence of the proposed hypothesis that DM plays the role of the variations carrier.

Among 19 S-stars (see Table 1 of \cite{Gualandris2009}), the most precisely measured properties are those of S2: it is bright (14.2 in the K-band) and has a short orbital period of 16.2 years. According to \cite{Boehle2016}, they have been able to track its motion since Keck observations of the GC began in 1995, so that the observations now cover more than one full orbit of this star. In addition, when the new analysis technique is combined with the first complete re-reduction of Keck GC speckle images using speckle holography, it will be possible to track the short-period star S38 (K-band magnitude = 17, orbital period = 19 yr) through the speckle years (see Fig.~\ref{fig-stellar-orbits}a,b).

On the other hand, it is known that the periods of 19 S-stars at the GC lie in the range $11.5 ~yr \leqslant P_{S-star} \leqslant 400~yr$ (see \cite{Gillessen2009,Genzel2010,Gillessen2013,Gillessen2017}). The orbits of 19 remaining S-stars do not lie in the disc: they are consistent with a random distribution in space (see Fig.~\ref{fig-stellar-orbits}a). The period uncertainties for $11.5 ~yr \leqslant P_{S-star} \leqslant 60~yr$ are rather low \citep{Gillessen2009} because of the high measurement precision of Keck observations of the GC that began in 1995 (see e.g. Fig.~5 in \cite{Boehle2016}) and number $\sim 22$ year points to 2017. This means that the estimates for 13 of 19 S-stars at the GC give the accurate period values for $\sim$70\% of S-stars. They can also serve as an evidence for the nonrandom coincidence of these periods with the cycles of e.g. sunspots, geomagnetic field and temperature on the Earth (see Fig.~\ref{fig-stellar-orbits}b,c). The longer orbital periods, $60 ~yr \leqslant P_{S-star} \leqslant 400~yr$  (see \cite{Gillessen2009,Genzel2010,Gillessen2013,Gillessen2017}), which currently have high uncertainties, may be improved to 2030-2050 thanks to the higher number of yearly points measured for $\sim$95\% of S-stars.

Let us consider the question of how the ADM density variations, and ADM luminosity variations $L_{\chi}$, work as a “clock” regulating the tempo of the solar cycle. As we understand, Dikce's clock (Sect.~\ref{sec-sun-chronometer}) is a consequence of the ADM density modulation, which correlates with the baryonic matter, in the Sun and in S-stars \citep{Ghez2008,Gillessen2009,Gillessen2013} orbiting the SMBH
(of the radius $\leqslant 0.0024 pc$, see e.g. Fig.~\ref{fig-stellar-orbits}a and \cite{Gillessen2017}) in the Milky Way center  ($\approx 4 \cdot 10^ 6 M_{Sun}$) (see \cite{Genzel2010}), where the capture and annihilation of ADM occur in punctuated stages, clearly correlated with the orbital period (see \cite{Scott2009}). The S-star cluster lying between the BH and the clockwise disk, is one of the most mysterious components of the GC: most of the S-stars are early-type stars and could not form in situ, with a pericenter so close to the SMBH (this is the so-called “paradox of youth” \citep{Ghez2003}). Among all S-stars ($\leqslant 30$ \citep{Mapelli2016}) the closest one is S102, which has a period of only 11.5 years \citep{Meyer2012}.

A striking coincidence of revolution periods of S-stars orbiting a SMBH at the GC of the Milky Way and oscillation periods of such solar and terrestrial observables as the sunspot number, the geomagnetic field Y-component and the global temperature is established on the basis of the corresponding experimental data. Rejecting randomness of this discovered coincidence, we put forward a hypothesis that modulation of DM flows in the Milky Way by the S-stars is responsible for such a frequency transfer from the GC to the Solar System.

Let us estimate the modulations of the ADM around the BH and their possible experimental observations.

\subsection{Active galactic nuclei variability as a trigger of the matter accretion onto a supermassive black hole}
            
One of the signature properties of AGNs is their variability. AGNs are the most powerful sources of constant light in the Universe (with the bolometric luminosities as high as $10^{48} ~erg/s$ or $10^{14} L_{Sun}$ (see e.g. \cite{Kozlowski2016a} and Refs. therein)). Their variability during the periods of months to several years about 10\% of the total light must be enormous. There are numerous evidences that AGNs are fed from the matter accretion onto SMBHs (\cite{Salpeter1964,LyndenBell1969,Shakura1973,Soltan1982,Rees1984,Richstone1998,Silk1998}).
The common picture of an AGN is that the material is pulled by the supermassive ($\sim 4 \cdot 10^6$ solar masses) BH in the center of a galaxy. Due to angular momentum conservation, this material forms the accretion disc around the BH. Some part of the material streams towards the BH boundary and accretes, and another part is accelerated by strong magnetic fields and forms the relativistic jet perpendicular to the disc (\mbox{Fig.~\ref{fig-accretion-disk}a}).

For example, the chromatic microlensing of the strongly gravitationally lensed AGN yields the sizes ($\sim$1~AU in the optical band) and temperature profiles of the accretion disks \citep{Czerny1994,Kochanek2004,Dai2010,Morgan2010,Blackburne2011,Blackburne2014}. The masses of SMBHs in the local Universe are known to be related to some properties of their home galaxies  \citep{Magorrian1998,Marconi2003,McConnell2013,Kormendy2013,Lasker2014}. The physical mechanisms responsible for such intense accretion episodes in galactic nuclei and their relation to the cosmological evolution of the galaxies are still unknown.

Since the discovery of AGNs numerous studies of their variability appeared, both theoretical and observational. Now it is known that AGNs are the variable sources on all wavelengths (see e.g. \cite{Kozlowski2017} and Refs. therein), and the accretion disk instability simulation \citep{Kawaguchi1998} demonstrates the best match to the observations in the optical band \citep{ChenTaam1995,VandenBerk2004,Kozlowski2016a}. The exact process driving the variability is unknown though. The typical AGN variability has the stochastic nature (see e.g. \cite{Kelly2009,Andrae2013,Zu2013,Kozlowski2016a,Kozlowski2016b}). It shows the flat spectrum in the low frequencies (white noise, PSD~$\propto \nu ^0$) and the red noise (PSD~$\propto \nu ^{-2}$) or even steeper dependence (PSD~$\propto \nu ^{-3}$) in the high frequencies (see e.g. \cite{Mushotzky2011,Kasliwal2015,Simm2016}).

An interesting semi-analytic model by \cite{Fanidakis2011}, which shows the formation and evolution of the galaxies in a CDM Universe, seems to be most plausible. According to \cite{Fanidakis2011}, the BH and galaxy formation models are coupled: during the evolution of the host galaxy, the hot and cold gas is added to the SMBH by the flows from the cooling gas halo, the disk instabilities and galaxy mergers. This provides the mass and spin of the BH, and the resulting accretion power regulates the consequent gas cooling and star formation. Next, the accretion flow is believed to form the geometrically thin cool disk if the accretion rate is higher than $0.01 \dot{M}_{Edd}$, and geometrically thick, radiation-inefficient hot disk if the accretion rate is lower than that. The origin of such dichotomy is yet unknown, but many studies suggest to make the first step towards the explanation why some AGNs fire prominent jets while others do not, and eventually to the explanation of the nature of central driver
of the accreting BH (see e.g. \cite{Falcke2004,Schawinski2015,Oh2015}). Understanding of the connection between galaxies with AGN, powered by matter accretion onto SMBHs at the centers of galaxies \citep{Silk1998}, and DM halo they reside in, is the key to the processes of initiation and
feeding of the BHs (see \cite{Fanidakis2013,Leauthaud2015}).

One of the most advanced ideas of the AGN initialization on the basis of galaxy interactions and disk instabilities is described in the paper by \cite{Gatti2016}.  The ultimate goal is to single out the key properties of the clustering in two modes, which can serve as the reliable probes for the dominating mechanism of SMBH feeding.

Among the remarkable results of \cite{Gatti2016} we would pick the most intriguing one. Their analysis suggests the presence of both a mild luminosity and a more consistent redshift dependence in the AGN clustering, with AGNs inhabiting progressively less massive DM halos as the redshift increases. In other words, less luminous AGNs are biased towards lower DM halo masses. At high redshift, the average halo mass sensibly moves towards lower values.

This means that these AGNs fed by the matter accretion onto the SMBHs, are observed with high redshift as quasars among the most luminous objects in the Universe. The brightest infrared galaxies really possess both the strong AGN activity and the star formation, which suggests these two phenomena are closely related. The details of whether AGN feeding and outflows drive star formation or vice versa remain unclear, and remain to be elucidated by future observations, especially at far-infrared and X-ray frequencies (see e.g. \cite{Silk2007}).

From here we derive our main assumption that the luminous AGN modulation, related to periodic variability, is determined by the modulation of matter accretion by the SMBH. The assumption about the modulation of baryon accretion is directly connected to the modulation of the DM density in the SMBH.

The Ockham's razor will strengthen such scenario if the DM density modulation succeeds in solving the problems related to the known observations. We discuss some of them below.

\subsubsection{AdS/CFT correspondence and the thermomagnetic Ettingshausen-Nernst effect}

In theoretical physics the anti-de Sitter -- conformal field theory correspondence, sometimes referred to as “the Maldacena duality” or “the gauge/gravity duality” (see \cite{Maldacena1999}), represents a hypothetical connection between two different types of physical theories -- the gravitation theory and the quantum field theory. The AdS spaces used in the quantum gravitation theory formulated in terms of the string theory or M-theory \citep{Becker2008} are on the one hand. The conformal field theories (CFTs), which are the quantum field theories including those similar to the Yang-Mills theories (see e.g. \cite{tHooft1972,tHooft2005}) describing the elementary particles, are on the other hand. Such equivalence is an example of the holographic duality -- a representation of the 3D space on the 2D photographic emulsion (see \cite{tHooft1993,Susskind1995,Hanada2014}). Thus, although these two types of theories seem to be very different, they are mathematically identical \citep{Maldacena1999,Gubser1998,Witten1998}).

In their ingenious paper \cite{Hartnoll2007} turn to the physics of a BH in the 3D AdS space, which carries both the electric and magnetic charges. This lets them derive the transport equations directly from the quantum field theory. They showed that these theoretical results apparently correspond to the known features of the thermomagnetic EN~effect – the perpendicular heat and charge fluxes in the presence of the magnetic field (see \cite{Spitzer1962,Spitzer2006,Wang2006,Hartnoll2007,Rusov2015}).

This is however darkened by the fact that the AdS/CFT has been proven to work in certain special cases only, such as the important theory of the EN~effect near the quantum phase transitions in the condensed matter, or the dyonic BHs (see \cite{Hartnoll2007,Hartnoll2016}). This correspondence must be applicable to the more general cases as well, but one cannot show this explicitly so far. The progress in this direction is also often complicated by the absence of the full control over both sides of this duality (see \cite{Lee2012,Lee2014,Lee2016}).

Let us recall that the basic link between the field theories on the surface and the gravitation theories in the volume is established via the renormalization group (RG), where the volume variables represent the scale-dependent couplings and the radial direction represents the length scale (see \cite{Skenderis2002,Papadimitriou2016}). The AdS/CFT correspondence maps the N-dimensional field theory into the (N+1)-dimensional gravitation theory. The extra dimension is usually accepted as a length scale. The conversion of the theories from one scale to another is exactly the thing that the RG does (more precisely, the quantum RG and holography). That is why it is believed to be the key to any proof of the AdS/CFT hypothesis (see \cite{Lee2014,Lee2016}).

It is also understood that any version of the “holographic renormalization”, trying to bridge the Planck-scale quantum gravitation theories with  the observations, must be associated with the minimum measurable length or the maximum measurable momentum. These are predicted by various quantum gravitation theories (such as the string theory or double special relativity) as well as the physics of BHs. The search of these limits is related to the so-called generalized uncertainty principle (GUP) or modified commutation relations between the coordinate and momentum (see \cite{Amati1989,Maggiore1993,Magueijo2005,DasVagenas2008}). The only GUP consistent with the symmetries and index structure of the modified commutator between the coordinate $x$ and momentum $p$, providing $[x_i,x_j] = 0 = [p_i,p_j]$ (Jacobi identity) at the same time, is \citep{Ali2009,Ali2011}:

\begin{equation}
[x_i,p_j] = i \hbar \left[ \delta _{ij} - \alpha (p \delta_{ij} + p_i p_j / p) + 
            \alpha^2 (p^2 \delta_{ij} + 3 p_i p_j) \right] ,
\label{eq07-121}
\end{equation}

\begin{equation}
\Delta x \Delta p \geqslant \frac{\hbar}{2} \left[ 1 - 2 \alpha \langle p \rangle + 4 \alpha ^2 p^2 \right]
\geqslant \frac{\hbar}{2} \left[ 1 + \left( [ \alpha / \sqrt{\langle p^2 \rangle} ] + 4 \alpha ^2 \right)
\Delta p^2 + 4 \alpha ^2 \langle p \rangle ^2 - 2 \alpha \sqrt{\langle p^2 \rangle} \right] .
\label{eq07-122}
\end{equation}

This yields the minimum measurable length and the maximum measurable momentum

\begin{equation}
\Delta x \geqslant (\Delta x)_{min} \approx \alpha _0 l_{Pl} ,
\label{eq07-123}
\end{equation}

\begin{equation}
\Delta p \leqslant (\Delta p)_{max} \approx M_{Pl} c / \alpha_0  ,
\end{equation}

\noindent
where $\alpha_0$ is a dimensionless number, $\alpha = \alpha_0 / M_{Pl} c = \alpha_0 l_{Pl} / \hbar$, $ M_{Pl}\approx 22 ~\mu g$ (Planck mass), $l_{Pl} \equiv \sqrt{\hbar G /c^3} = 1.6 \cdot 10^{-35} ~m$ (Planck length), and $ M_{Pl} c^2 = 1.2 \cdot 10^{19} ~GeV$ (Planck energy). According to this approach, the momentum must be modified as \citep{Ali2009,Ali2011,Moussa2014}

\begin{equation}
p \rightarrow p (1 - \alpha p + 2\alpha ^2 p^2) .
\end{equation}

If one assumes that the parameter $\alpha _0$ is equal to unity, the terms with $\alpha$ are significant for the energies (momenta) of the Planck energy order and the lengths of the Planck length order (see \cite{Ali2009}).

It was recently shown (see \cite{DasVagenas2008,Ali2009}) that the GUP leads to the corrections  to the Schr\"{o}dinger and Dirac equations affecting  both the relativistic and nonrelativistic quantum Hamiltonians. \cite{Ali2011,Das2011} also determined the bound on $\alpha$ demonstrating that the GUP may be detected in the low-energy systems like the Landau levels, Lamb shift, quantum Hall effect and the anomalous magnetic moment of the muon. I.e. the GUP is also applicable to the low energy scales. The most important consequence of this bound is that the space must be discrete, and all measurable lengths must be quantized with the fundamental measurable length of $\alpha \hbar = \alpha _0 l_p$ which cannot exceed the scale of the electroweak interaction \citep{Ali2011,Das2011,Das2010}.

This rises an interesting problem of the GUP quantum gravity corrections to various quantum phenomena, in particular, to the thermomagnetic EN~effect in the tachocline at low energies.

\subsubsubsection{Quantum gravity and the generalized thermomagnetic Ettingshausen-Nernst effect  in tachoclines of the Sun and magnetic white dwarfs}

Let us consider some solutions of the problem of GUP quantum gravity, emerging in response to the question: “What kind of the mysterious nature of the solar tachocline gives birth to the repulsive magnetic field through the thermomagnetic EN~effect?”

According to \cite{Ali2009,Ali2011}, it can be shown that

\begin{equation}
x_i = x_{0i} , ~~~ p_i = p_{0i} (1 - \alpha p_0 + 2\alpha ^2 p_0 ^2) ,
\label{eq07-126}
\end{equation}

\noindent
where $x_{0i}$, $p_{0j}$ satisfy the canonical commutation relations $[x_{0i}, p_{0j}]=i \hbar \delta _{ij}$, and $p_0 = -i \hbar \partial / \partial x_0$, satisfy Eq.~(\ref{eq07-121}).

Through Eq.~(\ref{eq07-126}) it was shown in \cite{DasVagenas2008,Ali2009,DasVagenas2009} that any nonrelativistic Hamiltonian of the form $H = p^2/2m + V(r)$ may be given as

\begin{equation}
H = p^2 / 2m - (\alpha / m) p_0 ^3 + (5 \alpha ^2 / 2m) p_0 ^4 + V(r) + O(\alpha ^2) ,
\end{equation}

\noindent
and the modified Schr\"{o}dinger equation is

\begin{equation}
\left[ - \frac{\hbar ^2}{2m} \nabla ^2 + \frac{i \alpha \hbar}{m} \nabla ^3 + \frac{5 \alpha ^2 \hbar ^4}{2m} \right] \psi = i \hbar \frac{\partial \psi}{\partial t} .
\label{eq07-128}
\end{equation}

Let us consider the terms with $\alpha$ and $\alpha ^2$ as the perturbations, although this higher order Schr\"{o}dinger equation has the non-perturbative solutions of the form $\psi \sim e^{ix / 2 \alpha \hbar}$, which may lead to the interesting physical consequences \citep{Ali2009}. Some phenomenological consequences of the modified Hamiltonian of such GUP model were addressed in \cite{DasVagenas2009,Das2010}.

So, we examined the Planck scale effects in some low energy system via the GUP, which is a clear prediction of the quantum gravity theories, and found the effect to be small yet nonzero for the fundamental quantum flux in the thermomagnetic EN~effect in the tachocline.

\textit{The thermomagnetic EN~effect in the tachocline of the Sun.} Let us note that using Eqs.~(\ref{eq07-121})-(\ref{eq07-122}) and (\ref{eq07-126}), we applied this GUP version to the problem of the thermomagnetic EN~effect, which is actually an almost perfect analogy of the superconductivity problem (see Sect.~1 in \cite{Das2011}). Making use of this analogy, we involved a Schr\"{o}dinger current (from Eq.~(\ref{eq07-128})) through the density of states of the Landau levels and the substitution $2e \rightarrow e$, since the charge carriers are electrons in this case, and not the Cooper pairs. These effects have not been observed so far, but it is possible to estimate the upper bound on the GUP parameter

\begin{equation}
\alpha _0 < \frac{10^{-n/2}}{\sqrt{2.5}} \frac{M_{Pl} c^2}{e BL} \sim 3 \cdot 10^6 ,
\label{eq07-129}
\end{equation}

\noindent
assuming the experimental error of 1 per $10^n$, e.g. for $n = 2$, $B \approx 4.1 \cdot 10^3 ~T \approx 8 \cdot 10^{-13} ~GeV^{2}$, $L_{tacho} \approx 4 \cdot 10^4 ~km \approx 2 \cdot 10^{23} ~GeV^{-1}$ (see \cite{Rusov2015}). The total bound for $l_n \geqslant (\Delta x)_{min} = \alpha_0 l_{Pl}$ (see Eq.~(\ref{eq07-123})) reads:

\begin{equation}
\alpha_0 l_{Pl} \leqslant l_n < 3 \cdot 10^{-28} ~m\, .
\label{eq07-130}
\end{equation}

It should be emphasized that the obtained upper bound on the GUP parameter is more strict than those obtained in recent experiments comparing the nonlocal relativistic effective field theory with the 8TeV LHC data (see $l \leqslant 10^{-19} ~m$ in \cite{Biswas2015}), as well as the low energy macroscopic optomechanical oscillators (see $l \sim 10^{-18} ~m$  in \cite{Marin2013}; $l \sim 10^{-22} \div 10^{-26} ~m$ in \cite{Belenchia2017}; $l \sim 10^{-29} ~m$ in \cite{Bawaj2015}). This bound is smaller than the electroweak interaction scale, which may indicate the existence of the intermediate length scale between the electroweak and Planck scales. Although the figures are 7 orders of magnitude larger than $l_{Pl} = 1.6 \cdot 10^{-35} ~m$, they can serve as a sign of the quantum gravity corrections to various quantum phenomena including the thermomagnetic EN~effect in the tachocline.

\textit{The generalized thermomagnetic EN~effect by analogy to the solar tachocline in the magnetic white dwarfs: the stably burning hydrogen shell on the helium core.} Now we understand that it is not “some kind of the mysterious nature” that gives birth to the thermomagnetic EN~effect and, consequently, to the repulsive magnetic field in the solar tachocline (see Eq.~(\ref{eq06-16})). It is the quantum gravity that does. The question is how exactly does the quantum gravity induce the repulsive magnetic field?

As it was noted above, the fundamental minimum length scale cannot exceed that of the electroweak interactions. Therefore GUP has to modify the density of states, which impacts the statistical and thermodynamic  properties of the ideal fermion and boson quantum systems. According to \cite{Ali2011}, the number of quantum states in a certain volume of the momentum space should change in the following way:

\begin{equation}
\frac{V}{(2 \pi \hbar)^3} \int d^3 p \rightarrow \frac{V}{(2 \pi \hbar)^3} \int \frac{d^3 p}{(1- \alpha p)^4} .
\end{equation}

Leaving aside the complex calculations of the modified density of states (see \cite{Camacho2006,Moussa2014}), related to the distribution function, the number of particles, the total energy and pressure of the gas, it is important to understand that quantum gravity induces repulsive forces between fermions and bosons in their ground state, thus increasing  the pressure in quantum systems.

Consequently, while increasing pressure in the ground state, quantum gravity induces repulsive magnetic fields e.g. in the solar tachocline (see Eqs.~(\ref{eq06-15})-(\ref{eq06-16})).

It is interesting that the quantum gravitation behaves differently if the fermions are in the degenerate state \citep{Camacho2006,Moussa2014}. In this case the repulsive force is smaller, and the pressure is smaller too. The smaller pressure of the ground state of fermions implies the smaller radius-to-mass ratio in white dwarfs as a test for GUP. For example, according to \cite{Camacho2006,Ali2013,Moussa2014}, the radius contraction is proportional to the mass of the white dwarf and is divergent with the mass when it reaches the Chandrasekhar limit. This result is consistent with the current observations indicating that the white dwarfs have smaller radii that it was theoretically predicted \citep{Provencal2002,Mathews2006}.

The most interesting phenomenon is, of course, the generalized thermomagnetic EN~effect in the tachoclines of stars, e.g. the magnetic white dwarfs (see \citet{Valyavin2014}). Let us start approaching this problem by estimating the magnetic field in the tachocline. In the quasi-steady magnetic field the thermomagnetic current generated in the fully ionized hydrogen plasma is (see Eqs.~(4-1) in \cite{Spitzer1962}):

\begin{equation}
{j} _{\perp} = {j} _y = - \frac{3 n k_B c}{4 B} \frac{dT}{dz} ,
\label{eq07-v4-136}
\end{equation}

\noindent where

\begin{equation}
T^{1/4} n = const .
\label{eq07-v4-137}
\end{equation}

It may be shown that Eq.~(\ref{eq07-v4-136}) which leads to the magnetic
equilibrium (see Eqs.~(4-1) in \cite{Spitzer1962})

\begin{equation}
j_y = - \frac{c}{B} \nabla p = - \frac{c}{B} \frac{dp}{dz} ,
\label{eq07-132}
\end{equation}

\noindent
where the pressure $p$ corresponds to the overshoot tachocline (at the bottom of the convection zone).

From the Maxwell equation $4 \pi j_{\perp} / c = 4 \pi j_y / c = \nabla \times \vec{B}$ we get

\begin{equation}
j_y = \frac{c}{4 \pi} \frac{dB}{dz} .
\label{eq07-133}
\end{equation}

By equating (\ref{eq07-132}) and (\ref{eq07-133}) and integrating in the limits $[B_{tacho}, 0]$ on the left and $[0, p_{tacho}]$ on the right, we find the repulsive magnetic field induced by the generalized thermomagnetic EN~effect:

\begin{equation}
\frac{B_{tacho}^2}{8 \pi} = p_{tacho} ,
\label{eq07-134}
\end{equation}

\noindent
where the magnetic field of the generalized thermomagnetic current in the overshoot tachocline ``neutralizes'' the magnetic field of the core in magnetic white dwarf.

We adopt a model of the stably burning hydrogen shell on the helium core obtained by solving the equations of hydrostatic balance, heat transfer, energy generation and mass conservation \citep{Robinson1995,Steinfadt2010}. However, the most important feature between the H and He layers for our purposes is not the chemical composition, but the neutralization of the magnetic fields in the core on its boundary.

Taking the density in the tachocline 
$\rho = \rho _{periph} \leqslant 10^2 ~g/cm^3$ and $T \sim 10^6 ~K$
(see \citep{Bhatia2001}), let us estimate the pressure of the
nonrelativistic electron plasma (\ref{eq07-134}):

\begin{equation}
\frac{B_{tacho}^2}{8\pi} = p_{tacho} \leqslant 10^{15} ~erg/cm^3,
\end{equation}

\noindent
where the toroidal magnetic field in the tachocline

\begin{equation}
B_{tacho} \leqslant 1.6 \cdot 10^4 ~T = 1.6 \cdot 10^8 ~G.
\end{equation}

By analogy to Eq.~(\ref{eq07-129}) for the solar tachocline, we can estimate an important upper bound of the GUP parameter:

\begin{equation}
\alpha _0 < \frac{10^{-n/2}}{\sqrt{2.5}} \frac{M_{Pl} c^2}{e BL} \sim 10^7 ,
\end{equation}

\noindent
again, assuming experimental precision of 1 part per $10^n$, e.g. for $n = 2$, where we used $B = B_{tacho} \approx 4.1 \cdot 10^3 ~T \approx 8 \cdot 10^{-12} ~GeV^{2}$, $L = L_{tacho} \leqslant 10^2 ~km \approx 10^{21} ~GeV^{-1}$, where $L_{tacho}$ is the thickness of the tachocline in the magnetic white dwarf \citep{Kissin2015}. The total bound on $l_n \geqslant (\Delta x)_{min} = \alpha _0 l_{Pl}$ (see Eq.~(\ref{eq07-123})) reads:

\begin{equation}
\alpha _0 l_{Pl} \leqslant l _{n=2} < 10^{-28} ~m\, ,
\label{eq07-140}
\end{equation}

\noindent
which again indicates an intermediate length scale between the electroweak interactions and the
Planck length. The quantum gravity corrections lead to the restraint of the star collapse if the GUP
parameter $\alpha = \alpha_0 l_{Pl} / \hbar$ takes the values between the Planck length and the electroweak interactions scale (see Eqs.~(\ref{eq07-130}) and Eq.~(\ref{eq07-140})).

It means that quantum gravity induces the repulsive magnetic fields remarkably well in both the solar (see Eqs.~(\ref{eq06-15})-(\ref{eq06-16})) and white dwarfs tachoclines (see Eq.~(\ref{eq07-134})). Recalling in addition the EN~effect theory for the quantum phase transitions in dyonic BHs \citep{Hartnoll2007,Hartnoll2016}, one may say that the quantum gravitation and the induced repulsive magnetic field in the generalized thermomagnetic EN~effect play an important role in the astrophysical phenomena and the structure of the Universe. This is a separate problem though, and we hope to obtain some results in this direction in the future.

Meanwhile, the part related to the generalized thermomagnetic EN~effect by analogy to the solar tachocline near the boundary of the BH is presented below.

\subsubsubsection{Thermomagnetic Ettingshausen-Nernst effect near quantum phase transitions in dyonic black holes}

BHs in the centers of galaxies are surrounded by a strong magnetic field due to charged matter around them  (see e.g. \cite{Aliev2002,Konoplya2006,Konoplya2007,Konoplya2008,Jamil2015,Garofalo2017}). As noted by \cite{Johnson2015}, “Understanding these magnetic fields is critical. Nobody has been able to resolve magnetic fields near the event horizon until now”. The authors measured polarization near the event horizon of Sgr A*, the SMBH at the center of our galaxy. The polarization is believed to be a signature of ordered magnetic fields generated in the accretion disk around the BH. The results are expected to explain how BHs accrete gas and launch jets of material into their surroundings.

Such explanation may be related to the known solution of the perturbative Einstein-Maxwell equations depending on the BH mass $M$, magnetic field $B$, and the parameter $K$, which characterizes the above surrounding structure (see \cite{Konoplya2006}). Hence, the presence of the gravitational tidal force from surroundings considerably changes the parameters of the test particle motion (see e.g. \cite{Konoplya2006}): when the magnetic field is weak (strong), it increases (decreases) the radius of circular orbits of particles and the binding energy of massive particles going from a given circular orbit to the innermost stable orbit near the BH. In addition, it increases (decreases) the distance of minimal approach, time delay and bending angle for a ray of light propagating near the BH (see e.g. \cite{Gammaldi2016,Belikov2016,Lacroix2016}).

On the other hand, taking into account the DM existence 
and the possible modifications to general relativistic gravitational wave patterns (see \cite{Abbott2016a,Abbott2016b,Barcelo2017}), let us consider some important thermodynamic features of this BH space-time, which are solutions of the Einstein-Maxwell theory in 3+1 dimensions, and include the quantum gravity effects beyond general relativity.

The matter is the AdS/CFT correspondence \citep{Maldacena1999}, which demonstrates a close connection between Einstein’s general theory of relativity and quantum physics.

\begin{figure*}[tbp]
  \begin{center}
    \includegraphics[width=16.5cm]{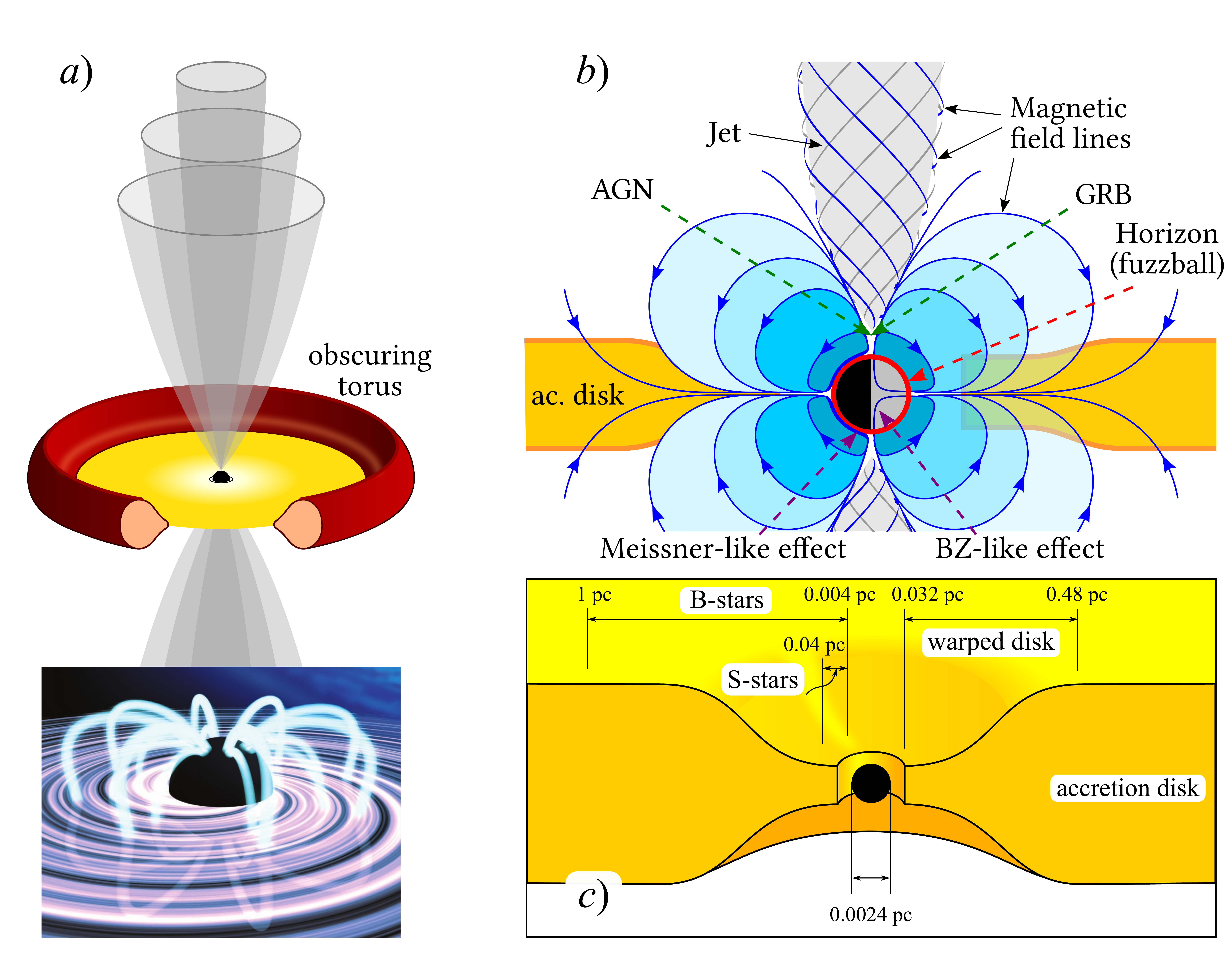}
  \end{center}
\caption{Anatomy of a spinning BH.
\textbf{(a)} Generic accretion disk around the BH and AGN jet. 
SMBH in the center of a galaxy (black ball) with the accretion disc around (light-yellow torus). Blue lines show the magnetic fields.
Inset: The magnetic field is amplified inside the disk and erupts to form a corona. The magnetic field penetrating the ergosphere (see Fig.~1a in \cite{Begelman2003}) extracts energy from the BH, which can accelerate a jet (if the field lines escape into space) or enhance the emission from the disk (if the field lines connect to the disk, as shown here). Adopted from \cite{Begelman2003}.
\textbf{(b)} BH magnetosphere, accretion disk and AGN jets, which describe the mapping of the hydrodynamic and AdS/CFT results under particle-vortex duality (or the bulk electromagnetic duality (see Figs.~5.45-5.46 in \cite{Hartnoll2007})). 
As shown here, the AGN jets are collimated outflows with a roughly helical magnetic field structure. They require three major components: a source of power (either matter accreting onto a compact object, or the spinning object itself), rotation, and magnetic fields (see the analog in \cite{Polko2013}).
\textbf{(c)} The regions of the disk having different physical conditions. Here we present the new data on B-stars with projected radii $0.1'' < p < 25''$ ($\sim 0.004 - 1 ~pc$) from the massive BH in the GC \citep{Madigan2014}. So far, dozens of B-type stars (so-called “S-star cluster”, or “S-cluster”) have been discovered in the region as close as $< 1''$  ($1'' \simeq 0.04 ~pc$ at the GC) from Sgr A* (see \cite{Eisenhauer2005,Ghez2005,Gillessen2009,Schodel2003}). Young B-stars and old red giants in the GC constitute the nuclear star cluster of the Milky Way. There is another dynamically distinct structure centered on Sgr A* and spanning a radial range of (0.032 -- 0.480) pc, known as the “mini disk” (warped disk \citep{Pfuhl2014}), which consists of $O(10^2)$ young ($6 \pm 2 ~Myr$) and massive ($>20 M_{Sun}$) Wolf-Rayet (WR) and O-type stars, in a configuration of a (possibly two) mildly thick disk(s) (see \cite{ChenAmaro2015}).}
\label{fig-accretion-disk}
\end{figure*}

Here we are interested in the system at finite temperature $T$, charge density
$\rho$ and with a background magnetic field $B$. The finite temperature is realized
in $\text{AdS}_4/\text{CFT}_3$ by allowing the space-time to contain a BH \citep{Witten1998}, where the temperature in the field theory is just the
Hawking temperature of the BH and the finite temperature dissipation in the
field theory is dual to bulk matter fields falling into the BH. The
important result of this is the fact that the crosswise flow of heat,
associated with the finite temperature dissipation in the field theory, and charge
currents in the presence of the magnetic field are obtained by allowing the black
hole to carry the thermomagnetic EN~effect 
(see \cite{Spitzer1962,Spitzer2006,Rusov2015}), which in turn produces the
large-scale magnetic fields near the BH (see inset in 
Fig.~\ref{fig-accretion-disk}a, and also Eqs.~(\ref{eq06-15})-(\ref{eq06-16})).

Magnetism, produced by the generalized thermomagnetic EN~effect (see the essence of quantum gravity for dyonic BHs by \cite{Hartnoll2007}), plays a fundamental role in the generation of the large-scale magnetic fields of the BH. It pushes the gas away from the BH (see e.g. similarities and differences in the papers by \cite{Fukumura2017} and \cite{Contopoulos2015,Contopoulos2018}).
Therefore, we understand that quantum gravity is not only the base for the GUP and the dyonic BH solution \citep{Hartnoll2007,Hartnoll2016}, but also gives birth to the repulsive magnetic field induced in the “tachocline” by the generalized thermomagnetic EN~effect in compact objects (see e.g. \cite{Shapiro1983}): our Sun, magnetic white dwarfs, accreting neutron stars (similar to white dwarfs; see the review by \cite{Wang2016}) and dyonic BHs.

A number of intriguing questions appear at this point (see Fig.~\ref{fig-accretion-disk}). How do the large-scale magnetic fields of the thermomagnetic EN~effect extract the energy from the central BH? Or, to put it differently, how does the ergosphere cause the magnetosphere inside it to rotate, making the outgoing flux of angular momentum result in extraction of energy from the BH? How can the large-scale magnetic fields of the thermomagnetic EN~effect be explained with the magnetic connection between the BH and the accretion disk? If there is a magnetic link, then how does the BH exert a torque on the accretion disk, thus transferring energy and angular momentum between the BH and the disk? How do the cycles  of accretion from a thin disk and the associated spin-up of the BH alternate with the periods of no accretion and magnetic transfer of energy from the BH to the disk? How is the jet launched as a magnetically dominated outflow with strong magnetic fields accelerating the flow to relativistic velocities?

One can ask a lot of other questions, but first of all, it is necessary to understand and describe in detail the thermomagnetic EN~effect -- a fundamental phenomenon of crucial importance for powering relativistic outflows of jets from BHs (see \cite{Blandford1977,MacDonald1982,Thorne1986,Punsly1990}).

In Fig.~\ref{fig-accretion-disk}b one can see how the magnetic flux is pushed out of the quickly rotating BH horizon. The nature of this phenomenon is similar to the Meissner effect \citep[see e.g.][]{Bicak2007,Komissarov2007,Penna2014a,Penna2014b,Bicak2015,Gurlebeck2017,Hirsch2016,Hirsch2017}, i.e. displacement of magnetic fields in superconductors. Later \citet{Bicak1984} \citep[see also][]{Bicak2007} showed that this was true for all axisymmetric stationary vacuum solutions of the BH electrodynamics. They concluded that their discovery  undermines the widely known electromagnetic mechanism of \cite{Blandford1977}, which, in contrast to the vacuum Meissner effect, is concerned with plasma-filled magnetospheres. \cite{Bicak1984,Bicak2007} believe that this BH Meissner effect could quench jet power at high spins, but we think the actual answer is “yes and no”. Let us show why.

\begin{figure*}[tbp]
  \begin{center}
    \includegraphics[width=10cm]{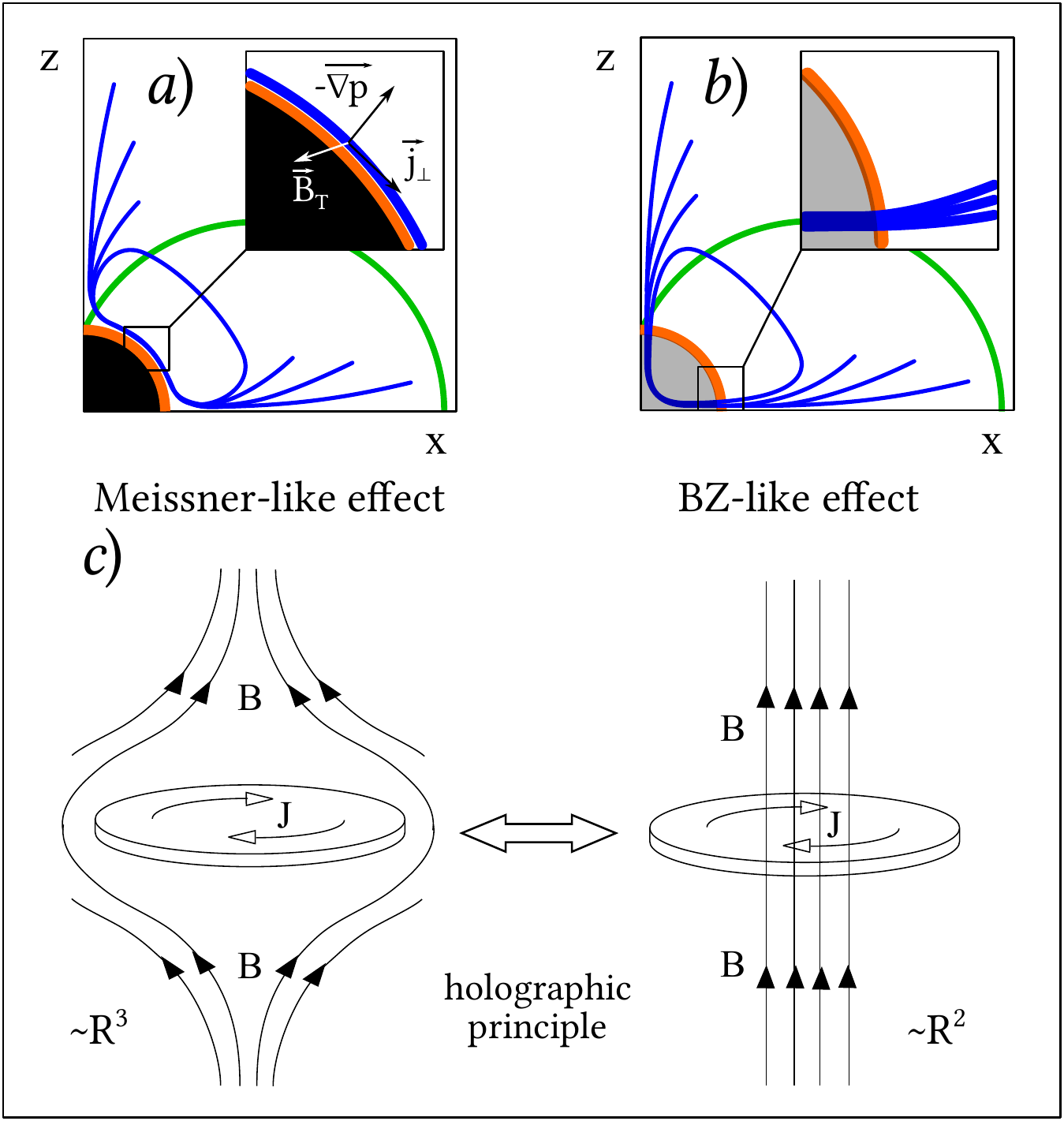}
  \end{center}
\caption{\textbf{(a)} The magnetic flux is expelled from the event horizon, thus illustrating the Meissner-like effect, which arises also for the dyonic BH -- Reissner-Nordstrom AdS BH with both electric and magnetic charges and no scalar hair (see \cite{Hartnoll2007,Hartnoll2008a}). The thick green line shows the BH ergosphere (with maximal hole’s angular momentum per unit mass $a \equiv J/M = 1.0$). Inset: For a fully ionized plasma the thermomagnetic EN~effect leads to the current density $j_{\perp} = (c/B_T) dp / dz$ (see Eqs.~(5-49) in \cite{Spitzer1962,Spitzer2006}), where $p$ is the gas pressure, $B_T$ is the toroidal magnetic field. 
\textbf{(b)} The magnetic flux is pulled back to the horizon, thus illustrating the Blandford-Znajek-like (BZ-like) effect. 
\textbf{(c)} Possible transition of Meissner effect to the BZ-like effect and back. In order to prevent the flux from penetrating the holographic superconductor of area $R^2$, the currents would have to do enough work to expel the field from a volume of size $R^3$, as shown in the left figure. If the external magnetic field varies for some reason, e.g. because of the baryon density variations within $R^3$, this work cannot be supplied by the free energy  gain (see Sect.~5.2 in \cite{Hartnoll2008a}) of the thin superconducting film (i.e. of the area $R^2$).
To restore the Meissner effect, it is necessary to exclude variations of the magnetic field and, thereby, to push out the external magnetic field with the help of a current (see e.g. Fig.~(a)), the force of which must neutralize the strength of the opposite field inside the three-dimensional sample. On the other hand, if neutralization is not complete, the varying flux always penetrates into the thin film (see e.g. Fig.~(b)). This argument is illustrated in Fig.~(c).}
\label{fig-meissner-effect}
\end{figure*}

The EN thermomagnetic mechanism, which generates the Meissner effect at Hawking's BH temperature, is considered one of the most effective ways of extracting the net energy from rotating BHs, for example, a dyonic BH (see \cite{Spitzer1962,Spitzer2006,Hartnoll2007,Hartnoll2008a,Hartnoll2008b,Rusov2015}). Our view of this mechanism is as follows. The spinning BH distorts the poloidal magnetic field $B_P$ and induces the poloidal electric field $E_P$ and toroidal magnetic field $B_T$ (see inset in Fig.~\ref{fig-meissner-effect}a), which generate an outward Poynting flux $E_P \times B_T$ along the magnetic field lines threading the spinning BH. The rotation energy of the spinning dyonic BH is extracted in the form of the Poynting flux. The formal physics of this idea almost completely coincides with the well-known idea of \cite{Komissarov2009} and \cite{Beskin2010}, but, surprisingly, it describes not only the BZ-like effect (see e.g. \cite{Pan2017,Contopoulos2017} and Refs. therein), but also the thermomagnetic EN~effect (see Fig.~\ref{fig-accretion-disk}b and Fig.~\ref{fig-meissner-effect}b). To proceed further, it is natural to ask why doesn't the relativistic magnetohydrodynamic (MHD) simulation of jets of BHs give any evidence (see e.g. \cite{Penna2014a,Penna2014b}) that jets are extinguished by the Meissner effect of the BH. In general, why should the Meissner effect exist along with the BZ-like effect, from the point of view of the BH jets existence? Below we demonstrate the possibility of overcoming the contradictions between these effects.

It is important to understand this possibility and the essence of the existence of holographic superconductors. This is due to the fact that in determining whether our model is Type I or Type II superconductors, we are confronted with the limitation that the currents (see the right panel in Fig.~\ref{fig-meissner-effect}c) in the model are not a source of electromagnetic fields (see \cite{Hartnoll2008a}). As the material does not produce its own magnetic fields, the applied magnetic field is the actual magnetic field outside the horizon and we can set $B_T$. The major benefit of the London equation is the ability to explain the Meissner effect \citep{Meissner1933}, when the material exponentially pushes out the internal magnetic fields after passing the superconducting threshold -- the horizon. However, the most important is the fact that the holographic superconductors (Type II) generate the currents required to expel magnetic fields (the London equation and the associated London magnetic penetration depth) and that the theory can consistently be weakly gauged (see \cite{Hartnoll2008a}).

So, our starting point is the holographic superconductor as a dyonic BH background (see \cite{Hartnoll2008a,Hartnoll2008b}). This leads to the existence of the thermomagnetic EN~effect at low temperatures (the Hawking temperature), otherwise known as the BH Meissner effect, which extracts pure energy near the horizon by the BH spin or by an accretion flow, feeding on dyonic BH jets. Note that this is not the BZ~mechanism, but the EN~mechanism (see Fig.~\ref{fig-meissner-effect}a), which, strangely enough, is one of the most effective ways to extract the rotation energy from spinning BHs. And this argument is reasonable for accreting BH systems, such as AGNs and X-ray binaries, but not in gamma-ray bursts (GRBs) events by the BZ~mechanism  (see Fig.~\ref{fig-meissner-effect}b), which we will discuss below.

If the thermomagnetic EN~effect, which gives rise to the Meissner effect at low temperatures (see Fig.~\ref{fig-meissner-effect}a), transforms the variable magnetic fields within the holographic superconductor (see \cite{Hartnoll2008a}), it means that if the strength of the external magnetic field changes for some reason (e.g. because of the chaotic variations of the baryon density and, consequently, the ADM density; see Fig.~\ref{fig-meissner-effect}c), the variations in baryon density would lead to a rather quick transition from the Meissner effect (Fig.~\ref{fig-meissner-effect}a) to the BZ~effect (Fig.~\ref{fig-meissner-effect}b). 
Hence, we understand that the BZ~effect excludes the external magnetic field of the Meissner effect near the horizon, since, unlike the Meissner effect, it changes the shape of the magnetic field and thereby reveals a mechanism, how the spin energy of a rotating BH may be extracted electromagnetically through the magnetic field that threads the BH horizon (see e.g. \cite{Nathanail2014,Penna2014a,Penna2014b,Contopoulos2017}). 
Therefore, the chaotic variations of the flux always penetrate into the thin film (see e.g. Fig.~\ref{fig-meissner-effect}b on the right). It is very important that the BZ~effect also shows that it is possible to hold this magnetic flux for the duration of the GRB even without accretion onto the BH (see the original Penrose process of \cite{Penrose1971}, \cite{Contopoulos1984} and the analogous one of \cite{Contopoulos2017}).

So, on the one hand, the duration of a GRB depends closely on the magnetic flux accumulated on the event horizon, and on the other hand, that is actually why we observe a clear exponential decay in the fraction of GRB events (see \cite{Contopoulos2017}). And vice versa, in order to restore the Meissner effect, it is necessary to 
include the non-chaotic variations of the magnetic field, and thereby to push out the external magnetic field with the help of a current (see e.g. Fig.~\ref{fig-meissner-effect}a), the force of which must ``neutralize'' the strength of the opposite field inside the three-dimensional sample. If neutralization is not complete, then the chaotic variations of the flux always penetrate into the thin film (see e.g. Fig.~\ref{fig-meissner-effect}b on the right). This argument is illustrated in Fig.~\ref{fig-meissner-effect}c (adopted from \cite{Hartnoll2008a}). As a result, we have a possible transition from the Meissner effect to the BZ~effect and back. This means that the appearance or disappearance of the chaotic modulating density of baryons (and correspondingly, the density of DM) on the horizon of a BH is the main basis for the manifestation of the BZ or Meissner effects, respectively.

Ignoring many other questions of the Meissner effect in the BH, we focus on the main result of the theoretical-experimental proof of the fundamental existence of the repulsive magnetic field induced in the ``tachocline'' by the generalized thermomagnetic EN~effect of quantum gravity, which can be evaluated on compact objects: our Sun, magnetic white dwarfs, accreting neutron stars and dyonic BHs.

In this context, let us consider the physics of the so-called holographic principle of quantum gravity. The outstanding paper by \cite{Hanada2014} presents an intriguing argument for the superstring theory as a specific implementation of the holographic principle \citep{Bekenstein1973,Hawking1973,tHooft1993,Susskind1995,Maldacena1999}. Simply speaking, the hologram is a 2D object, but if we look at it under certain light, we can see the image of a 3D object encoded within this 2D surface (see \cite{Bekenstein1981,Bekenstein1997,Bekenstein2003,Bekenstein2007,Brown2016}). According to these new theories, the whole Universe (or an individual BH, or a white dwarf, or a neutron star) may be a sort of a hologram. It may be thought of as a 3D object with all internal complex systems consisting of ``quarks'' and ``gluons'' being described by the laws based on gravity, or as a 2D surface within which the ``quarks'' and ``gluons'' are subject to completely different laws like quantum mechanics, and no gravity on the surface.

There is however a problem related to the fact of Hawking radiation from BHs, which would lead to their evaporation and the paradox of information loss \citep{Hawking1976}. At first it may seem to violate the laws of quantum mechanics. On the other hand, quantum mechanics  imposes a strong rule forbidding the information loss – the unitarity principle, which is closely related to other inviolable physical laws such as the energy conservation (see e.g. \cite{Banks1984,Giddings2013,Giddings1992,Giddings2012b,Giddings2013a,Modak2015}). And there is a controversy in understanding of the quantum mechanics with and without the unitary processes (see e.g. \cite{ChenP2015}). One party suggests that the statement by \cite{Banks1984}, according to which the unitarity violation would violate both the locality and the energy-momentum conservation, is a strong argument against non-unitarity of Hawking radiation \citep{Hawking1976}. Another party relies on the fact that \cite{Nikolic2015} recently showed that the alleged problem with non-unitary Hawking radiation, the problem of violation of either locality or conservation of energy-momentum, does not really exist. This is because the non-unitary time evolution of Hawking radiation is not in contradiction with a generalized concept of unitarity (see \cite{Vafa2014,Nikolic2015}).

The remarkable paper by \cite{Vafa2014} strongly supports the fact that the unitarity is not a necessary consequence of holography, and that the existence of non-unitary holography can probably be established using the known 2D non-unitary models which should lead to non-unitary AdS$^3$ holographic duals (see \cite{Hawking1976,Vafa2014}).

We are struck by the fact that non-unitarity is consistent with the idea of holography, wherein quantum gravity is to be constructed in terms of degrees of freedom that are highly nonlocal from the bulk point of view (see \cite{Almheiri2013a,Almheiri2013b}). A possible unexpected effect is that the properties of non-unitary holography can be experimentally measured precisely on the basis of the fundamental physics of the holographic principle of quantum gravity.

$\bullet$ First of all, we are interested in the properties of the holographic principle of quantum gravity, which make it possible to measure the ratio between the magnetic fields of the 2D surface and the 3D volume of compact objects: our Sun, magnetic white dwarfs, accreting neutron stars and dyonic BHs. 

Using the thermomagnetic EN~effect (see \cite{Spitzer1962,Spitzer2006,Rusov2015} and Fig.~\ref{fig-lower-heating}), the magnetic pressure of the ideal gas inside e.g. the solar tachocline may be estimated as

\begin{equation}
\left( \frac{B_{tacho}^2}{8 \pi} \right)_{Sun} = n_{tacho} k_B T_{tacho} =
6.5 \cdot 10^{13} ~erg/cm^3 = \left( \frac{B_{core}^2}{8 \pi} \right)_{Sun} ,
\end{equation}

\noindent
which indirectly shows that, according to the holographic principle of quantum gravity, the repulsive magnetic field of the tachocline exactly ``compensates'' the magnetic field of the solar core (see Fig.~\ref{fig-R-MagField})
where the projections of the magnetic fields in the tachocline and the core
have equal values but opposite directions:

\begin{equation}
(B_{tacho})_{Sun} = 4.1 \cdot 10^7 ~G = -(B_{core})_{Sun} .
\label{eq07-142}
\end{equation}

The pressure of the nonrelativistic electron plasma in the tachocline of cold magnetic white dwarfs may also be estimated:

\begin{equation}
\left( \frac{B_{tacho}^2}{8 \pi} \right)_{WD} = p_{tacho} \leqslant
10^{15} ~erg/cm^3 \approx \left( \frac{B_{core}^2}{8 \pi} \right)_{WD} .
\end{equation}

Accordingly, the repulsive magnetic field of the tachocline exactly
``neutralizes'' the magnetic field in  white dwarfs (see the analog of
\cite{Kissin2015}):

\begin{equation}
(B_{tacho})_{WD} = 1.6 \cdot 10^4 ~T = 1.6 \cdot 10^8 ~G \approx -(B_{core})_{WD} .
\label{eq07-144}
\end{equation}

A rough estimate of the pressure of degenerate relativistic electron plasma in the tachoclines of accreting neutron stars and dyonic BHs is also possible:

\begin{equation}
\left( \frac{B_{tacho}^2}{8 \pi} \right)_{NS,BH} = K_1 \rho_{tacho}^{4/3} = 
\left( \frac{B_{core}^2}{8 \pi} \right)_{NS,BH} .
\end{equation}

\noindent
The corresponding ``neutralizing'' magnetic field in accreting neutron stars and dyonic BHs

\begin{equation}
(B_{tacho})_{NS,BH} = -(B_{core})_{NS,BH} \geqslant 10^{13} ~G ,
\end{equation}

\noindent
where in a fully ionized hydrogen plasma with $Z = 1$ the generalized thermomagnetic EN~effect leads (see Fig.~\ref{fig-meissner-effect}a) to the current density given by

\begin{equation}
j_{\perp} = - \frac{c}{B} \frac{dp}{dz} .
\end{equation}

So our semi-phenomenological result consists in the fact that the repulsive magnetic field of the tachocline, produced by the thermomagnetic EN~effect (see e.g. Fig.~\ref{fig-meissner-effect}a) on compact objects, is the fundamental consequence of the holographic principle of quantum gravity in the Universe.

$\bullet \bullet$ Second, we are interested in the connection between the thermomagnetic EN~effect and the fuzzball effect near the BH boundary.

As we know, according to the AdS/CFT correspondence \citep{Maldacena1999} or, generally, the gauge/gravity correspondence \citep{Horowitz2006}, the quantum gravitation theory in the volume is mathematically equivalent to the quantum field theory on its border. Therefore if the thermomagnetic EN~effect on the border of the BH is physically equivalent to the Meissner effect (see Fig.~\ref{fig-meissner-effect}a), the EN~effect under low Hawking temperatures (or the unitary Meissner effect) is a consequence of the so-called ‘‘energetic curtain’’ \citep{Braunstein2009,Braunstein2013} or ``fuzzball'' \citep{Almheiri2013a,Almheiri2013b,Marolf2013,tHooft2017a} formation upon the so-called fuzzball (see \cite{Mathur2005,Almheiri2013b,tHooft2017b,Barcelo2017,Chakraborty2017} and Refs. therein), which blocks the non-unitary Hawking radiation by the BZ-like effect (see Fig.~\ref{fig-meissner-effect}b). This is because the non-unitary time evolution of Hawking radiation (see \cite{Hawking1976,Vafa2014,Nikolic2015}) does not contradict the generalized unitarity principle (see \cite{Hartle1998,Nikolic2009,Nikolic2012,Nikolic2014,Nikolic2015}), and thus does not contradict the unitary time evolution of the holographic Meissner effect (see Fig.~\ref{fig-meissner-effect}).

Note that in contrast to the firewall \citep{Almheiri2013a,Almheiri2013b,Marolf2013,tHooft2017a} argument with the help of the so-called `validity of effective field theory' \citep{Almheiri2013a,Almheiri2013b,Marolf2013}, an infalling shell-object of fuzzball never falls into the trap of its own horizon, and thus avoids any problem with causality, as opposed to the firewall \citep{Mathur2009a,Mathur2017a,Mathur2017b}.

The surprising result is that it excludes the information loss due to non-unitarity completely,  thus undermining the very logic used for the paradox formulation (see \cite{Braunstein2013}). One should also keep in mind that our indirect experimental results of the thermomagnetic EN~effect in tachoclines of the Sun (see Eq.~(\ref{eq07-142})) and white dwarfs (see Eq.~(\ref{eq07-144})) may become a direct proof if, for example, the new research succeeds in getting the information on the structure of the BH fuzzball from the observations of gravitational waves (see e.g. \cite{Abedi2017,Zhang2017} and Refs. therein). According to our holographic model of the fuzzball, it is possible to discover the gravitational echo-effects as a result of the macroscopic quantum gravity effects in BHs beyond general relativity (see \cite{Barcelo2017} and Refs. therein).

\phantomsection
\label{topic-THIRD}
$\bullet \bullet \bullet$ Third, we are interested in the connection between the holographic Meissner-like effect near the BH boundary, which produces the “fuzzball” phenomenon, and the Higgs phenomenon in particle physics.

We start with understanding that the holographic Meissner-like effect near the BH boundary -- fuzzball surface (see Fig.~\ref{fig-meissner-effect}a), which is essentially identical to the EN~effect at Hawking's BH temperature, is a consequence of quantum gravity by the gauge/gravity correspondence. It is then easy to show that the EN~effect is physically identical to the Meissner-like effect not only for low temperatures, but for any (gradients of) temperature. It exactly compensates the magnetic fields between the 2D surface of the ``tachocline'' and the 3D volume of the core of compact objects. Since the exact neutralization of the 2D and 3D fields (see Fig.~\ref{fig-meissner-effect}c) is determined by one of the properties of the holographic principle of quantum gravity (spontaneous breaking of electroweak symmetry), it leads to a surprising result: the EN~effect is absolutely identical to the Higgs phenomenon in particle physics (see \cite{Wilczek2000,Wilczek2005}).

\phantomsection
\label{topic-FOURTH}
$\bullet~\bullet~\bullet~\bullet$ Forth, we are interested in the connection between the Higgs phenomenon in particle physics (identical to the thermomagnetic EN~effect) and the quantum phenomenon of DM particles.

\begin{figure*}[tbp]
  \begin{center}
      \includegraphics[width=16cm]{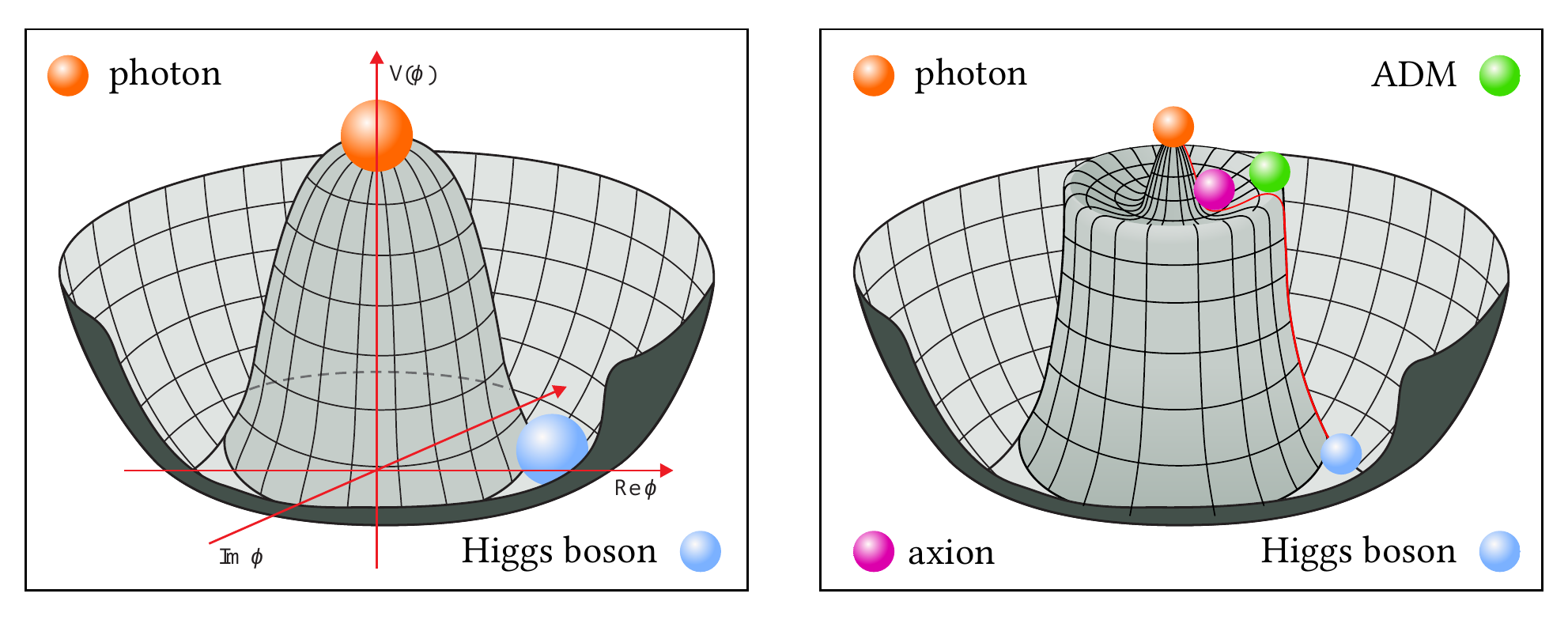}
  \end{center}
\caption{\textbf{(a)} The Mexican-hat potential energy density considered by Jeffrey Goldstone in his seminal paper \citep{Goldstone1961}. The energy density is a function of the real (Re) and imaginary (Im) values of a spinless field $\phi$. In the context of the electroweak theory developed later in the decade, the orange ball at the top of the hat would represent the symmetric solution for the potential, in which the photon, W bosons, and Z boson are all massless. The blue ball in the trough represents the solution after symmetry breaking. In that solution the W and Z bosons are massive and the photon remains massless. The steepness of the trough is related to the mass of the Higgs boson. Adopted from \cite{Lykken2013}. 
\textbf{(b)} Analog of the Mexican-hat potential energy density considered by \cite{Merkotan2017}, where Higgs-like particles do not represent the extra fundamental interaction, but carry the visible part of matter  Higgs-like baryon particles, and the DM component -- Higgs-like DM as ADM. The  axion here is an independent particle of DM, which is reproduced via the inflation due to quantum potential \citep{EingornRusov2015}, similarly to the Higgs-like particles.}
\label{fig-mexican-hat}
\end{figure*}

The Higgs discovery showed the scientists a new top-priority direction in solving one of the major puzzles in cosmology – the nature of DM. The experimental results pushed physicists away from the Z-bozon, so the only particle able to interact with DM directly was the Higgs particle (see \cite{Lykken2013}). Taking into account the cogitative history of the Higgs mechanism (see e.g. \cite{Lykken2013,Ivanov2017} and Refs. therein), the Higgs boson (see Fig.~\ref{fig-mexican-hat}a) is naturally a hot topic of new physics beyond the Standard Model through the Higgs window, with the strongest indications coming from its capacity to accommodate DM and provide a viable explanation of the baryon asymmetry of the Universe.

\textit{\textbf{Higgs mechanism problem.}} In our view, the theory of electroweak interactions \citep{Weinberg1967,Salam1964,Glashow1961} and the related Higgs mechanism \citep{Higgs1964a,Higgs1964b} contain a number of problems, revealed, among other things, by some inconsistencies with the recent experimental results (see \cite{Chatrchyan2012,Andersen2013,Aad2016}). In particular, the cited papers mention the experimental observations of the Higgs boson decay channels gathered in the report by Particle Data Group \citep{Patrignani2016}. As it is seen from these results, the total weak isospin of the particles in the final state of the decay may be integer only. At the same time, the Higgs field in the Standard Model \citep{Weinberg1967,Patrignani2016} is transformed according to the fundamental representations of the SU(2) group, and is two-component. So, according to the Standard Model, the Higgs boson has the weak isospin of 1/2 (see e.g. \cite{Peskin1995,Ryder1996}). The analysis of the $\beta$-decay as well as various channels of lepton decays \citep{Patrignani2016} leads to a conclusion that the weak interaction conserves the weak isospin. Since the carriers of other interactions have zero weak isospin, it must conserve in the processes related to other interactions as well.

On the theory side, we believe that the problem of the Standard Model is the introduction of the “new” non-gauge interactions. Particularly, the nonzero vacuum mean of the Higgs field is achieved due to the non-gauge $\varphi^4$ interaction. The only manifestation of such interaction being discussed is the spontaneous breaking of symmetry \citep{Goldstone1962}. So it is unclear how the existence of such interaction may be confirmed experimentally. The same is applicable to the Yukawa interaction of fermion fields with the Higgs field \citep{Weinberg1967,Peskin1995,Ryder1996}, which provides them with mass. This is not a result of some symmetry localization, and thus cannot be reduced to one of the known interactions!

Another point is the “unnatural” sign of the Lagrangian term quadratic in Higgs field. Although it does not lead to nonphysical results, the question remains of why there is only one field with such properties. And the only argument in favour of such notation seems to be the appearance of mass for the gauge particles (see e.g. \cite{Grojean2007,Quigg2009,Troitsky2012,Grojean2014}).

While considering the multiparticle fields 
(see \cite{Volkotrub2015,Chudak2016,Merkotan2017}), it was noticed that the
dynamic equation of the two-particle gauge field looks similar to the dynamic
equation of the $\varphi^3$-theory. The difference is that instead of the
squared mass there is an operator which may have non-negative eigenvalues
under certain conditions. The two-particle gauge field operators describe the
creation and annihilation of particles, which are the bound states of the gauge
bosons. Therefore, the self-action of the two-particle gauge field is the
manifestation of the interaction between quanta of the non-Abelian gauge field
and does not require the addition of an extra interaction. So in the model of
multiparticle fields we have both components necessary for the spontaneous
violation of symmetry. The Higgs boson is then (like in some other models) not
an elementary particle, but a bound state. This possibility was also pointed
out by Peter Higgs \citep{Higgs1964a,Higgs1964b}. However, he expected the
scalar field, which violates the symmetry, to consist of fermion fields, and
not of gauge bosons. \cite{Hoh2016a,Hoh2016b} considers the Higgs boson as a
bound state of gauge bosons, which are in the state of confinement. In the papers by
\cite{Volkotrub2015,Chudak2016} the two-particle gauge field describes the
gluon and quark confinement. The field with spontaneous breaking of symmetry
describes the creation and annihilation of the bound states of gauge bosons
without confinement, i.e. the bound states with a finite binding energy. Still,
although such fields have nonzero vacuum mean, they cannot lead to the 
appearance of mass for the gauge bosons, since the scalar representation
of the internal symmetry group is realized upon them. As a result, they cannot
interact with the single-particle gauge field and produce its mass.

For this reason in the present paper the aim is to consider the two-particle gauge field with the vector representation of the internal symmetry group. Since we are interested in the Higgs mechanism, we choose the SU(2) group \citep{Merkotan2017}.

\textit{\textbf{Multiparticle fields and the Higgs mechanism -- a brief discussion of the main results}} (see \cite{Merkotan2017}).
\phantomsection
\label{topic-MultiparticleFields}
In contrast to the (incomplete) Standard Model, the Higgs boson here is supposed to have integer weak isospin, which is consistent with the experimental data \citep{Patrignani2016} on its decay channels. It is considered as a bound state of W$^+$ and W$^-$ bosons -- the particles with weak isospin of 1. Such a bound state may possess a weak isospin of 0 or 1 or 2, depending on three terms in expansion into irreducible tensors (see Eq.~(7) in \cite{Merkotan2017}). The channel of decay into two photons (see \cite{Patrignani2016}) means that the Higgs boson may be in the state with weak isospin of 0. The channels of decay into two particles with 1/2 weak isospin, e.g. into electron and positron, add the possibility to observe the Higgs boson in the state with weak isospin of 1. The four lepton channels add the possible value of 2.

So, the known decay channels indicate that the Higgs boson is not an eigenstate of the weak isospin, and its measurement may yield only the integer values of 0, 1 or 2. This property is represented by Eq.~(7) in \cite{Merkotan2017}. It is essential that this model considers the self-action of the Higgs field (which provides the nonzero vacuum value) not as an independent non-gauge interaction, but as a manifestation of the non-Abelian gauge field. Namely, since the Higgs boson is considered to be a bound state of  W$^+$ and W$^-$ bosons, the interaction of Higgs bosons is a consequence of the non-Abelian weak SU(2) interaction of gauge bosons. This is formally seen from the derivation of equations for the two-particle gauge field in Sect.~3 in \cite{Merkotan2017}. The authors acknowledge that this procedure is somewhat artificial, but it lets one describe some important experimental details. In particular, the use of the similar procedure in the papers by \cite{Volkotrub2015,Chudak2016} made it possible to describe the quark and gluon confinement. So it lets one derive the Lagrangian term with “unnatural” sign of the quadratic mass, instead of simply introducing it, as it is done in the common Standard Model.

Unfortunately, the suggested model does not solve the problem of non-gauge introduction of the Yukawa interaction between fermion fields and the Higgs field into the Standard Model. From the physical point of view, fermions interact with W bosons, and the Higgs boson consists of the W bosons in this framework, so such interaction should indeed take place. It is unclear how to introduce it on the basis of the gauge principle though. All fundamental fermion fields interact with W bosons, so they can interact with each other through the Higgs field. On the one hand, it explains the neutrino mixing, which leads to neutrino oscillations (see \cite{Fukuda1998,Ahmad2001,Ahmad2002}). On the other hand, it is unclear why such oscillations exist for neutrinos only, i.e. why there are no oscillations between electron and muon, both of which may interact with the Higgs field through W bosons. This field should be “mixing” them just like the corresponding neutrinos.

Another problem not solved in the present paper is that it is impossible to divide the gauge fields into the charged field of W bosons and the uncharged field of Z boson in the gauge-invariant way. So in order to move to a new gauge, it is necessary to switch to the real fields $A_{a_1, g_1} (x)$ (see Sect.~2 in \cite{Merkotan2017}), do the gauge transformations and then separate the fields of W and Z bosons. This is the view of the W$^+$, W$^-$ and Z fields in the present paper.

According to Eq.~(7) in \cite{Merkotan2017}, in addition to the discussed antisymmetric and symmetric traceless irreducible terms, the two-particle gauge field contains a scalar part. According to Eqs.~(15)-(42) in \cite{Merkotan2017}, this part also has nonzero vacuum value, which appears as a result of spontaneous breaking of symmetry. So it is natural to call such a field the Higgs field, scalar with respect to interior indices. However, this field cannot interact with the gauge field and provide the mass for its components. The particles of this field carry no electric charge, because the field is real. They have no interior SU(2) indices, so they do not participate in weak interactions. Since they are considered as the bound states of gauge SU(2) bosons, such particles cannot participate in strong interactions. At the same time, such a scalar field contributes into the gravitational field through the energy-momentum tensor. Thus, the Higgs bosons, scalar with respect to interior indices, may be considered a candidate for a part of DM!

\phantomsection
\label{topic-FIFTH}
$\bullet \bullet \bullet \bullet \bullet$ And finally, fifth, we are interested in the link between the physics of the emission of gravitational waves and the generalized thermomagnetic EN~effect of ADM around a BH.

In this case we are looking for the possible signs of the quantum gravity alternatives near the BHs horizons through the gravitational waves emission from the BH mergers observed by the advanced Laser Interferometer Gravitational-Wave Observatory (aLIGO) \citep{Abbott2016a,Abbott2016b,Giddings2016}. Interestingly, by creating a phenomenological pattern for the consecutive echoes from exotic quantum mirrors, expected in the firewall or fuzzball paradigms, \cite{Abedi2017} obtained the first preliminary evidence of such echo signals in aLIGO data on BH mergers (false detection probability was 0.011, and the significance level was 2.5$\sigma$).

In our view, these quantum reflected echo signals (see Fig.~1 in \cite{Abedi2017}) are possible not only from the BH mergers, but also from the single SMBHs, where the ADM (and accordingly, baryons) modulation determines the formation and existence of the galactic gravitational waves (see Eq.~(2.1) of \cite{Allen1999}; Eq.~(2) of \cite{Abbott2016b}; Eqs.~(4.1) and (4.3) of \cite{Raidal2017}).

\begin{figure*}[tbp]
  \begin{center}
     \includegraphics[width=14cm]{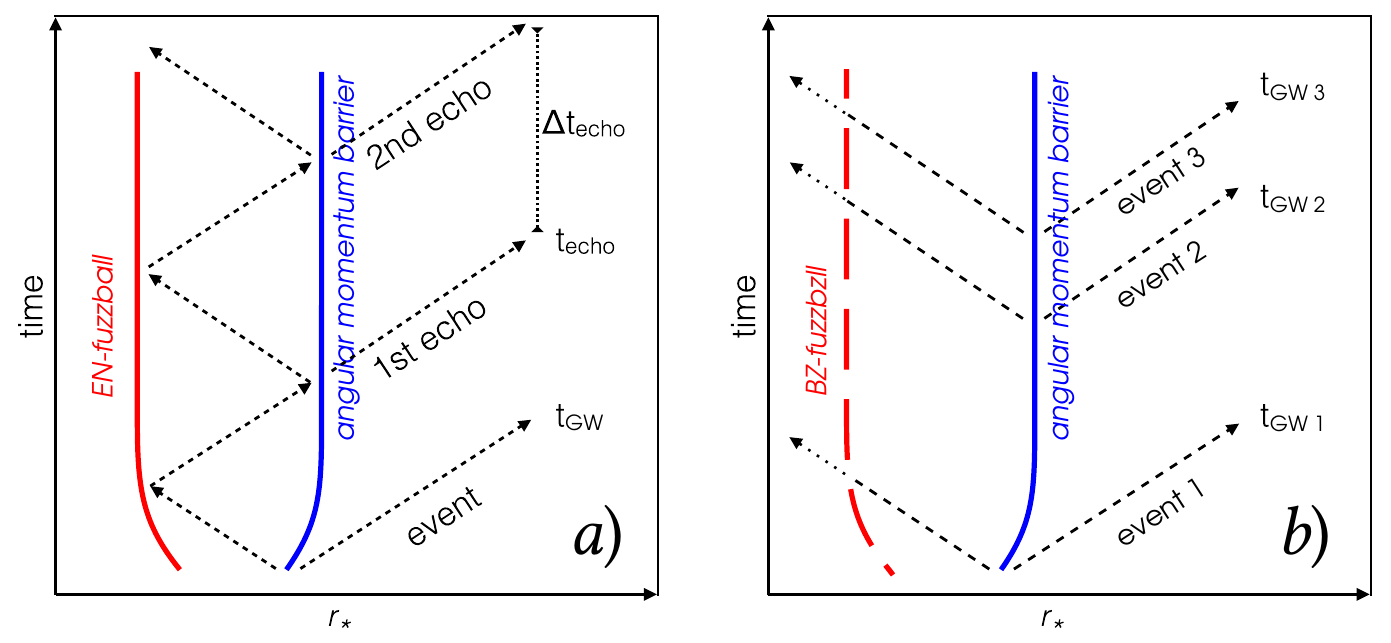}
  \end{center}
\caption{The gravitational waves signals as the signatures of quantum gravity near the BH boundary. 
\textbf{(a)} Space-time depiction of gravitational wave echoes from the EN~fuzzball (Meissner-like fuzzball (see Fig.~\ref{fig-meissner-effect}a)) on the stretched BZ~horizon, following a BH accretion event (adopted from \cite{Abedi2017}). Echoes are the gravitational waves trapped between the structure near the horizon and the angular momentum barrier (see Fig.~1 of \cite{Abedi2017}), where the EN~fuzzball is the result of the generalized thermomagnetic EN~effect on BHs (see Fig.~\ref{fig-meissner-effect}a).
The space-time ends in the string theory EN~fuzzball outside $r=2GM$, with no horizon. Unitary Hawking radiation at the energy $E \sim kT$ is emitted from the EN~fuzzball surface (see Fig.~\ref{fig-fuzzball}b).
\textbf{(b)} Gravitational waves ``evaporation'', penetrating through the surface of BZ~fuzzball, is predetermined by account for any BZ~fuzzball complementarity through collective oscillations generated by hard impacts of $E \gg kT$ quanta (see Fig.~\ref{fig-fuzzball}c). Adopted from \cite{Mathur2014}. The full proof of the physics of fuzzball complementarity is presented in Sect.~\ref{sec-conclusion}.}
\label{fig-gravitational-echoes}
\end{figure*}

But here it is very important to understand that the modulations of ADM (and,
accordingly, baryons) predetermine not only the emergence of galactic
gravitational waves, but also the modulation of gravitational waves. If the
changes in the gravitational field are non-chaotic, then the EN~fuzzball is
formed near the BH boundary (see Figs.~\ref{fig-gravitational-echoes}a, 
\ref{fig-meissner-effect}a, \ref{fig-fuzzball}b). Let us note that the EN, 
Meissner and Higgs effects are absolutely identical here!
On the other hand, if the changes in the gravitational field are chaotic, then
the surface of the BZ~fuzzball is determined by the BZ-like effect (see 
Fig.~\ref{fig-meissner-effect}b), the EN~fuzzball disappears instantly, in the
absence of which the BZ~fuzzball becomes ``visible'', e.g. through the gamma
ray burst (GRB) from the Higgs decay (see 
Figs.~\ref{fig-gravitational-echoes}b, \ref{fig-meissner-effect}b,
\ref{fig-fuzzball}c).

Since the holographic principle of quantum gravity, consisting in the relation between the magnetic fields of the 2D surface of ``tachocline'' and 3D volume of compact objects, is the consequence of the generalized thermomagnetic EN~effect on BHs, the physics of the gravitational waves modulation is determined by the physics of the EN or BZ~fuzzball surface near the BH boundary.
Simply put, this means that if these magnetic fields are exactly
``holographically'' equal (see Fig.~\ref{fig-meissner-effect}c), the ``mirror''
of the EN~fuzzball appears near the BH boundary with a completely extruded magnetic field from the EN~fuzzball surface (see Fig.~\ref{fig-meissner-effect}a and Fig.~\ref{fig-gravitational-echoes}a). If the holographic magnetic fields do not compensate each other, the BZ~fuzzball surface appears as a ``mirror with holes'' instead (see Figs.~\ref{fig-gravitational-echoes}c, \ref{fig-fuzzball}b), and it cannot push the magnetic field out of the BH (see Figs.~\ref{fig-meissner-effect}b and \ref{fig-gravitational-echoes}b). This simple and clear model assumes that the ``mirror'' in the form of the EN~fuzzball surface predetermines the appearance of Hawking radiation and signals reflecting the quantum echo signal (see Fig.~\ref{fig-gravitational-echoes}a), while the BZ~fuzzball surface near the BH boundary causes ``evaporation'' gravitational waves (see Fig.~\ref{fig-gravitational-echoes}b). This is not so far-fetched, because the ``evaporation'' of the gravitational waves that appear during the transition from the EN to BZ~fuzzball surface is predetermined by taking into account any complementarity of the BZ~fuzzball through their collective dynamics (see Figs.~\ref{fig-gravitational-echoes}b, \ref{fig-fuzzball}b), in which the problem of the information paradox, known as the obvious contradiction between the general theory of relativity and quantum mechanics, disappears completely!

\textbf{\textit{ADM modulation and magnetic fields near the BH boundary.}}
\phantomsection
\label{par-DM-modulation}
Taking into account the properties of dyonic BHs \citep{Hartnoll2007}, we find the repulsive magnetic field induced by the generalized thermomagnetic EN effect, where the wind from the BH accretion disk is governed by magnetic pressure, which within the wind must be as follows (see analogous equations by \cite{Shakura1973,Proga2003,Miller2016}):

\begin{equation}
\frac{B_{tacho}^2 (r,t)}{8 \pi} = n_{tacho} (r,t) k_B T_{tacho} ,
 ~~~ [T(r,t)]^{1/4} n(r,t) = const ,
\label{eq07-148}
\end{equation}

\noindent
where $B_{tacho}$ is the magnetic field of baryon matter, $T_{tacho}$ is the absolute temperature and  $n_{tacho} = \rho / m$ is the baryon number density.

\begin{figure*}[tbp]
  \begin{center}
     \includegraphics[width=15cm]{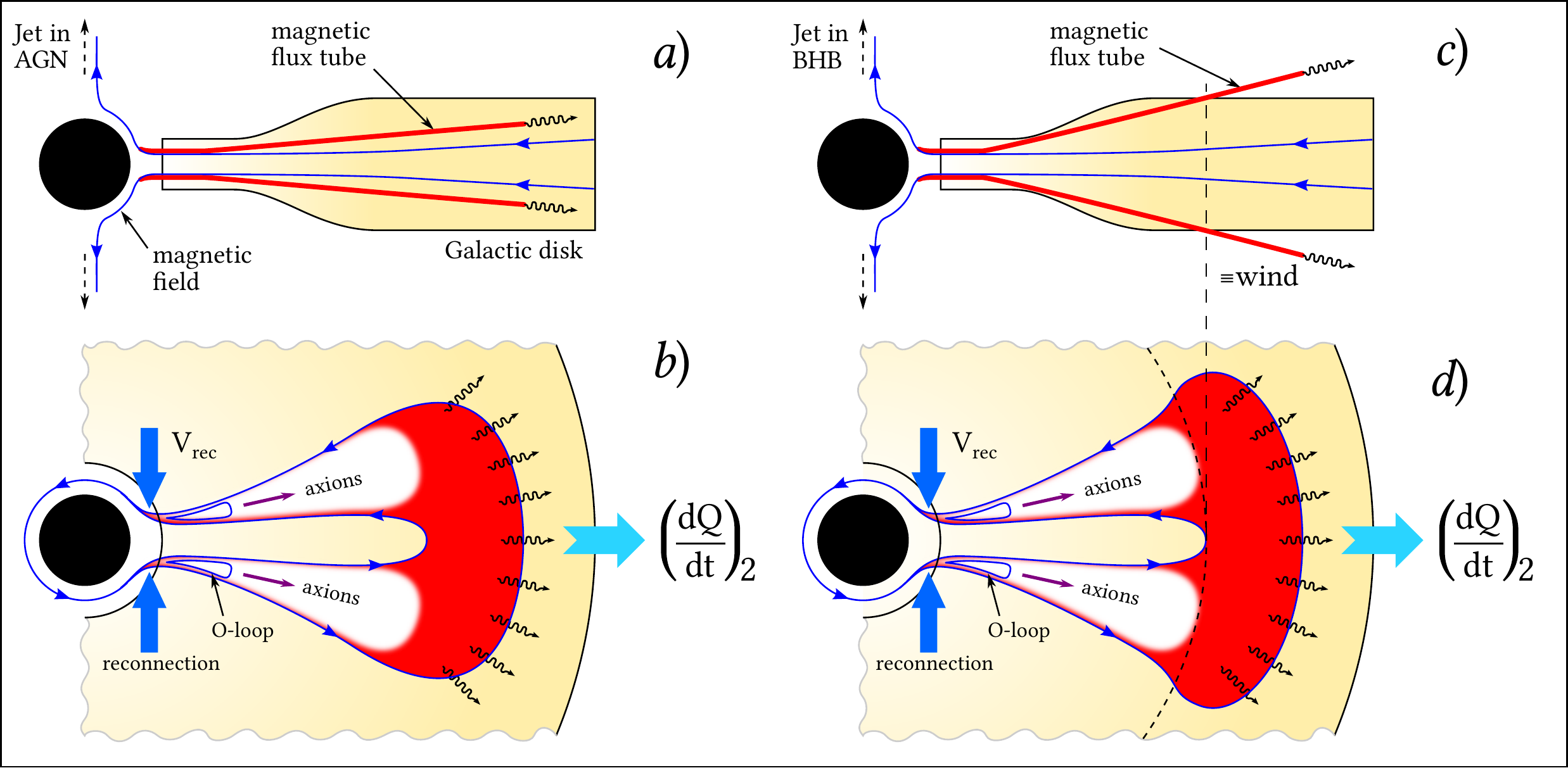}
  \end{center}
\caption{Schematic representation not to scale: Generalized thermomagnetic
EN effect and virtually empty MFT born
anchored to the BH tachocline and risen to the disk surface by the
neutral buoyancy ($\rho_{ext} = \rho_{int}$; see also Figs.~\ref{fig-axion-compton}
and~\ref{fig-lower-heating}). If the virtually empty magnetic tubes are born
with strong fields \textbf{(a)}, then the flux tubes practically do not reach
outside of the disk, while with less strong fields \textbf{(c)}, the flux tubes
go out of the disk at a small angle. So the visible (c,d) or invisible (a,b)
buoyant flux tubes are the analog of the variation or disappearance of the
wind through the various states of BH X-ray binaries (BHBs), which are
interpreted as a variation in the driving mechanism of the wind
(see \cite{Chakravorty2016} and Refs. therein). Based on the generalized
thermomagnetic EN effect, the buoyancy of MFTs
with the toroidal magnetic field $\gg 10^5 ~G$ in the BH tachocline,
ultimately, through $\nabla \rho$-pumping with the magnetic field $\sim 10^5 ~G$ and the dominant
Coriolis force creates an upward curved magnetic loop with a tilt angle from
the Joy's law (see (a), (c); see also Sect.~\ref{sec-tilt} and 
Fig.~\ref{fig-meridional-cut-tilt}a). The slope at low latitudes is near the
disk plane (see (a), (c); see also Fig.~\ref{fig-magtube-tilt}a,b). Here the keV
photons ((a)-(d); see analogous Fig.~\ref{fig-lower-heating}a), coming into the
tachocline from the solar-like radiation zone, are turned (b,d) into axions
by means of the horizontal magnetic field of the O-loop ((b,d); see analogous 
Figs.~\ref{fig-lampochka}a, \ref{fig-axion-compton}a and 
\ref{fig-lower-heating}a). Some small photon flux can still pass through the
``ring'' between the O-loop and the tube walls (see Figs.~\ref{fig-lampochka}a,
\ref{fig-axion-compton}a) and reach the solar-like penumbra of the MFT in the disk (see Fig.~\ref{fig-axion-compton} and 
Sect.~\ref{sec-empty-tubes}). The physics and the possible estimate of the
radiative heating $(dQ/dt)_2$ (see e.g. Fig.~\ref{fig-lower-heating}a) passing
through the ``ring'' of the magnetic tube (b,d), the speed of the reconnection
$V_{rec}$ (see analogous Fig.~\ref{fig-lower-reconnection2}), O-loops (see 
Figs.~\ref{fig-lampochka}a, \ref{fig-axion-compton}a, 
\ref{fig-lower-reconnection}d, \ref{fig-lower-heating}a), and the lifetime of
the magnetic tube are presented in Sect.~\ref{sec-radiative-heating}.
}
\label{fig-GalacticDisk-MagTube}
\end{figure*}

This means that the existence of the repulsive magnetic field induced in the “tachocline” near the BH boundary (as well as on the other compact objects, see Figs.~\ref{fig-accretion-disk}b, \ref{fig-GalacticDisk-MagTube}) plays an essential role in the acceleration and collimation of the wind, as well as the accretion disk dynamics and evolution \citep[see e.g.][]{Miller2017}. Let us explain ``where the disk magnetic field comes from”.

The accretion disks have to transfer the angular momentum in order for the matter
to move radially towards a compact object. The inner viscosity of the magnetic processes
\citep[see][]{Shakura1973,Balbus1991,Balbus1998,Hawley1995,Pariev2007p1,Pariev2007p2,Colgate2014,Colgate2015},
and disk winds \citep[see][and Refs. therein]{Blandford1982,Miller2006a,Miller2008,Miller2017,Fukumura2017}
may in principle transfer the angular moment, but we still lack for the proof that it happens.

It becomes clearer with our alternative (see Figs.~\ref{fig-accretion-disk}b, \ref{fig-GalacticDisk-MagTube}), which is substantially different from the other candidate processes, e.g. \cite{Blandford1982}, where the gas might escape along the toroidal magnetic field lines, transferring the angular momentum and allowing the mass transfer through the disk and then accreting to the BH (see differences between Fig.~1 of \cite{Pariev2007p1,Pariev2007p2} and Fig.~\ref{fig-accretion-disk}b). In the generally accepted Shakura-Sunyaev disk model \citep{Shakura1973}, depending on the initial conditions, the method of matter compression and the degree of ordering of the magnetic field, the energy of matter is $\varepsilon = 3 \rho (kT / m) + bT^4 = \rho (v_s^2 / 2)$, as well as can sufficiently exceed it, reaching the magnetic pressure value $(B^2 / 8\pi) \sim \rho v_s^2 / 2 = \rho (GM / R)$ ($energy/cm^3$). In this last case, the stress ($w_{r \varphi} = \vert B_r \times B_{\varphi} / 8 \pi \vert$) and the efficiency of the angular momentum transport are so high that the radial accretion will occur (see \cite{Shakura1973} and Refs. therein).

The essence of fundamental magnetic processes associated with quantum gravity and the generalized thermomagnetic EN effect in the “tachocline” near the BH boundary may be described rather simply as follows. According to our understanding, one of the fundamental effects of the holographic principle of quantum gravity is the existence of ``tachoclines'' in all stellar objects in the Universe, including all galaxies and, of course, our Sun, magnetic white dwarfs, neutron stars and BHs of the Milky Way (see Sect.~\ref{sec-toroidal-field}).

Let us recall that the solar radiation zone rotates approximately like a solid,
and the convection zone has a differential rotation. This leads to the
formation of a very strong shear layer between these two zones, called the
tachocline. Similar physics of the tachocline exists in a BH. As a
result, the tachocline shear layers produce virtually empty MFTs
(see Sect.~\ref{parker-biermann}), anchored in the BH tachocline
and rising to the surface of the disk (see 
Fig.~\ref{fig-GalacticDisk-MagTube}b,d). Since it is known that disk parts
rotate around a BH at different speeds that reinforce the fields
(Fig.~\ref{fig-GalacticDisk-MagTube}a), this turns the accretion disk into
a vortex, pulling the substance into a BH and fueling winds that blow
some of it outwards (see Fig.~\ref{fig-GalacticDisk-MagTube}b).

So, if the physics of winds in the disk is identical to the physics of
practically empty MFTs, which start from the tachocline and rise
to the outer (Fig.~\ref{fig-GalacticDisk-MagTube}c,d) or inner
(Fig.~\ref{fig-GalacticDisk-MagTube}a,b) part of the disk, then it means that
the generalized thermomagnetic EN effect produces the magnetic
tubes or the so-called winds and, on the one hand, the substance current
flowing in the direction of the pole of the poloidal (meridional) field of the
tachocline in the form of jets.

If we recall that the variations of the toroidal magnetic field and ADM
density in the BH tachocline exactly anticorrelate with each other
(see Eq.~(\ref{eq07-151})), this means that in strong fields the magnetic tubes
rise only inside the disk, which is equivalent to the disappearance of winds,
and thus they slow down the speed of the accretion. And the opposite, in weak
fields the magnetic tubes rise out of the disk surface (see 
Fig.~\ref{fig-GalacticDisk-MagTube}c), which is equivalent to observing the
disk winds, and thus they increase the speed of the accretion.

Here a remarkable connection between the density $\rho_{accr}$ of the
BH accretion material and the toroidal magnetic field $B_{tacho}$ of
the tachocline appears:

\begin{equation}
j_{\perp} \sim \frac{1}{B_{tacho}} \sim \rho_{accr}^{5/2} ,
\end{equation}

\noindent
where $j_{\perp}$ is the density of the poloidal (meridional) current towards
the BH pole, which is perpendicular to the toroidal magnetic field
$B_{tacho}$ in the tachocline due to the thermomagnetic EN
effect (see Eq.~(\ref{eq06-09}), inset in Fig.~\ref{fig-meissner-effect}a, and
also Fig.~\ref{fig-lampochka}a).

When the magnetic pressure is large at low densities and
high temperatures, the low densities of the BH accretion material and
high magnetic fields in the tachocline lead to the formation of highly
collimated spectrally-hard jets, but without a ``visible'' magnetic tube
(jet in AGN in Fig.~\ref{fig-GalacticDisk-MagTube}c). And vice versa, when the 
magnetic pressure is relatively weak at high densities and 
low temperatures, the high density of the BH accretion material and
relatively low magnetic fields in the tachocline lead to the formation of
strongly collimated spectrally-soft jets and a visible magnetic tube, i.e. 
a visible disk ``wind'' (jet in BHB in Fig.~\ref{fig-GalacticDisk-MagTube}c,d).

So, the resulting conclusion looks quite clear: the physical nature of the
generalized thermomagnetic EN effect is the cause of both the
formation of magnetic tubes rising from the BH tachocline to the disk
and the formation of meridional currents in the direction of the BH
pole, which generate jets, for example, in AGN or BHB.

Since the physics of magnetic tubes, rising from the BH ``tachocline''
to the disk, is almost identical, in our view, to the physics of visible or
``invisible'' disk winds, below we consider the common properties of the
generalized thermomagnetic EN effect and the known models of
disk winds. Curiously enough, the latter are associated with the mechanism of
DM variations around a BH.

Such winds may arise from various processes, which makes their sources disputable (see e.g. \cite{Fukumura2017} and Refs. therein).
However, the X-ray spectroscopic data and analysis of the wind associated with the X-ray binary (XRB) GRO J1655-40 \citep{Miller2006a,Miller2006b,Miller2008,Miller2012,Kallman2009,Luketic2010} argued in favour of the magnetic origin, excluding all candidate processes except for the following two: the semi-analytic MHD wind model of \cite{Fukumura2017} and the MHD outflow model of \cite{Chakravorty2016}. Plus our model of the generalized thermomagnetic EN effect.

The observations indicate that disk winds and jets in X-ray binaries are anticorrelated \citep{Miller2006b,Miller2008,Miller2012,Neilsen2009,King2012a,King2013,Ponti2012}. This indicates a link between disk properties, magnetic field configurations and outflow modes, the repulsive magnetic field induced by the generalized thermomagnetic EN effect (see Eq.~(\ref{eq07-148})), which produces the variations of the magnetic fields near the BH boundary (see e.g. \cite{Eatough2013,Zamaninasab2014,Johnson2015}),

\begin{equation}
\frac{\left[ B_{tacho}^2 (r,t) \right]_{max-cycle}}{\left[ B_{tacho}^2 (r,t) \right]_{min-cycle}} = 
\frac{\left[n_{tacho}^{min} (r,t) \cdot T_{tacho}^{max} (r,t)\right]_{max-cycle}}
{\left[n_{tacho}^{max} (r,t) \cdot T_{tacho}^{min} (r,t)\right]_{min-cycle}} ,
 ~~~ [T (r,t)]^{1/4} n (r,t) = const ,
\end{equation}

\noindent
as well as the variations of the ADM density $n_{tacho} ^{ADM}$ and baryon matter $n_{tacho}$,

\begin{equation}
\frac{\left[ n_{tacho}^{min} (r,t) \right]_{max-cycle}}
{\left[ n_{tacho}^{max} (r,t) \right]_{min-cycle}} \approx
\frac{\left[ n_{tacho}^{ADM-min} (r,t) \right]_{max-cycle}}
{\left[ n_{tacho}^{ADM-max} (r,t) \right]_{min-cycle}}\, ,
\label{eq07-150}
\end{equation}

\noindent
determined by the modulations of the ADM density $n_{tacho}^{ADM}$.

For these magnetic field configurations, let us point out some important features of the density variations of the dark matter $n_{tacho}^{ADM}$, that can distinguish between ``cold'' and ``warm'' wind solutions from near-Keplerian accretion disks.

It is known (see \cite{Chakravorty2016}) that the X-ray spectra of BH X-ray binaries (BHBs) contain blueshifted absorption lines, which means the presence of outflowing winds. The observations also show that the disk winds are equatorial and they mostly occur in the Softer (disk dominated) states of the outburst, being less in the Harder (power-law dominated) states.

The properties of the accretion disk are used to infer the driving mechanism of the winds (see e.g. \cite{Neilsen2012} and Refs. therein). And more or less prominent winds through the various states of the BHB have been interpreted as a variation in the magnetic driving mechanism of the wind \citep{Miller2006a,Miller2006b,Kallman2009,Neilsen2012}.

In our case the ADM density $n_{tacho}^{ADM}$ modulations, associated with the generalized thermomagnetic EN effect (see Eqs.~(\ref{eq07-150}), (\ref{eq07-151})), lead to the variations of the toroidal magnetic fields in the accretion disk (see Fig.~\ref{fig-GalacticDisk-MagTube}). In order to understand the main motivation for the toroidal magnetic field variations in the accretion disk, forming the ``cold'' and ``warm'' wind, it is necessary to discuss the difference between the winds and jets from accretion disks.

Given Eq.~(\ref{eq07-150}) and the high (low) magnetic pressure, the density of ADM is relatively low (high), and the gas temperature is high (low). This means that with the high density of ADM and low temperature, the magnetic driving mechanism produces the accretion disk wind, which is equatorial (see Fig.~\ref{fig-accretion-disk}c, and the analogous physics of the X-rays of axion origin in Figs.~\ref{fig-axion-compton}a, \ref{fig-lampochka}a, and \ref{fig-GalacticDisk-MagTube}b,d) and occurs in the soft states of the BHB outbursts (see e.g. the analogous model by \cite{Chakravorty2016}). Alternatively, with the low density of ADM and high temperature, the magnetic pressure is rather high and, consequently, the magnetic driving mechanism produces the weak or less prominent wind in the hard states (AGN jets).

Either way, the nonrelativistic disk winds and relativistic jets are anticorrelated, since the relativistic AGN jet, induced by the vertical toroidal magnetic field (see Fig.~\ref{fig-GalacticDisk-MagTube})
and collisions between ADM and nuclei
in the close vicinity of a SMBH (see e.g. the analogous model by \cite{Lacroix2016}), is determined by the modulations of the ADM density (see Eq.~(\ref{eq07-150})):

\begin{equation}
\frac{\left[ B_{tacho} \right]_{max-cycle}}
     {\left[ B_{tacho} \right]_{min-cycle}} =
\left \lbrace 
\frac{\left[ \rho _{tacho}^{ADM-max} (r,t) \right]_{min-cycle}}
     {\left[ \rho _{tacho}^{ADM-min} (r,t) \right]_{max-cycle}}
\right \rbrace ^{3/2} .
\label{eq07-151}
\end{equation}

This eventually means that the strong magnetic fields near the BH boundary are caused by quantum gravity of the dyonic BH \citep{Hartnoll2007}, which determines the existence of the generalized thermomagnetic EN effect (see Fig.~\ref{fig-meissner-effect}b and \cite{Spitzer1962,Spitzer2006,Hartnoll2007,Rusov2015}). 
So we understand that the major effect of quantum gravity here is that the
initial acceleration (deceleration) of the disk winds and BHB flares (AGN jets)
originate from less (more) strong magnetic fields in the
accretion disk near the BH boundary (see also Eq.~(\ref{eq07-151}), Fig.~\ref{fig-GalacticDisk-MagTube} and inset in Fig.~\ref{fig-meissner-effect}a), which predetermine the modulations of the ADM density -- the process connected with the variability of the accretion flows, nonrelativistic disk winds and relativistic BHB or AGN jets.

\subsubsection{Who generates and controls the modulation of dark matter in the supermassive black hole?}
\label{sec-who-generates}

The basic idea is that the modulation of the luminous AGN, associated with the periodical variability, is supposed to be the modulated accretion of baryon matter onto the SMBH, which modulates ADM of baryon nature in the SMBH. Here are some points in support of such a scenario through Occam's razor:

$\bullet$ It is known that ``collisions and interactions between gas-rich galaxies are thought to be pivotal stages in their formation and evolution, causing the rapid production of new stars, and possibly serving as a mechanism for fueling supermassive black holes'' (\citet{Goulding2018}; see also \citet{DiMatteo2005,Gatti2015,Koss2018,Steinborn2018,Storchi2019}).

$\bullet$ A significant result of the last decade is the discovery that the ``...the mass of the central black holes (BHs) and properties of the host galaxies, notably the stellar bulge mass or central stellar velocity dispersion''\citep{Bogdan2015}, as well as
the total gravitational mass of the host galaxy, or the mass of the ADM halo, closely correlate with each other (see e.g. \cite{Magorrian1998,Gebhardt2000,Ferrarese2000,Tremaine2002,Gultekin2009,McConnell2013,Bogdan2015,Zahid2018,Arguelles2019}).

$\bullet$ The idea of the galaxies spending most of their life in star formation and in the main sequence (see \cite{Elbaz2011,Stanley2017}) also suggests that the star formation and BH accretion are closely related (see \cite{Alexander2012,Chen2013,Hickox2014,Asmus2014,Thacker2014,Stanley2017,Harrison2017,Cresci2018, Yang2019}).

$\bullet$ Another possible evidence of the causal link between the rising of the AGN activity and the quenching of a galaxy can be obtained from the observed correlation of the star formation rate with the AGN luminosity (see \cite{Shao2010,Lutz2010,Bonfield2011,Harrison2012,Hickox2014,Gurkan2015,Bernhard2016,Stanley2018,
Dai2018,Brown2019,Aird2019}).

$\bullet$ The most important point is the fact that ``understanding the relationship between the galaxies in which the active galactic nuclei (AGNs) are located and the halo of dark matter in which they are located is the key to limiting how black hole refueling is triggered and regulated''\citep{Leauthaud2015,Netzer2013,Gatti2016,Ballantyne2017,Padovani2017,Georgakakis2018,Outmezguine2018}.

These points are connected with the physics of the baryon ADM modulation in the SMBH. In order to solve this problem, it should be remembered that the question of how the AGN feedback controls the star formation and the growth of luminous galaxies (see \cite{Asmus2015}) is solved in a rather straightforward way for the modulation of the luminous AGN (see e.g. \cite{McNamara2007,Fanidakis2013}). The heating and cooling of gas in galactic clusters coupled by AGN feedback, which suppresses the star formation and luminous galaxies growth, necessarily using the AGN clustering (see \cite{Fanidakis2013,RetanaMontenegro2017}), may be interpreted in terms of their distribution in host ADM haloes (see \cite{Fanidakis2013,Hickox2016}).

A rather natural question arises here: if the modulations of the ADM density at the GC give rise to the modulations of the ADM halo density near the solar surface -- through the density waves (see e.g. \cite{Lin1964,Toomre1977,Bertin1996,Dobbs2014,Shu2016}), crossing the spiral arms (see \cite{Binney2008,Baba2013,Sellwood2014}), is there any ``experimental'' link between the ADM density modulation in the solar interior and the sunspot number?

\begin{figure}[tbp]
  \begin{center}
    \includegraphics[width=12cm]{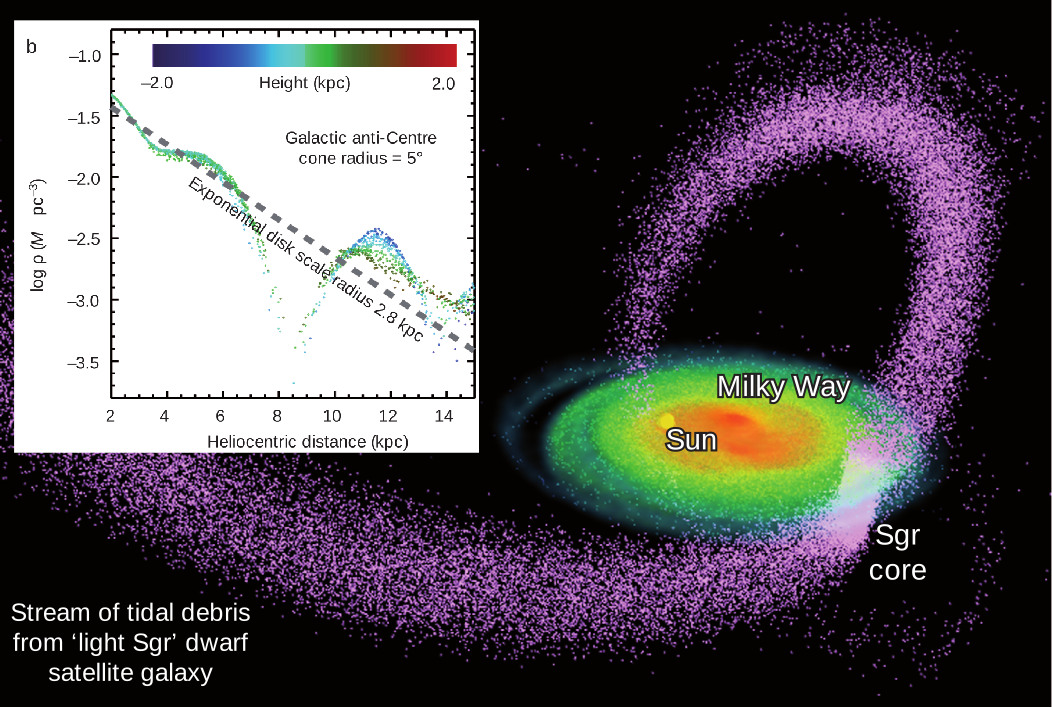}
  \end{center}
\caption{Global rendering of the Sagittarius dwarf galaxy (Sgr) tidal debris and the Milky Way disk \citep{Purcell2011}. All simulations (of DM and stellar components in both the Milky Way and the Sgr progenitor) used the parallel N-body tree code ChaNGa with the gravitational softening length of one parsec and followed the evolution of 30 million particles with masses in the range $1.1-1.9 \cdot 10^4 ~M_{Sun}$. Adopted from Fig.~1 of \citet{Purcell2011}.
Inset: Endstate disk overdensities in the ‘heavy Sgr’ simulation \citep{Purcell2011}. The local stellar density in a thin cone directed from the solar neighbourhood towards the Galactic anti-center, as a function of the heliocentric distance along the cone. In both panels, off-plane overdensities resemble ‘multiple tributaries’ observed in the Milky Way \citep{Grillmair2006}, and the spiral arm wrapping at a distance of around 10~kpc is strikingly similar to the Monoceros ring \citep{Penarrubia2005}. Adopted from Fig.~4b of \citet{Purcell2011}.}
\label{fig-dwarf-galaxy}
\end{figure}

Until now, ``an implicit assumption in modeling the tracer density profile was that the local neighborhood is axisymmetric, and long-lived spiral patterns in stellar disks are in dynamic equilibrium''~\citep{Buch2019}. However, according to~\citet{Buch2019}, the growing evidence in \citet{Gaia2016,Gaia2018} data for asymmetry in the vertical number counts \citep{Widrow2012,Bennett2018}, vertical waves in the disk at the Sun position \citep{Purcell2011,Gomez2013,Carlin2013,Widrow2012,Widrow2014,Xu2015,Gomez2017,Carrillo2018,Carrillo2019,Laporte2018a,Laporte2018b} and the kinematic substructure \citep{Antoja2018,Myeong2018,Necib2018} warrants a closer look at sources of disequilibrium in the solar neighborhood.
Additionally, it is necessary to include the indicators that connect the movements of tracers in different directions, for example, vertical density waves propagating through the spiral arms of our galaxy (see e.g. inset in Fig.~\ref{fig-dwarf-galaxy}).

From here we understand that the ADM halo density
variability is a consequence of the existence of sources of disequilibrium in
the solar neighborhood based on the vertical density waves in the Galactic disk
(see \cite{Purcell2011,Gomez2013,Carlin2013,Widrow2014,Carrillo2018,
Carrillo2019,Laporte2018a,Laporte2018b}). The passing satellite galaxy or a ADM sub-halo excites vertical, radial and azimuthal coherent oscillations of
the stellar disk of the Milky Way from the  solar neighborhood towards the
Galactic anti-center, where the vertical wave perturbations act in the direction
perpendicular to the Galactic plane (see inset in Fig.~\ref{fig-dwarf-galaxy};
see also Fig.4 in \cite{Purcell2011}).

This asymmetry as a source of disequilibrium becomes more evident when
compared with the well-known model of the Milky Way disk response to the tidal
interaction with the Sagittarius dwarf spheroidal galaxy
(Sgr; see \cite{Purcell2011}), that leads to the galactic north-south
asymmetry (in both the number and the mean vertical velocity distribution of
stars \citep{Widrow2012}) and the vertical wave-like behavior of modes
(see e.g. Fig.~4 in \cite{Purcell2011}) penetrating in the Milky Way disk. Not
surprisingly, it is qualitatively consistent with the observations
(see \cite{Purcell2011,Widrow2012,Gomez2013}).

Note that the vertical perturbations in the galactic disk are the result of the
orbiting Sgr satellite $\rightarrow$ host DM halo $\rightarrow$ disk
interaction. The reason for this has been explained in great detail by
\cite{Vesperini2000}, and also by \cite{Gomez2016}. The main driving force
behind this perturbation is low-mass, low-velocity fly-by encounter, which
intersects the plane of the disk at a relatively large galactocentric distance.
Although the Sgr satellite is not massive enough to directly perturb the galaxy
disk, the density field of the host DM halo responds to this
interaction, and consequently, the asymmetric features of the density
perturbations are transmitted to the internal parts of the primary system,
acting on a deeply submerged galactic disk. Such vertical perturbations as a
natural consequence of dynamic friction
\citep{Chandrasekhar1943,Weinberg1985,Weinberg1986}, which excites the DM inside the host halo, can cause the formation of vertical
structures of the galactic disk \citep{Weinberg1995,Weinberg1998,Vesperini2000}.

Considering the latest $\sim$2~Gyr evolution of the Milky Way, according to
\cite{Laporte2018a,Laporte2018b}, it may be particularly affected not only by
the Sgr satellite, but also by the Large Magellanic Cloud (LMC) at the same
time. Skipping the complex but remarkable work of
\cite{Laporte2018a,Laporte2018b}, one can simply justify that the physical
cause is the same as in the works by
\cite{Weinberg1995,Weinberg1998,Vesperini2000}: the addition of force from the
DM halo induced by the resonant interaction between the DM halo of the
Milky Way and both the Sgr satellite and the Magellanic clouds
(see e.g. \cite{Weinberg1998}) may be sufficient to explain the observed warp
(see \cite{WeinbergBlitz2006}), 
which causes the observed vertical density perturbation in the direction of the
Milky Way disk (see analogous inset in Fig.~\ref{fig-dwarf-galaxy}). In fact, 
\cite{Laporte2018a,Laporte2018b,Laporte2019} showed that the nonlinear
combination of the effects caused by LMC and Sgr remains dominant due to the
Sgr effect, which is the most convincing initiator and the main architect of
the spiral structure and vertical perturbation of the Milky Way stellar disk.

As will be described in detail, in addition to the large
timescale influence of the Sgr satellite and LMC towards the center of the
Milky Way, we are interested in the small timescale interaction of the orbital
satellite of AGNs $\rightarrow$ bar/host DM
halo $\rightarrow$ accretion disk.

Let us start with the condition of birth and variability of AGNs. On the one
hand, it is believed that radiation from the AGN is the result of matter accretion
onto a SMBH (see \cite{Shakura1973,Rees1984}) in the center of its host galaxy.
On the other hand, it is known that the luminosities of quasars and other
AGNs vary from X-rays to radio wavelengths and at time
scales from several hours to many years. It is clear that the matter of a
quasar emits light not by command, but by virtue of the processes occurring in
it. The fact of synchronism in all points of the region, that is
simultaneity of change in conditions and magnitude of the radiation, indicates
the compactness of this quasi-star object. It should be also kept in mind that
the light from distant quasars comes to us very ``reddened''. Funny enough, the
wavelength increased due to the redshift, as if deliberately to pass through the
Earth atmosphere and be registered in the instruments. In this regard we are
interested in the evolution of the ultraviolet luminosity function of AGNs
between redshifts z~=~4 and 7 using the X-ray/near infrared selection criterion
(see \cite{Kolodzig2013,Giallongo2015,Kulkarni2019}), especially in the
evolution of the luminosity function of the Magellan quasars
(see e.g. \cite{Kozlowski2009,Ivanov2016}).

Most important is the fact that the observation of the AGN radiation
variability is the result of the accretion rate variability in the SMBH at the
center of its host galaxy. Among the most well-known scenarios of the basic
structure of the accretion disk, which can explain the AGN strong and rapid
variability, are the models of magnetically elevated (or “thick”) disks
(see \cite{Dexter2014,Begelman2015,Begelman2017,Dexter2018}), the transitions
of the accretion state (see \cite{Noda2018,Ruan2019,Graham2019}), instability
arising as a result of the magnetic moment near the inner stable circular orbit
around the accretion disk (see \cite{Stern2018,Ross2018}) and the misaligned
disks (see \cite{Nixon2012,Nixon2013,Nealon2015,Nixon2016,Pounds2017,
Pounds2018,King2018}) and disk wind
models (see \cite{Elitzur2006,Elitzur2009,Elitzur2014,MacLeod2019}).

Our disk wind model 
is very different from the known scenarios of the above-mentioned models.

We have shown that the existence of toroidal magnetic processes in the accretion
disk is the result of the fundamental holographic principle of quantum gravity
and, consequently, the existence of the generalized thermomagnetic 
EN effect in the BH ``tachocline''.
It is known that a toroidal (i.e. azimuthal) magnetic field is subject
to both Parker (see \cite{Parker1966,Parker1967,Parker1969b,Parker1979a}) and 
rotational shearing instabilities 
(differential rotation; see \cite{Velikhov1959,Chandrasekhar1960,Balbus1991}).
The most important fact is that both amplification mechanisms operate consistently
and at the same time (see e.g. \cite{Foglizzo1994,Foglizzo1995}), depending
on such differential rotation forces, which turn the accretion disk into a
whirl, drawing matter into a BH, and rousing winds that blow some
of it out.

This means (see Eq.~(\ref{eq07-151})) that at the high density of DM the
magnetic driving mechanism creates an accretion disk wind, which is equatorial
(see Fig.~\ref{fig-accretion-disk}c) and occurs in soft states of BHB bursts
(see e.g. a similar model by \cite{Chakravorty2016}). Alternatively, at the low
density of ADM, the magnetic pressure is rather high and, consequently,
the magnetic driving mechanism creates a weak or less noticeable wind in hard
states, which characterizes the appearance of jets from a BH.

Hence, we understand that the force of the magnetic driving mechanism in the disk,
connected by hydrostatic equilibrium, is weak or strong when the vertical size
of the horizontal magnetic field from the center of the disk is inversely
proportional to the density (i.e. the sum of gravitation, thermal gas pressure,
CR gas pressure and DM halo density 
(see e.g. \cite{Rodrigues2015})) of
the gas in the disk (see e.g. Fig.~2 in \cite{Parker1966}, Fig.~3 in
\cite{Parker1967}). In other words, the larger the toroidal field, the smaller
the vertical size of the horizontal magnetic field from the middle of the
disk, and vice versa.

This raises the question of how the observational data on AGN (or jet)
variability, which theoretically anticorrelates with a variation of the
toroidal magnetic field in the accretion disk, will be related to the
observational data on variations in the accretion rate or e.g. the magnetic
disk winds? In our opinion, despite the elusiveness of direct observations,
some ideas, for example, of \cite{Rodrigues2015},
about the possible observation of Parker loops from synchrotron radiation of 
galaxies near their edges, or from the Faraday rotation, or the idea of 
\cite{Chakravorty2016} on the possible observable difference between winds and
jets from accretion disks can provide the indirect observational support.

In this sense, we are very interested in obtaining the indirect observations of
anticorrelation between winds and jets from accretion disks. The point is that
the modulation of a ADM halo leads to the situation when the high density
of ADM corresponds to the disk wind and high accretion rate, while
the low density of ADM corresponds to the low accretion rate and, as a
result, to the jets from the BH. As we understand, in order to get a
generalized picture of the connections between winds and jets from accretion
disks, it is necessary to obtain the indirect observation, which physically
explains the remarkable connection between all mentioned systems: the 
modulation of the ADM halo at the center of the galaxy $\rightarrow$ waves
of the vertical gravitational density from the disk to the solar neighborhood 
$\rightarrow$ variations (cycles) of sunspots $\rightarrow$ the variability of
the local positions of orbital S-stars near the BH.

Our plan is the following. The main goal is to show that the observed
variability of the local positions of the S-stars orbiting near the BH
is an indicator (or dynamic probe) of the disk wind speeds variation or,
equivalently, of the accretion rate variations of the BH. 
The intermediate goal is to briefly discuss the formation of the elliptically
orbital distributions and periodic cycles of the S-stars, which are located
in 0.13 ly $\approx$ 0.04 pc (see Fig.~\ref{fig-accretion-disk}c)
from the BH.

Here we present a detailed analysis of the effects of DM capture
and energy transport on the structure and evolution of main-sequence B-stars,
specifically those which might exist at the GC. First we need to
highlight some important modulation properties of the S-stars and ADM around
the BH:

$\bullet$
The greatest
capture happens when the star is farthest from the center of the galaxy, at
apoapsis. This is because it slows down relative to the ADM halo and achieves a
significant capture rate for a time
before turning back towards the BH. By the time it reaches
periapsis, the star is moving so quickly that the capture is essentially zero,
regardless of how high the ADM density is.

$\bullet$ When the S-stars, and especially, for example, S102 and S2, approach
(retreat from) the center of the Milky-Way, at periapsis (apoapsis), the
ADM density increases (decreases) and, thus, increases (decreases) the baryonic
sector in the subparsec region near the SMBH at the GC. As a
result, the variability of the ADM density and the baryonic matter
density are determined by the variability of the gravitational potential around
the BH, which is controlled by tidal interactions with other galaxies 
in the form of the Virgo-like cluster (see e.g. \citet{Semczuk2016}).

\begin{figure}[tbp!]
  \begin{center}
    \includegraphics[width=9cm]{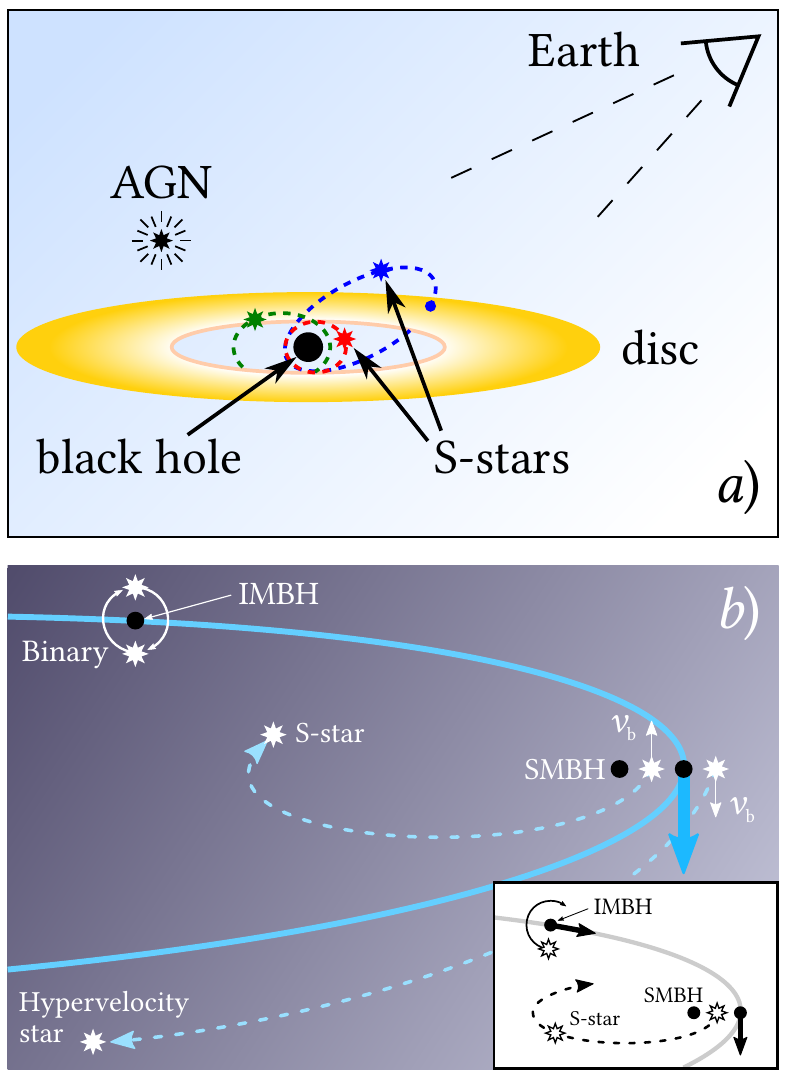}
  \end{center}
\caption{S-stars between the inner edge of the disk and the BH
``tachocline''.
\textbf{(a)} Simplified sketch of the observed variability of local positions of orbital
S-stars near the BH.
\textbf{(b)} Here is how the solution of the problem of energy conservation on the basis
of four bodies works: a single body (SMBH) exchanges partners of binary
stars orbiting around the IMBH, and through an
extreme gravitational tidal field, one star is captured (by the SMBH) and loses
energy, while the other runs away, gets all this energy and (through
hypervelocity) just flies out of the galaxy (see \citet{Hills1988,Hills1991,
BrownEtAl2005,BrownEtAl2014,BrownEtAl2018,Brown2015,Brown2016b,SubrHaas2016,
Kenyon2018,Rasskazov2019}). This is the so-called double slingshot.
\textbf{Inset:} If two objects -- an IMBH and a star rotating around it -- are
approaching a SMBH, then with three-body interactions the
gravitational tidal field can be so extreme that it can separate the star from the
IMBH. The capture or ejection of the star depends on the direction of motion of
this star relative to the pair of BHs (SMBH-IMBH). Most likely, the
star is captured by the SMBH. The resulting torus-like configuration is
determined by the Kozai-Lidova eccentric mechanism in binary SMBHs (see e.g. analogous works by 
\citet{Naoz2014,Naoz2016,SubrHaas2016,Rasskazov2019}).
}
\label{fig-hypervelocity}
\end{figure}

We have found that, in contrast to the notable paper by \citet{Merritt2009},
the presence of the intermediate mass BH
(IMBH; see evidence by \citet{Takekawa2019}) orbiting inside a nuclear star
cluster at the GC can effectively help (by means of the AR-CHAIN
​​code \citep{Mikkola2008} and the ``indirectly observable'' ADM
halo density modulation in the center of the Milky Way) to randomize the orbits of S-stars near the SMBH,
converting the initially thin co-rotating disk (see the warped or
mini-disk in 
Fig.~\ref{fig-accretion-disk}c; see also \citet{ChenAmaro2014,ChenAmaro2015}) into the
almost isotropic distribution of stars moving on eccentric orbits around the
SMBH (see Fig.~\ref{fig-hypervelocity}). Here the word ``almost'', as we
believe, practically overturns the essence of understanding of the physics of
randomizing the orbits of S-stars near the SMBH, since the initial distribution of
stars is somewhat ad hoc, and the evolution to the distribution of thermal
eccentricity itself occurs at the even more special time scale.

First of all, the randomization of the orbital planes requires
$\leqslant 20~Myr$ (see \citet{Subr2019}) if the IMBH mass is 
$(3.2 \pm 0.6) \times 10^4 ~M_{Sun}$ (see \citet{Takekawa2019}) and if the
orbital eccentricity is $\sim 0.7$ or greater. So, this means that in our view
of the S-stars randomization, the final distribution of the main semiaxes of star
orbits does not depend on the estimated size of the IMBH orbit
(see \citet{Merritt2009}) or, for example, on the special orientation of the
binary orbit (see \citet{Rasskazov2017}), but depends on the ``indirectly observable''
density of the DM halo at the center of the Milky Way.
Its time scale modulation is
determined by the time scale of the S-stars periods
through the
variability of the gravitational potential at the GC, which is
controlled by means of the ``observed'' disk wind rates variations or, 
equivalently, the accretion rate variations of the BH.

Let us now return to the question posed above. If we take into account the
evolution of the SMBH-IMBH binaries through the case of the ejection of
high-velocity S-stars (Fig.~\ref{fig-hypervelocity}a) and B-type hypervelocity
stars (Fig.~\ref{fig-hypervelocity}b), i.e. the so-called IMBH slingshot,
then how is the time scale of the S-stars periods
(through the
variability of the gravitational potential at the center of the galaxy) driven
and controlled by the ``observable'' variations of the disk wind speeds or,
equivalently, the variations of the BH accretion rate?

It works as follows. Among the chaotically oriented orbits of S-stars, some of
them which are oriented near the fundamental plane of the GC
have a direction along the accretion to the SMBH. When an
elliptically orbital star moves from the BH periapsis (at the high
speed the capture of ADM is almost zero) to apoapsis (when the speed is
lower, the capture of ADM is not too small), it means that the
point of apoapsis is determined by the condition of hydrostatic balance, at
which the deceleration of the S-star must be identical (in the absolute value) to
the deceleration of the disk wind, or equivalently, the lower BH
accretion rate. Conversely, the increase in the speed of an elliptical-orbit
star from apoapsis to periapsis assumes the higher speed of the disk wind, or
equivalently, the higher rate of the BH accretion. As a result, the periods of 
variability in the disk wind speed or accretion rate are an indicator of the
ADM variability, and consequently, an indicator of the periods of 
S-stars, e.g. S102 and S2 with periods of about 11 and 16 years 
(see Fig.~\ref{fig-stellar-orbits}a,b).

It also means that among the isotropically distributed S-stars there are such
stars that, although they have random orientations, do not have random 
velocities and periods moving along eccentric orbits near the fundamental plane
of the GC (see Fig.~\ref{fig-hypervelocity}a).

A unique result of our model is the fact that the periods, velocities and
modulations of S-stars are a fundamental indicator of the modulation of the ADM halo density at the center of the Galaxy, which closely correlates with
the density modulation of the baryon matter near the SMBH. If the modulations of the
ADM halo at the GC lead to modulations of the ADM
density on the surface of the Sun (through vertical density waves
from the disk to the solar neighborhood), then there is an ``experimental'' 
anticorrelation connection between the modulation of the ADM density in
the solar interior and the number of sunspots. Therefore, this is also true for
the relationship between the periods of S-star cycles and the sunspot cycles
(see ``experimental'' anticorrelated data in Fig.~\ref{fig-stellar-orbits}a,b).

\section{Summary and Outlook}
\label{sec-conclusion}

In the given paper we present a self-consistent model of the axion mechanism of
the Sun luminosity variations, in the framework of which we estimate the values of the axion
mass ($m_a \sim 3.2 \cdot 10^{-2} ~eV$) and the axion coupling constant to
photons ($g_{a \gamma} \sim 4.4 \cdot 10^{-11} ~GeV^{-1}$).

This result for the first time (see \cite{Rusov2015}) suggested that the existence of photons of axion origin in a virtually empty MFT, which is anchored to the tachocline, is the consequence of the Parker-Biermann cooling effect, which causes the strong magnetic pressure in the twisted magnetic tubes (see Fig.~\ref{fig-twisted-tube}b and Fig.~\ref{fig04-3}). This result also suggests that the existence of the magnetic field of the virtually empty anchored flux tube should be a consequence of the existence of the strong magnetic field in the tachocline, which, in contrast to the solar dynamo action, is much larger than $10^5~G$.

This means that the existence of the tachocline, on the one hand, must be predetermined by the existence of the holographic principle of quantum gravity, and on the other hand, it gives rise to the BL holographic mechanism (see Fig.~\ref{fig-solar-dynamos}b), which, unlike the solar dynamo models, generates a strong toroidal magnetic field by the thermomagnetic EN~effect in the tachocline ($\sim 10^7 ~G$ (Eq.~(\ref{eq06-16})); see also Fig.~\ref{fig-R-MagField}).

Leaving aside some conclusions from the interesting solved problems (the nature of the toroidal magnetic field of the tachocline in the interior of the Sun (see Sect.~\ref{sec-toroidal-field}), the universal model of antidynamo flux tubes and the phenomenon of DM solar axions (see Sect.~\ref{sec-radiative-heating}), the phenomenon of magnetic reconnection in the lower layers of the magnetic tube and observable tendencies of the tilt angle of Joy's Law (see Sect.~\ref{sec-tilt}), the solar axion and coronal heating problem solution (see Sect.~\ref{sec-coronal-heating}), the chronometer of DM hidden in the Sun core (see Sect.~\ref{sec-sun-chronometer}), the modulation of the ADM density as a ``clock'' regulating the tempo of the solar cycle and the mechanism of ADM variations around the BH (see Sect.~\ref{sec-dark-matter})), we focus on the fundamental connection between the Meissner-like and BZ-like effects on the BH boundary, which resolves the information loss paradox, a long-standing problem in theoretical physics.

Understanding the physics of this topic reinforces a very surprising and intriguing question about whether the theory of the complete restoration of information loss is different or not in our model and the known Hawking model \citep{Hawking2015b,Hawking2016}. Our answer is yes and no!

At the conference in Stockholm on August 28, 2015, Stephen Hawking announced that he had solved the information paradox. According to Hawking \citep{Hawking2015a,Hawking2015b,Hawking2016}, the AdS/CFT correspondence showed no loss of information, where all information about matter and energy in the three-dimensional volume of the BH is encoded in the form of a hologram on its two-dimensional surface of the event horizon (see \cite{tHooft1993}, \cite{Susskind1995}, \cite{Maldacena1999}, \cite{Hanada2014}). It resembles the analog of the famous photo-hologram by \cite{Gabor1948}, which is a photo of a certain type that generates a three-dimensional image when it is well-lit; all information describing a three-dimensional scene is encoded in a pattern of light and dark areas inscribed on a two-dimensional film. This means that the possible description and solution of the ``encoded'' holographic two-dimensional “film-plate” is predetermined by the fact that the problem of the information paradox, known as the apparent contradiction between the general theory of relativity and quantum mechanics, completely disappears. This approach ultimately leads to the holographic principle of quantum gravity, which can be generalized for any physical system occupying space-time. For example, in the holograms of the Universe, a BH, white dwarfs, neutron stars or, surprisingly, in the hologram of the Sun tachocline!

Hawking's main result is the fact that the paradox of information loss can be resolved by supertranslations of the horizon of BHs forming a hologram of incoming particles \citep{Hawking2015a,Hawking2015b}, where the information itself can be later completely restored, albeit in the chaotic form because of the radiation emitted during the quantum evaporation of the BH \citep{Hawking2015a,Hawking2015b,Hawking2016} predicted forty years ago by Hawking. This is due to the fact that the concept of the asymptotic isometry of the AdS space (see \cite{BrownHenneaux1986}), which consists only of the supertranslation of time for all Bondi-van der Burg-Metzner-Sachs models (\cite{Bondi1962}, \cite{Sachs1962}), has long attracted considerable attention from the point of view of the AdS/CFT correspondence (\cite{Maldacena1999}), in which supergravity in the AdS space is equivalent to the CFT at the boundary and gives a lot of information about the entropy problem in BHs and the strong connection between the gauge theory and the string theory (see \cite{StromingerVafa1996}, \cite{Witten1998}, \cite{Strominger1998}; \cite{Strominger2016}).

But the most surprising is the fact that the information stored in supertranslation with the help of a shift in the horizon caused by ingoing particles is returned via outgoing particles in a scrambled  chaotically useless form. In other words, we can say that the volume of recovered information is a complete ``book'' of information, but, unfortunately, it is not readable.

Oddly enough, this raises the question of whether a complete ``book'' of information can somehow be read. The answer is very simple: ``yes''!

In this regard, we consider a rather unexpected connection between the theory of the EN~effect, the second law of thermodynamics, and the paradigm of fuzzball complementarity on the horizon of BH events.

It is known that the AdS/CFT correspondence brought the holographic principle to the central stage of the string theory and, thus, can help to understand the nature of BHs. On the other hand, we know that several years ago, Mathur and his group discovered 
(see e.g. \cite{MathurTurton2012,MathurTurton2014a,MathurTurton2014b,MathurTurton2018} and Refs. therein; see also 
\cite{Guo2018,Mathur2005,Mathur2008,Mathur2009a,Mathur2010,Mathur2011,
Mathur2012a,Mathur2012b,Mathur2012c,Mathur2015,Mathur2017a}) that in the string theory BHs have a completely different structure: instead of vacuum and the singularity of a BH, in which the region just outside the horizon is the Rindler space (see Fig.~3a in \cite{Mathur2017a}), there are microstates of the string theory that are known as fuzzballs of mass $M$ (with surface at $r = 2 GM + \varepsilon \equiv r_b$); they do not have a horizon or singularity, and do not collapse the spheres into a BH. It should be remembered that because of the altered vacuum fluctuations, the region near the fuzzball  boundary is called the pseudo-Rindler space (see Fig.~3b in \cite{Mathur2017a}).

The most interesting is the fact that this fuzzball, according to 
\cite{Mathur2009a,MathurTurton2014a,MathurTurton2014b,Mathur2017a,Mathur2018a,Mathur2018b} is radiated at the same temperature as the Hawking BH. But the radiation emerges from the fuzzball surface in the same way as for any other normal body (like burning a lump of coal (see \cite{Mathur2009a,Mathur2009b,AlonsoSerranoVisser2016}), and therefore the radiation (and its thermal spectrum) carries the information present on this surface. This eliminates the information paradox 
(see e.g. \cite{Mathur2009b,Mathur2017a,Mathur2017b,AlonsoSerranoVisser2018}).

\begin{figure}[tb!]
  \begin{center}
      \includegraphics[width=16cm]{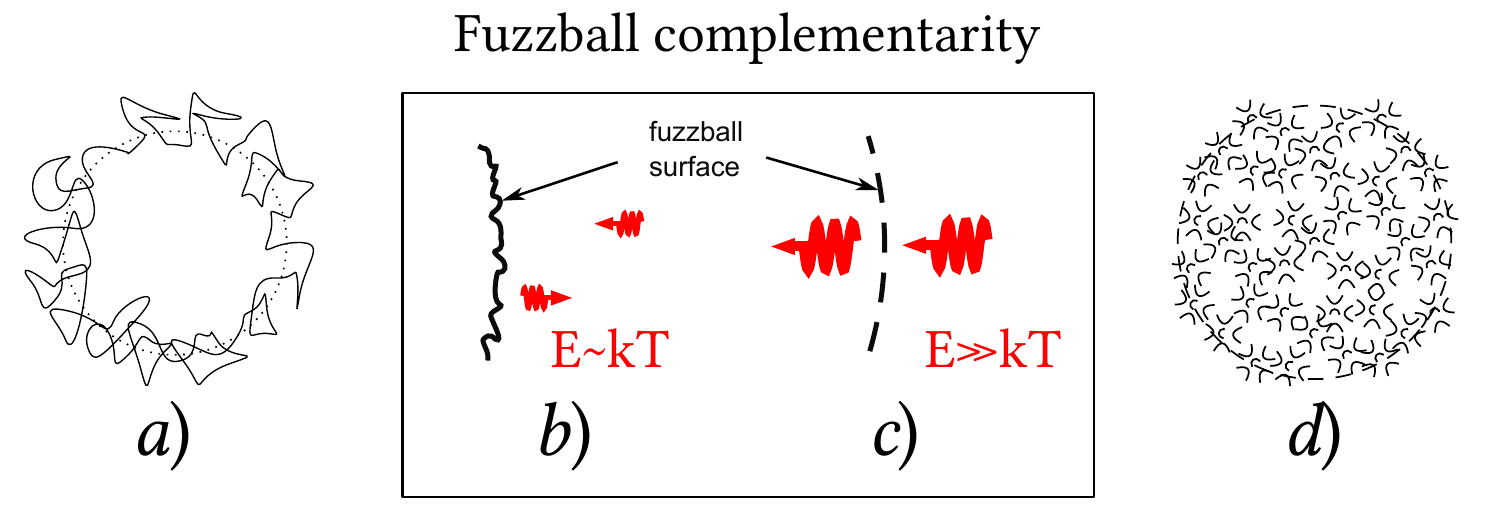}
  \end{center}
\caption{Fuzzball complementarity.
\textbf{(a)} Schematic description of a microstate solution of Einstein's equations: the fuzzball structure is fluctuating on a very fine scale for generic microstates. There are ‘local ergoregions’ with rapidly changing direction of frame dragging near the horizon. The geometry closes off without having an interior horizon or singularity due to its peculiar topological structure. Generic microstates are very quantum, and so this figure just gives a schematic description of the rapid fluctuations near the fuzzball surface. Adopted from \cite{Mathur2011}.
\textbf{(b)} The space-time ends in string theory sources outside $r=2GM$, with no horizon. Unitary Hawking radiation ($E \sim T$) is emitted from the fuzzball surface. 
\textbf{(c)} $E \gg T$ quanta excite collective dynamics of the fuzzball. This dynamics has an approximate complementary description in terms of smooth infall. Adopted from \cite{MathurTurton2014a}. 
\textbf{(d)} In the string theory fuzzball reaches the radius of the ball, and the wave function of the ball spreads over this space with no horizon or singularity. Adopted from \cite{Mathur2018b}.}
\label{fig-fuzzball}
\end{figure}

Strengthening the paradigm of fuzzball, i.e. the microstates of BHs (see e.g. Fig.~\ref{fig-fuzzball}a), we note the fact that unlike the well-known Hawking's problem of quantum corrections (see e.g. Hawking's argument as 'theorem' \citep{Mathur2009a,Mathur2017b}) and the firewall argument with the help of the so-called `validity of effective field theory' (\cite{Almheiri2013b}), an infalling shell-object never falls into the trap of its own horizon, and thus avoids any problem with causality.

Hence, it follows that, according to \cite{MathurTurton2014a,MathurTurton2014b}, the information paradox (and causality) is solved with the help of the construction of fuzzball in the string theory, which, without resorting to new physics, gives a real degree of freedom on the fuzzball surface and not a virtual degree of freedom on the horizon. These degrees of freedom should play two roles: 1) radiate Hawking quanta at $E \sim T$ (see Fig.~\ref{fig-fuzzball}b) and 2) account for any complementarity through collective oscillations generated by hard impacts of $E \gg T$ quanta (see Fig.~\ref{fig-fuzzball}c), where $E$ is the energy related to the physical process and $T$ is the Hawking temperature, both measured at infinity.

The key point is that in the string theory the fuzzball surface differs from the microstate to the microstate (see Fig.~\ref{fig-fuzzball}a) or from the wave function of the ball to the wave function (see Fig.~\ref{fig-fuzzball}a) that propagates through this space without a horizon or singularity and thus gives birth to two processes: (i) the Hawking radiation ($E \sim T$), which is very sensitive to microstates and carries the microstate information, and (ii) only complementarily excited collective modes ($E \gg T$), which mimic free infall and, as a consequence, are relatively insensitive to the choice of microstates and can have any possible complementary description with an information-free horizon, where, according to \cite{Mathur2015}, the spectrum of excitations of the fuzzball surface agrees, to a good approximation, with the spectrum of infalling modes in the traditional BH. This solves the information paradox.

It also means that Hawking radiation ($E \sim T$), which is very sensitive to
microstates, will carry information about the microstate through the
thermomagnetic EN effect in the tachocline of the black hole. From this we
understand that it is necessary to replace the fuzzball surface with the 
EN-fuzzball surface, which is now clear in Fig.~\ref{fig-gravitational-echoes}a.

Unlike the information about the BH horizon by \cite{Hawking2015b,Hawking2016}, the recovered information of the BH fuzzball found by Mathur and his group (see e.g. \cite{MathurTurton2014a,MathurTurton2014b,Mathur2011,Mathur2015,Mathur2017a,Mathur2017b,Guo2018} is surprisingly a completely readable ``book''. However, although at this stage these two groups 
cannot prove that the recovered information is true, we need to understand how one of these information models of a BH can be associated with some estimates of the already known measured signals of deviations from a traditional BH.

In this regard, we are interested in the observable astrophysical signs of quantum gravity near BHs (see e.g. \cite{Cardoso2017}) in which some features of the data already obtained can be interpreted as evidence of a gravitational echo (see \cite{Cardoso2016,Abedi2017}). This means that in our case, in addition to measurable signals exhibiting significant deviations from classical BH evolution in the near-horizon region, it can give unique observable signs of the formation of fuzzball, where the collapsing shell suffer ``entropic enhanced tunneling'' into fuzzballs \citep{Mathur2017a}, which usually breaks symmetry and is expected to result in bursts of gravitational and electromagnetic radiation (see \cite{Hertog2017}).

In our view, these quantum reflected echo signals (see Fig.~1 in \cite{Abedi2017}) are possible not only from the BH mergers, but also from the single SMBHs, where the DM (and, accordingly, baryons) modulation determines the formation and existence of the galactic gravitational waves (see Eq.~(2.1) in \cite{Allen1999}; Eq.~(2) in \cite{Abbott2016b}; Eqs.~(4.1) and (4.3) in \cite{Raidal2017}; \cite{Romano2017}; \cite{Kovetz2017}).

It is very important to understand that the modulations of DM (and, accordingly, baryons) predetermine not only the emergence of galactic gravitational waves, but also the modulation of gravitational waves, in which gravitational theories predict the possibility of creating matter associated with the geometry-matter connection in cosmology and astrophysics 
(see \cite{Parker1968,Parker1969,Zeldovich1970,Zeldovich1972,Fulling1974,
ParkerToms2009,Parker2009}).

It means that the particle creation corresponds to the irreversible energy flow from the gravitational field to the created matter constituents, with the second law of thermodynamics requiring that space-time transforms into matter  (see \cite{Harko2014}). We are interested in the irreversible thermodynamic process that is a direct consequence of the so-called non-minimal coupling of curvature-matter, induced in this case by quantum fluctuations of the gravitational metric (see \cite{Harko2014,HarkoLobo2013,Harko2015,LiuXing2016}), as initiated by \cite{Yang2016,DzhunushalievEtAl2015,Dzhunushaliev2015}.

The essence of the remarkable idea of ​​Dzhunushaliev and his group is as follows. They show that if the metric in quantum gravity can be represented as a sum of the classical and quantum parts (see Eq.~(9) in \cite{DzhunushalievEtAl2014}), then such a gravitating physical system looks like modified gravity (see Eq.~(21) in \cite{DzhunushalievEtAl2014}). This is, on the one hand, due to the nonzero mathematical expectation of the quantum part of the metric, and on the other hand, there is a non-minimal interaction between matter and gravity. The general approach of quantum corrections arises both from the nonperturbative quantization of the metric \citep{DzhunushalievFolomeev2014} and from the perturbative quantization of various types of fields (scalar, electromagnetic, QCD) \citep{DzhunushalievFolomeev2015}.

Based on the idea of Dzhunushaliev and his group, Harko and his group made a very important assumption 
(see \cite{Harko2014,HarkoLobo2013,Harko2015,LiuXing2016}) that the coupling coefficient between the metric and the average value of the quantum fluctuation tensor is a scalar field with a non-vanishing self-interaction potential and a simple scalar function. One of the most important results of this assumption is the fact (see \cite{LiuXing2016}) that  it is the self-interaction potential scalar field that was assumed to be of Higgs type \citep{Aad2015}.

Hence, there is an intriguing problem of the Higgs-like boson. In our work (see the topic 
\hyperref[topic-MultiparticleFields]{``Multiparticle fields and the Higgs 
mechanism''} in Sect.~\ref{sec-dark-matter}), we have shown that, given the experimental data of \cite{Patrignani2016}, it is possible to evaluate the existence of a scalar Higgs field in internal indices that do not interact for electroweak and strong interactions. At the same time, such a scalar field can contribute to the gravitational field through the energy-momentum tensor. Thus, the Higgs bosons, scalar in internal indices, can be considered a candidate for a part of DM!

As a result, we understand that the Higgs-like particles do not have any additional fundamental interaction and contain the part of VM -- the Higgs-like baryon particles, and the part of DM -- Higgs-like DM (see Fig.~\ref{fig-mexican-hat}b). That is why the effects of Higgs-like particles can, on the one hand, be identical to the effects of EN-fuzzball (see Fig.~\ref{fig-gravitational-echoes}a), and, on the other hand, when they decay into two photons, they will be identical to the effects of BZ-fuzzball, which is now also clear from Fig.~\ref{fig-gravitational-echoes}b.

Here we are interested in the problem of the Higgs-like particles decay channel into two photons (see ATLAS Collaboration \citep{Aad2015} and \cite{Patrignani2016}) and the link (see the topic 
\hyperref[topic-FIFTH]{``•••••''} in Sect.~\ref{sec-dark-matter}) between the physics of radiation of gravitational waves and the generalized thermomagnetic EN~effect of DM around a BH. Since we know (see the topics \hyperref[topic-THIRD]{``•••''} and \hyperref[topic-FOURTH]{``••••''} in Sect.~\ref{sec-dark-matter}) that the EN~effect near the ``fuzzball'' surface of a BH is absolutely identical to the Higgs phenomenon in particle physics (see \cite{Wilczek2000,Wilczek2005}), i.e. quantum phenomenon of particles of DM, there appears a wonderful way to solve the problem of the Higgs-like particles decay channel near the ``BZ~fuzzball'' surface of a BH.

In the beginning, we demonstrate the simple physics of this problem. (1) If the Higgs-like particles are near the surface of the ``fuzzball'' of a BH, then we call it EN~fuzzball, at which unitary Hawking radiation is emitted ($E \sim T$; see Fig.~\ref{fig-fuzzball}b). (2) If the decay channel of Higgs-like particles into two photons appears near the surface of a BH fuzzball, then we call it EN~fuzzball, in which photons with energy $E \gg T$ excite the collective dynamics of the fuzzball. This is the solution of the problem of BZ~fuzzball complementarity (see Fig.~\ref{fig-fuzzball}c).
  
Here the question arises as to how the theoretical solution of the fuzzball complementarity problem (see Fig.~\ref{fig-fuzzball}) is related to the known experimental data, for example, using signals from gravitational waves as signatures of quantum gravity towards the horizon of BHs. In this sense, we are interested in the well-known observation of the space-time image of the echo signals of gravitational waves on the EN~fuzzball surface (or the identical Meissner-like fuzzball (see Fig.~\ref{fig-meissner-effect}a)) on the stretched horizon after the BH accretion event (see Fig.~\ref{fig-gravitational-echoes}a; adopted from \cite{Abedi2017}).

Based on the model of the fuzzball complementarity (see Fig.~\ref{fig-gravitational-echoes}), we see that the echo waves, i.e. the gravitational waves trapped between the structure near the horizon and the angular momentum barrier (see Fig.~1 in \cite{Abedi2017}), are reflected on the EN~fuzzball surface (see Fig.~\ref{fig-gravitational-echoes}a), which is the result of the generalized thermomagnetic EN~effect on BHs (see Fig.~\ref{fig-meissner-effect}a) or, equivalently, the phenomenon of Higgs-like particles. On the other hand, if we assume that the gravitational waves penetrate the BZ~fuzzball surface (see Fig.~\ref{fig-gravitational-echoes}b) as a consequence of the BZ~effect (see Fig.~\ref{fig-meissner-effect}b), this means that due to the decay of the Higgs particles, photons with energy $E \gg T$, exciting the collective dynamics of the BZ~fuzzball surface, effectively penetrate into it, and thus the gravitational waves also pass there.

Here comes the second question about what the physics of the BZ-like effect is. Until now, it is believed that the central BH engine, which is the source of GRBs, includes the electromagnetic-magnetic extraction of the BH rotational energy, proposed by \cite{Blandford1977}. Unlike the well-known discussions in the GRB community 
(see e.g. \cite{Lee2000,Wang2002,McKinney2005,KomissarovBarkov2009,
Nagataki2009,Contopoulos2017}), we use the very interesting interpretation of \cite{Salafia2017} about the properties of a particular short GRB (SGRB), whose progenitor is the merger of two neutron stars in the form of a BH, or the BH and neutron star, and the remarkable first observation of gravitational waves from the fusion of a binary neutron star with a SGRB-GW connection \citep{Abbott2017a,Abbott2017b,Abbott2017c}. The two phenomena are thus inseparable (see e.g. \cite{Salafia2017}).

Skipping the complex but very beautiful task of the first observations of a kilonova (which follows the initial fuzzball), the UV/Optical/Infrared emission from the expanding material ejected during the merger stages and coalescence driven by the nuclear decay of unstable nuclei synthesized by the $r$-process (see \cite{Salafia2017} and Refs. therein) and/or the creation of new (dark) matter related to the geometry-substance coupling induced by the quantum fluctuations of the gravitational metric (see \cite{LiuXing2016} and Refs. therein), we note that the physics of two undivided observations of SGRB and GW is the theoretical basis of the fuzzball complement model (see Fig.~\ref{fig-gravitational-echoes}), which as a consequence of holographic quantum gravity is predetermined by the BZ-like effect (see Fig.~\ref{fig-meissner-effect}b) on BHs.

We will discuss the implications of our findings in a forthcoming publication.

And finally, let us emphasize one essential and most painful point of this paper. This is the key problem of the holographic principle of quantum gravity, on the basis of which the thermomagnetic EN~effect predetermines the possibility of observational measurements of magnetic fields between the two-dimensional surface of the tachocline and the three-dimensional volume of cores in compact objects -- our Sun, magnetic white dwarfs, accreting neutron stars and BHs. For example, with the help of the thermomagnetic EN~effect, a simple estimate of the magnetic pressure of an ideal gas in the tachocline of e.g. the Sun can indirectly prove that by using the holographic principle of quantum gravity, the toroidal magnetic field of the tachocline accurately ``neutralizes'' the magnetic field in the Sun core. This means that the holographic BL~mechanism is the main process of regenerating the primary toroidal field in the tachocline and, as a consequence, the formation of buoyant toroidal tubes of magnetic flux at the base of the convective zone, which then rise to the surface of the Sun. Hence, two phenomena -- DM in the form of ADM and axions -- ground the understanding of the nature of the holographic principle, starting, for example, from the solution of the problem of antidynamo, sunspots and coronal heating and, oddly enough, to the complete restoration of information loss of the BH, which, although it differs from the well-known Hawking model, does not contradict the experimental observations.

Here the question arises: is it true, is the reason simple or not? Or is it just the guessed rules of calculation that do not reflect any real nature of things, or “…when we would backward see from what region of remoteness the whatness of our whoness hath fetched his whenceness…” (James Joyce. Ulysses, Episode 14)?

\appendix
\numberwithin{equation}{section}
\numberwithin{figure}{section}
\numberwithin{table}{section}

\

\section{X-ray coronal luminosity variations}
\label{appendix-luminosity}

\begin{figure*}[htpb!]
\noindent
\centerline{\includegraphics[width=15cm]{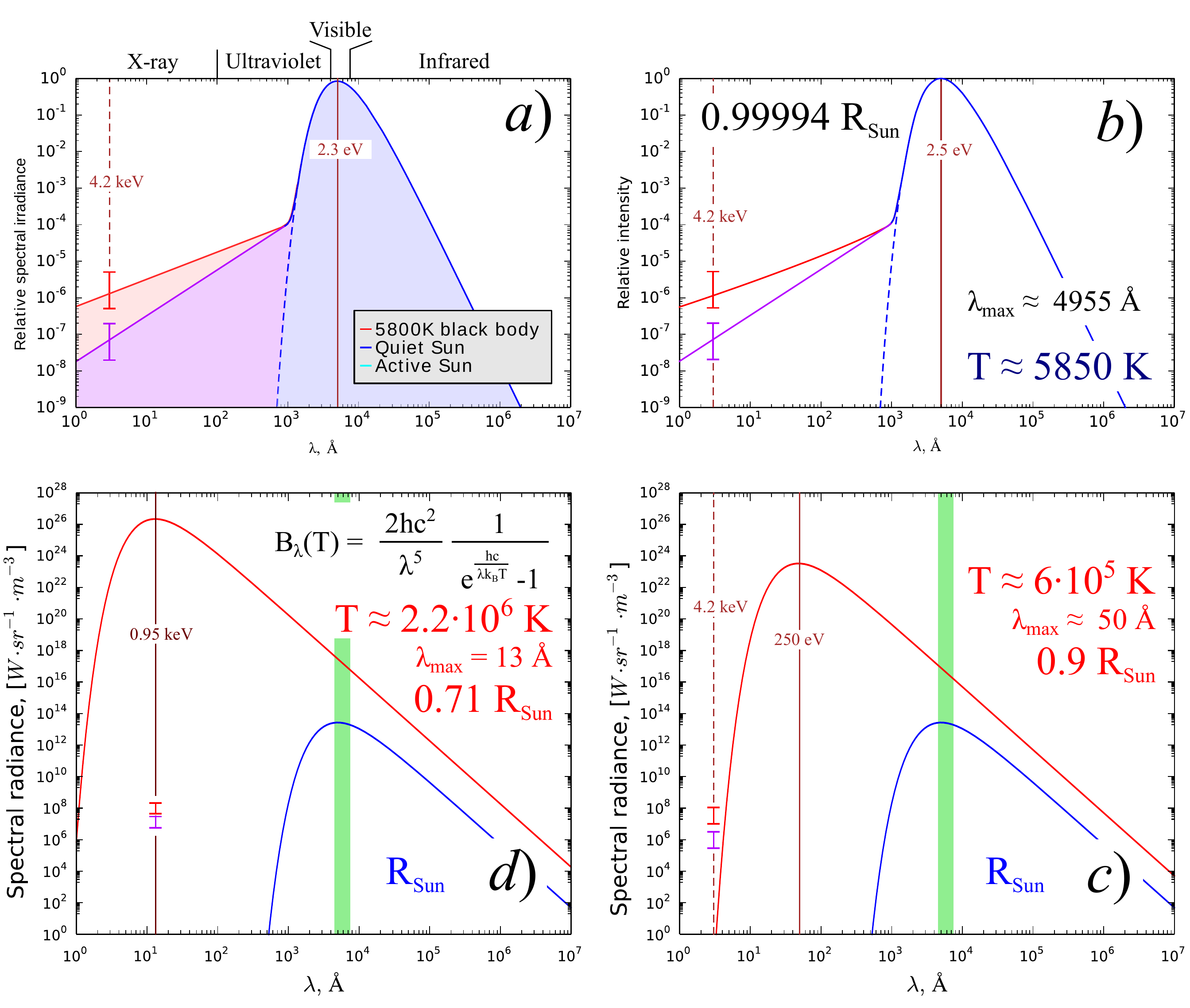}}
\caption{a) The smoothed solar spectrum corresponds to a black body with
         temperature $5.77 \cdot 10^3$~K at $R_{Sun}$ (see Fig.~11
         in~\cite{Zioutas2009}). 
         The red and violet bars represent
         the high and low contributions from the X-rays of axion origin with mean energy of 
         4.2~keV born in the ``magnetic steps'' at $\sim 0.72 R_{Sun}$,
         roughly estimated for 
         the solar cycle 22. It is possible to perform similar 
         estimations for any other solar cycle using Eqs.~(\ref{eq06-08}), (\ref{eq7.5-sep-11}), (\ref{app-b-eq03})
         and the data from \cite{Dikpati2008,Pevtsov2014,Lockwood2001}.
         Note that these X-ray contributions exist within the magnetic tubes 
         only.
         \newline
         b) A spectrum of a black body with temperature 5850~K at 
         0.99994$R_{Sun}$ (see Fig.~\ref{fig-Bz}).
         The X-ray luminosity (high intensity red and low intensity violet lines) 
         is determined by the Compton scattering of the mentioned 4.2~keV X-rays.
         \newline
         c) The red line is a spectrum of a black body with temperature $6 \cdot 10^5$~K
         at 0.9$R_{Sun}$ (see Fig.~\ref{fig-Bz} and \cite{Bahcall1992}). 
         The high and low X-ray luminosity (red and violet bars) is determined by the 4.2~keV
          X-rays which propagate along the cool region of the magnetic tube without scattering 
         (Fig.~\ref{fig-lampochka}a). The blue line corresponds to the black-body spectrum
         of the solar surface.
         \newline
         d) A spectrum of a black body with temperature
         $2.22 \cdot 10^6$~K at 0.71$R_{Sun}$ (see Fig.~\ref{fig-Bz} and 
         \cite{Bahcall1992}). The X-ray luminosity
         is determined by thermal photons with average energy $\sim\,$0.95~keV 
         only (in the tachocline \citep{Bailey2009}). These 
         photons are converted into axions in the ``magnetic steps'' at 
         $\sim 0.72 R_{Sun}$, and thus do not constitute the spectra of the 
         higher layers. As above, the blue line corresponds to the black-body 
         spectrum of the solar surface.}
\label{app-b-fig01}
\end{figure*}

The X-ray luminosity of the solar corona during the active phase of the solar
cycle is defined by the following expression:
\begin{align}
\left( L_{corona} ^{X} \right)_{max} = \int \limits_{X-ray} \frac{d\Phi_{corona}^{max} (E)}{dE} E dE =  \int
\limits_{X-ray} \frac{d W_{corona}^{max} (E)}{dE} dE\, .
\label{app-b-eq01}
\end{align}

In the quiet phase it can be written as
\begin{align}
\left( L_{corona} ^{X} \right)_{min} = \int \limits_{X-ray} \frac{d\Phi_{corona}^{min} (E)}{dE} E dE =  \int
\limits_{X-ray} \frac{d W_{corona}^{min} (E)}{dE} dE\, .
\label{app-b-eq02}
\end{align}

Then, integrating the blue curve in Fig.~\ref{app-b-fig01} for
$\left(L_{corona}^X \right) _{max}$ and the cyan curve for
$\left(L_{corona}^X \right)_{min}$, we obtain

\begin{equation}
\frac{\left( L_{corona}^X \right) _{min}}{L_{Sun}} \sim 7 \cdot 10^{-8} ; ~~~~~
\frac{\left( L_{corona}^X \right) _{max}}{L_{Sun}} \sim 1.2 \cdot 10^{-6} .
\label{app-b-eq03}
\end{equation}

So it may be derived from here that the Sun luminosity is quite low in
X-rays~(\ref{app-b-eq03}), typically (see analogue in \citet{Rieutord2014})

\begin{equation}
7 \cdot 10^{-8} L_{Sun} \leqslant L_{corona}^X \leqslant 1.2 \cdot 10^{-6} L_{Sun}, 
\label{app-b-eq04}
\end{equation}

\noindent but it varies with the cycle (see blue and cyan lines in
Fig.~\ref{app-b-fig01}a) as nicely shown by the pictures obtained with the
Yohkoh satellite (see analogous Fig.~4 in~\citet{Rieutord2014}).

And finally, it may be supposed that X-rays, propagating from the tachocline
towards the photosphere, interact with the charged particles via the Compton
scattering, but only outside the magnetic tubes. The axion-originated X-ray
radiation channeling inside the magnetic tubes does not experience the Compton
scattering up to the photosphere (Figure~\ref{app-b-fig01}).

\section{Explanation of the high X-ray intensity bands widening near the Yohkoh image edges}
\label{appendix-widening}

It is interesting to note that the bands of high X-ray intensity on Yohkoh
images deviate from the solar parallels 
(Fig.~13b in \citet{Zioutas2009}). 
This is
especially the case near the edges of the visible solar disk.

\begin{figure*}
\centerline{\includegraphics[width=15cm]{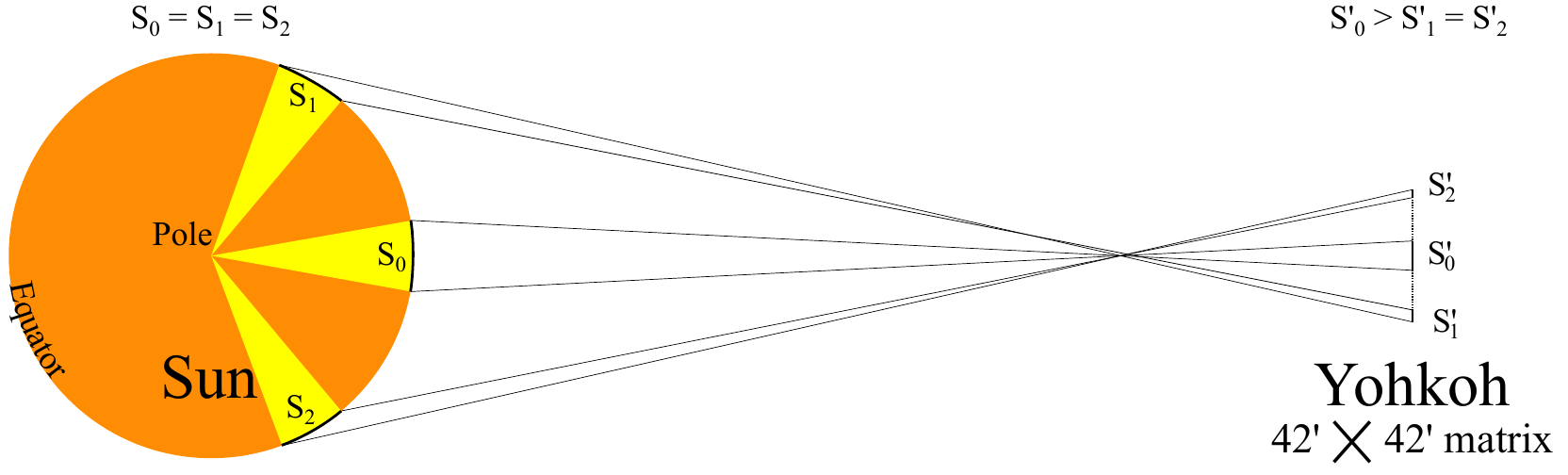}}

\caption{A sketch of the Sun image formation on the Yohkoh matrix. \label{app-a-fig05}}
\end{figure*}

\begin{figure*}[htb!]
\centerline{\includegraphics[width=9cm]{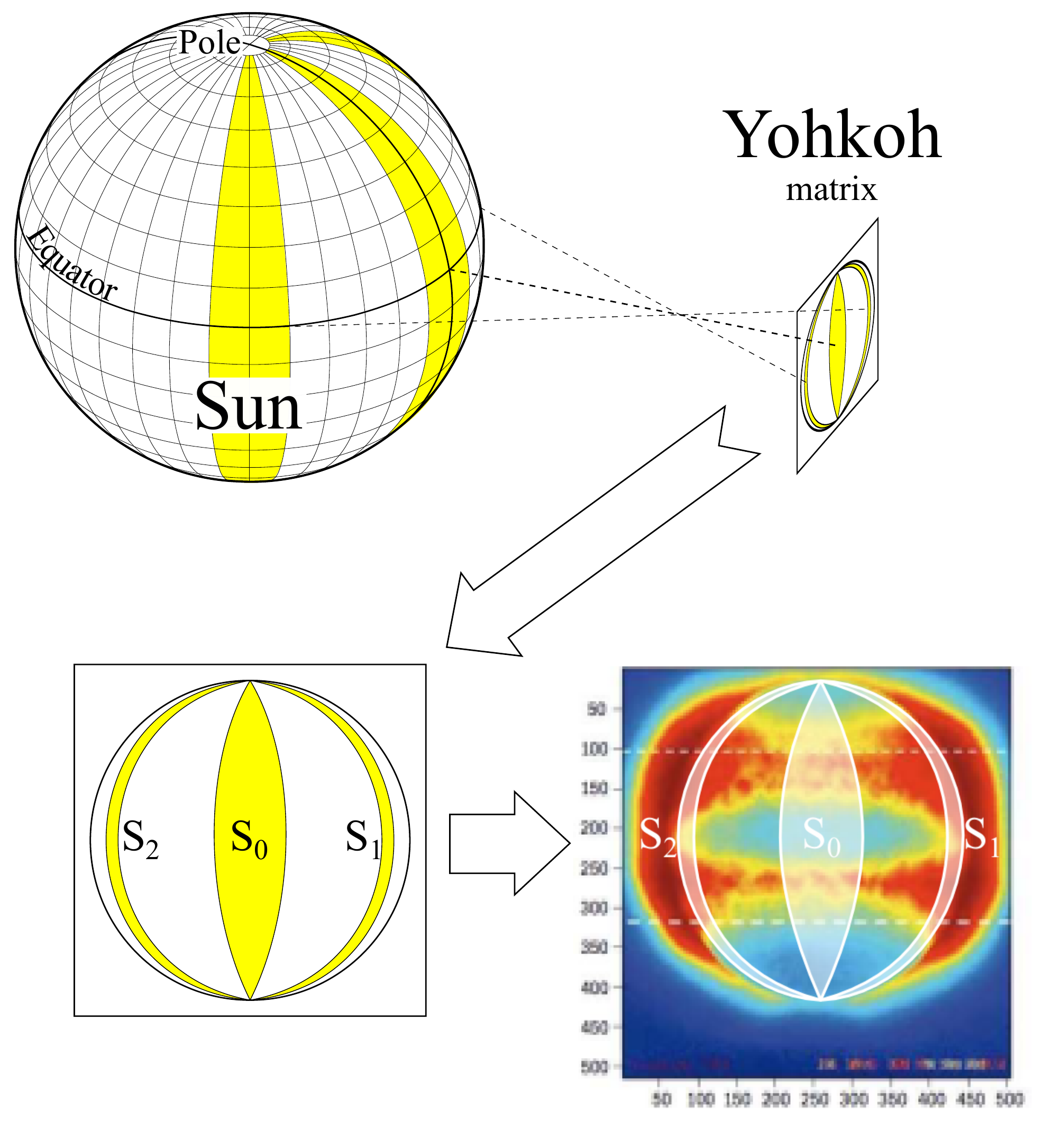}}

\caption{A sketch of the Sun image formation on the Yohkoh matrix. \label{app-a-fig04}}
\end{figure*}

This effect can be explained graphically by means of Figs.~\ref{app-a-fig05}
and~\ref{app-a-fig04}. These figures show the schematic concept of the Sun
image formation on the Yohkoh matrix. Fig.~\ref{app-a-fig05} shows the Sun from
its pole.

Let us choose three sectors of equal size on the surface of the Sun
(Figs.~\ref{app-a-fig05}, \ref{app-a-fig04}). The areas of the photosphere cut
by these sectors are also equal ($S_0 = S_1 = S_2$). However, as it is easily
seen in the suggested scheme, the \emph{projections} of these sectors on the
Yohkoh matrix are not of equal area (Figs.~\ref{app-a-fig05},
\ref{app-a-fig04}). The $S'_1$ and $S'_2$ projections areas are much less than
that of the $S'_0$ projection (Fig.~\ref{app-a-fig05}). This means that the
radiation emitted by the sectors $S_1$ and $S_2$ of the solar photosphere and
captured by the satellite camera will be concentrated within \emph{lesser} areas
(near the edges of the solar disk) than the radiation coming from the $S_0$
sector (in the center of the solar disk). As a result, the satellite shows
higher intensity near the image edges than that in the center, in spite of the
obvious fact that the real radiation intensity is equal along the parallel of
the Sun.

Therefore, because of the system geometry, the satellite tends to ``amplify''
the intensity near the image edges, and the areas that correspond to the yellow
and green areas in the center 
(Fig.~13b in \citet{Zioutas2009})
become red near the
edges, thus leading to the visible widening of the high intensity bands.

The particularly high radiation intensity near the very edges of the visible
solar disk, observed even during the quiet phase of the Sun
Fig.~13a in \citet{Zioutas2009}, indicates a rather ``wide'' directional radiation
pattern of the solar X-rays.

Let us make a simple computational experiment. We choose a sphere of unit radius and spread the points over its surface in such a way that their
density changes smoothly according to some dependence on the \emph{polar} angle
($\theta$). The \emph{azimuth} angle does not influence the density of these
points. For this purpose any function that provides a smooth change of the
density will do. For example, this one:
\begin{equation}
\rho (\theta) = [\rho _0 + \rho _{max} \cdot \cos(2 \cos\theta)]^{-1}\, . \label{app-a-eq01}
\end{equation}

\begin{figure*}[tb!]
\centerline{\includegraphics[width=9cm]{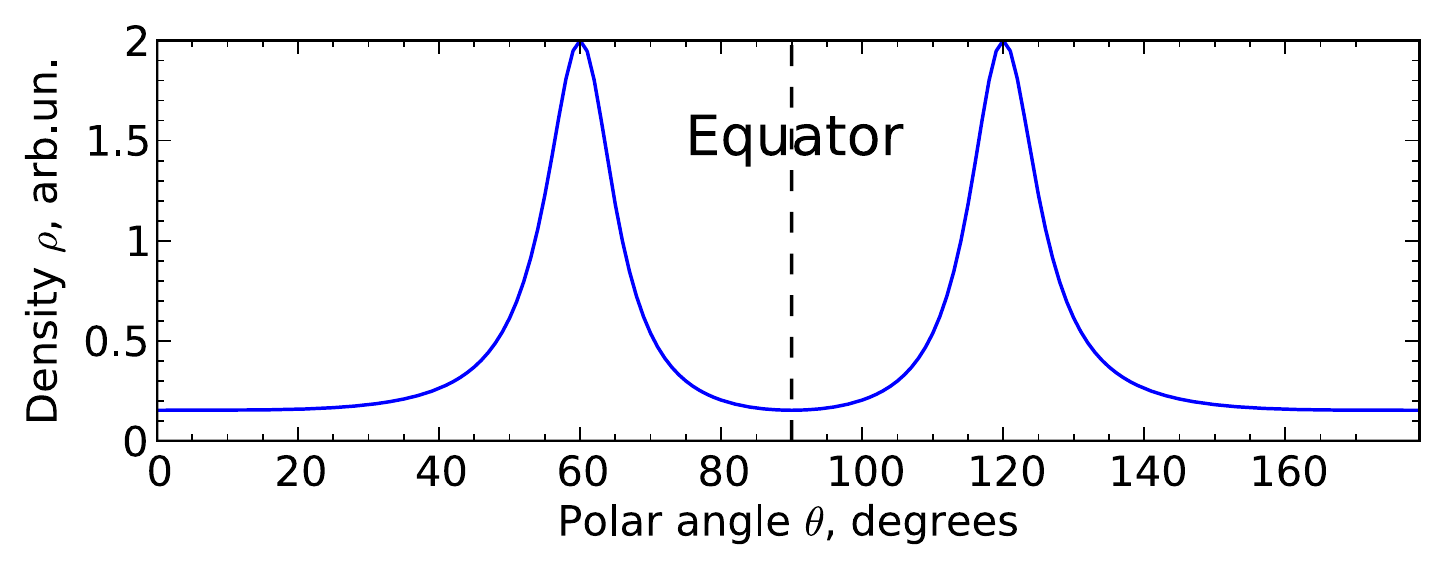}}

\caption{Graphical representation of Eq.~(\ref{app-a-eq01}). \label{app-a-fig06}}
\end{figure*}

Here we take $\rho_0=3.5$ and $\rho_{max}=3$ in arbitrary units. The graphical
representation of this dependence is shown in Fig.~\ref{app-a-fig06}. It yields the minimum density of points near the poles and the equator,
and the maximum density of points near $\theta = 60^{\circ}$ and
$\theta = 120^{\circ}$.

The polar angle $\theta$ was set in the range $[0, 180^{\circ}]$ with the step
of $1^{\circ}$. The azimuth angle $\varphi$ was set in the range
$[0,180^{\circ}]$ (one hemisphere) with a variable step $\Delta \varphi$
representing the variable density, since the points density is inversely
proportional to the step between them ($\Delta \varphi \sim 1 / \rho$). We
assume that
\begin{equation}
\Delta \varphi (\theta) = \Delta \varphi _0 + \Delta \varphi _{max} \cdot \cos(2 \cos\theta) ~~[deg] .
\label{app-a-eq01a}
\end{equation}

The values of $\Delta \varphi _0 = 3.5^{\circ}$ and
$\Delta \varphi _{max} = 3^{\circ}$ were chosen arbitrarily. So,
\begin{align}
\Delta \varphi (\theta) = 3.5 + 3 \cdot \cos(2 \cos\theta) ~~[deg] .
\label{app-a-eq03}
\end{align}

\begin{figure*}[tb!]
  \begin{center}
  \begin{minipage}[h]{0.49\linewidth}
    \center{a) \includegraphics[width=5cm]{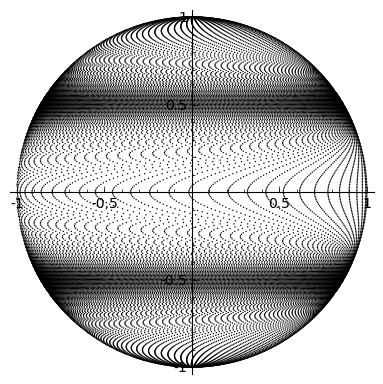}}
  \end{minipage}
  \begin{minipage}[h]{0.49\linewidth}
    \center{b) \includegraphics[width=6cm]{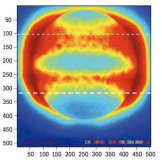}}
  \end{minipage}
  \caption{a) Simulation of the high intensity bands formation on the 2D
           projection of the sphere.
  \newline b) Sun X-ray image from Yohkoh satellite during the active
  phase of the Sun.}
  \label{app-a-fig07}
  \end{center}
\end{figure*}

From Eq.~(\ref{app-a-eq03}) it is clear that the minimum step was 0.5$^{\circ}$
and the maximum step was 6.5$^{\circ}$. Apparently, the more is the step, the
less is the density (near the poles and the equator) and vice versa, the less
is the step, the more is the points density. This was the way of providing the
smooth change of the points density by latitude (along the solar meridians).

Obviously, this forms the ``belts'' of the high density of points along the
parallels. The projection of such a sphere on any plane perpendicular to its
equator plane will have the form shown in Fig.~\ref{app-a-fig07}a. As it is
seen in this figure, although the density does not depend on the azimuth angle,
there are high density bands widening near the edges of the \emph{projected}
image. These bands are similar to those observed on the images of the Sun in
Fig.~\ref{app-a-fig07}b.

Let us emphasize that the exact form of the dependence (\ref{app-a-eq01a}) as
well as the exact values of its parameters were chosen absolutely arbitrarily
for the sole purpose of the qualitative effect demonstration. They have no
relation to the actual latitudinal X-ray intensity distribution over the
surface of the Sun.

\bibliographystyle{agsm}        

\bibliography{Rusov-AxionSunLuminosity}

@Article{ref01,
author = {Raffelt, G. G.},
title = {Axions and other very light bosons: {Part II} (astrophysical constraints)},
journal = {Phys. Lett.},
year = {2004},
volume = {592},
pages = {391},
}

@Article{ref02,
author = {Raffelt, G. G.},
title = {Astrophysical axion bounds},
journal = {Lect. Notes Phys.},
year = {2008},
volume = {741},
pages = {51-71},
note = {\href{http://arxiv.org/abs/hep-ph/0611350}{arXiv:hep-ph/0611350}},
}

@Article{ref03,
author = {De Angelis, Alessandro and Galanti, Giorgio and Roncadelli, Marco},
title = {Relevance of axion-like particles for very-high-energy astrophysics},
journal = {Phys. Rev. D},
year = {2011},
volume = {84},
pages = {105030},
}

@Article{ref04,
author = {Masso, E.},
title = {Axions and Their Relatives},
journal = {Lect. Notes Phys.},
year = {2008},
volume = {741},
pages = {83},
}

@Article{ref05,
author = {Fairbairn, M. and Rashba, T. and Troitsky, S.},
title = {Photon-axion mixing and ultra-high-energy cosmic rays from {BL Lac} type objects -- Shining light through the Universe},
journal = {Phys. Rev. D},
year = {2011},
volume = {84},
pages = {125019},
}

@Article{ref06,
author = {Csaki, C. and Kaloper, N. and Terning, J.},
title = {Dimming supernovae without  cosmic  acceleration},
journal = {Phys. Rev. Lett.},
year = {2002},
volume = {88},
pages = {161302},
note = {\href{http://arxiv.org/abs/astro-ph/0111311}{arXiv:astro-ph/0111311}},
}

@Article{Mirizzi2005,
author = {Mirizzi, A. and Raffelt, G. G. and Serpico, P. D.},
title = {Photon axion conversion as a mechanism for supernova dimming: Limits from {CMB} spectral distortion},
journal = {Phys. Rev. D},
year = {2005},
volume = {72},
pages = {023501},
}

@Article{ref08,
author = {Adler, S. and Gamboa, J. and Mendez, F. and Lopez-Sarrior, J.},
title = {Axions and "Light Shining Through a Wall": A Detailed Theoretical Analysis},
journal = {Annals Phys.},
year = {2008},
volume = {32},
pages = {2851-2872},
}

@Unpublished{ref09,
author = {Redondo, J. and Ringwald, A.},
title = {Light shining through walls},
year = {2010},
note = {\href{http://arxiv.org/abs/1011.3741}{arXiv:1011.3741 [hep-ph]}},
}

@Article{ref10,
author = {Vysotsskii, M. I. and Zel'dovich, Ya. B. and Khlopov, M. Yu. and Chechetkin, V. M.},
title = {Some astrophysical limitations on the axion mass},
journal = {JETP letters},
year = {1978},
volume = {27},
pages = {533-536},
}

@article{ref11,
  title = {Photo-Production of Neutral Mesons in Nuclear Electric Fields and the Mean Life of the Neutral Meson},
  author = {Primakoff, H.},
  journal = {Phys. Rev.},
  volume = {81},
  issue = {5},
  pages = {899--899},
  year = {1951},
  month = {Mar},
  doi = {10.1103/PhysRev.81.899},
  url = {http://link.aps.org/doi/10.1103/PhysRev.81.899},
  publisher = {American Physical Society}
}

@Book{ref12,
author = {Hoyle, F.},
title = {Some Recent Researches in Solar Physics},
publisher = {Cambridge University Press},
year = {1949},
}

@Article{ref13,
author = {Chitre, S.M.},
title = {The structure of sunspots},
journal = {Monthly Notices Roy Astron. Soc.},
year = {1963},
volume = {126},
pages = {431-443},
}

@Article{ref14,
author = {Zwaan, C.},
title = {On the appearance of magnetic flux in the solar photosphere},
journal = {Solar Phys.},
year = {1978},
volume = {60},
pages = {213-240},
}

@Article{ref15,
author = {Spruit, H.C. and Roberts, B.},
title = {Magnetic flux tubes on the {Sun}},
journal = {Nature},
year = {1983},
volume = {304},
pages = {401-406},
}

@Article{Parker1955b,
author = {Parker, E. N.},
title = {Hydromagnetic dynamo models},
journal = {Astrophysics J.},
year = {1955},
volume = {122},
pages = {293-314},
}

@Article{ref17,
author = {Hassan, S.S.},
title = {Magnetic Flux Tubes and Activity on the {Sun}, Lectures on Solar Physics},
journal = {Lection Notes in Physics},
year = {2003},
volume = {619},
pages = {173-201},
}

@Article{ref18,
author = {Spruit, H.C.},
title = {Theories of the Solar Cycle and Its Effect on Climate},
journal = {Progr. Theor. Physics Supplement},
year = {2012},
volume = {195},
pages = {185-200},
}

@article{Ilonidis2011,
	author = {Ilonidis, Stathis and Zhao, Junwei and Kosovichev, Alexander},
	title = {Detection of Emerging Sunspot Regions in the Solar Interior},
	volume = {333},
	number = {6045},
	pages = {993--996},
	year = {2011},
	doi = {10.1126/science.1206253},
	publisher = {American Association for the Advancement of Science},
	issn = {0036-8075},
	journal = {Science}
}

@Article{ref20,
author = {Caligari, P. and Schuessler, M. and Moreno-Insertis, F.},
title = {Emerging flux tubes in the solar convective zone II. The influence of initial conditions},
journal = {Astrophys. J.},
year = {1981},
volume = {243},
pages = {309-316},
}

@Article{Benz2008,
author = {Benz, A.O.},
title = {Flare Observations},
journal = {Living Rev. Solar Phys.},
year = {2008},
volume = {5},
pages = {1-62},
}

@Article{ref23,
author = {Ruzmaikin, A.},
title = {Can we get the bottom {B}?},
journal = {Solar Phys.},
year = {2000},
volume = {192},
pages = {49-57},
note = {(Invited Review)},
}

@Article{ref24,
author = {Fisher, G.H. and Fan, Y. and Longcope, D.W. and Linton, M.G. and Pevtsov, A.A.},
title = {The solar dynamo and emerging flux},
journal = {Solar phys.},
year = {2000},
volume = {192},
pages = {119-139},
note = {(Invited Review)},
}

@Article{ref26,
author = {Golub, L. and Rosner, R. and Vaiana, G.S. and Weiss, N.O.},
title = {Solar magnetic fields: the generation of emerging flux},
journal = {Astrophys. J.},
year = {1981},
volume = {243},
pages = {309-316},
}

@ARTICLE{MorenoInsertis1983,
   author = {{Moreno-Insertis}, F.},
    title = "{Rise times of horizontal magnetic flux tubes in the convection zone of the sun}",
  journal = {Astronomy and Astrophysics},
     year = 1983,
    month = jun,
   volume = 122,
    pages = {241-250},
   adsurl = {http://adsabs.harvard.edu/abs/1983A%26A...122..241M}
}

@Article{ref28,
author = {Spruit, H.C.},
title = {Propagation speeds and acoustic damping of waves in magnetic flux tubes},
journal = {Solar Phys.},
year = {1982},
volume = {75},
pages = {3-17},
}

@InCollection{ref29,
author = {Hollweg, J.V.},
title = {{MHD} waves on solar magnetic flux tubes},
booktitle = {Physics of Magnetic Flux Ropes},
series = {Geophys. Monograph},
number = {58},
year = {1990},
publisher = {American Geophysical Union},
editor = {Russell, C.T. and Priest, E.R. and Lee, L.C.},
address = {Washington, 23},
}

@Article{ref30,
author = {Roberts, B.},
title = {Waves in Solar Atmosphere},
journal = {Geophys. Astrophys. Fluid Dyn.},
year = {1991},
volume = {62},
pages = {83},
}

@incollection{ref31,
year={1997},
isbn={978-3-540-63072-2},
booktitle={Solar and Heliospheric Plasma Physics},
volume={489},
series={Lecture Notes in Physics},
editor={Simnett, GeorgeM. and Alissandrakis, ConstantineE. and Vlahos, Loukas},
doi={10.1007/BFb0105671},
title={Dynamics of flux tubes in the solar atmosphere: Theory},
url={http://dx.doi.org/10.1007/BFb0105671},
publisher={Springer Berlin Heidelberg},
author={Roberts, B. and Ulmschneider, P.},
pages={75-101},
language={English}
}

@Book{Stix2004,
author = {Stix, M.},
title = {The {Sun}: An Introduction},
publisher = {Springer},
year = {2004},
address = {Berlin},
}

@Article{ref33,
author = {Fawzy, D.F. and Cuntz, M. and Rammacher, W},
title = {Solar magnetic flux tube simulations with time-dependent ionization},
journal = {Mon. Not. R. Astron. Soc.},
year = {2012},
volume = {426},
pages = {1916-1927},
}

@Article{ref34,
author = {Fawzy, D.F. and Cuntz, M.},
title = {Generation of longitudinal flux tube waves in theoretical main-sequence stars: effects of model parameters},
journal = {Astron. Astrophys.},
year = {2011},
volume = {521},
pages = {A91},
}

@Article{Zioutas2009,
author = {Zioutas, K. and Tsagri, M. and Semertzidis, Y. and Papaevangelou, T. and Dafni, T. and Anastassopoulos, T.},
title = {Axion Searches with Helioscopes and astrophysical signatures for axion(-like) particles},
journal = {New J. Physics},
year = {2009},
volume = {11},
pages = {105020},
note = {\href{http://arxiv.org/abs/0903.1807}{arXiv:0903.1807 [astro-ph.SR]}},
}

@Article{ref46,
author = {Kim, J.},
title = {Weak Interaction Singlet and Strong {CP} Invariance},
journal = {Phys. Rev. Lett.},
year = {1979},
volume = {43},
pages = {103},
}

@Article{ref46a,
author = {Shifman, M.A. and Vainshtein, A.I. and Zakharov, V.I.},
title = {Can Confinement Ensure Natural {CP} Invariance of Strong Interactions?},
journal = {Nucl. Phys. B},
year = {1980},
volume = {166},
pages = {493},
}

@Article{ref47,
author = {Zhitnisky, A. P.},
title = {On Possible Suppression of the Axion Hadron Interactions (In {Russian})},
journal = {Sov. J. Nucl. Phys.},
year = {1980},
volume = {31},
pages = {260},
}

@Article{ref50,
author = {S. J. Asztalos and G. Carosi and C. Hagmann and D. Kinion and and K. van Bibber and J. Hoskins and J. Hwang and P. Sikivie and D. B. Tanner and R. Bradley and J. Clarke},
title = {{SQUID}-Based Microwave Cavity Search for Dark-Matter Axions},
journal = {Phys. Rev. Lett.},
year = {2010},
volume = {104},
pages = {041301},
}

@Article{ref51,
author = {W. U. Wuensch and  S. De Panfilis-Wuensch and Y. K. Semertzidis and  J. T. Rogers and A. C. Melissinos and H.  J. Halama and B. E. Moskowitz and A. G. Prodell and W. B. Fowler and F. A. Nezrick},
title = {Results of a laboratory search for cosmic axions and other weakly coupled light particles},
journal = {Phys. Rev. D},
year = {1989},
volume = {40},
number = {10},
pages = {3153-3167},
}

@Article{Andriamonje2007,
author = {Andriamonje, S. and others},
collaboration = {{CAST}},
title = {An improved limit on the axion-photon coupling from the {CAST} experiment},
journal = {J. Cosmol. Astropart. Phys.},
year = {2007},
volume = {10},
pages = {0702},
note = {\href{http://arxiv.org/abs/hep-ex/0702006}{arXiv:hep-ex/0702006}},
}

@Article{Peres2000,
author = {Peres, G. and Orlando, S. and Reale and Rosner, R. and Hudson, H.},
title = {The {Sun} as an {X}-ray Star II. Using the {Yohkoh/Soft} {X}-ray Telescope-derived solar emission measure versus temperature to interpret stellar {X}-ray observations},
journal = {The Astrophys. J.},
year = {2000},
volume = {528},
pages = {537-551},
}

@Article{Bahcall2004,
author = {Bahcall, J. N. and Pinsonneault, M. H.},
title = {What do we (not) know theoretically about solar neutrino fluxes?},
journal = {Phys. Rev. Lett.},
year = {2004},
volume = {92},
pages = {121301},
note = {\href{http://arxiv.org/abs/astro-ph/0402114}{arXiv:astro-ph/0402114}},
}

@article{ref72,
  author={{Arik, E. \textit{et al.} (CAST collaboration)}},
  title={Probing eV-scale axions with CAST},
  journal={Journal of Cosmology and Astroparticle Physics},
  volume={2009},
  number={02},
  pages={008},
  url={http://stacks.iop.org/1475-7516/2009/i=02/a=008},
  year={2009},
}

@Article{ref34-3,
author = {Stein, R.F.},
title = {Solar Surface Magneto-Convection},
journal = {Living Rev.Solar Phys.},
year = {2012},
volume = {9},
pages = {4},
}

@Article{Fan2009,
author = {Fan, Y.},
title = {Magnetic Fields in the Solar Convective Zone},
journal = {Living Rev.Solar Phys.},
year = {2009},
volume = {6},
pages = {4},
}

@Article{Parker2009,
author = {Parker, E. N.},
title = {Solar Magnetism: The State of Our Knowledge and Ignorance},
journal = {Space Sci. Rev.},
year = {2009},
volume = {122},
pages = {15-24},
}

@Article{Parker1955a,
author = {Parker, E. N.},
title = {The formation of sunspot from solar toroidal field},
journal = {Astrophys. J.},
year = {1955},
volume = {121},
pages = {491-507},
}

@Article{Biermann1941,
author = {Biermann, L.},
title = {Der gegenw\"{a}artige Stand der Theorie konvektiver Sonnenmodelle},
journal = {Vierteljahrsschrift Astron. Gesellsch.},
year = {1941},
volume = {76},
pages = {194-200},
}

@Article{Parker1979b,
author = {Parker, E. N.},
title = {Sunspots and the physics of magnetic flux tubes. {I}. The general nature of the sunspot},
journal = {Astrrophys. J.},
year = {1979},
volume = {230},
pages = {905-913},
}

@Article{Bahcall1992,
author = {Bahcall, J.N. and Pinsonneault, M.N.},
title = {Standard solar models, with and without helium diffusion, and the solar neutrino problem},
journal = {Rev. Mod. Phys.},
year = {1992},
volume = {64},
pages = {885},
}

@Article{Fan1996,
author = {Fan, Y. and Fisher, G.H.},
title = {Radiative heating and the buoyant rise of magnetic flux tubes in the solar interior},
journal = {Solar Phys.},
year = {1996},
volume = {166},
pages = {17-41},
}

@Book{Priest2000,
author = {Priest, E. and Forbes, T.},
title = {Magnetic reconnection. {MHD} theory amd applications},
publisher = {Cambridge University Press},
year = {2000},
}

@Article{Parker1994,
author = {Parker, E. N.},
title = {Theoretical properties of {$\Omega$}-loops in the connection zone of the Sun. {I}. Emerging bipolar magnetic regions},
journal = {The Astrophysics Journal},
year = {1994},
volume = {433},
pages = {867-874},
}

@article{Spruit1974,
author = {Spruit, H.C.},
title = {A model of the solar convective zone},
journal = {Solar Physics},
year = {1974},
volume = {34},
pages = {277-290},
}

@Article{Pevtsov2011,
  author={Alexei A. Pevtsov and Yury A. Nagovitsyn and Andrey G. Tlatov and Alexey L. Rybak},
  title={Long-term Trends in Sunspot Magnetic Fields},
  journal={The Astrophysical Journal Letters},
  volume={742},
  number={2},
  pages={L36},
  url={http://stacks.iop.org/2041-8205/742/i=2/a=L36},
  year={2011},
}

@ARTICLE{Pevtsov2014,
   author = {{Pevtsov}, A.~A. and {Bertello}, L. and {Tlatov}, A.~G. and 
	{Kilcik}, A. and {Nagovitsyn}, Y.~A. and {Cliver}, E.~W.},
    title = "{Cyclic and Long-Term Variation of Sunspot Magnetic Fields}",
  journal = {Solar Physics},
archivePrefix = "arXiv",
   eprint = {1301.5935},
 primaryClass = "astro-ph.SR",
     year = 2014,
    month = feb,
   volume = 289,
    pages = {593-602},
      doi = {10.1007/s11207-012-0220-5},
   adsurl = {http://adsabs.harvard.edu/abs/2014SoPh..289..593P}
}

@Article{Parker1993,
author = {Parker, E. N.},
title = {A solar dynamo surface wave at the interface between convection and nonuniform},
journal = {The Astrophysical Journal},
year = {1993},
volume = {408},
pages = {707},
}

@Article{Frandsen2010,
author = {Frandsen, M.T. and Sarkar, S.},
title = {Asymmetric Dark Matter and the Sun},
journal = {Phys. Rev. Lett.},
year = {2010},
volume = {105},
pages = {011301},
}

@article{Gillessen2009,
  author={S. Gillessen and F. Eisenhauer and S. Trippe and T. Alexander and R. Genzel and F. Martins and T. Ott},
  title={Monitoring Stellar Orbits Around the Massive Black Hole in the Galactic Center},
  journal={The Astrophysical Journal},
  volume={692},
  number={2},
  pages={1075},
  url={http://stacks.iop.org/0004-637X/692/i=2/a=1075},
  year={2009},
}

@Article{DSilva1993,
author = {D'Silva, S.},
title = {Can equipartition fields produce the tilts of bipolar magnetic regions?},
journal = {The Astrophysical Journal},
year = {1993},
volume = {407},
pages = {385},
}

@Article{Dikpati2008,
author = {Dikpati, M. and Gilman, P.A. and de Toma, Giuliana},
title = {THE WALDMEIER EFFECT: AN ARTIFACT OF THE DEFINITION OF WOLF SUNSPOT NUMBER?},
journal = {The Astrophysical Journal},
year = {2008},
volume = {673},
pages = {L99-L101},
month = {January},
}

@incollection{Gough2010,
year={2010},
isbn={978-3-642-02858-8},
booktitle={Magnetic Coupling between the Interior and Atmosphere of the Sun},
series={Astrophysics and Space Science Proceedings},
editor={Hasan, S. and Rutten, R. J.},
doi={10.1007/978-3-642-02859-5_4},
title={Vainu Bappu Memorial Lecture: What is a Sunspot?},
url={http://dx.doi.org/10.1007/978-3-642-02859-5_4},
publisher={Springer Berlin Heidelberg},
author={Gough, D.O.},
pages={37-66},
language={English}
}

@Article{Raffelt-Stodolsky1988,
author = {Raffelt, G. and  Stodolsky, L.},
title = {Mixing  of the  photon with low-mass particles},
journal = {Phys. Rev. D},
year = {1988},
volume = {37},
pages = {1237},
}

@Article{Hochmuth2007,
author = {Hochmuth, K.A. and Sigl, G.},
title = {Effects of axion-photon mixing on gamma-ray spectra from magnetized astrophysical sources},
journal = {Phys.Rev. D},
year = {2007},
volume = {76},
pages = {123011},
}

@techreport{Irastorza2013,
      author        = "Irastorza, Igor G",
      title         = "{The International Axion Observatory IAXO. Letter of
                       Intent to the CERN SPS committee}",
      institution   = "CERN",
      collaboration = "IAXO Collaboration",
      address       = "Geneva",
      number        = "CERN-SPSC-2013-022. SPSC-I-242",
      month         = "Aug",
      year          = "2013",
}

@Unpublished{Carosi2013,
author = {Carosi, G. and Friedland, A. and Giannotti, M. and Pivovarov, M.J. and Ruz, J. and Vogel, J.K.},
title = {Probing the axion-photon coupling: phenomenological and experimental perspectives. A snowmass white paper},
year = {2013},
note = {\href{http://arxiv.org/abs/1309.7035}{arXiv:1309.7035}},
}

@Article{Baer2011,
author = {Baer, H. and Lessa, A. and Rajagopalan, S. and Sreethawong, W.},
title = {Mixed axion/neutralino cold dark matter in supersymmetric models},
journal = {JCAP},
year = {2011},
volume = {31},
}

@Article{Arias2012,
author = {Arias, P. and Cadamuro, D. and Goodsell, M. and Jaeckel, J. and Redondo, J. and others},
title = {{WISP}y Cold Dark Matter},
journal = {JCAP},
year = {2012},
volume = {13},
note = {\href{http://arxiv.org/abs/1201.5902}{arXiv:1201.5902}},
}

@Unpublished{Arik2013,
author = {Arik, M. and Aune, S. and Barth, K. and Belov, A. and Borghi, S. and others},
title = {{CAST} solar axion search with ${^3}${He} buffer gas: Closing the hot dark matter gap},
year = {2013},
note = {\href{http://arxiv.org/abs/1307.1985}{arXiv:1307.1985}},
}

@Article{Genzel2010,
author = {Genzel, R. and Eisenhauer, F. and Gillessen, S.},
title = {The Galactic Center massive black hole and nuclear star cluster},
journal = {Reviews of Modern Physics},
year = {2010},
volume = {82},
pages = {3121},
}

@Article{Gillessen2013,
author = {Gillessen, S. and Genzel, R. and Fritz, T. K. and Eisenhauer, F. and Pfuhl, O. and Ott, O.},
title = {New observation of the gas cloud {G2} in the Galactic Center},
journal = {The Astrophysical Journal},
year = {2013},
volume = {78},
pages = {763},
}

@Misc{WDC2007,
author = {{WDC for Geomagnetic}},
organization = {World Data Centre for Geomagnetic ({Edinburg})},
title = {2007 worldwide observatory annual means, data of the observatory {Eskdalemuir} ({England}).},
howpublished = {\url{http://www.geomag.bgs.ak.uk./gifs/annual_means.shtml}},
year = {2007},
}

@Article{Bintanja2008,
author = {Bintanja, R. and van de Wal, R. S. W.},
title = {North American ice-sheet dynamics and the onset of 100,000-year glacial cycles},
journal = {Nature},
year = {2008},
volume = {454},
pages = {869-872},
}

@InProceedings{Rieutord2014,
author = {Rieutord, M.},
title = {Magnetohydrodynamics and solar physics},
booktitle = {Soci\'{e}t\'{e} Francaise d'Astronomie et d'Astrophysique (SF2A)},
year = {2014},
note = {\href{http://arxiv.org/abs/1410.3725}{arXiv:1410.3725}},
}

@article{CAST2011,
  title = {Search for Sub-eV Mass Solar Axions by the CERN Axion Solar Telescope with $^{3}\mathrm{He}$ Buffer Gas},
  author = {{Arik, E. \textit{et al.} (CAST collaboration)}},
  collaboration = {CAST Collaboration},
  journal = {Phys. Rev. Lett.},
  volume = {107},
  issue = {26},
  pages = {261302},
  numpages = {5},
  year = {2011},
  month = {Dec},
  publisher = {American Physical Society},
  doi = {10.1103/PhysRevLett.107.261302},
  url = {http://link.aps.org/doi/10.1103/PhysRevLett.107.261302}
}

@Article{Rempel2011,
author = {Matthias Rempel and Rolf Schlichenmaier},
title = {Sunspot Modeling: From Simplified Models to Radiative {MHD} Simulations},
journal = {Living Rev. Solar Phys.},
year = {2011},
volume = {8},
}

@ARTICLE{Fowler1955,
   author = {Fowler, W. A. and Burbidge, G. R. and Burbidge, E.~M.},
    title = "Nuclear Reactions and Element Synthesis in the Surface of Stars.",
  journal = {Astrophysical Journal Supplement},
     year = 1955,
    month = dec,
   volume = 2,
    pages = {167},
      doi = {10.1086/190020},
   adsurl = {http://adsabs.harvard.edu/abs/1955ApJS....2..167F}
}

@article{Couvidat2003,
  author={S. Couvidat and S. Turck-Chi\`{e}ze and A. G. Kosovichev},
  title={Solar Seismic Models and the Neutrino Predictions},
  journal={The Astrophysical Journal},
  volume={599},
  number={2},
  pages={1434},
  url={http://stacks.iop.org/0004-637X/599/i=2/a=1434},
  year={2003},
}

@Book{Schwarzschild1958,
author = {Martin Schwarzschild},
title = {Structure and evolution of the stars},
publisher = {Princeton, Princeton University Press},
year = {1958},
}

@Unpublished{Winterberg2015a,
author = {Friedwardt Winterberg},
year = {2015},
title = {Emission of Gravitational Waves from a Magnetohydrodynamic Dynamo},
note = {\href{http://arxiv.org/abs/1503.03003}{arXiv:1503.03003 [physics.gen-ph]}},
}

@article{Winterberg2016, title={Thermonuclear dynamo inside ultracentrifuge with supersonic plasma flow stabilization}, volume={23}, ISSN={1070-664X, 1089-7674}, DOI={10.1063/1.4938196}, number={1}, journal={Physics of Plasmas}, author={Winterberg, F.}, year={2016}, month={Jan}, pages={012502} }

@Article{Ettingshausen1886,
author = {A. Von Ettingshausen and W. Nernst},
title = {Ueber das Auftreten electromotorischer Kr\"{a}fte in Metallplatten, welche von einem W\"{a}rmestrome durchflossen werden und sich im magnetischen Felde befinden},
journal = {Wied. Ann.},
year = {1886},
volume = {29},
pages = {343-347},
}

@Article{Sondheimer1948,
author = {E. H. Sondheimer},
title = {The Theory of the Galvanomagnetic and Thermomagnetic Effects in Metals},
journal = {Proceedings of the Royal Society of London. Series A, Mathematical and Physical Sciences},
year = {1948},
volume = {193},
pages = {484-512},
}

@Book{Spitzer1956,
author = {Spitzer, Lyman Jr.},
title = {Physics of Fully Ionized Gases},
publisher = {John Wiley \& Sons, Inc.},
year = {1956},
address = {New York},
}

@InCollection{Kim1969,
author = {Y. B. Kim and M. J. Stephen},
title = {},
booktitle = {Superconductivity},
publisher = {Dekker},
year = {1969},
editor = {R.D. Parks},
volume = {2},
address = {New York},
}

@Book{Spitzer1962,
author = {Spitzer, Lyman Jr.},
title = {Physics of Fully Ionized Gases},
publisher = {Interscience Publishers, John Wiley \& Sons},
year = {1962},
address = {New York},
edition = {2nd},
}

@Book{Spitzer2006,
author = {Spitzer, Lyman Jr.},
title = {Physics of Fully Ionized Gases},
publisher = {Dover Publications, Inc.},
year = {2006},
address = {Mineola, New York},
edition = {2nd},
}

@Unpublished{Zioutas2007,
author = {Zioutas, K. and Tsargi, M. and Semertzidis, Y. and Papaevangelov, T. and Nordt, A. and Anastassopoulos, V.},
title = {Overlooked astrophysical signatures of axion(-like) particles},
year = {2007},
note = {\href{http://arxiv.org/abs/astro-ph/0701627}{arXiv:astro-ph/0701627}},
}

@Book{Parker1979a,
author = {Parker, E. N.},
title = {Cosmical magnetic fields: Their origin and their activity},
publisher = {Oxford, Clarendon Press},
year = {1979},
pages = {858},
}

@article{Stein2012,
  title    = {Solar Surface Magneto-Convection},
  author   = {Robert F. Stein},
  journal  = {Living Reviews in Solar Physics},
  year     = {2012},
  number   = {4},
  volume   = {9},
  doi      = {10.12942/lrsp-2012-4},
  url      = {http://www.livingreviews.org/lrsp-2012-4}
}

@article{Gold1960,
author = {Gold, T. and Hoyle, F.}, 
title = {On the Origin of Solar Flares},
volume = {120}, 
number = {2}, 
pages = {89-105}, 
year = {1960}, 
doi = {10.1093/mnras/120.2.89}, 
journal = {Monthly Notices of the Royal Astronomical Society} 
}

@article{Sturrock2001,
  author={Peter A. Sturrock and Mark Weber and Michael S. Wheatland and Richard Wolfson},
  title={Metastable Magnetic Configurations and Their Significance for Solar Eruptive Events},
  journal={The Astrophysical Journal},
  volume={548},
  number={1},
  pages={492},
  url={http://stacks.iop.org/0004-637X/548/i=1/a=492},
  year={2001},
}

@article{Nelson2014,
  author={N J Nelson and M S Miesch},
  title={Generating buoyant magnetic flux ropes in solar-like convective dynamos},
  journal={Plasma Physics and Controlled Fusion},
  volume={56},
  number={6},
  pages={064004},
  url={http://stacks.iop.org/0741-3335/56/i=6/a=064004},
  year={2014},
}

@article{Rempel2011a,
  author={Matthias Rempel},
  title={Subsurface Magnetic Field and Flow Structure of Simulated Sunspots},
  journal={The Astrophysical Journal},
  volume={740},
  number={1},
  pages={15},
  url={http://stacks.iop.org/0004-637X/740/i=1/a=15},
  year={2011},
}

@ARTICLE{Vincent2015a,
   author = {{Vincent}, A.~C. and {Scott}, P. and {Serenelli}, A.},
    title = "{Possible Indication of Momentum-Dependent Asymmetric Dark Matter in the Sun}",
  journal = {Physical Review Letters},
archivePrefix = "arXiv",
   eprint = {1411.6626},
 primaryClass = "hep-ph",
     year = 2015,
    month = feb,
   volume = 114,
   number = 8,
      eid = {081302},
    pages = {081302},
      doi = {10.1103/PhysRevLett.114.081302},
   adsurl = {http://adsabs.harvard.edu/abs/2015PhRvL.114h1302V}
}

@ARTICLE{Vincent2015b,
   author = {{Vincent}, A.~C. and {Serenelli}, A. and {Scott}, P.},
    title = "{Generalised form factor dark matter in the Sun}",
  journal = {Journal of Cosmology and Astroparticle Physics},
archivePrefix = "arXiv",
   eprint = {1504.04378},
 primaryClass = "hep-ph",
     year = 2015,
    month = aug,
   volume = 8,
      eid = {040},
    pages = {040},
      doi = {10.1088/1475-7516/2015/08/040},
   adsurl = {http://adsabs.harvard.edu/abs/2015JCAP...08..040V}
}

@Article{Alfven1942,
author = {Alfv\'{e}n, H.},
title = {Existence of Electromagnetic-Hydrodynamic Waves},
journal = {Nature},
year = {1942},
volume = {150},
pages = {405-406},
month = {October},
}

@article{Ayala2014,
  title = {Revisiting the Bound on Axion-Photon Coupling from Globular Clusters},
  author = {Ayala, Adrian and Dom\'{i}nguez, Inma and Giannotti, Maurizio and Mirizzi, Alessandro and Straniero, Oscar},
  journal = {Phys. Rev. Lett.},
  volume = {113},
  issue = {19},
  pages = {191302},
  numpages = {5},
  year = {2014},
  month = {Nov},
  publisher = {American Physical Society},
  doi = {10.1103/PhysRevLett.113.191302},
  url = {http://link.aps.org/doi/10.1103/PhysRevLett.113.191302}
}

@Article{ChenF2015,
author = {Chen, F. and Peter, H. and Bingert, S. and Cheung, M. C. M.},
title = {Magnetic jam in the corona of the Sun},
journal = {Nature Physics},
year = {2015},
volume = {11},
pages = {492-495},
url = {http://dx.doi.org/10.1038/nphys3315},
}

@ARTICLE{Parker1988,
   author = {Parker, E. N.},
    title = "{Nanoflares and the solar X-ray corona}",
  journal = {Astrophysical Journal},
     year = 1988,
    month = jul,
   volume = 330,
    pages = {474-479},
      doi = {10.1086/166485},
   adsurl = {http://adsabs.harvard.edu/abs/1988ApJ...330..474P}
}

@article{Shibata2011,
  title    = {Solar Flares: Magnetohydrodynamic Processes},
  author   = {Kazunari Shibata and Tetsuya Magara},
  journal  = {Living Reviews in Solar Physics},
  year     = {2011},
  number   = {6},
  volume   = {8},
  doi      = {10.12942/lrsp-2011-6},
  url      = {http://www.livingreviews.org/lrsp-2011-6}
}

@article{Friedland2013,
  title = {Constraining the Axion-Photon Coupling with Massive Stars},
  author = {Friedland, Alexander and Giannotti, Maurizio and Wise, Michael},
  journal = {Phys. Rev. Lett.},
  volume = {110},
  issue = {6},
  pages = {061101},
  numpages = {5},
  year = {2013},
  month = {Feb},
  publisher = {American Physical Society},
  doi = {10.1103/PhysRevLett.110.061101},
  url = {http://link.aps.org/doi/10.1103/PhysRevLett.110.061101}
}

@article{Dine1981,
title = "A simple solution to the strong \{CP\} problem with a harmless axion ",
journal = "Physics Letters B ",
volume = "104",
number = "3",
pages = "199 - 202",
year = "1981",
note = "",
issn = "0370-2693",
doi = "http://dx.doi.org/10.1016/0370-2693(81)90590-6",
url = "http://www.sciencedirect.com/science/article/pii/0370269381905906",
author = "Michael Dine and Willy Fischler and Mark Srednicki"
}

@ARTICLE{Bak1987,
   author = {Bak, P. and Tang, C. and Wiesenfeld, K.},
    title = "{Self-organized criticality - An explanation of 1/f noise}",
  journal = {Physical Review Letters},
     year = 1987,
    month = jul,
   volume = 59,
    pages = {381-384},
      doi = {10.1103/PhysRevLett.59.381},
   adsurl = {http://adsabs.harvard.edu/abs/1987PhRvL..59..381B}
}

@article{Bak1988,
  title = {Self-organized criticality},
  author = {Bak, Per and Tang, Chao and Wiesenfeld, Kurt},
  journal = {Phys. Rev. A},
  volume = {38},
  issue = {1},
  pages = {364--374},
  numpages = {0},
  year = {1988},
  month = {Jul},
  publisher = {American Physical Society},
  doi = {10.1103/PhysRevA.38.364},
  url = {http://link.aps.org/doi/10.1103/PhysRevA.38.364}
}

@article{Bak1989,
title = "The physics of fractals ",
journal = "Physica D: Nonlinear Phenomena ",
volume = "38",
number = "1–3",
pages = "5 - 12",
year = "1989",
note = "",
issn = "0167-2789",
doi = "http://dx.doi.org/10.1016/0167-2789(89)90166-8",
url = "http://www.sciencedirect.com/science/article/pii/0167278989901668",
author = "Per Bak and Kan Chen",
}

@Book{Bak1996,
author = {Per Bak},
title = {How Nature Works: the science of self-organized criticality},
publisher = {Springer New York},
year = {1996},
url = {http://dx.doi.org/10.1007/978-1-4757-5426-1}
}

@Book{Aschwanden2011,
author = {Markus Aschwanden},
title = {Self-Organized Criticality in Astrophysics},
publisher = {Springer-Verlag Berlin Heidelberg},
year = {2011},
pages = {416},
}

@article{Charbonneau2001,
year={2001},
issn={0038-0938},
journal={Solar Physics},
volume={203},
number={2},
doi={10.1023/A:1013301521745},
title={Avalanche models for solar flares (Invited Review)},
url={http://dx.doi.org/10.1023/A%3A1013301521745},
publisher={Kluwer Academic Publishers},
author={Charbonneau, Paul and McIntosh, ScottW. and Liu, Han-Li and Bogdan, ThomasJ.},
pages={321-353},
language={English}
}

@article{Aschwanden2014,
  author={Markus J. Aschwanden},
  title={A Macroscopic Description of a Generalized Self-organized Criticality System: Astrophysical Applications},
  journal={The Astrophysical Journal},
  volume={782},
  number={1},
  pages={54},
  url={http://stacks.iop.org/0004-637X/782/i=1/a=54},
  year={2014},
}

@ARTICLE{Lu1993,
   author = {{Lu}, E.~T. and {Hamilton}, R.~J. and {McTiernan}, J.~M. and 
	{Bromund}, K.~R.},
    title = "{Solar flares and avalanches in driven dissipative systems}",
  journal = {Astrophysical Journal},
     year = 1993,
    month = aug,
   volume = 412,
    pages = {841-852},
      doi = {10.1086/172966},
   adsurl = {http://adsabs.harvard.edu/abs/1993ApJ...412..841L}
}

@ARTICLE{Lu1991,
   author = {{Lu}, E.~T. and {Hamilton}, R.~J.},
    title = "{Avalanches and the distribution of solar flares}",
  journal = {Astrophysical Journal Letters},
     year = 1991,
    month = oct,
   volume = 380,
    pages = {L89-L92},
      doi = {10.1086/186180},
   adsurl = {http://adsabs.harvard.edu/abs/1991ApJ...380L..89L}
}

@article{Uchaikin2013,
  author={V V Uchaikin},
  title={Fractional phenomenology of cosmic ray anomalous diffusion},
  journal={Physics-Uspekhi},
  volume={56},
  number={11},
  pages={1074},
  url={http://stacks.iop.org/1063-7869/56/i=11/a=1074},
  year={2013}
}

@Article{Haxton2014,
author = {Haxton, Wick},
title = {Neutrino physics: What makes the Sun shine},
journal = {Nature},
year = {2014},
volume = {512},
pages = {378-380},
url = {http://dx.doi.org/10.1038/512378a},
}

@article{Raffelt1986,
title = "Axion constraints from white dwarf cooling times",
journal = "Physics Letters B",
volume = "166",
number = "4",
pages = "402 - 406",
year = "1986",
note = "",
issn = "0370-2693",
doi = "http://dx.doi.org/10.1016/0370-2693(86)91588-1",
url = "http://www.sciencedirect.com/science/article/pii/0370269386915881",
author = "Georg G. Raffelt",

}

@Book{Bhatia2001,
author = {Bhatia, V. B.},
title = {Textbook of astronomy and astrophysics with elements of cosmology},
publisher = {Pangbourne, India : Alpha Science International Ltd},
year = {2001},
}

@ARTICLE{Robinson1995,
   author = {{Robinson}, E.~L. and {Mailloux}, T.~M. and {Zhang}, E. and 
	{Koester}, D. and {Stiening}, R.~F. and {Bless}, R.~C. and {Percival}, J.~W. and 
	{Taylor}, M.~J. and {van Citters}, G.~W.},
    title = "{The pulsation index, effective temperature, and thickness of the hydrogen layer in the pulsating DA white dwarf G117-B15A}",
  journal = {Astrophysical Journal},
     year = 1995,
    month = jan,
   volume = 438,
    pages = {908-916},
      doi = {10.1086/175132},
   adsurl = {http://adsabs.harvard.edu/abs/1995ApJ...438..908R}
}

@article{Steinfadt2010,
  author={Justin D. R. Steinfadt and Lars Bildsten and Phil Arras},
  title={Pulsations in Hydrogen Burning Low-mass Helium White Dwarfs},
  journal={The Astrophysical Journal},
  volume={718},
  number={1},
  pages={441},
  url={http://stacks.iop.org/0004-637X/718/i=1/a=441},
  year={2010},
}

@Article{Valyavin2014,
author = {G. Valyavin and D. Shulyak and G. A. Wade and K. Antonyuk and S. V. Zharikov and G. A. Galazutdinov and S. Plachinda and S. Bagnulo and L. Fox Machado and M. Alvarez and D. M. Clark and J. M. Lopez and D. Hiriart and Inwoo Han and Young-Beom Jeon and C. Zurita and R. Mujica and T. Burlakova and T. Szeifert and A. Burenkov},
title = {Suppression of cooling by strong magnetic fields in white dwarf stars},
journal = {Nature},
year = {2014},
volume = {515},
pages = {88-91},
}

@article{Kissin2015,
  author={Yevgeni Kissin and Christopher Thompson},
  title={Spin and Magnetism of White Dwarfs},
  journal={The Astrophysical Journal},
  volume={809},
  number={2},
  pages={108},
  url={http://stacks.iop.org/0004-637X/809/i=2/a=108},
  year={2015},
}

@article{Rogers1994,
author = {Rogers, Forrest J. and Iglesias, Carlos A.}, 
title = {Astrophysical Opacity},
volume = {263}, 
number = {5143}, 
pages = {50-55}, 
year = {1994}, 
doi = {10.1126/science.263.5143.50}, 
journal = {Science} 
}

@ARTICLE{Ferguson2005,
   author = {{Ferguson}, J.~W. and {Alexander}, D.~R. and {Allard}, F. and 
	{Barman}, T. and {Bodnarik}, J.~G. and {Hauschildt}, P.~H. and 
	{Heffner-Wong}, A. and {Tamanai}, A.},
    title = "{Low-Temperature Opacities}",
  journal = {Astrophysical Journal},
   eprint = {astro-ph/0502045},
     year = 2005,
    month = apr,
   volume = 623,
    pages = {585-596},
      doi = {10.1086/428642},
   adsurl = {http://adsabs.harvard.edu/abs/2005ApJ...623..585F},
}

@ARTICLE{Bailey2009,
   author = {{Bailey}, J.~E. and {Rochau}, G.~A. and {Mancini}, R.~C. and 
	{Iglesias}, C.~A. and {Macfarlane}, J.~J. and {Golovkin}, I.~E. and 
	{Blancard}, C. and {Cosse}, P. and {Faussurier}, G.},
    title = "{Experimental investigation of opacity models for stellar interior, inertial fusion, and high energy density plasmas}",
  journal = {Physics of Plasmas},
     year = 2009,
    month = may,
   volume = 16,
   number = 5,
    pages = {058101},
      doi = {10.1063/1.3089604},
   adsurl = {http://adsabs.harvard.edu/abs/2009PhPl...16e8101B}
}

@article{Stamatellos2007,
	author = {{D. Stamatellos} and {A. P. Whitworth} and {T. Bisbas} and {S. Goodwin}},
	title = {Radiative transfer and the energy equation   in SPH simulations of star formation},
	DOI= "10.1051/0004-6361:20077373",
	url= "http://dx.doi.org/10.1051/0004-6361:20077373",
	journal = {A\&A},
	year = 2007,
	volume = 475,
	number = 1,
	pages = "37-49",
	month = "",
}

@unpublished{Cranmer2015,
author = {Steven R. Cranmer},
title = {University of Colorado ASTR-3760 course, Solar and Space Physics},
year = {2015},
url = {http://lasp.colorado.edu/~cranmer/astr_3760_sp2015.html},
}

@article {Lockwood2001,
author = {Lockwood, M.},
title = {Long-term variations in the magnetic fields of the Sun and the heliosphere: Their origin, effects, and implications},
journal = {Journal of Geophysical Research: Space Physics},
volume = {106},
number = {A8},
issn = {2156-2202},
url = {http://dx.doi.org/10.1029/2000JA000115},
doi = {10.1029/2000JA000115},
pages = {16021--16038},
year = {2001},
}

@Article{Hathaway2015,
author="Hathaway, David H.",
title="The Solar Cycle",
journal="Living Reviews in Solar Physics",
year="2015",
month="Sep",
day="21",
volume="12",
number="1",
pages="4",
issn="1614-4961",
doi="10.1007/lrsp-2015-4",
url="https://doi.org/10.1007/lrsp-2015-4"
}

@InBook{Cowling1953,
author = {Cowling, T. G.},
title = {The solar system},
chapter = {The Sun},
publisher = {The University of Chicago Press},
year = {1953},
volume = {1},
address = {Chicago},
pages = {532},
}

@Article{Hale1908,
author = {Hale, George E.},
title = {On the Probable Existence of a Magnetic Field in Sun-Spots},
journal = {Astrophysical Journal},
year = {1908},
volume = {28},
pages = {315-343},
month = {11},
}

@ARTICLE{Hale1919,
   author = {{Hale}, G.~E. and {Ellerman}, F. and {Nicholson}, S.~B. and 
	{Joy}, A.~H.},
    title = "{The Magnetic Polarity of Sun-Spots}",
  journal = {Astrophysical Journal},
     year = 1919,
    month = apr,
   volume = 49,
    pages = {153},
      doi = {10.1086/142452},
   adsurl = {http://adsabs.harvard.edu/abs/1919ApJ....49..153H}
}

@article{Evershed1909,
author = {Evershed, J.},
title = {Radial movement in sun-spots},
journal = {Monthly Notices of the Royal Astronomical Society},
volume = {69},
number = {5},
pages = {454-458},
year = {1909},
doi = {10.1093/mnras/69.5.454},
}

@Article{Solanki2003,
author="Solanki, Sami K.",
title="Sunspots: An overview",
journal="The Astronomy and Astrophysics Review",
year="2003",
month="Apr",
day="01",
volume="11",
number="2",
pages="153--286",
issn="1432-0754",
doi="10.1007/s00159-003-0018-4",
url="https://doi.org/10.1007/s00159-003-0018-4"
}

@Article{Borrero2011,
author="Borrero, Juan M.
and Ichimoto, Kiyoshi",
title="Magnetic Structure of Sunspots",
journal="Living Reviews in Solar Physics",
year="2011",
month="Sep",
day="09",
volume="8",
number="1",
pages="4",
issn="1614-4961",
doi="10.12942/lrsp-2011-4",
url="https://doi.org/10.12942/lrsp-2011-4"
}

@article{Tiwari2015,
	author = {Tiwari, Sanjiv K. and van Noort, Michiel and Solanki, Sami K. and Lagg, Andreas},
	title = {Depth-dependent global properties of a sunspot observed  by Hinode using the Solar Optical Telescope/Spectropolarimeter},
	DOI= "10.1051/0004-6361/201526224",
	url= "https://doi.org/10.1051/0004-6361/201526224",
	journal = {A\&A},
	year = 2015,
	volume = 583,
	pages = "A119",
	month = "",
}

@article{Pozuelo2015,
  author={S. Esteban Pozuelo and L. R. Bellot Rubio and J. de la Cruz Rodr\'{i}guez},
  title={Lateral Downflows in Sunspot Penumbral Filaments and their Temporal Evolution},
  journal={The Astrophysical Journal},
  volume={803},
  number={2},
  pages={93},
  url={http://stacks.iop.org/0004-637X/803/i=2/a=93},
  year={2015},
}

@article{Heinemann2007,
  author={T. Heinemann and \r{A}. Nordlund and G. B. Scharmer and H. C. Spruit},
  title={MHD Simulations of Penumbra Fine Structure},
  journal={The Astrophysical Journal},
  volume={669},
  number={2},
  pages={1390},
  url={http://stacks.iop.org/0004-637X/669/i=2/a=1390},
  year={2007},
}

@Article{Miesch2015,
author="Miesch, Mark
and Matthaeus, William
and Brandenburg, Axel
and Petrosyan, Arakel
and Pouquet, Annick
and Cambon, Claude
and Jenko, Frank
and Uzdensky, Dmitri
and Stone, James
and Tobias, Steve
and Toomre, Juri
and Velli, Marco",
title="Large-Eddy Simulations of Magnetohydrodynamic Turbulence in Heliophysics and Astrophysics",
journal="Space Science Reviews",
year="2015",
month="Nov",
day="01",
volume="194",
number="1",
pages="97--137",
issn="1572-9672",
doi="10.1007/s11214-015-0190-7",
url="https://doi.org/10.1007/s11214-015-0190-7"
}

@ARTICLE{Choudhuri1995,
   author = {Choudhuri, A. R. and Sch\"ussler, M. and Dikpati, M.},
    title = "The solar dynamo with meridional circulation.",
  journal = {Astronomy and Astrophysics},
     year = 1995,
    month = nov,
   volume = 303,
    pages = {L29},
   adsurl = {http://adsabs.harvard.edu/abs/1995A%26A...303L..29C},
}

@article{Guerrero2011,
	author = {Guerrero, G. and K\"apyl\"a, P. J.},
	title = {Dynamo action and magnetic buoyancy in convection simulations
          with vertical shear},
	DOI= "10.1051/0004-6361/201116749",
	url= "https://doi.org/10.1051/0004-6361/201116749",
	journal = {A\&A},
	year = 2011,
	volume = 533,
	pages = "A40",
	month = "",
}

@article{Nelson2013,
  author={Nicholas J. Nelson and Benjamin P. Brown and Allan Sacha Brun and Mark S. Miesch and Juri Toomre},
  title={Magnetic Wreaths and Cycles in Convective Dynamos},
  journal={The Astrophysical Journal},
  volume={762},
  number={2},
  pages={73},
  url={http://stacks.iop.org/0004-637X/762/i=2/a=73},
  year={2013},
}

@article{Kapyla2013,
  author={Petri J. K\"apyl\"a and Maarit J. Mantere and Elizabeth Cole and J\"orn Warnecke and Axel Brandenburg},
  title={Effects of Enhanced Stratification on Equatorward Dynamo Wave Propagation},
  journal={The Astrophysical Journal},
  volume={778},
  number={1},
  pages={41},
  url={http://stacks.iop.org/0004-637X/778/i=1/a=41},
  year={2013},
}

@article{Birch2013,
  author={A. C. Birch and D. C. Braun and K. D. Leka and G. Barnes and B. Javornik},
  title={Helioseismology of Pre-emerging Active Regions. II. Average Emergence Properties},
  journal={The Astrophysical Journal},
  volume={762},
  number={2},
  pages={131},
  url={http://stacks.iop.org/0004-637X/762/i=2/a=131},
  year={2013},
}

@article {Birch2016,
	author = {Birch, Aaron C. and Schunker, Hannah and Braun, Douglas C. and Cameron, Robert and Gizon, Laurent and L{\"o}ptien, Bj{\"o}rn and Rempel, Matthias},
	title = {A low upper limit on the subsurface rise speed of solar active regions},
	volume = {2},
	number = {7},
	year = {2016},
	doi = {10.1126/sciadv.1600557},
	publisher = {American Association for the Advancement of Science},
	journal = {Science Advances}
}

@article{Hanasoge2010,
  author={Shravan M. Hanasoge and Thomas L. Duvall Jr and Marc L. DeRosa},
  title={Seismic Constraints on Interior Solar Convection},
  journal={The Astrophysical Journal Letters},
  volume={712},
  number={1},
  pages={L98},
  url={http://stacks.iop.org/2041-8205/712/i=1/a=L98},
  year={2010},
}

@article{Hanasoge2012,
author = {Hanasoge, Shravan M. and Duvall, Thomas L. and Sreenivasan, Katepalli R.}, 
title = {Anomalously weak solar convection},
volume = {109}, 
number = {30}, 
pages = {11928-11932}, 
year = {2012}, 
doi = {10.1073/pnas.1206570109}, 
journal = {Proceedings of the National Academy of Sciences} 
}

@Article{Hanasoge2015,
author="Hanasoge, S.
and Miesch, M. S.
and Roth, M.
and Schou, J.
and Sch{\"u}ssler, M.
and Thompson, M. J.",
title="Solar Dynamics, Rotation, Convection and Overshoot",
journal="Space Science Reviews",
year="2015",
month="Dec",
day="01",
volume="196",
number="1",
pages="79--99",
issn="1572-9672",
doi="10.1007/s11214-015-0144-0",
url="https://doi.org/10.1007/s11214-015-0144-0"
}

@article{Gizon2012,
author = {Gizon, Laurent and Birch, Aaron C.}, 
title = {Helioseismology challenges models of solar convection},
volume = {109}, 
number = {30}, 
pages = {11896-11897}, 
year = {2012}, 
doi = {10.1073/pnas.1208875109}, 
journal = {Proceedings of the National Academy of Sciences} 
}

@ARTICLE{Spiegel1980,
   author = {Spiegel, E. A. and Weiss, N. O.},
    title = "{Magnetic activity and variations in solar luminosity}",
  journal = {Nature},
     year = 1980,
    month = oct,
   volume = 287,
    pages = {616},
      doi = {10.1038/287616a0},
   adsurl = {http://adsabs.harvard.edu/abs/1980Natur.287..616S}
}

@ARTICLE{Rosner1980,
   author = {Rosner, R.},
    title = "{Stellar Coronae - Interpretation and Modeling of Stellar Activity}",
  journal = {SAO Special Report},
     year = 1980,
   volume = 389,
    pages = {79},
   adsurl = {http://adsabs.harvard.edu/abs/1980SAOSR.389...79R},
}

@ARTICLE{Glatzmaier1985,
   author = {Glatzmaier, G. A.},
    title = "{Numerical simulations of stellar convective dynamos. II - Field propagation in the convection zone}",
  journal = {Astrophysical Journal},
     year = 1985,
    month = apr,
   volume = 291,
    pages = {300-307},
      doi = {10.1086/163069},
   adsurl = {http://adsabs.harvard.edu/abs/1985ApJ...291..300G},
}

@inbook{Weiss1994,
  author={Weiss, N. O.},  
  title={Solar and Stellar Dynamos},
  editor={Proctor, M. R. E. and Gilbert, A. D.Editors},
  place={Cambridge},
  series={Publications of the Newton Institute},
  DOI={10.1017/CBO9780511624025.004},
  booktitle={Lectures on Solar and Planetary Dynamos},
  publisher={Cambridge University Press},
  year={1994},
  pages={59–96},
  collection={Publications of the Newton Institute}
}

@article{Schou1998,
  author={J. Schou and H. M. Antia and S. Basu and R. S. Bogart and R. I. Bush and S. M. Chitre and J. Christensen-Dalsgaard and M. P. Di Mauro and W. A. Dziembowski and A. Eff-Darwich and D. O. Gough and D. A. Haber and J. T. Hoeksema and R. Howe and S. G. Korzennik and A. G. Kosovichev and R. M. Larsen and F. P. Pijpers and P. H. Scherrer and T. Sekii and T. D. Tarbell and A. M. Title and M. J. Thompson and J. Toomre},
  title={Helioseismic Studies of Differential Rotation in the Solar Envelope by the Solar Oscillations Investigation Using the Michelson Doppler Imager},
  journal={The Astrophysical Journal},
  volume={505},
  number={1},
  pages={390},
  url={http://stacks.iop.org/0004-637X/505/i=1/a=390},
  year={1998},
}

@article{Zaqarashvili2010,
  author={Teimuraz V. Zaqarashvili and Marc Carbonell and Ramón Oliver and Jos\'{e} Luis Ballester},
  title={Magnetic Rossby Waves in the Solar Tachocline and Rieger-Type Periodicities},
  journal={The Astrophysical Journal},
  volume={709},
  number={2},
  pages={749},
  url={http://stacks.iop.org/0004-637X/709/i=2/a=749},
  year={2010},
}

@article{Zaqarashvili2015,
  author={Teimuraz V. Zaqarashvili and Ramon Oliver and Arnold Hanslmeier and Marc Carbonell and Jose Luis Ballester and Tamar
Gachechiladze and Ilya G. Usoskin},
  title={Long-term variation in the Sun's activity caused by magnetic Rossby waves in the tachocline},
  journal={The Astrophysical Journal Letters},
  volume={805},
  number={2},
  pages={L14},
  url={http://stacks.iop.org/2041-8205/805/i=2/a=L14},
  year={2015},
}

@article{Guerrero2016,
  author={G. Guerrero and P. K. Smolarkiewicz and E. M. de Gouveia Dal Pino and A. G. Kosovichev and N. N. Mansour},
  title={On the Role of Tachoclines in Solar and Stellar Dynamos},
  journal={The Astrophysical Journal},
  volume={819},
  number={2},
  pages={104},
  url={http://stacks.iop.org/0004-637X/819/i=2/a=104},
  year={2016},
}

@article{Schussler2006,
  author={M. Sch\"ussler and A. V\"ogler},
  title={Magnetoconvection in a Sunspot Umbra},
  journal={The Astrophysical Journal Letters},
  volume={641},
  number={1},
  pages={L73},
  url={http://stacks.iop.org/1538-4357/641/i=1/a=L73},
  year={2006},
}

@article{Chandrasekhar1952,
author = {Chandrasekhar, F. R. S.},
title = {XLVI. On the inhibition of convection by a magnetic field},
journal = {The London, Edinburgh, and Dublin Philosophical Magazine and Journal of Science},
volume = {43},
number = {340},
pages = {501-532},
year = {1952},
doi = {10.1080/14786440508520205},
URL = {http://dx.doi.org/10.1080/14786440508520205},
eprint = {http://dx.doi.org/10.1080/14786440508520205}
}

@article{Tremblay2015,
  author={P.-E. Tremblay and G. Fontaine and B. Freytag and O. Steiner and H.-G. Ludwig and M. Steffen and S. Wedemeyer and P. Brassard},
  title={On the Evolution of Magnetic White Dwarfs},
  journal={The Astrophysical Journal},
  volume={812},
  number={1},
  pages={19},
  url={http://stacks.iop.org/0004-637X/812/i=1/a=19},
  year={2015},
}

@article{Miesch2012,
	author = {Miesch, Mark S.},
	title = {The solar dynamo},
	volume = {370},
	number = {1970},
	pages = {3049--3069},
	year = {2012},
	doi = {10.1098/rsta.2011.0507},
	publisher = {The Royal Society},
	issn = {1364-503X},
	journal = {Philosophical Transactions of the Royal Society of London A: Mathematical, Physical and Engineering Sciences}
}

@Article{Spruit1987,
author="Spruit, H. C.
and Title, A. M.
and Van Ballegooijen, A. A.",
title="Is there a weak mixed polarity background field? Theoretical arguments",
journal="Solar Physics",
year="1987",
month="Mar",
day="01",
volume="110",
number="1",
pages="115--128",
issn="1573-093X",
doi="10.1007/BF00148207",
url="https://doi.org/10.1007/BF00148207"
}

@Article{Wilson1990,
author="Wilson, P. R.
and McIntosh, P. S.
and Snodgrass, H. B.",
title="The reversal of the solar polar magnetic fields",
journal="Solar Physics",
year="1990",
month="May",
day="01",
volume="127",
number="1",
pages="1--9",
issn="1573-093X",
doi="10.1007/BF00158510",
url="https://doi.org/10.1007/BF00158510"
}

@ARTICLE{Gaizauskas1983,
   author = {Gaizauskas, V. and Harvey, K. L. and Harvey, J. W. and Zwaan, C.},
    title = "{Large-scale patterns formed by solar active regions during the ascending phase of cycle 21}",
  journal = {Astrophysical Journal},
     year = 1983,
    month = feb,
   volume = 265,
    pages = {1056-1065},
      doi = {10.1086/160747},
   adsurl = {http://adsabs.harvard.edu/abs/1983ApJ...265.1056G}
}

@Article{Petrovay1997,
author="Petrovay, K.
and van Driel-Gesztelyi, L.",
title="Making Sense of Sunspot Decay. I. Parabolic Decay Law and Gnevyshev--Waldmeier Relation",
journal="Solar Physics",
year="1997",
month="Dec",
day="01",
volume="176",
number="2",
pages="249--266",
issn="1573-093X",
doi="10.1023/A:1004988123265",
url="https://doi.org/10.1023/A:1004988123265"
}

@ARTICLE{Spruit1982,
   author = {Spruit, H. C. and van Ballegooijen, A. A.},
    title = "{Stability of toroidal flux tubes in stars}",
  journal = {Astronomy and Astrophysics},
     year = 1982,
    month = feb,
   volume = 106,
    pages = {58-66},
   adsurl = {http://adsabs.harvard.edu/abs/1982A%26A...106...58S}
}

@article{Nandy2002,
	author = {Nandy, Dibyendu and Choudhuri, Arnab Rai},
	title = {Explaining the Latitudinal Distribution of Sunspots with Deep Meridional Flow},
	volume = {296},
	number = {5573},
	pages = {1671--1673},
	year = {2002},
	doi = {10.1126/science.1070955},
	publisher = {American Association for the Advancement of Science},
	issn = {0036-8075},
	journal = {Science}
}

@ARTICLE{Parker1975,
   author = {Parker, E. N.},
    title = "{The generation of magnetic fields in astrophysical bodies. X - Magnetic buoyancy and the solar dynamo}",
  journal = {Astrophysical Journal},
     year = 1975,
    month = may,
   volume = 198,
    pages = {205-209},
      doi = {10.1086/153593},
   adsurl = {http://adsabs.harvard.edu/abs/1975ApJ...198..205P}
}

@ARTICLE{Ballegooijen1982,
   author = {van Ballegooijen, A. A.},
    title = "The overshoot layer at the base of the solar convective zone and the problem of magnetic flux storage",
  journal = {Astronomy and Astrophysics},
     year = 1982,
    month = sep,
   volume = 113,
    pages = {99-112},
   adsurl = {http://adsabs.harvard.edu/abs/1982A%26A...113...99V}
}

@ARTICLE{BohmVitense1958,
   author = {B{\"o}hm-Vitense, E.},
    title = "{\"U}ber die Wasserstoffkonvektionszone in Sternen verschiedener Effektivtemperaturen und Leuchtkr{\"a}fte. Mit 5 Textabbildungen",
  journal = {Zeitschrift fuer Astrophysik},
     year = 1958,
   volume = 46,
    pages = {108},
   adsurl = {http://adsabs.harvard.edu/abs/1958ZA.....46..108B}
}

@phdthesis{Spruit1977,
  title={Magnetic Flux Tubes and Transport of Heat},
  author={Spruit, H. C.},
  year={1977},
  school={Ph. D. Thesis, Utrecht University}
}

@article{Brun2011,
  author={Allan Sacha Brun and Mark S. Miesch and Juri Toomre},
  title={Modeling the Dynamical Coupling of Solar Convection with the Radiative Interior},
  journal={The Astrophysical Journal},
  volume={742},
  number={2},
  pages={79},
  url={http://stacks.iop.org/0004-637X/742/i=2/a=79},
  year={2011},
}

@ARTICLE{Smolec2008,
   author = {Smolec, R. and Moskalik, P.},
    title = "Convective Hydrocodes for Radial Stellar Pulsation. Physical and Numerical Formulation",
  journal = {Acta Astronomica},
archivePrefix = "arXiv",
   eprint = {0809.1979},
     year = 2008,
    month = sep,
   volume = 58,
    pages = {193-232},
   adsurl = {http://adsabs.harvard.edu/abs/2008AcA....58..193S}
}

@article{Smolec2010,
	author = {Smolec, R. and Moskalik, P.},
	title = {Non-linear modelling of beat Cepheids: resonant  and non-resonant models},
	DOI= "10.1051/0004-6361/201014494",
	url= "https://doi.org/10.1051/0004-6361/201014494",
	journal = {A\&A},
	year = 2010,
	volume = 524,
	pages = "A40",
	month = "",
}

@article{Christensen1995,
author = {Christensen-Dalsgaard, Jørgen and Monteiro, Mário J. P. F. G. and Thompson, Michael J.},
title = {Helioseismic estimation of convective overshoot in the Sun},
journal = {Monthly Notices of the Royal Astronomical Society},
volume = {276},
number = {1},
pages = {283-292},
year = {1995},
doi = {10.1093/mnras/276.1.283},
}

@ARTICLE{Weber2015,
   author = {Weber, M. A. and Fan, Y.},
    title = "Effects of Radiative Diffusion on Thin Flux Tubes in Turbulent Solar-like Convection",
  journal = {Solar Physics},
archivePrefix = "arXiv",
   eprint = {1503.08034},
 primaryClass = "astro-ph.SR",
     year = 2015,
    month = may,
   volume = 290,
    pages = {1295-1321},
      doi = {10.1007/s11207-015-0674-3},
   adsurl = {http://adsabs.harvard.edu/abs/2015SoPh..290.1295W}
}

@article{Weber2016,
  author={Maria A. Weber and Matthew K. Browning},
  title={Modeling the Rise of Fibril Magnetic Fields in Fully Convective Stars},
  journal={The Astrophysical Journal},
  volume={827},
  number={2},
  pages={95},
  url={http://stacks.iop.org/0004-637X/827/i=2/a=95},
  year={2016},
}

@article{Christensen2011,
author = {Christensen-Dalsgaard, J. and Monteiro, M. J. P. F. G. and Rempel, M. and Thompson, M. J.},
title = {A more realistic representation of overshoot at the base of the solar convective envelope as seen by helioseismology},
journal = {Monthly Notices of the Royal Astronomical Society},
volume = {414},
number = {2},
publisher = {Blackwell Publishing Ltd},
issn = {1365-2966},
url = {http://dx.doi.org/10.1111/j.1365-2966.2011.18460.x},
doi = {10.1111/j.1365-2966.2011.18460.x},
pages = {1158--1174},
year = {2011},
}

@article{Marsh2016,
title = "Axion cosmology",
journal = "Physics Reports",
volume = "643",
number = "",
pages = "1 - 79",
year = "2016",
note = "Axion cosmology",
issn = "0370-1573",
doi = "http://dx.doi.org/10.1016/j.physrep.2016.06.005",
url = "http://www.sciencedirect.com/science/article/pii/S0370157316301557",
author = "David J.E. Marsh",
}

@Article{Gudel2004,
author="G{\"u}del, Manuel",
title="X-ray astronomy of stellar coronae",
journal="The Astronomy and Astrophysics Review",
year="2004",
month="Sep",
day="01",
volume="12",
number="2",
pages="71--237",
issn="1432-0754",
doi="10.1007/s00159-004-0023-2",
url="https://doi.org/10.1007/s00159-004-0023-2"
}

@article{Ilonidis2012,
	author = {Ilonidis, Stathis and Zhao, Junwei and Kosovichev, Alexander},
	title = {Response to Comment on {\textquotedblleft}Detection of Emerging Sunspot Regions in the Solar Interior{\textquotedblright}},
	volume = {336},
	number = {6079},
	pages = {296--296},
	year = {2012},
	doi = {10.1126/science.1215539},
	publisher = {American Association for the Advancement of Science},
	issn = {0036-8075},
	journal = {Science}
}

@article{Ilonidis2013,
  author={Stathis Ilonidis and Junwei Zhao and Thomas Hartlep},
  title={Helioseismic Investigation of Emerging Magnetic Flux in the Solar Convection Zone},
  journal={The Astrophysical Journal},
  volume={777},
  number={2},
  pages={138},
  url={http://stacks.iop.org/0004-637X/777/i=2/a=138},
  year={2013},
}

@ARTICLE{Kosovichev2016,
   author = {Kosovichev, A. G. and Zhao, J. and Ilonidis, S.},
    title = "{Local Helioseismology of Emerging Active Regions: A Case Study}",
  journal = {ArXiv e-prints},
archivePrefix = "arXiv",
   eprint = {1607.04987},
 primaryClass = "astro-ph.SR",
     year = 2016,
    month = jul,
      url = {https://arxiv.org/abs/1607.04987},
   adsurl = {http://adsabs.harvard.edu/abs/2016arXiv160704987K}
}

@Unpublished{Rusov2015,
author = {V.D. Rusov and M.V. Eingorn and I.V. Sharph and V.P. Smolyar and M.E. Beglaryan},
title = {Thermomagnetic {Ettingshausen-Nernst} effect in tachocline and axion mechanism of solar luminosity variations},
note = {arXiv:1508.03836 [astro-ph.SR]},
year = {2015},
url = {https://arxiv.org/abs/1508.03836}
}

@article{Hertz1976,
  title = {Quantum critical phenomena},
  author = {Hertz, John A.},
  journal = {Phys. Rev. B},
  volume = {14},
  issue = {3},
  pages = {1165--1184},
  numpages = {0},
  year = {1976},
  month = {Aug},
  publisher = {American Physical Society},
  doi = {10.1103/PhysRevB.14.1165},
  url = {https://link.aps.org/doi/10.1103/PhysRevB.14.1165}
}

@article{Coleman2005,
author = {Coleman, Piers and Schofield, Andrew J.},
title = {Quantum criticality},
journal = {Nature},
year = {2005},
volume = {433},
pages = {226},
doi = {10.1038/nature03279},
}

@article{Michaeli2009,
  author={K. Michaeli and A. M. Finkel'stein},
  title={Fluctuations of the superconducting order parameter as an origin of the Nernst effect},
  journal={EPL (Europhysics Letters)},
  volume={86},
  number={2},
  pages={27007},
  url={http://stacks.iop.org/0295-5075/86/i=2/a=27007},
  year={2009},
}

@ARTICLE{Xu2000,
   author = {{Xu}, Z.~A. and {Ong}, N.~P. and {Wang}, Y. and {Kakeshita}, T. and 
	{Uchida}, S.},
    title = "{Vortex-like excitations and the onset of superconducting phase fluctuation in underdoped La$_{2-x}$Sr$_{x}$CuO$_{4}$}",
  journal = {Nature},
     year = 2000,
    month = aug,
   volume = 406,
    pages = {486-488},
      doi = {10.1038/35020016},
   adsurl = {http://adsabs.harvard.edu/abs/2000Natur.406..486X}
}

@article{Wang2006,
  title = {Nernst effect in high-${T}_{c}$ superconductors},
  author = {Wang, Yayu and Li, Lu and Ong, N. P.},
  journal = {Phys. Rev. B},
  volume = {73},
  issue = {2},
  pages = {024510},
  numpages = {20},
  year = {2006},
  month = {Jan},
  publisher = {American Physical Society},
  doi = {10.1103/PhysRevB.73.024510},
  url = {https://link.aps.org/doi/10.1103/PhysRevB.73.024510}
}

@article{Bergman2010,
  title = {Theory of Dissipationless Nernst Effects},
  author = {Bergman, Doron L. and Oganesyan, Vadim},
  journal = {Phys. Rev. Lett.},
  volume = {104},
  issue = {6},
  pages = {066601},
  numpages = {4},
  year = {2010},
  month = {Feb},
  publisher = {American Physical Society},
  doi = {10.1103/PhysRevLett.104.066601},
  url = {https://link.aps.org/doi/10.1103/PhysRevLett.104.066601}
}

@ARTICLE{Zaanen2007,
   author = {{Zaanen}, J.},
    title = "{Theoretical physics: A black hole full of answers}",
  journal = {Nature},
     year = 2007,
    month = aug,
   volume = 448,
    pages = {1000-1001},
      doi = {10.1038/4481000a},
   adsurl = {http://adsabs.harvard.edu/abs/2007Natur.448.1000Z}
}

@article{Hartnoll2007,
  title = {Theory of the Nernst effect near quantum phase transitions in condensed matter and in dyonic black holes},
  author = {Hartnoll, Sean A. and Kovtun, Pavel K. and M\"uller, Markus and Sachdev, Subir},
  journal = {Phys. Rev. B},
  volume = {76},
  issue = {14},
  pages = {144502},
  numpages = {23},
  year = {2007},
  month = {Oct},
  publisher = {American Physical Society},
  doi = {10.1103/PhysRevB.76.144502},
  url = {https://link.aps.org/doi/10.1103/PhysRevB.76.144502}
}

@article{Skenderis2002,
  author={Kostas Skenderis},
  title={Lecture notes on holographic renormalization},
  journal={Classical and Quantum Gravity},
  volume={19},
  number={22},
  pages={5849},
  url={http://stacks.iop.org/0264-9381/19/i=22/a=306},
  year={2002},
}

@InBook{Papadimitriou2016,
author = {Papadimitriou, I.},
editor = {Kallosh, R. and Orazi, E.},
title = {Theoretical Frontiers in Black Holes and Cosmology},
chapter = {Lectures on Holographic Renormalization},
publisher = {Springer, Cham},
year = {2016},
volume = {176},
series = {Springer Proceedings in Physics},
doi = {10.1007/978-3-319-31352-8_4},
}

@article{Maldacena2005,
author = {Juan Maldacena},
title = {The Illusion of Gravity},
journal = {Scientific American},
year = {2005},
volume = {293},
number = {5},
}

@article{Maldacena1999,
      author         = "Maldacena, Juan Martin",
      title          = "{The Large N limit of superconformal field theories and
                        supergravity}",
      journal        = "Int. J. Theor. Phys.",
      volume         = "38",
      year           = "1999",
      pages          = "1113-1133",
      doi            = "10.1023/A:1026654312961, 10.4310/ATMP.1998.v2.n2.a1",
      note           = "[Adv. Theor. Math. Phys.2,231(1998)]",
      eprint         = "hep-th/9711200",
      archivePrefix  = "arXiv",
      primaryClass   = "hep-th",
      reportNumber   = "HUTP-97-A097, HUTP-98-A097",
      SLACcitation   = "%%CITATION = HEP-TH/9711200;%%"
}

@article{tHooft1993,
      author         = "{'t Hooft}, Gerard",
      title          = "{Dimensional reduction in quantum gravity}",
      booktitle      = "{Conference on Highlights of Particle and Condensed
                        Matter Physics (SALAMFEST) Trieste, Italy, March 8-12,
                        1993}",
      journal        = "Conf. Proc.",
      volume         = "C930308",
      year           = "1993",
      pages          = "284-296",
      eprint         = "gr-qc/9310026",
      archivePrefix  = "arXiv",
      primaryClass   = "gr-qc",
      reportNumber   = "THU-93-26",
      SLACcitation   = "%%CITATION = GR-QC/9310026;%%"
}

@article{Susskind1995,
author = {Susskind,Leonard },
title = {The world as a hologram},
journal = {Journal of Mathematical Physics},
volume = {36},
number = {11},
pages = {6377-6396},
year = {1995},
doi = {10.1063/1.531249},
URL = { https://doi.org/10.1063/1.531249 },
eprint = { https://doi.org/10.1063/1.531249}
}

@article {Hanada2014,
	author = {Hanada, Masanori and Hyakutake, Yoshifumi and Ishiki, Goro and Nishimura, Jun},
	title = {Holographic description of a quantum black hole on a computer},
	volume = {344},
	number = {6186},
	pages = {882--885},
	year = {2014},
	doi = {10.1126/science.1250122},
	publisher = {American Association for the Advancement of Science},
	issn = {0036-8075},
	journal = {Science}
}

@article{Cowling1933,
author = {Cowling, T. G.},
title = {The Magnetic Field of Sunspots},
journal = {Monthly Notices of the Royal Astronomical Society},
volume = {94},
number = {1},
pages = {39-48},
year = {1933},
doi = {10.1093/mnras/94.1.39},
}

@ARTICLE{Sanchez2014,
   author = {{Sanchez}, S. and {Fournier}, A. and {Aubert}, J.},
    title = "{The Predictability of Advection-dominated Flux-transport Solar Dynamo Models}",
  journal = {Astrophysical Journal},
     year = 2014,
    month = jan,
   volume = 781,
      eid = {8},
    pages = {8},
      doi = {10.1088/0004-637X/781/1/8},
   adsurl = {http://adsabs.harvard.edu/abs/2014ApJ...781....8S}
}

@ARTICLE{Fan1993,
   author = {{Fan}, Y. and {Fisher}, G.~H. and {Deluca}, E.~E.},
    title = "{The origin of morphological asymmetries in bipolar active regions}",
  journal = {Astrophysical Journal},
     year = 1993,
    month = mar,
   volume = 405,
    pages = {390-401},
      doi = {10.1086/172370},
   adsurl = {http://adsabs.harvard.edu/abs/1993ApJ...405..390F}
}

@article{Zwaan1987,
author = {Zwaan, Cornelis},
title = {Elements and Patterns in the Solar Magnetic Field},
journal = {Annual Review of Astronomy and Astrophysics},
volume = {25},
number = {1},
pages = {83-111},
year = {1987},
doi = {10.1146/annurev.aa.25.090187.000503},
URL = { https://doi.org/10.1146/annurev.aa.25.090187.000503},
}

@PhdThesis{Jabbari2016,
author = {Sarah Jabbari},
title = {Origin of solar surface activity and sunspots},
school = {Stockholm University, Faculty of Science, Department of Astronomy. NORDITA.},
year = {2016},
note = {URN: urn:nbn:se:su:diva-128774, OAI: oai:DiVA.org:su-128774, DiVA, id: diva2:916647},
ISBN = {978-91-7649-370-0}
}

@article{JabbariEtAl2016,
author = {Jabbari, S. and Brandenburg, A. and Mitra, Dhrubaditya and Kleeorin, N. and Rogachevskii, I.},
title = {Turbulent reconnection of magnetic bipoles in stratified turbulence},
journal = {Monthly Notices of the Royal Astronomical Society},
volume = {459},
number = {4},
pages = {4046-4056},
year = {2016},
doi = {10.1093/mnras/stw888},
}

@ARTICLE{Ossendrijver2003,
   author = {{Ossendrijver}, M.},
    title = "{The solar dynamo}",
  journal = {Astronomy and Astrophysicsr},
     year = 2003,
   volume = 11,
    pages = {287-367},
      doi = {10.1007/s00159-003-0019-3},
   adsurl = {http://adsabs.harvard.edu/abs/2003A%26ARv..11..287O}
}

@ARTICLE{Beresnyak2017,
   author = {{Beresnyak}, A.},
    title = "{Three-dimensional Spontaneous Magnetic Reconnection}",
  journal = {Astrophysical Journal},
archivePrefix = "arXiv",
   eprint = {1301.7424},
 primaryClass = "astro-ph.SR",
     year = 2017,
    month = jan,
   volume = 834,
      eid = {47},
    pages = {47},
      doi = {10.3847/1538-4357/834/1/47},
      url = {https://arxiv.org/abs/1301.7424},
   adsurl = {http://adsabs.harvard.edu/abs/2017ApJ...834...47B}
}

@article{Loureiro2009,
author = {Loureiro, N. F. and Uzdensky, D. A. and Schekochihin, A. A. and Cowley, S. C. and Yousef, T. A.},
title = {Turbulent magnetic reconnection in two dimensions},
journal = {Monthly Notices of the Royal Astronomical Society: Letters},
volume = {399},
number = {1},
pages = {L146-L150},
year = {2009},
doi = {10.1111/j.1745-3933.2009.00742.x},
}

@article{Huang2010,
author = {Huang,Yi-Min  and Bhattacharjee,A. },
title = {Scaling laws of resistive magnetohydrodynamic reconnection in the high-Lundquist-number, plasmoid-unstable regime},
journal = {Physics of Plasmas},
volume = {17},
number = {6},
pages = {062104},
year = {2010},
doi = {10.1063/1.3420208},
URL = { https://doi.org/10.1063/1.3420208},
eprint = { https://doi.org/10.1063/1.3420208}
}

@Unpublished{Beresnyak2013,
 author = {Beresnyak, A},
 title = {Three-dimensional Spontaneous Magnetic Reconnection},
 note = {arXiv:1301.7424v1 [astro-ph.SR]},
 OPTkey = {•},
 OPTmonth = {•},
 year = {2013},
 OPTannote = {•},
 url = {https://arxiv.org/abs/1301.7424}
 }

@article{Baggaley2009,
  title = {Reconnecting flux-rope dynamo},
  author = {Baggaley, Andrew W. and Barenghi, Carlo F. and Shukurov, Anvar and Subramanian, Kandaswamy},
  journal = {Phys. Rev. E},
  volume = {80},
  issue = {5},
  pages = {055301},
  numpages = {4},
  year = {2009},
  month = {Nov},
  publisher = {American Physical Society},
  doi = {10.1103/PhysRevE.80.055301},
  url = {https://link.aps.org/doi/10.1103/PhysRevE.80.055301}
}

@ARTICLE{Kolmogorov1941,
   author = {Kolmogorov, A.},
    title = "{The Local Structure of Turbulence in Incompressible Viscous Fluid for Very Large Reynolds' Numbers}",
  journal = {Akademiia Nauk SSSR Doklady},
     year = 1941,
   volume = 30,
    pages = {301-305},
   adsurl = {http://adsabs.harvard.edu/abs/1941DoSSR..30..301K}
}

@article{Kolmogorov1968,
  author={A. N. Kolmogorov},
  title={LOCAL STRUCTURE OF TURBULENCE IN AN INCOMPRESSIBLE VISCOUS FLUID AT VERY HIGH REYNOLDS NUMBERS},
  journal={Soviet Physics Uspekhi},
  volume={10},
  number={6},
  pages={734},
  url={http://stacks.iop.org/0038-5670/10/i=6/a=R02},
  year={1968}
}

@article {Kolmogorov1991,
	author = {A. N. Kolmogorov},
	title = {The local structure of turbulence in incompressible viscous fluid for very large Reynolds numbers},
	volume = {434},
	number = {1890},
	pages = {9--13},
	year = {1991},
	doi = {10.1098/rspa.1991.0075},
	publisher = {The Royal Society},
	issn = {0962-8444},
	journal = {Proceedings of the Royal Society of London A: Mathematical, Physical and Engineering Sciences}
}

@ARTICLE{DSilvaChoudhuri1993,
   author = {{D'Silva}, S. and {Choudhuri}, A.~R.},
    title = "{A theoretical model for tilts of bipolar magnetic regions}",
  journal = {Astronomy and Astrophysics},
     year = 1993,
    month = may,
   volume = 272,
    pages = {621},
   adsurl = {http://adsabs.harvard.edu/abs/1993A%26A...272..621D}
}

@ARTICLE{Choudhuri1987,
   author = {{Choudhuri}, A.~R. and {Gilman}, P.~A.},
    title = "{The influence of the Coriolis force on flux tubes rising through the solar convection zone}",
  journal = {Astrophysical Journal},
     year = 1987,
    month = may,
   volume = 316,
    pages = {788-800},
      doi = {10.1086/165243},
   adsurl = {http://adsabs.harvard.edu/abs/1987ApJ...316..788C}
}

@ARTICLE{Choudhuri1989,
   author = {{Choudhuri}, A.~R.},
    title = "{The evolution of loop structures in flux rings within the solar convection zone}",
  journal = {Solar Physics},
     year = 1989,
    month = sep,
   volume = 123,
    pages = {217-239},
      doi = {10.1007/BF00149104},
   adsurl = {http://adsabs.harvard.edu/abs/1989SoPh..123..217C}
}

@article{Deluca1986,
author = { Edward E.   Deluca  and  Peter A.   Gilman },
title = {Dynamo theory for the interface between the convection zone and the radiative interior of a star: Part I model equations and exact solutions},
journal = {Geophysical \& Astrophysical Fluid Dynamics},
volume = {37},
number = {1-2},
pages = {85-127},
year  = {1986},
publisher = {Taylor \& Francis},
doi = {10.1080/03091928608210092},
URL = { https://doi.org/10.1080/03091928608210092},
eprint = { https://doi.org/10.1080/03091928608210092}
}

@ARTICLE{Browning2016,
   author = {{Browning}, M.~K. and {Weber}, M.~A. and {Chabrier}, G. and 
	{Massey}, A.~P.},
    title = "{Theoretical Limits on Magnetic Field Strengths in Low-mass Stars}",
  journal = {Astrophysical Journal},
archivePrefix = "arXiv",
   eprint = {1512.05692},
 primaryClass = "astro-ph.SR",
     year = 2016,
    month = feb,
   volume = 818,
      eid = {189},
    pages = {189},
      doi = {10.3847/0004-637X/818/2/189},
   adsurl = {http://adsabs.harvard.edu/abs/2016ApJ...818..189B}
}

@ARTICLE{DasiEspuig2010,
   author = {Dasi-Espuig, M. and {Solanki}, S.~K. and {Krivova}, N.~A. and 
	{Cameron}, R. and {Pe{\~n}uela}, T.},
    title = "{Sunspot group tilt angles and the strength of the solar cycle}",
  journal = {Astronomy and Astrophysics},
archivePrefix = "arXiv",
   eprint = {1005.1774},
 primaryClass = "astro-ph.SR",
     year = 2010,
    month = jul,
   volume = 518,
      eid = {A7},
    pages = {A7},
      doi = {10.1051/0004-6361/201014301},
   adsurl = {http://adsabs.harvard.edu/abs/2010A%26A...518A...7D}
}

@article{DasiEspuig2013,
	author = {Dasi-Espuig, M. and {Solanki, S. K.} and {Krivova, N. A.} and {Cameron, R.} and {Pe\~nuela, T.}},
	title = {Sunspot group tilt angles and the strength of the solar cycle (Corrigendum)},
	DOI= "10.1051/0004-6361/201014301e",
	url= "https://doi.org/10.1051/0004-6361/201014301e",
	journal = {A\&A},
	year = 2013,
	volume = 556,
	pages = "C3",
}

@ARTICLE{Ivanov2012,
   author = {{Ivanov}, V.~G.},
    title = "{Joy's law and its features according to the data of three sunspot catalogs}",
  journal = {Geomagnetism and Aeronomy},
     year = 2012,
    month = dec,
   volume = 52,
    pages = {999-1004},
      doi = {10.1134/S0016793212080130},
   adsurl = {http://adsabs.harvard.edu/abs/2012Ge%26Ae..52..999I}
}

@ARTICLE{McClintock2013,
   author = {{McClintock}, B.~H. and {Norton}, A.~A.},
    title = "{Recovering Joy's Law as a Function of Solar Cycle, Hemisphere, and Longitude}",
  journal = {Solar Physics},
archivePrefix = "arXiv",
   eprint = {1305.3205},
 primaryClass = "astro-ph.SR",
     year = 2013,
    month = oct,
   volume = 287,
    pages = {215-227},
      doi = {10.1007/s11207-013-0338-0},
   adsurl = {http://adsabs.harvard.edu/abs/2013SoPh..287..215M}
}

@ARTICLE{Tlatova2015,
   author = {{Tlatova}, K.~A. and {Vasil'eva}, V.~V. and {Pevtsov}, A.~A.
	},
    title = "{Long-term variations in the sunspot magnetic fields and bipole properties from 1918 to 2014}",
  journal = {Geomagnetism and Aeronomy},
     year = 2015,
    month = dec,
   volume = 55,
    pages = {896-901},
      doi = {10.1134/S0016793215070233},
   adsurl = {http://adsabs.harvard.edu/abs/2015Ge%26Ae..55..896T}
}

@article{ Pavai2015,
	author = {{Senthamizh Pavai, V.} and {Arlt, R.} and {Dasi-Espuig, M.} and {Krivova, N. A.} and {Solanki, S. K.}},
	title = {Sunspot areas and tilt angles for solar cycles 7-10},
	DOI= "10.1051/0004-6361/201527080",
	url= "https://doi.org/10.1051/0004-6361/201527080",
	journal = {A\&A},
	year = 2015,
	volume = 584,
	pages = "A73",
	month = "",
}

@article{Baranyi2015,
author = {Baranyi, T.},
title = {Comparison of Debrecen and Mount Wilson/Kodaikanal sunspot group tilt angles and the Joy's law},
journal = {Monthly Notices of the Royal Astronomical Society},
volume = {447},
number = {2},
pages = {1857-1865},
year = {2015},
doi = {10.1093/mnras/stu2572},
}

@ARTICLE{Wang2015,
   author = {Wang, Y.-M. and {Colaninno}, R.~C. and {Baranyi}, T. and {Li}, J.
	},
    title = "{Active-region Tilt Angles: Magnetic versus White-light Determinations of Joy's Law}",
  journal = {Astrophysical Journal},
archivePrefix = "arXiv",
   eprint = {1412.2329},
 primaryClass = "astro-ph.SR",
     year = 2015,
    month = jan,
   volume = 798,
      eid = {50},
    pages = {50},
      doi = {10.1088/0004-637X/798/1/50},
   adsurl = {http://adsabs.harvard.edu/abs/2015ApJ...798...50W}
}

@ARTICLE{Wang2017,
   author = {Wang, Y.-M.},
    title = "{Surface Flux Transport and the Evolution of the Sun's Polar Fields}",
  journal = {Space Science Reviews},
     year = 2017,
    month = sep,
   volume = 210,
    pages = {351-365},
      doi = {10.1007/s11214-016-0257-0},
   adsurl = {http://adsabs.harvard.edu/abs/2017SSRv..210..351W}
}

@ARTICLE{Goldreich1995,
   author = {Goldreich, P. and {Sridhar}, S.},
    title = "{Toward a theory of interstellar turbulence. 2: Strong alfvenic turbulence}",
  journal = {Astrophysical Journal},
     year = 1995,
    month = jan,
   volume = 438,
    pages = {763-775},
      doi = {10.1086/175121},
   adsurl = {http://adsabs.harvard.edu/abs/1995ApJ...438..763G}
}

@article{Loureiro2017,
  title = {Role of Magnetic Reconnection in Magnetohydrodynamic Turbulence},
  author = {Loureiro, Nuno F. and Boldyrev, Stanislav},
  journal = {Phys. Rev. Lett.},
  volume = {118},
  issue = {24},
  pages = {245101},
  numpages = {6},
  year = {2017},
  month = {Jun},
  publisher = {American Physical Society},
  doi = {10.1103/PhysRevLett.118.245101},
  url = {https://link.aps.org/doi/10.1103/PhysRevLett.118.245101}
}

@BOOK{Childress1995,
   author = {Childress, S. and Gilbert, A. D.},
    title = "{Stretch, Twist, Fold}",
booktitle = {The Fast Dynamo, XI, 406 pp..~ Springer-Verlag Berlin Heidelberg New York.~ Also Lecture Notes in Physics, volume 37},
     year = 1995,
   adsurl = {http://adsabs.harvard.edu/abs/1995stf..book.....C}
}

@article{Blackman1996,
  title = {Overcoming the Backreaction on Turbulent Motions in the Presence of Magnetic Fields},
  author = {Blackman, Eric G.},
  journal = {Phys. Rev. Lett.},
  volume = {77},
  issue = {13},
  pages = {2694--2697},
  numpages = {0},
  year = {1996},
  month = {Sep},
  publisher = {American Physical Society},
  doi = {10.1103/PhysRevLett.77.2694},
  url = {https://link.aps.org/doi/10.1103/PhysRevLett.77.2694}
}

@article{ Archontis2003a,
	author = {Archontis, V. and Dorch, S. B. F. and Nordlund, \AA{}.},
	title = {Numerical simulations of kinematic dynamo action },
	DOI= "10.1051/0004-6361:20021568",
	url= "https://doi.org/10.1051/0004-6361:20021568",
	journal = {A\&A},
	year = 2003,
	volume = 397,
	number = 2,
	pages = "393-399",
}

@article{ Archontis2003b,
	author = {Archontis, V. and Dorch, S. B. F. and Nordlund, \AA{}.},
	title = {Dynamo action in turbulent flows},
	DOI= "10.1051/0004-6361:20031293",
	url= "https://doi.org/10.1051/0004-6361:20031293",
	journal = {A\&A},
	year = 2003,
	volume = 410,
	number = 3,
	pages = "759-766",
}

@article{Baggaley2010,
author = {Baggaley, A.W. and Shukurov, A. and Barenghi, C.F. and Subramanian, K.},
title = {Fluctuation dynamo based on magnetic reconnections},
journal = {Astronomische Nachrichten},
volume = {331},
number = {1},
pages = {46-62},
year = {2010},
doi = {10.1002/asna.200911298},
}

@article {Testa2014,
	author = {Testa, P. and De Pontieu, B. and Allred, J. and Carlsson, M. and Reale, F. and Daw, A. and Hansteen, V. and Martinez-Sykora, J. and Liu, W. and DeLuca, E. E. and Golub, L. and McKillop, S. and Reeves, K. and Saar, S. and Tian, H. and Lemen, J. and Title, A. and Boerner, P. and Hurlburt, N. and Tarbell, T. D. and Wuelser, J. P. and Kleint, L. and Kankelborg, C. and Jaeggli, S.},
	title = {Evidence of nonthermal particles in coronal loops heated impulsively by nanoflares},
	volume = {346},
	number = {6207},
	year = {2014},
	doi = {10.1126/science.1255724},
	publisher = {American Association for the Advancement of Science},
	issn = {0036-8075},
	journal = {Science}
}

@article{Isik2015,
  author={Emre Isik},
  title={A Mechanism for the Dependence of Sunspot Group Tilt Angles on Cycle Strength},
  journal={The Astrophysical Journal Letters},
  volume={813},
  number={1},
  pages={L13},
  url={http://stacks.iop.org/2041-8205/813/i=1/a=L13},
  year={2015},
}

@article{Karak2017,
  author={Bidya Binay Karak and Mark Miesch},
  title={Solar Cycle Variability Induced by Tilt Angle Scatter in a Babcock–Leighton Solar Dynamo Model},
  journal={The Astrophysical Journal},
  volume={847},
  number={1},
  pages={69},
  url={http://stacks.iop.org/0004-637X/847/i=1/a=69},
  year={2017},  
}

@INPROCEEDINGS{Schussler2002,
   author = {{Sch{\"u}ssler}, M. and {Rempel}, M.},
    title = "{Structure of the magnetic field in the lower convection zone}",
booktitle = {From Solar Min to Max: Half a Solar Cycle with SOHO},
     year = 2002,
   series = {ESA Special Publication},
   volume = 508,
   editor = {{Wilson}, A.},
    month = jun,
    pages = {499-506},
   adsurl = {http://adsabs.harvard.edu/abs/2002ESASP.508..499S}
}

@article{Krivodubskij2005,
author = {Krivodubskij, V. N.},
title = {Turbulent dynamo near tachocline and reconstruction of azimuthal magnetic field in the solar convection zone},
journal = {Astronomische Nachrichten},
volume = {326},
number = {1},
pages = {61-74},
year = {2005},
doi = {10.1002/asna.200310340},
}

@ARTICLE{Ballegooijen1988,
   author = {{van Ballegooijen}, A.~A. and {Choudhuri}, A.~R.},
    title = "{The possible role of meridional flows in suppressing magnetic buoyancy}",
  journal = {Astrophysical Journal},
     year = 1988,
    month = oct,
   volume = 333,
    pages = {965-977},
      doi = {10.1086/166805},
   adsurl = {http://adsabs.harvard.edu/abs/1988ApJ...333..965V}
}

@ARTICLE{Khaibrakhmanov2017,
   author = {{Khaibrakhmanov}, S.~A. and {Dudorov}, A.~E. and {Parfenov}, S.~Y. and 
	{Sobolev}, A.~M.},
    title = "{Large-scale magnetic field in the accretion discs of young stars: the influence of magnetic diffusion, buoyancy and Hall effect}",
  journal = {Monthly Notices of the RAS},
archivePrefix = "arXiv",
   eprint = {1609.03969},
 primaryClass = "astro-ph.SR",
     year = 2017,
    month = jan,
   volume = 464,
    pages = {586-598},
      doi = {10.1093/mnras/stw2349},
   adsurl = {http://adsabs.harvard.edu/abs/2017MNRAS.464..586K}
}

@misc{Roberts2001,
title = "Solar photospheric magnetic flux tubes: theory",
author = "Bernard Roberts and P Murdin",
note = "Encyclopedia of Astronomy and Astrophysics, Institute of Physics, Nature Publishing Group, London",
year = "2001",
language = "English",
type = "Other",
}

@ARTICLE{Stix1990,
   author = {Stix, M. and Skaley, D.},
    title = "{The equation of state and the frequencies of solar P modes}",
  journal = {Astronomy and Astrophysics},
     year = 1990,
    month = jun,
   volume = 232,
    pages = {234-238},
   adsurl = {http://adsabs.harvard.edu/abs/1990A%26A...232..234S}
}

@ARTICLE{Kichatinov1991,
   author = {Kichatinov, L. L.},
    title = "Turbulent transport of magnetic fields in a highly conducting rotating fluid and the solar cycle",
  journal = {Astronomy and Astrophysics},
     year = 1991,
    month = mar,
   volume = 243,
    pages = {483-491},
   adsurl = {http://adsabs.harvard.edu/abs/1991A%26A...243..483K}
}

@Article{Zeldovich1956,
author = {Ya. B. Zel'dovich},
title = {Magnetic Field at Two-Dimensional Motion of Conducting Liquid},
journal = {Sov. Phys. JETP},
year = {1956},
volume = {31},
pages = {154},
}

@Article{Zeldovich1957,
author = {Ya. B. Zel'dovich},
title = {The Magnetic Field in Two-dimensional Motion of a Conducting Turbulent Liquid},
journal = {Sov. Phys. JETP},
year = {1957},
volume = {4},
pages = {460},
}

@article{Radler1968a,
  title={Zur Elektrodynamik turbulent bewegter leitender Medien. I. Grundz\"{u}ge der Elektrodynamik der mittleren Felder},
  author={R{\"a}dler, K-H},
  journal={Zeitschrift f{\"u}r Naturforschung A},
  volume={23},
  number={11},
  pages={1841--1851},
  year={1968},
  publisher={Verlag der Zeitschrift f{\"u}r Naturforschung}
}

@article{Radler1968b,
  title={Zur Elektrodynamik turbulent bewegter leitender Medien. II. Turbulenzbedingte Leitf\"{a}higkeits- und Permeabilit\"{a}ts\"{a}nderungen},
  author={R{\"a}dler, K-H},
  journal={Zeitschrift f{\"u}r Naturforschung A},
  volume={23},
  number={11},
  pages={1841--1851},
  year={1968},
  publisher={Verlag der Zeitschrift f{\"u}r Naturforschung}
}

@Book{Vainshtein1980,
author = {Vainshtein, S. I. and Zel'dovich, Ya. B. and Ruzmaikin, A. A.},
title = {The Turbulent Dynamo in Astrophysics},
publisher = {Nauka},
year = {1980},
address = {Moscow},
}

@article{VainshteinKichatinov1983,
author = { S. I.   Vainshtein  and  L. L.   Kichatinov },
title = {The macroscopic magnetohydrodynamics of inhomogeneously turbulent cosmic plasmas},
journal = {Geophysical \& Astrophysical Fluid Dynamics},
volume = {24},
number = {4},
pages = {273-298},
year  = {1983},
publisher = {Taylor \& Francis},
doi = {10.1080/03091928308209069},
URL = { https://doi.org/10.1080/03091928308209069 },
eprint = { https://doi.org/10.1080/03091928308209069 }
}

@BOOK{Stix1989,
   author = {Stix, M.},
    title = "{The Sun. an Introduction}",
booktitle = {The Sun.~ An Introduction, XIII, 390 pp.~192 figs..~Springer-Verlag Berlin Heidelberg New York.~ Also Astronomy and Astrophysics Library},
     year = 1989,
    pages = {192},
   adsurl = {http://adsabs.harvard.edu/abs/1989sun..book.....S}
}

@ARTICLE{Kichatinov1992,
   author = {Kichatinov, L. L. and Ruediger, G.},
    title = "Magnetic-field advection in inhomogeneous turbulence",
  journal = {Astronomy and Astrophysics},
     year = 1992,
    month = jul,
   volume = 260,
    pages = {494-498},
   adsurl = {http://adsabs.harvard.edu/abs/1992A%26A...260..494K}
}

@article{Kitchatinov2016,
title = "Diamagnetic pumping in a rotating convection zone",
journal = "Advances in Space Research",
volume = "58",
number = "8",
pages = "1554 - 1559",
year = "2016",
note = "Solar Dynamo Frontiers",
issn = "0273-1177",
doi = "https://doi.org/10.1016/j.asr.2016.04.014",
url = "http://www.sciencedirect.com/science/article/pii/S0273117716301491",
author = "L. L. Kitchatinov and A. A. Nepomnyashchikh"
}

@article{Drobyshevski1974, 
	title={Topological pumping of magnetic flux by three-dimensional convection}, 
	volume={65}, 
	DOI={10.1017/S0022112074001236}, 
	number={1}, 
	journal={Journal of Fluid Mechanics}, 
	publisher={Cambridge University Press}, 
	author={Drobyshevski, E. M. and Yuferev, V. S.}, 
	year={1974},
	pages={33–44}}

@Book{Vainshtein1983,
author = {Vainshtein, S.I.},
title = {Magnetic Fields in Space},
publisher = {Nauka},
year = {1983},
address = {Moscow},
}

@article{ Ossendrijver2002,
	author = {Ossendrijver, M. and Stix, M. and Brandenburg, A. and R\"udiger, G.},
	title = {Magnetoconvection and dynamo coefficients - II. Field-direction dependent pumping of magnetic field},
	DOI= "10.1051/0004-6361:20021224",
	url= "https://doi.org/10.1051/0004-6361:20021224",
	journal = {A\&A},
	year = 2002,
	volume = 394,
	number = 2,
	pages = "735-745",
}

@ARTICLE{Kitchatinov2012,
   author = {Kitchatinov, L. L. and Olemskoy, S. V.},
    title = "Solar Dynamo Model with Diamagnetic Pumping and Nonlocal {$\alpha$}-Effect",
  journal = {Solar Physics},
archivePrefix = "arXiv",
   eprint = {1108.3138},
 primaryClass = "astro-ph.SR",
     year = 2012,
    month = feb,
   volume = 276,
    pages = {3-17},
      doi = {10.1007/s11207-011-9887-2},
   adsurl = {http://adsabs.harvard.edu/abs/2012SoPh..276....3K}
}

@ARTICLE{Dudorov1985,
   author = {Dudorov, A. E. and Kirillov, A. K.},
    title = "{Dynamics of magnetic flux tubes in the convection zone of the sun.}",
  journal = {Byulletin Solnechnye Dannye Akademie Nauk SSSR},
     year = 1986,
   volume = 1985,
    pages = {85-93},
   adsurl = {http://adsabs.harvard.edu/abs/1986BSolD1985...85D}
}

@ARTICLE{Wang1991,
   author = {Wang Y.-M. and Sheeley, Jr., N. R. and Nash, A. G.},
    title = "{A new solar cycle model including meridional circulation}",
  journal = {Astrophysical Journal},
     year = 1991,
    month = dec,
   volume = 383,
    pages = {431-442},
      doi = {10.1086/170800},
   adsurl = {http://adsabs.harvard.edu/abs/1991ApJ...383..431W}
}

@ARTICLE{Caligari1995,
   author = {Caligari, P. and Moreno-Insertis, F. and Schussler, M.	},
    title = "Emerging flux tubes in the solar convection zone. 1: Asymmetry, tilt, and emergence latitude",
  journal = {Astrophysical Journal},
     year = 1995,
    month = mar,
   volume = 441,
    pages = {886-902},
      doi = {10.1086/175410},
   adsurl = {http://adsabs.harvard.edu/abs/1995ApJ...441..886C}
}

@article{Kitchatinov2008,
author = {Kitchatinov, L. L. and R\"{u}diger, G.},
title = {Diamagnetic pumping near the base of a stellar convection zone},
journal = {Astronomische Nachrichten},
volume = {329},
number = {4},
year = {2008},
pages = {372-375},
doi = {10.1002/asna.200810971},
}

@BOOK{Krause1980,
   author = {{Krause}, F. and {Raedler}, K.-H.},
    title = "{Mean-field magnetohydrodynamics and dynamo theory}",
booktitle = {Organic Photonics and Photovoltaics},
     year = 1980,
   adsurl = {http://adsabs.harvard.edu/abs/1980opp..bookR....K}
}

@ARTICLE{Bradshaw1974,
   author = {Bradshaw, P.},
    title = "Possible origin of Prandt's mixing-length theory",
  journal = {Nature},
     year = 1974,
    month = may,
   volume = 249,
    pages = {135-136},
      doi = {10.1038/249135b0},
   adsurl = {http://adsabs.harvard.edu/abs/1974Natur.249..135B}
}

@ARTICLE{Gough1977a,
   author = {Gough, D. O.},
    title = "{Mixing-length theory for pulsating stars}",
  journal = {Astrophysical Journal},
     year = 1977,
    month = may,
   volume = 214,
    pages = {196-213},
      doi = {10.1086/155244},
   adsurl = {http://adsabs.harvard.edu/abs/1977ApJ...214..196G}
}

@InProceedings{Gough1977b,
	author = {Gough, D.},
    title = "The current state of stellar mixing-length theory",
 booktitle = {Problems of Stellar Convection},
     year = 1977,
   series = {Lecture Notes in Physics, Berlin Springer Verlag},
   volume = 71,
   editor = {{Spiegel}, E.~A. and {Zahn}, J.-P.},
      doi = {10.1007/3-540-08532-7},
   adsurl = {http://adsabs.harvard.edu/abs/1977LNP....71.....S}
}

@article{Barker2014,
  author={Adrian J. Barker and Adam M. Dempsey and Yoram Lithwick},
  title={Theory and Simulations of Rotating Convection},
  journal={The Astrophysical Journal},
  volume={791},
  number={1},
  pages={13},
  url={http://stacks.iop.org/0004-637X/791/i=1/a=13},
  year={2014},
}

@article{Brandenburg2016,
  author={Axel Brandenburg},
  title={Stellar Mixing Length Theory with Entropy Rain},
  journal={The Astrophysical Journal},
  volume={832},
  number={1},
  pages={6},
  url={http://stacks.iop.org/0004-637X/832/i=1/a=6},
  year={2016},  
}

@article{Gough1976,
author = {Gough, D. O. and Weiss, N. O.},
title = {The Calibration of Stellar Convection Theories},
journal = {Monthly Notices of the Royal Astronomical Society},
volume = {176},
number = {3},
pages = {589-607},
year = {1976},
doi = {10.1093/mnras/176.3.589},
}

@Article{Karak2014,
author="Karak, Bidya Binay
and Jiang, Jie
and Miesch, Mark S.
and Charbonneau, Paul
and Choudhuri, Arnab Rai",
title="Flux Transport Dynamos: From Kinematics to Dynamics",
journal="Space Science Reviews",
year="2014",
month="Dec",
day="01",
volume="186",
number="1",
pages="561--602",
issn="1572-9672",
doi="10.1007/s11214-014-0099-6",
url="https://doi.org/10.1007/s11214-014-0099-6"
}

@article{ Kapyla2014,
	author = {{K\"apyl\"a, P. J.} and {K\"apyl\"a, M. J.} and {Brandenburg, A.}},
	title = {Confirmation of bistable stellar differential rotation profiles},
	DOI= "10.1051/0004-6361/201423412",
	url= "https://doi.org/10.1051/0004-6361/201423412",
	journal = {A\&A},
	year = 2014,
	volume = 570,
	pages = "A43",
	month = "",
}

@article{Brandenburg2005,
title = "Astrophysical magnetic fields and nonlinear dynamo theory",
journal = "Physics Reports",
volume = "417",
number = "1",
pages = "1 - 209",
year = "2005",
issn = "0370-1573",
doi = "https://doi.org/10.1016/j.physrep.2005.06.005",
url = "http://www.sciencedirect.com/science/article/pii/S037015730500267X",
author = "Axel Brandenburg and Kandaswamy Subramanian"
}

@article{Kapyla2011,
author = {K\"{a}pyl\"{a}, P.J.},
title = {On global solar dynamo simulations},
journal = {Astronomische Nachrichten},
volume = {332},
number = {1},
pages = {43-50},
year = {2011},
doi = {10.1002/asna.201012345},
}

@article{KapylaEtAl2011,
author = {K\"{a}pyl\"{a}, P.J. and Mantere, M.J. and Brandenburg, A.},
title = {Effects of stratification in spherical shell convection},
journal = {Astronomische Nachrichten},
volume = {332},
number = {9-10},
pages = {883-890},
year = {2011},
doi = {10.1002/asna.201111619},
}

@article{Cowling1981,
author = {Cowling, T. G.},
title = {The Present Status of Dynamo Theory},
journal = {Annual Review of Astronomy and Astrophysics},
volume = {19},
number = {1},
pages = {115-135},
year = {1981},
doi = {10.1146/annurev.aa.19.090181.000555},
URL = { https://doi.org/10.1146/annurev.aa.19.090181.000555 },
eprint = { https://doi.org/10.1146/annurev.aa.19.090181.000555 }
}

@Inbook{Gilman1986,
author="Gilman, Peter A.",
editor="Sturrock, Peter A.
and Holzer, Thomas E.
and Mihalas, Dimitri M.
and Ulrich, Roger K.",
title="The Solar Dynamo: Observations and Theories of Solar Convection, Global Circulation, and Magnetic Fields",
bookTitle="Physics of the Sun: The Solar Interior",
year="1986",
publisher="Springer Netherlands",
address="Dordrecht",
pages="95--160",
isbn="978-94-009-5253-9",
doi="10.1007/978-94-009-5253-9_5",
url="https://doi.org/10.1007/978-94-009-5253-9_5"
}

@ARTICLE{Spruit1981,
   author = {Spruit, H. C.},
    title = "{Motion of magnetic flux tubes in the solar convection zone and chromosphere}",
  journal = {Astronomy and Astrophysics},
     year = 1981,
    month = may,
   volume = 98,
    pages = {155-160},
   adsurl = {http://adsabs.harvard.edu/abs/1981A%26A....98..155S}
}

@Article{Roberts1979,
author="Roberts, B.
and Webb, A. R.",
title="Vertical motions in an intense magnetic flux tube",
journal="Solar Physics",
year="1979",
month="Nov",
day="01",
volume="64",
number="1",
pages="77--92",
issn="1573-093X",
doi="10.1007/BF00151117",
url="https://doi.org/10.1007/BF00151117"
}

@ARTICLE{FerrizMas1989,
   author = {Ferriz-Mas, A. and Schuessler, M. and Anton, V.},
    title = "{Dynamics of magnetic flux concentrations - The second-order thin flux tube approximation}",
  journal = {Astronomy and Astrophysics},
     year = 1989,
    month = feb,
   volume = 210,
    pages = {425-432},
   adsurl = {http://adsabs.harvard.edu/abs/1989A%26A...210..425F}
}

@ARTICLE{Fan1994,
   author = {Fan, Y. and Fisher, G. H. and McClymont, A. N.},
    title = "{Dynamics of emerging active region flux loops}",
  journal = {Astrophysical Journal},
     year = 1994,
    month = dec,
   volume = 436,
    pages = {907-928},
      doi = {10.1086/174967},
   adsurl = {http://adsabs.harvard.edu/abs/1994ApJ...436..907F}
}

@ARTICLE{Gough1988,
   author = {Gough, D.},
    title = "{Deep roots of solar cycles}",
  journal = {Nature},
     year = 1988,
    month = dec,
   volume = 336,
    pages = {618},
      doi = {10.1038/336618a0},
   adsurl = {http://adsabs.harvard.edu/abs/1988Natur.336..618G}
}

@Article{Dicke1988,
author="Dicke, R. H.",
title="The phase variations of the solar cycle",
journal="Solar Physics",
year="1988",
month="Mar",
day="01",
volume="115",
number="1",
pages="171--181",
issn="1573-093X",
doi="10.1007/BF00146238",
url="https://doi.org/10.1007/BF00146238"
}

@ARTICLE{Dicke1978,
   author = {Dicke, R. H.},
    title = "{Is there a chronometer hidden deep in the sun}",
  journal = {Nature},
     year = 1978,
    month = dec,
   volume = 276,
    pages = {676-680},
      doi = {10.1038/276676b0},
   adsurl = {http://adsabs.harvard.edu/abs/1978Natur.276..676D}
}

@ARTICLE{Dicke1979,
   author = {{Dicke}, R.~H.},
    title = "{Solar luminosity and the sunspot cycle}",
  journal = {Nature},
     year = 1979,
    month = jul,
   volume = 280,
    pages = {24-27},
      doi = {10.1038/280024a0},
   adsurl = {http://adsabs.harvard.edu/abs/1979Natur.280...24D}
}

@Article{Dicke1982,
author="Dicke, R. H.",
title="A magnetic core in the Sun? The solar rotator",
journal="Solar Physics",
year="1982",
month="May",
day="01",
volume="78",
number="1",
pages="3--16",
issn="1573-093X",
doi="10.1007/BF00151138",
url="https://doi.org/10.1007/BF00151138"
}

@InCollection{Gough1980,
author = {Gough, D},
title = {On the seat of the solar cycle},
booktitle = {Variations of the Solar Constant},
pages = {185-206},
publisher = {NASA Publ.},
year = {1980},
number = {Report 2191},
}

@Article{Gough1981,
author="Gough, D. O.",
title="Solar interior structure and luminosity variations",
journal="Solar Physics",
year="1981",
month="Nov",
day="01",
volume="74",
number="1",
pages="21--34",
issn="1573-093X",
doi="10.1007/BF00151270",
url="https://doi.org/10.1007/BF00151270"
}

@Article{DeMortel2015,
author = {De Moortel, Ineke and Browning, Philippa},
title = {Recent advances in coronal heating},
journal = {Philosophical transactions. Series A, Mathematical, physical, and engineering sciences},
year = {2015},
volume = {373},
pages = {2042},
doi = {10.1098/rsta.2014.0269}
}

@article{DeMortel2016,
  author={I De Moortel and D J Pascoe and A N Wright and A W Hood},
  title={Transverse, propagating velocity perturbations in solar coronal loops},
  journal={Plasma Physics and Controlled Fusion},
  volume={58},
  number={1},
  pages={014001},
  url={http://stacks.iop.org/0741-3335/58/i=1/a=014001},
  year={2016},
}

@ARTICLE{Gudel2009,
   author = {{G{\"u}del}, M. and {Naz{\'e}}, Y.},
    title = "{X-ray spectroscopy of stars}",
  journal = {Astronomy and Astrophysicsr},
archivePrefix = "arXiv",
   eprint = {0904.3078},
 primaryClass = "astro-ph.SR",
     year = 2009,
    month = sep,
   volume = 17,
    pages = {309-408},
      doi = {10.1007/s00159-009-0022-4},
   adsurl = {http://adsabs.harvard.edu/abs/2009A%26ARv..17..309G}
}

@article{ Kariyappa2011,
	author = {Kariyappa, R. and DeLuca, E. E. and Saar, S. H. and Golub, L. and Dam\'e, L. and Pevtsov, A. A. and Varghese, B. A.},
	title = {Temperature variability in X-ray bright points  observed with Hinode/XRT},
	DOI= "10.1051/0004-6361/201014878",
	url= "https://doi.org/10.1051/0004-6361/201014878",
	journal = {A\&A},
	year = 2011,
	volume = 526,
	pages = "A78",
	month = "",
}

@ARTICLE{Cirtain2013,
   author = {Cirtain, J. W. and Golub, L. and Winebarger, A. R. and 
	de Pontieu, B. and Kobayashi, K. and Moore, R. L. and 
	{Walsh}, R.~W. and {Korreck}, K.~E. and {Weber}, M. and {McCauley}, P. and 
	{Title}, A. and {Kuzin}, S. and {Deforest}, C.~E.},
    title = "{Energy release in the solar corona from spatially resolved magnetic braids}",
  journal = {Nature},
     year = 2013,
    month = jan,
   volume = 493,
    pages = {501-503},
      doi = {10.1038/nature11772},
   adsurl = {http://adsabs.harvard.edu/abs/2013Natur.493..501C}
}

@article {Peter2014,
	author = {Peter, H. and Tian, H. and Curdt, W. and Schmit, D. and Innes, D. and De Pontieu, B. and Lemen, J. and Title, A. and Boerner, P. and Hurlburt, N. and Tarbell, T. D. and Wuelser, J. P. and Mart{\'\i}nez-Sykora, Juan and Kleint, L. and Golub, L. and McKillop, S. and Reeves, K. K. and Saar, S. and Testa, P. and Kankelborg, C. and Jaeggli, S. and Carlsson, M. and Hansteen, V.},
	title = {Hot explosions in the cool atmosphere of the Sun},
	volume = {346},
	number = {6207},
	year = {2014},
	doi = {10.1126/science.1255726},
	publisher = {American Association for the Advancement of Science},
	issn = {0036-8075},
	journal = {Science}
}

@ARTICLE{Reale2014,
   author = {Reale, F.},
    title = "{Coronal Loops: Observations and Modeling of Confined Plasma}",
  journal = {Living Reviews in Solar Physics},
     year = 2014,
    month = jul,
   volume = 11,
      eid = {4},
    pages = {4},
      doi = {10.12942/lrsp-2014-4},
   adsurl = {http://adsabs.harvard.edu/abs/2014LRSP...11....4R}
}

@ARTICLE{Laming2015,
   author = {Laming, J. M.},
    title = "{The FIP and Inverse FIP Effects in Solar and Stellar Coronae}",
  journal = {Living Reviews in Solar Physics},
archivePrefix = "arXiv",
   eprint = {1504.08325},
 primaryClass = "astro-ph.SR",
     year = 2015,
    month = sep,
   volume = 12,
      eid = {2},
    pages = {2},
      doi = {10.1007/lrsp-2015-2},
   adsurl = {http://adsabs.harvard.edu/abs/2015LRSP...12....2L}
}

@ARTICLE{Peter2015,
   author = {Peter, H.},
    title = "{What can large-scale magnetohydrodynamic numerical experiments tell us about coronal heating?}",
  journal = {Philosophical Transactions of the Royal Society of London Series A},
     year = 2015,
    month = apr,
   volume = 373,
    pages = {20150055-20150055},
      doi = {10.1098/rsta.2015.0055},
   adsurl = {http://adsabs.harvard.edu/abs/2015RSPTA.37350055P}
}

@ARTICLE{Morgan2017,
   author = {Morgan, H. and Taroyan, Y.},
    title = "{Global conditions in the solar corona from 2010 to 2017}",
  journal = {Science Advances},
     year = 2017,
    month = jul,
   volume = 3,
    pages = {e1602056},
      doi = {10.1126/sciadv.1602056},
   adsurl = {http://adsabs.harvard.edu/abs/2017SciA....3E2056M}
}

@article{Alfven1947,
author = {Alfv\'{e}n, Hannes and Lindblad, B.},
title = {Granulation, Magneto-Hydrodynamic Waves, and the Heating of the Solar Corona},
journal = {Monthly Notices of the Royal Astronomical Society},
volume = {107},
number = {2},
pages = {211-219},
year = {1947},
doi = {10.1093/mnras/107.2.211},
}

@ARTICLE{Schatzman1949,
   author = {Schatzman, E.},
    title = "{The heating of the solar corona and chromosphere}",
  journal = {Annales d'Astrophysique},
     year = 1949,
    month = jan,
   volume = 12,
    pages = {203},
   adsurl = {http://adsabs.harvard.edu/abs/1949AnAp...12..203S}
}

@ARTICLE{Schatzman1962,
   author = {{Schatzman}, E.},
    title = "{A theory of the role of magnetic activity during star formation}",
  journal = {Annales d'Astrophysique},
     year = 1962,
    month = feb,
   volume = 25,
    pages = {18},
   adsurl = {http://adsabs.harvard.edu/abs/1962AnAp...25...18S}
}

@ARTICLE{Parker1964,
   author = {{Parker}, E.~N.},
    title = "{A Mechanism for Magnetic Enhancement of Sound-Wave Generation and the Dynamical Origin of Spicules.}",
  journal = {Astrophysical Journal},
     year = 1964,
    month = oct,
   volume = 140,
    pages = {1170},
      doi = {10.1086/148014},
   adsurl = {http://adsabs.harvard.edu/abs/1964ApJ...140.1170P}
}

@ARTICLE{Callebaut1994,
   author = {Callebaut, D. K. and Tsintsadze, N. L.},
    title = "{Heating of plasma by Alfv{\'e}n waves envelope}",
  journal = {Physica Scripta},
     year = 1994,
    month = sep,
   volume = 50,
    pages = {283-289},
      doi = {10.1088/0031-8949/50/3/011},
   adsurl = {http://adsabs.harvard.edu/abs/1994PhyS...50..283C}
}

@article {Jess2009,
	author = {Jess, David B. and Mathioudakis, Mihalis and Erd{\'e}lyi, Robert and Crockett, Philip J. and Keenan, Francis P. and Christian, Damian J.},
	title = {Alfv{\'e}n Waves in the Lower Solar Atmosphere},
	volume = {323},
	number = {5921},
	pages = {1582--1585},
	year = {2009},
	doi = {10.1126/science.1168680},
	publisher = {American Association for the Advancement of Science},
	issn = {0036-8075},
	journal = {Science}
}

\end{document}